\documentclass[letterpaper,titlepage,11pt]{article}
\usepackage{hyperref}
\usepackage{amssymb,amsmath,amsfonts}
\usepackage{graphicx}
\usepackage{epsfig}
\usepackage{subfigure}

\usepackage[nosort]{cite}

\usepackage[swedish,english]{babel} 
	\usepackage{bbm}
	\usepackage{mathbbol}

\newcommand{\matrto}[4]{\left( \begin{array}{cc} #1 & #2 \\
#3 & #4 \end{array} \right) }

\def\[{\begin{equation}}
\def\]{\end{equation}}
\newcommand{\be}{\begin{eqnarray}}
\newcommand{\ee}{\end{eqnarray}}
\newcommand{\nln}{\nonumber\\}
\newcommand{\nn}{\nonumber}

\newcommand{\spa}{\ , \ \ }

\newcommand{\p}{\partial}
\def\pbar{\bar\p}

\newcommand{\ads}{\mbox{AdS}}
\newcommand{\gym}{g_{\rm YM}}

\newcommand{\CO}{\mathcal{O}}
\newcommand{\CN}{\mathcal{N}}

\newcommand{\CH}{\mathcal{H}}

\newcommand{\Z}{\mathbb{Z}}
\newcommand{\C}{\mathbb{C}}
\newcommand{\R}{\mathbb{R}}

\newcommand{\CFT}{\mbox{CFT}}

\newcommand{\s}{\mbox{S}}

\newcommand{\Jflat}{\mathcal{J}}

\newcommand{\F}{\mathcal{F}}

\newcommand{\ul}{\underline}

\newcommand{\ula}{\underline{a}}
\newcommand{\ulb}{\underline{b}}

\newcommand{\Op}{\mathcal{O}}

\newcommand{\Tr}{\mathop{\mathrm{Tr}}}

\newcommand{\STr}{\mathop{\mathrm{STr}}}

\newcommand{\Reals}{\mathbbm{R}}

\newcommand{\Sphere}{S}  

\newcommand{\sphere}{\mathrm{S}}

\newcommand{\infinity}{\infty}

\ifx\genfrac\sdflkaj

\else

\fi
\newcommand{\sfrac}[2]{{\textstyle\frac{#1}{#2}}}
\newcommand{\half}{\sfrac{1}{2}}

\newcommand{\quarter}{\sfrac{1}{4}}

\newcommand{\Half}{\frac{1}{2}}

\newcommand{\Quarter}{\frac{1}{4}}

\newcommand{\alg}[1]{\mathfrak{#1}}
\newcommand{\grp}[1]{\mathrm{#1}}

\newcommand{\grSU}{\grp{SU}}
\newcommand{\grU}{\grp{U}}
\newcommand{\grSO}{\grp{SO}}

\newcommand{\grPSU}{\grp{PSU}}
\newcommand{\algSU}{\alg{su}}
\newcommand{\algSO}{\alg{so}}

\newcommand{\algPSU}{\alg{psu}}
\newcommand{\algU}{\alg{u}}

\newcommand{\al}{\alpha}
\newcommand{\bt}{\beta}
\newcommand{\si}{\sigma}
\newcommand{\dl}{\delta}           
\newcommand{\eps}{\epsilon}
\newcommand{\lm}{\lambda}    \newcommand{\Lm}{\Lambda}
\newcommand{\om}{\omega}   \newcommand{\Om}{\Omega}
\newcommand{\gm}{\gamma}  \newcommand{\Gm}{\Gamma}  
\newcommand{\Bsi}{\Upsilon}

\def\bth{\hat{\beta}}
\def\alh{\hat{\alpha}}

\newcommand{\fraz}{\frac{1}{2}}

\def\vbar{\bar{v}}

\def\gma{\gm_{\ula}}

\newcommand{\Smatrix}{S}  
\newcommand{\smatrix}{\mathsf{S}}          
\newcommand{\Amp}{\mathcal{A}}      
\newcommand{\Samp}{\mathcal{S}}
\newcommand{\Ramp}{\mathcal{R}}
\newcommand{\lAA}{{a}}

\newcommand{\laa}{{\alpha}}

\newcommand{\lBB}{{b}}

\newcommand{\lbb}{{\beta}}

\newcommand{\lCC}{{c}}

\newcommand{\lcc}{{\gamma}}

\newcommand{\lDD}{{d}}

\newcommand{\ldd}{{\delta}}

\newcommand{\pin}{\eta}    
\newcommand{\pou}{\bar{\eta}}

\newcommand{\prop}{\mathsf{I}}

\newcommand{\bubble}{\mathsf{B}}

\renewcommand{\vec}[1]{\mathbf{#1}}
\newcommand{\vecpin}{\boldsymbol{\eta}}

\newcommand{\lrbrk}[1]{\left(#1\right)}
\newcommand{\bigbrk}[1]{\bigl(#1\bigr)}

\newcommand{\lrsbrk}[1]{\left[#1\right]}
\newcommand{\bigsbrk}[1]{\bigl[#1\bigr]}

\newcommand{\biggsbrk}[1]{\biggl[#1\biggr]}

\newcommand{\ket}[1]{\mathopen{|}#1\mathclose{\rangle}}
\newcommand{\bra}[1]{\mathopen{\langle}#1\mathclose{|}}
\newcommand{\braket}[2]{\mathopen{\langle}#1|#2\mathclose{\rangle}}

\newcommand{\lrabs}[1]{\left|#1\right|}
\newcommand{\abs}[1]{{|#1|}}

\newcommand{\bigeval}[1]{#1\big|}

\newcommand{\levi}{\epsilon}

\begin{document}

\begin{titlepage}

\rightline{\vbox{\small\hbox{\tt NORDITA-2010-41} }}
 \vskip 1.8 cm

\centerline{\LARGE \bf{ On string integrability}}
\vskip 0.2cm
\centerline{ \large{{\bf A journey through the two-dimensional hidden symmetries in the AdS/CFT dualities}}}

\vskip 1.5cm

\centerline{\large {\bf Valentina
Giangreco Marotta Puletti} }

\vskip 0.5cm

\begin{center}
 NORDITA\\
Roslagstullsbacken 23,
SE-106 91 Stockholm,
Sweden\\
\end{center}
\vskip 0.5cm

\centerline{\small\tt 
valentina@nordita.org}

\vskip 1.5cm
 \centerline{\bf Abstract} 
\vskip 0.2cm
 \noindent 
One of the main topics in the modern String Theory are the AdS/CFT dualities. Proving such conjectures is extremely difficult since the gauge and string theory perturbative regimes do not overlap. In this perspective, the discovery of infinitely many conserved charges, i.e. the integrability, in the planar AdS/CFT has allowed us to reach immense progresses in understanding and confirming the duality. 
We review the fundamental concepts  and properties of integrability in two-dimensional $\sigma$-models and in the AdS/CFT context. The first part is focused on the $\mbox{AdS}_5/\mbox{CFT}_4$ duality, especially the classical and quantum integrability of the type IIB superstring on $\mbox{AdS}_5\times \mbox{S}^5$ are discussed in both pure spinor and Green-Schwarz formulations. The second part is dedicated to the $\mbox{AdS}_4/\mbox{CFT}_3$ duality with  particular attention to the type IIA superstring on $\mbox{AdS}_4\times \mathbb{C} P^3$ and its integrability. This review is based on a shortened and revised version of the author's PhD thesis, discussed at Uppsala University in September 2009.
 
\end{titlepage}

\small
\tableofcontents
\normalsize
\setcounter{page}{1}


\section{Introduction: Motivations, Overview and Outline}
\label{chapter:Intro}

\begin{figure}%
	\begin{center}%
		\includegraphics[scale=0.3]{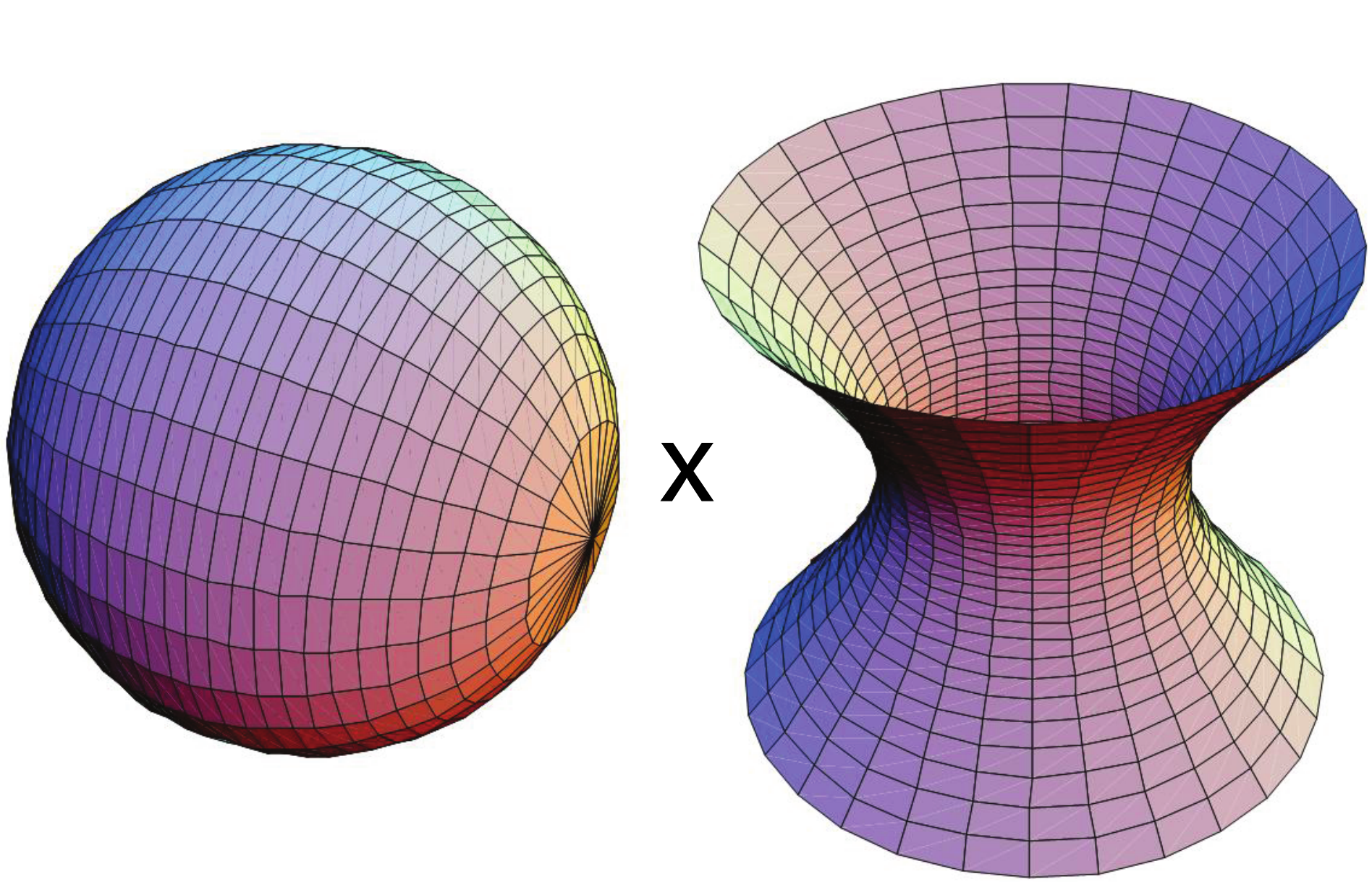}%
	\end{center}%
	\caption[vvv]{$\ads_5\times \sphere^5$. The five-dimensional anti-De Sitter space is represented as a hyperboloid on the right hand side, while the five-dimensional sphere is drawn on the left hand side.}%
	\label{fig:adss}%
\end{figure}

In 1997, Maldacena conjectured that type IIB superstrings on $\ads_5 \times \sphere^5$ describe the {\it same} physics of the supersymmetric $\grSU(N)$ Yang-Mills theory in four dimensions~\cite{Maldacena:1997re} ($\ads_5/\CFT_4$). The background where the string lives ($\ads_5 \times \sphere^5$) is built of a five-dimensional anti-De Sitter space (AdS), a space with constant negative curvature, times a five-dimensional sphere (S), cf. figure \ref{fig:adss}. 
In 2008 Aharony, Bergman, Jafferis and Maldacena proposed the existence of a further gauge/gravity duality between a theory of M2-branes in eleven dimensions and a certain three-dimensional gauge theory~\cite{Aharony:2008ug} ($\ads_4/\CFT_3$).  
The eleven-dimensional M2-theory can be effectively described by type IIA superstrings when the string coupling constant is very small. For a reason that will be clear later, I will consider only the type IIA as the gravitational dual in the $\ads_4/\CFT_3$ correspondence, but the reader should keep in mind that this is just a particular regime of the full correspondence. The background where the type IIA strings live is a four-dimensional anti-De Sitter space times a six-dimensional projective space ($\C P^3$).
\footnote{For a very recent analysis on lower-dimensional examples of AdS/CFT dualities we refer the reader to~\cite{Babichenko:2009dk}.}
%
%

The conformal field theories contained in the AdS/CFT dualities, namely $\mathcal N=4$ super Yang-Mills (SYM) in the $\ads_5/ \CFT_4$ case and the supersymmetric $\mathcal N=6$ Chern-Simons (CS) theory in the $\ads_4/ \CFT_3$ case, are rather difficult to solve. 
A general approach to quantum field theory is to compute quantities such as cross sections, scattering amplitudes and correlation functions.
In particular, for conformal field theories the correlation functions are constrained by the conformal symmetry.%
\footnote{Actually this is also true for the scattering amplitudes as it turns out in recent developments, but we will not focus on these aspects of the conformal field theories.}
For a certain class of operators (the conformal primary operators) their two-point function has a characteristic behavior: in the configuration space it is an inverse power function of the distance. The specific behavior, namely the specific power (the so called scaling dimension) depends on the nature of the operators and of the theory we are considering. It reflects how this operator transforms under conformal symmetry, in particular for the scaling dimension it reflects how the conformal primary operator transforms under the action of the dilatation operator.
At high energy, the scaling dimensions acquire quantum corrections, i.e. the anomalous dimension.%
\footnote{In conformal field theories there are special classes of operators, the chiral primary operators, whose scaling dimension does not receive quantum corrections.}
In conformal field theories, the anomalous dimension encodes the physical information about the behavior of the operators under the renormalization process. I will expand this point in section \ref{chapter:AdSIntro}. 
For the moment it is enough to note that collecting the spectrum of the correlation functions, namely the spectrum of the anomalous dimensions, gives an outstanding insight of the theory. However, in general it is a very hard task to reach such a knowledge for a quantum field theory. 

For this purpose the gauge/string dualities can play a decisive role. Let me explain why. Both correspondences are strong/weak-coupling dualities: the strongly coupled gauge theory corresponds to a free non-interacting string and vice versa fully quantum strings are equivalent to weakly interacting particles. The two perturbative regimes on the string and on the gauge theory side do not overlap. Technical difficulties usually prevent to depart from such regimes. This implies that it is incredibly difficult to compare directly observable computed on the string and on the gauge theory side, and thus to prove the dualities. However, there is a positive aspect of such a weak/strong coupling duality: in this way it is possible to reach the non-perturbative gauge theory once we acquire enough knowledge of the classical string theory. 

Ironically, we are moving on a circle. In 1968, String Theory has been developed with the purposes to explain the strong nuclear interactions. Thus it started as a theory for particle physics. With the advent of the Quantum Chromo Dynamics (QCD) namely the quantum field theory describing strong nuclear forces, String Theory was abandoned and only later in 1974 it has been realized that the theory necessarily contained gravity. The AdS/CFT dualities give us the possibility to reach a better insight and knowledge of SYM (and hopefully of the CS theory) {\it by means} of String Theory. In this sense, String Theory is turning back to a particle physics theory. In this scenario the long-term and ambitious hope is that also QCD might have a dual string description which might give us a deeper theoretical understanding of its non-perturbative regime.

At this point I will mostly refer to the $\ads_5/\CFT_4$ correspondence, I will explicitly comment on the new-born duality at the end of the section. On one side of the correspondence, the $\ads_5\times\sphere^5$ type IIB string is described by a quantum two-dimensional $\sigma$-model in a very non-trivial background. On the other side, we have a quantum field theory, the SYM theory, which is also a rather complicated model. Some simplifications come from considering the planar limit, namely when in the gauge theory the number of colors $N$ of the gluons is very large, or equivalently in the string theory when one does not consider higher-genus world-sheet. In this limit both gauge and string theories show their {\it integrable} structure, which turns out to be an incredible tool to explore the duality. 

What does ``integrable'' mean? We could interpret such a word as ``solvable'' in a first approximation. However, this definition is not precise enough and slightly unsatisfactory. 
Integrable theories posses infinitely many (local and non-local) conserved charges which allow one to solve completely the model. Such charges generalize the energy and momentum conservation which is present in all the physical phenomena as for example the particle scatterings. Among all the integrable theories, those which live in two-dimensions are very special: in this case, the infinite set of charges manifests its presence by severely constraining the dynamics of the model through selection rules and through the factorization, cf. section \ref{chapter:Intro_integrability}. In order to fix the ideas, let me consider the scattering of $n$ particles in two-dimensions. The above statement means that for an integrable two-dimensional field theory, a general $n$-particle scattering will be reduced to a sequence of two-particle scattering. The set of necessary information to solve the model is then restricted in a dramatic way: we only need to solve the two-body problem to have access to the full model! This is indeed the ultimate power of integrability.

The impressing result (which has been historically the starting point of the exploit of integrability in the AdS/CFT context) has been the discovery of a relation between the SYM gauge theory and certain spin chain models. In 2002 Minahan and Zarembo understood that the single trace operators (which are the only relevant ones in the planar limit) could be represented as spin chains~\cite{Minahan:2002ve}: each field in the trace becomes a spin in the chain. This is not only a pictorial representation: the equivalence is concretely extended also to the dilatation operator whose eigenvalues are the anomalous dimensions and to the spin chain Hamiltonian. The key-point is that such a spin chain Hamiltonian is integrable, ``solvable''. On the gravity side, the integrability of the $\ads_5\times \sphere^5$ type IIB string has been rigorously proved only at classical level, which, in general, does not imply that the infinite conserved charges survive at quantum level.
However, 
the {\it assumption} of an exact quantum integrability on both sides of $\ads_5/\CFT_4$ has allowed one to reach enormous progresses in testing and in investigating the duality, thanks to the S-matrix program and to the entire Bethe Ansatz machinery, whose construction relies on such a hypothesis. Nowadays nobody doubts about the existence of integrable structures underlying the gauge and the gravity side of the $\ads_5/\mbox{CFT}_4$ correspondence. There have been numerous and reliable manifestations, even though indirect. Despite of such remarkable developments one essentially {\it assumes} that the $\ads_5\times \Sphere^5$ type IIB superstring theory is quantum integrable%
\footnote{It is correct to say that on the gauge theory side the quantum integrability relies on more robust basis, cf. section \ref{chapter:AdSIntro}.}.
And on general ground, proving integrability at quantum level is a very hard task as much as proving the correspondence itself. For this reason, there have been very few {\it direct} checks of quantum integrability in the string theory side. 
These are the main motivations of the present work: give some direct and explicit evidence for the quantum integrability of the AdS superstring. 

For the ``younger'' $\ads_4/\CFT_3$ duality, valuable results have been already obtained, cf. section \ref{chapter:ABJM}. It is very natural to ask whether and when it is possible to expect the existence of similar infinite symmetries also in this case. Considering the impressing history of the last ten years in $\ads_5/\CFT_4$, one would like to reach analogous results also in this second gauge/string duality. Probably understanding which are the differences between these two dualities might provide another perspective of how we should think about the gauge/string dualities and their infinite ``hidden'' symmetries.
  

\paragraph{Outline.}

In section \ref{chapter:AdSIntro} I will briefly introduce the $\ads_5/\CFT_4$ correspondence and the $\mathcal N=4$ SYM theory. It contains also a description of the symmetry algebra, $\algPSU(2,2|4)$, which controls the duality. I will also explain the crucial relation between the anomalous dimension and the spin chain systems as well as the Bethe Ansatz Equations for a sub-sector of the full $\algPSU(2,2|4)$ algebra. 

Section \ref{chapter:Intro_integrability} is dedicated to two-dimensional integrable field theories, in particular to some prototypes for our string theory, such as the Principal Chiral Models and the Coset $\sigma$-models.
I will explain the definition of integrability in the first order formalism approach as well as its dynamical implications for a two-dimensional integrable theory. I will stress the importance of the distinction between classical and quantum integrability. 

In section \ref{chapter:GS_formalism} I will review the type IIB string theory on $\ads_5\times \sphere^5$: starting from the Green-Schwarz formalism, the Metsaev-Tseytlin formulation of the theory based on a coset approach and finally its classical integrability. 

In section \ref{chapter:PSformalism} it is presented an alternative formulation of the type IIB $\ads_5\times \sphere^5$ superstring based on the Berkovits formalism, also called Pure Spinor formalism, and I will focus on its relation with integrability topics, such as the construction of the BRST charges, the finiteness of the monodromy matrix and of its path deformation. 

In section \ref{chapter:NFS} I will come back to the Green-Schwarz formalism and discuss some important limits of the $\ads_5\times\sphere^5$ string theory such as the plane wave limit (also called BMN limit) and the near-flat-space limit. I will present the Arutyunov-Frolov-Staudacher dressing phase, sketch the construction of the world-sheet scattering matrix, also in the near-flat-space limit, and finally, I will illustrate its factorization.

Section \ref{chapter:ABJM} is entirely based on the $\ads_4/\CFT_3$ duality. I will retrace certain fundamental results of the $\ads_5/\CFT_4$ correspondence in the new context, with a special attention to the near-BMN corrections of string theory.

In the appendices some complementary material is reported. In the first appendix notation and conventions are summarized. The second one contains the full all-loop Bethe ansatz equations. The third one is devoted to the pure spinor formalism, in particular the results concerning the operator product expansion for the matter and Lorentz ghost currents are listed. The fourth appendix contains an example showing the three-body S-matrix factorization. Finally in the last one the geometrical set-up for the $\ads_4/\CFT_3$ is reported.


\paragraph{Note added.}
This work is a shortened and revised version of the author's PhD thesis, submitted to Uppsala University. It is based on the papers~\cite{Puletti:2006vb, Puletti:2007hq, Astolfi:2008ji, Puletti:2008ym}.


\section{The {\texorpdfstring{$\ads_5/\mbox{CFT}_4$}{\ads_5/\mbox{CFT}_4}  } duality}

\label{chapter:AdSIntro}


The first part of this section is an introduction to the $\ads_5/\CFT_4$ correspondence, based on the original works, which are cited in the main text, and on the following reviews~\cite{Maldacena:2003nj, Aharony:1999ti, D'Hoker:2002aw}. For the introductory part dedicated to the $\mathcal N=4$ SYM and to the Coordinate Bethe Ansatz, I mainly refer to Minahan's review~\cite{Minahan:2006sk}, Plefka's review~\cite{Plefka:2005bk} and Faddeev's review~\cite{Faddeev:1996iy}
and by N. Dorey at RTN Winter School (2008)~\cite{Dorey:2008zz}. Finally, I find very useful also the Ph.D. theses written by Beisert~\cite{Beisert:2004ry} and Okamura~\cite{Okamura:2008jm}. 

\subsection{Introduction}


The Maldacena correspondence~\cite{Maldacena:1997re, Gubser:1998bc, Witten:1998qj} conjectures an exact duality between the type IIB superstring theory on the curved space $\ads_5\times \Sphere^5$ and $\CN=4$ super Yang-Mills (SYM) theory on the flat four-dimensional space $\R^{3,1}$ with gauge group $\grSU(N)$. In order to briefly illustrate the content of the duality, we will start by recalling all the parameters which are present in both theories.

The geometrical background in which the string lives is supported by a self-dual Ramond-Ramond (RR) five-form $F_5$. In particular, the flux through the sphere is quantized, namely it is an integer $N$, multiple of the unit flux. Both the sphere and the anti-de Sitter space have the same radius $R$:
\[
ds_{IIB}^2= R^2 ds^2_{AdS_5}+ R^2 ds_{S^5}^2
\]
where $ds^2_{AdS_5}$ and $ds_{S^5}^2$ are the unit metric in $\ads_5$ and $\Sphere^5$ respectively. The string coupling constant is $g_s$ and the effective string tension is $T= {R^2\over 2 \pi \alpha'}$ with $\alpha'=l_s^2$. The string theory side thus has two parameters%
\footnote{It might seem that also $N$ is an independent parameter in the string theory context. Actually it is related to the target space radius $R$ by $R^4= 4 \pi g_s N {\alpha'}^2$. This relation follows from supergravity arguments. In particular $R$ is the radius of the $D3$-brane solutions and $\alpha'$ the Planck length and the equality gives the threshold for the validity of the supergravity approximation ($g_s N\gg 1$).}:
$T$, $g_s$.

On the other side, SYM is a gauge theory with gauge group $\grSU(N)$, thus $N$ is the number of colors. The theory is maximally supersymmetric, namely it contains the maximal number of global supersymmetries which are allowed in four dimensions ($\CN=4$)~\cite{Gliozzi:1976qd, Brink:1976bc}. Another important aspect is that SYM is scale invariant at classical and quantum level, which means that the coupling constant $\gym$ is not renormalized~\cite{Sohnius:1981sn, Seiberg:1988ur, Mandelstam:1982cb, Howe:1983sr, Brink:1982pd}. The theory contains two parameters, i.e. $N$ and $\gym$. One can introduce the 't Hooft coupling constant $\lambda= \gym^2 N$. Notice that $\lambda$ is a continuous parameter. Summarizing, the gauge theory side has two parameters, we choose $\lm$ and $N$.

The correspondence states an identification between the coupling constants in the two theories, i.e. 
\[
\label{rel_coupling_const}
\gym^2 = 4 \pi g_s 
\qquad 
T = {\sqrt{\lambda}\over 2\pi}
\]
(or in terms of $\lambda$: $g_s = {\lambda\over 4\pi N}$), and between the observables, i.e. between the string energy and the scaling dimension for local operators:
\[
\label{ads_cft_def}
E (\lambda, N ) \,=\, \Delta (\lambda, N)\,. 
\]
The conjecture is valid for any value of the coupling constant $\lambda$ and for any value of $N$.%
\footnote{This is the conjecture statement in its strongest version. However, there are weaker versions: e.g. it might be considered to hold only in the large $N$ limit ($N\rightarrow \infinity$) and for finite values of $\lm$, namely without considering $g_s$ corrections to the string theory, or even weaker, without $\alpha'$ corrections (i.e. large $N$ and $\lm$ limits). In this work we will always assume the strongest version, namely that the AdS/CFT correspondence is valid for any value of the string coupling constant $g_s$ and of the color number $N$. }

We can consider certain limits of the full general $\ads_5/\CFT_4$ duality, which are simpler to be treated but still extremely interesting.

Let us consider the limit where $N$ is very large and $\lambda$ is kept fixed, namely $\gym\rightarrow0$~\cite{'tHooft:1973jz}. In this limit, $N$ is a continuous parameter and the gauge theory admits a ${1\over N}$-expansion. In the large-$N$ regime (also called the 't Hooft limit) of the SYM theory only the planar diagrams survive, namely all the Feynman diagrams whose topology is a sphere.
The corresponding gravity dual is a free string propagating in a non-trivial background ($\ads_5\times \Sphere^5$). The string is non-interacting since now $g_s\rightarrow 0$ and the tension $T$ is kept fixed, cf. eq.~\eqref{rel_coupling_const}. 
Notice that even though we are suppressing $g_s$-corrections, so that the string is a free string on a curved background, it is still described by a non-linear sigma model whose target-space geometry is $\ads_5\times \Sphere^5$. This is a highly non-trivial quantum field theory: the string can have quantum fluctuations which are described by an $\alpha'$-expansion. 

Furthermore, we can also vary the smooth parameter $\lambda$ between the strong-coupling regime ($\lambda \gg 1$) and the weak-coupling regime ($\lambda \ll 1$). In the first case the gauge theory is strongly coupled, while the gravity dual can be effectively described by type IIB supergravity. Indeed, the radius of the background is very large ($R= \lambda^{1\over 4} l_s$), thus the string is in a classical regime ($T\gg 1$). 

Conversely, when $\lambda$ takes very small values ($\lambda \ll 1$), the gauge theory can be treated with a perturbative analysis, while the background where the string lives is highly curved. The string is still free, but now the quantum effects become important (i.e. $T\ll 1$). 

For what we have learned above, the Maldacena duality is also called a weak/strong-coupling correspondence. This is an incredibly powerful feature, since it allows one to reach strong coupling regimes through perturbative computations in the dual description. At the same time, proving such a correspondence becomes an extremely ambitious task, simply because it is hard to directly compare the relevant quantities. For a summary about the different regimes and parameters we refer the reader to the table \ref{tab:AdS/CFT_summary}.
\begin{table}
\begin{center}
\begin{tabular}{|c c c|}
\hline
                 Gauge theory & &String theory \\ \hline
                 Yang Mills Coupling $\gym$ & &String coupling $g_s$\\
                 Number of colors $N$   &             &     String tension $T\equiv {R^2\over 2\pi\alpha'}$ \\
                 't Hooft coupling $\lm\equiv \gym^2N$ & &$\ads_5\times\sphere^5$ radius R\\
\hline
            &  $\ads_5/\mbox{CFT}_4$    &   \\
   & $g_s= {\gym^2\over 4\pi}$ &  \\
    & $T={\sqrt{\lm}\over 2\pi}$&
     \\ \hline 
& 't Hooft limit &\\
$N\rightarrow \infinity$ $\quad \lm=\text{fixed}$ & & $g_s\rightarrow 0$ $\quad T=\text{fixed}$\\
planar limit & & non-interacting string\\ \hline
& Strong Coupling & \\
 $N\rightarrow \infinity$ $\quad \lm\gg 1$ & &     $g_s\rightarrow 0$ $\quad T\gg 1$\\
  & & classical supergravity\\ \hline
  & Weak coupling & \\
  $N\rightarrow \infinity$ $\quad \lm\ll 1$ & &     $g_s\rightarrow 0$ $\quad T\ll 1$\\
  perturbative SYM  & & \\ \hline     \end{tabular}
\end{center}
\caption{Summary of the contents and parameters involved in $\ads_5/\mbox{CFT}_4$ duality.}
\label{tab:AdS/CFT_summary}
\end{table}

We will only deal with the planar $\ads/\mbox{CFT}$, since it is in this regime that both theories have integrable structures. In particular, we are interested in the strong coupling regime ($\lm\gg1$), since the string theory side is reachable perturbatively (${1\over \sqrt{\lm}}$ expansion) in the large 't Hooft coupling limit (cf. table~\ref{tab:AdS/CFT_summary}). The present work is mainly devoted to this sector.

If the two theories are dual, then they should have the same symmetries. This is the theme of the next section, after a more detailed introduction to $\mathcal N=4$ SYM theory.

%
%
%
%
%

\subsection{\texorpdfstring{$\mathcal{N}=4$}{\mathcal{N}=4x} super Yang-Mills theory in 4d}

%
As already mentioned, the $\CN=4$ super Yang-Mills theory in four dimensions~\cite{Gliozzi:1976qd, Brink:1976bc} is a maximally supersymmetric and superconformal gauge theory. The theory is scale invariant at classical and quantum level and the $\beta$-function is believed to vanish to all orders in perturbation theory as well as non perturbatively~\cite{Sohnius:1981sn, Seiberg:1988ur, Mandelstam:1982cb, Howe:1983sr, Brink:1982pd}. The action can be derived by dimensional reduction from the corresponding $\CN =1$ $\grSU(N)$ gauge theory in ten dimensions:
\[
\label{LYM_10d}
\mathcal L_{YM}= {1\over g^2_{\rm 10}} \mathrm{Tr} \big( -\Half F_{MN} F^{MN} + i \bar \psi \Gamma^M D_M \psi \big) \,.
\]
$D_M$ is the covariant derivative, $D_M=\partial_M - i A_M$, where $A_M$ is the gauge field with $M$ the $\grSO(9,1)$ Lorentz index, $M=0,1,...,9$, and $F_{MN}$ the corresponding field strength, which is given by $F_{MN}= \partial_M A_N -\partial_N A_M - i  [A_M, A_N]$. The matter content $\psi$ is a ten-dimensional Majorana-Weyl spinor.
The gauge group is $\grSU(N)$ and the fields $A_M$ and $\psi$ transform in the adjoint representation of $\grSU(N)$.%
%
%

By dimensionally reducing the action \eqref{LYM_10d}, the ten-dimensional Lorentz group $\grSO(9,1)$ is broken to $\grSO(3,1)\times \grSO(6)$, where the first group is the Lorentz group in four dimensions and the second one remains as a residual global symmetry (\emph{R-symmetry}). Correspondingly, the Lorentz index splits in two sets: $M= (\mu, I)$, where $\mu=0,1,2,3$ and $I=1,...,6$. We need to require that the fields do not depend on the transverse coordinates $I$. Hence, the gauge field $A_M$ gives rise to a set of six scalars $\phi_I$ and to four gauge fields $A_\mu$.
Also the fermions split in two sets of four complex Weyl fermions $\psi_{a,\alpha}$ and $\bar \psi^{ \bar a,\dot \alpha}$ in four dimensions, where $a=1,...,4$ is an $\grSO(6)\cong\grSU(4)$ spinor index and $\alpha, \dot\alpha=1,2$ are both $\grSU(2)$ indices. 

The final action for $\CN=4$ SYM in four dimensions is
\[
\label{LYM_4d}
\mathcal L_{YM} = {1\over \gym^2} \mathrm{ Tr} \big( - \Half F_{\mu\nu} F^{\mu\nu}-  (D_\mu \phi_I)^2 + \Half [\phi_I, \phi_J]^2 
                                                                     + i \bar \psi \Gamma^\mu D_\mu \psi 
                                                                      +\bar\psi \Gamma^I [\phi_I,\psi]\big)\,.
\]


\subsection{The algebra}

We have already stressed that the theory has an $\grSU(N)$ gauge symmetry, thus the gauge fields are $\algSU(N)$-valued, and they also carry an index $i=1,...,N^2-1$, which is not explicit in the formulas above. %

The conformal group in four dimensions is%
\footnote{The symbol $\cong$ means that the two groups are locally isomorphic.}
$\grSO(4,2)\cong \grSU(2,2)$. The generators for the conformal algebra $\algSO(4,2)$ are the Lorentz transformation generators, which consist of three boosts and three rotations $M_{\mu\nu}$, the four translation generators $P_\mu$, coming from the Poincar\'{e} symmetry, the four special conformal transformation generators $K_\mu$ and the dilatation generator $D$. Hence in total we have fifteen generators. 

The theory is also invariant under the $\mathcal R$-symmetry, which plays the role of an internal flavor symmetry which can rotate the supercharges and the scalar fields. The $\mathcal R$-symmetry group is $\grSO(6) \cong \grSU(4)$ and it is spanned by fifteen generators, $R_{IJ}$.

The supersymmetry charges $Q^a_{\alpha}$, $\bar Q^{\bar a\dot\alpha}$, which transform under {\it R}-symmetry in the four-dimensional representations of $\grSU(4)$ ($\mathbf{4}$ and $\mathbf{\bar{4}}$ respectively), commute with the Poincar\'{e} generators $P_\mu$. They do not commute with the special conformal transformation generators $K_\mu$. However, their commutation relations give rise to a new set of supercharges. We denote this new set of supercharges with $S_{\alpha}^{\bar a}$ and $\bar S^{a\dot\alpha}$. They transform in the $\mathbf{\bar{4}}$ and $\mathbf{4}$ representation of $\grSU(4)$. Thus we have in total 32 real fermionic generators. 

The $\grSO(4,2)\times  \grSO(6)$ bosonic symmetry groups and the supersymmetries merge in a unique superconformal group $\grSU(2,2|4)$. Actually, due to the vanishing of central charge for SYM, the final symmetry group is $\grPSU(2,2|4)$, where P denotes the fact that we are removing \emph{ad hoc} the identity generators which can appear in the commutators.
Notice that in supersymmetric theories usually the anticommutators between the supercharges $Q$ and $S$ give an operator which commutes with all the rest, the so called \emph{central charge}. 

The relevant relations are
\be
\label{superconformal_algebra}
&& \left[ D\,,P_\mu\right] = -i P_\mu 
\qquad \left[ D\,, K_\mu\right]= i K_\mu
\qquad \left[ P_\mu,\, K_\nu\right] = 2 i (M_{\mu\nu} -\eta_{\mu\nu} D)
\nln
&& \left[ M_{\mu\nu},\, P_{\lm}\right] = i (\eta_{\lm \nu} P_\mu- \eta_{\mu\lm} P_\nu)
\qquad
\left[ M_{\mu\nu},\, K_{\lm}\right] = i (\eta_{\lm \nu} K_\mu- \eta_{\mu\lm} K_\nu)
\nln
&& \left[M_{\mu\nu}\,, M_{\lm\rho}\right]= -i \eta_{\mu\lm} M_{\nu\rho} +\text{cycl. perm.}\nln
&& 
\left \{ Q^a_\alpha,\bar Q^{\bar b}_{\dot\alpha}\right\}= \gamma_{\alpha\dot\alpha}^\mu \dl^{a\bar b} P_\mu 
\qquad 
\left \{ S^{\bar a}_\alpha,\bar S^{ b}_{\dot\alpha}\right\}= \gamma_{\alpha\dot\alpha}^\mu \dl^{\bar a b} K_\mu 
\nln
&& \left[ D\,, Q^a_\alpha\right]= -{i\over 2} Q^a_\alpha 
\qquad 
\left[ D\,, \bar Q^{\bar a}_{\dot\alpha}\right]= -{i\over 2} \bar Q^{\bar a}_{\dot \alpha} 
\nln
&& \left[ D\,, S^{\bar a}_\alpha\right]= {i\over 2} S^{\bar a}_\alpha
\qquad
\left[D\,, \bar S^{a \dot \alpha}\right]= {i\over 2}  \bar S^{a \dot \alpha}\nln
&& \left[ K^\mu, \, Q^a_\alpha\right]= \sigma^\mu_{\alpha\dot\alpha} \eps^{\dot\alpha\dot\beta} \bar S^a_{\dot\beta}
\qquad
\left[ K^\mu,\, \bar Q^{\bar a}_{\dot\alpha}\right]= \sigma^\mu_{\alpha\dot\alpha} \eps^{\alpha\beta} S^{\bar a}_\beta\nln
&& \left[ P_\mu,\, S^{\bar a}_\alpha\right] = (\sigma_\mu)_{\alpha\dot\alpha} \eps^{\dot\alpha\dot \beta} \bar Q^{\bar a}_{\dot \beta} 
\qquad
\left [P_\mu,\, \bar S^{a}_{\dot \alpha}\right]= (\sigma_\mu)_{\alpha\dot\alpha}\eps^{\alpha\beta} Q^a_\beta
\nln
&&\left[ M^{\mu\nu}, \, Q_\alpha^a\right]= i \sigma_{\alpha\beta}^{\mu\nu} \eps^{\beta\gamma} Q_\gamma^a
\qquad
\left[ M^{\mu\nu}, \, \bar Q_{\dot\alpha}^{\bar a}\right]= i \sigma_{\dot\alpha\dot\beta}^{\mu\nu} \eps^{\dot\beta\dot\gamma} \bar Q_{\dot\gamma}^{\bar a}
\nln
&&\left[ M^{\mu\nu},\, S^{\bar a}_{\alpha}\right]= i \sigma^{\mu\nu}_{\alpha \beta} \eps^{\beta\gamma} S^{\bar a}_\gamma
\qquad
\left[ M^{\mu\nu},\, \bar S^a_{\dot\alpha}\right]= i \sigma^{\mu\nu}_{\dot\alpha\dot\beta} \eps^{\dot\beta\dot\gamma} \bar S^a_{\dot\gamma}\nln
&& \left\{ Q_\alpha^a\,, S^{\bar b}_\beta \right\}
= -i \epsilon_{\alpha\beta} (\sigma^{IJ})^{a\bar b} R_{IJ} + \sigma_{\alpha\beta}^{\mu\nu}\delta^{a \bar b} M_{\mu\nu} -\eps_{\alpha\beta} \delta^{a\bar b} D
\nln
&& \left\{ \bar Q_{\dot\alpha}^{\bar a}\,, \bar S^{ b}_{\dot\beta} \right\}= -i \epsilon_{\dot\alpha\dot\beta} (\sigma^{IJ})^{\bar a b} R_{IJ} + \sigma_{\dot\alpha\dot\beta}^{\mu\nu}\delta^{\bar a b} M_{\mu\nu} -\eps_{\dot\alpha\dot\beta} \delta^{\bar a b} D\,.
\ee
The matrices $\sigma^\mu_{\alpha\dot\alpha}$ are the Dirac $2\times 2$ matrices and $(\sigma^{IJ})_{ a \bar b}$ are the antisymmetric product of the Dirac $4\times 4$ matrices.


\paragraph{Matrix realization.}

It is natural to reorganize the $\algSU(2,2|4)$ generators as $8\times 8$ super-matrices: 
\be
\label{supermatrix_generators}
M= \matrto{P_\mu,\, K_\mu, \, L_{\mu\nu},\,  D}{ Q^{\alpha a} , \,\bar S^{\dot \alpha a}}{S^{\bar a}_{\alpha} ,\,\bar Q^{\bar a}_{\dot \alpha}}{R^{IJ}}\,.
\ee
On the diagonal blocks we have the generators for two bosonic sub-sectors, $\algSU(2,2)$ and $\algSU(4)$, while on the off-diagonal blocks we have the fermionic generators. The super-algebra is realized by two conditions which naturally generalize the $\algSU(n,m)$ algebra. First, the super-trace%
\footnote{For any super-matrix 
\be
\label{supermatrix}
M= \matrto{A}{ X}{Y}{B}
\ee
where the block-diagonal are even matrices and off-block elements are odd, the super-trace is defined as $\STr M= \Tr A-\Tr B$.}
of the matrix \eqref{supermatrix} vanishes. Second it satisfies a reality condition 
\[
H M^\dagger -M H=0\,,
\]
where 
\be
\label{Hmatrix_def}
H= \matrto{\gamma_5}{0}{0}{\mathbf{\mathrm{1}}}\,.
\ee
The $4\times 4$ matrix $\gamma_5$ appears in the above condition because $\gm_5$ realizes the Hermitian conjugation in the $\grSU(2,2)\cong \grSO(4,2)$ sector. 

Actually, we want to consider the $\algPSU(2,2|4)$ algebra. The $8\times 8$ $\algSU(2,2|4)$ identity matrix trivially satisfies both properties of tracelessness and of Hermicity. This means that even though such a matrix is not among our set of initial generators of the $\algSU(2,2|4)$ algebra, at some point it will appear as a product of some commutators. This is analogous to what we have discussed above, where the anticommutator between Q and S might have a term proportional to the unit matrix.
In the SYM, the central charge is zero, thus we would like to remove the unit matrix. We therefore mod out the $\algU(1)$ factor \emph{ad hoc}. This is indeed the meaning of the $\mathfrak{p}$ in $\algPSU(2,2|4)$. Note that such an algebra cannot be realized in terms of matrices. 

The total rank for the $\grPSU(2,2|4)$ supergroup is 7. The unitary representation is labelled by the quantum numbers for the bosonic subgroup. This means that the fields of $\mathcal N=4$ SYM, or better, local gauge invariant operators, and the states of the $\ads_5\times \sphere^5$ string are characterized by 6 charges, which are the Casimirs of the group:
\[
\label{charges}
\left( \Delta= E, \, S_1,\, S_2,\, J_1,\, J_2,\, J_3\,\right)\, . 
\]
The equality for the first charge is really the expression of the AdS/CFT correspondence. Let us see in more detail what these quantum numbers are. 
Coming from the $\grSU(2,2)$ sector, since $\grSO(1,1)  \times  \grSO(3,1) \subset \grSO(4,2)$, we have the dilatation operator eigenvalue $\Delta$ (or the string energy E), which can take continuous values, and the two spin eigenvalues $S_1,\, S_2$, which can have half-integer values, and which are the charges related to the Lorentz rotations in $\grSO(3,1)$. Notice that $\Delta$ and $E$ depend on the coupling constant $\lambda$, cf. \eqref{ads_cft_def}.  
The other sector $\grSU(4) \cong \grSO(6)$ contributes with the ``spins'' $J_1,\, J_2,\, J_3$, which characterize how the scalars can be rotated. 


\paragraph{The string side.}

The isometry group of $\ads_5\times \Sphere^5$ is $\grSO(4,2)\times \grSO(6)$, which is nothing but the bosonic sector of $\grPSU(2,2,|4)$.
Thus on the string side the bosonic symmetries are realized as isometries of the background where the string lives. The superstring also contains fermionic degrees of freedom which will mix the two bosonic sectors corresponding to $\ads_5$ and $\Sphere^5$.
The string spectrum is labelled by the charges \eqref{charges}. In principle one can also have winding numbers to characterize the string state, in addition to \eqref{charges}. 
The string energy $E$ is the charge corresponding to global time translation in $\ads_5$, while $S_1,\, S_2$ correspond to the Cartan generators of rotations in $\ads_5$. The last three charges correspond to Cartan generators for $\Sphere^5$ rotations, since the five-dimensional sphere can be embedded in $\R^{6}$, so we have three planes the rotations.


\subsection{Anomalous dimension and spin chains}
\label{sec:spinchain}

In a conformal field theory the correlation functions between local gauge invariant operators contain most of the relevant dynamical information. There is a special class of local operators, the \emph{(super) conformal primary} operators, whose correlators are fixed by conformal symmetry. In particular, these are the operators annihilated by the special conformal generators $K$ and by the supercharges $S$, i.e. $K \mathcal O=0$ and $S\mathcal O=0$. 
Thus, representations corresponding to primary operators are classified by how the dilatation operator $D$ and the Lorentz transformation generators $M$ act on $\mathcal O$, i.e. by the 3-tuplet $(\Delta, S_1, S_2)$:
\[
D \mathcal O=\Delta \mathcal O\,, \qquad M \mathcal O=\Sigma_{S_1, S_2} \mathcal O\,.
\]
where $\Delta$ is the scaling dimension, namely the dilatation operator eigenvalues, and $\Sigma_{S_1,S_2}$ tells us how the operator $\mathcal O$ transforms under Lorentz transformations.
Since the special conformal transformation generator $K$  lowers the dimension by $1$ and the supercharge $S$ by $\half$, cf. \eqref{superconformal_algebra}, in a unitary field theory the primary operators correspond to those operators with lowest dimension. They are also called highest-weight states. 
All the other operators in the same multiplet can be obtained by applying iteratively the translation operator $P$ and the supercharges $Q$ (\emph{descendant conformal operators}).

The correlation functions of primary operators are highly restricted by the invariance under conformal transformations, and they are of the form:
\[
\label{2pt_funct_O}
\langle \mathcal O_m (x) \mathcal O_n (y)\rangle = {C \delta_{mn} \over |x-y|^{2\Delta}}\,. 
\]
In the scaling dimension there are actually two contributions: 
\[
\Delta= \Delta_0 +\gamma\,.
\]
$\Delta_0$ is the classical dimension and $\gamma$ is the so-called anomalous dimension. It is in general a non-trivial function of the coupling constant $\lambda$. It appears once one starts to consider quantum corrections, since in general the correlators will receive quantum corrections from their free field theory values.

When we move from the classical to the quantum field theory we also need to face the problem of renormalization. 
In general in quantum field theory the renormalization is multiplicative. The operators are redefined by a field strength function $Z$ according to
\[
\label{def_renorm_O}
\mathcal O_m = Z_m^n \mathcal O_{n,0}
\]
where the subscript $0$ denotes the bare operator, and Z depends on the physical scale $\mu$ (typically $Z\sim \mu^\gamma$). As an example, we can consider the correlators in eq. (\ref{2pt_funct_O}). Applying the Callan-Symanzik equation, recalling that the $\beta$-function vanishes and defining the so-called mixing matrix $\Gamma$ as
\[
\label{mixing_matrix}
\Gamma_m^k=  \sum_ n (Z^{-1})_m^n {\partial Z_n^k\over \partial \log{\mu}}\,,
\]
we see that when the operator $\Gamma$ acts on a basis $\{\mathcal O_m\}$, then the corresponding eigenvalues are indeed the anomalous dimensions $\gamma_m$:
\[
\Gamma \mathcal O_m= \gamma_m \mathcal O_m\,.
\]
Hence, $\Gamma$ provides the quantum correction to the scaling operator $D$, i.e. $D= D_0 + \Gamma$.


\subsubsection{The Coordinate Bethe Ansatz for the $\mathfrak{su}(2)$ sector}
\label{sec:BE}

In this section I will sketch the {\it Coordinate Bethe Ansatz}, also called {\it Asymptotic Bethe Equations} (ABE), for the bosonic closed $\grSU(2)$ sub-sector, as the title suggested, in order to get feeling of why such techniques are so important.
The ABE are the basic connection between integrability, SYM theory, spin chain and the S-matrix. 

As pointed out in the previous section, a lot of the relevant physical information are contained in the anomalous dimension of a certain class of gauge invariant operators. The fact that the operators are gauge invariant means that we have to contract the $\grSU(N)$ indices. This can be done by taking the trace. In general, we can have multi-trace operators. However, in the planar limit ($N\rightarrow\infinity$) the gauge invariant operators which survive are the single trace ones. Thus from now on, we are only dealing with single trace local operators (and with their anomalous dimension).

The incredible upshot of this section will be that the mixing matrix \eqref{mixing_matrix} {\it is} the Hamiltonian of an {\it integrable} (1+1) dimensional spin chain! There are two important points in the last sentence. First, it means that the eigenvalues of the mixing matrix are the eigenvalues of a spin chain Hamiltonian, {\it namely the corresponding anomalous dimensions are nothing but the solutions of the Schr\"{o}dinger equation of certain spin chain Hamiltonians.} I cannot say whether it is easier to compute $\gamma$, or to solve some quantum mechanical system such as a one-dimensional spin chain. 
But here it enters the second keyword used: {\it integrable}. The spin system has an infinite set of conserved charges, all commuting with the Hamiltonian (which is just one of the charges), which allows us to solve the model itself. In concrete terms, this means that we can compute the energies of the spin chain, namely the anomalous dimension (of a certain class) of $\mathcal N=4$ SYM operators! Here the advantage is not purely conceptual but also practical: we can exploit and/or export in a string theory context some methods and techniques usually used in the condensed matter physics for example. And this is what we will see in a moment.

We have just claimed that the anomalous dimensions (for a certain class of operators) can be computed via spin chain picture. We have to make this statement more precise. In particular, we need to specify when and how it is true. In order to illustrate how integrability enters in the gauge theory side, and its amazing implications, I have chosen to review in detail the simplest example: the closed bosonic $\grp{SU}(2)$ sub-sector of $\grSO(6)$. Historically, the connection between SYM gauge theory and spin chain was discovered by Minahan and Zarembo for the scalar $\grSO(6)$ sector of the planar $\grp{PSU}(2,2|4)$ group~\cite{Minahan:2002ve}.
This has been the starting point for all the integrability machinery in AdS/CFT.%
 \footnote{  
The appearance of integrable spin chains in QCD at high-energy was already discussed by Lipatov in~\cite{Lipatov:1993yb} and by Faddeev and Korchemsky in \cite{Faddeev:1994zg}. See also~\cite{Beisert:2004ry} and references therein. }

\begin{figure}
	\begin{center}
		\includegraphics[scale=0.3]{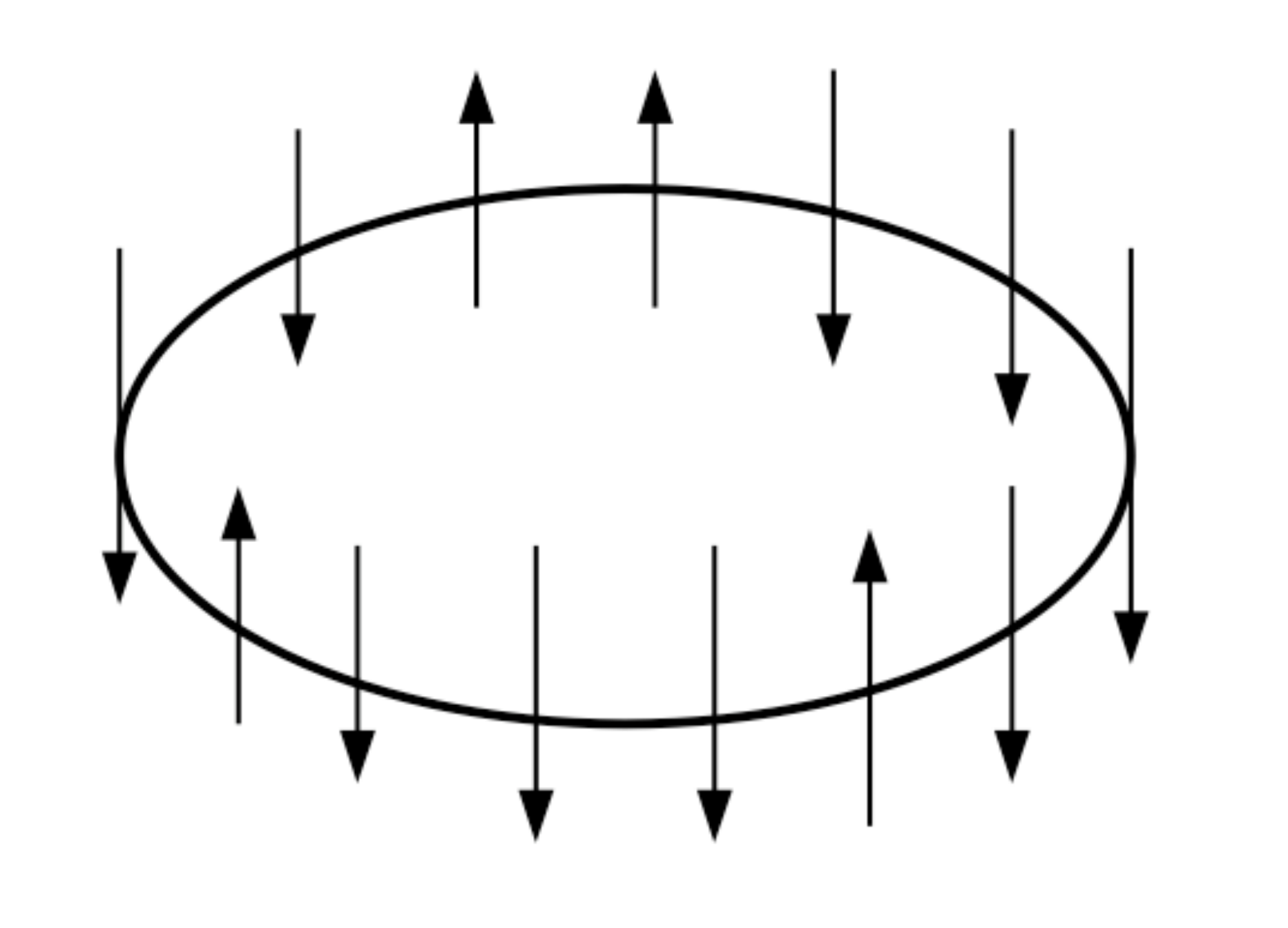}
	\end{center}
	\caption[vvv]{Example of a spin chain. The ``up'' arrow represents the field $Z$, while the down spin is represented by the field $W$.}
	\label{fig:spin_chain}
\end{figure}

The scalar fields $\phi_I$ with $I=1,\dots ,6$ can be rearranged in a complex basis. For example, we can write
\[
Z= \phi_1+i \phi_2
\qquad
W= \phi_3+i \phi_4
\qquad
Y=\phi_5+i \phi_6\,.
\]
The three complex fields $Z, \,W$ and $Y$ generate $\grSU(4)$. The $\grSU(2)$ subgroup is constructed by considering two of the three complex scalars. For example, we can take the fields $Z$ and $W$. We are considering gauge invariant operators of the type
\[
\label{su2_operator}
\mathcal O (x) = \Tr\left (WZWWZWWWWZZWW \right )_{|x}+\dots\,,
\]
where the dots indicate permutations of the fields and the subscript on the right hand side stresses the fact that these fields are all evaluated in the point $x$. If one identifies the fields in the following way 
\[
\label{field_spinchain}
Z\,=\, \uparrow 
 \qquad W\,=\, \downarrow\,,
\]
then the operator $\mathcal O$ in \eqref{su2_operator} can be represented by a spin chain. In particular, for the operator \eqref{su2_operator} the corresponding spin chain is represented in figure \ref{fig:spin_chain}. If we have L fields sitting in the trace of the operator $\mathcal O$, it means that we are considering a spin chain of length L, with L sites. Each site has assigned a spin, up or down, according to the identification \eqref{field_spinchain}.%

At one-loop the dilation operator for gauge invariant local operators which are $\mathfrak{su}(2)$ multiplets can be identified with the Hamiltonian of a Heisenberg spin chain, also denoted as a $XXX_{\half}$ spin chain. Note that this is a quantum mechanics system. 

The identification between the Heisenberg spin chain Hamiltonian and the $\grSU(2)$ one-loop dilatation operator can be seen by an explicit computation of such an operator \cite{Minahan:2002ve}. 
In particular, one has that
\[
\Gamma^{(1)}= {\lm\over 8 \pi^2} \sum_{l=1}^L H_{l,l+1}
\]
where $H_{l,l+1}$ is the operator acting on the sites $l$ and $l+1$, explicitly 
\be
\label{H_spinchain}
H &=&{\lm\over 8 \pi^2} \sum_{l=1}^L H_{l,l+1}
= {\lm\over 8\pi^2} \sum_{l=1}^L \left( \mathrm{I}_{l,l+1}- P_{l,l+1}\right) \nln
&=& {\lm\over 16 \pi^2} \sum_{l=1}^L \left( \mathrm{I}_{l,l+1}- \overrightarrow{\sigma}_l\cdot \overrightarrow{\sigma}_{l+1}\right)\,,
\ee 
where $P_{l,l+1}= \half (\mathrm{I}_{l,l+1}+\overrightarrow{\sigma}_l\cdot \overrightarrow{\sigma}_{l+1})$ is the permutation operator.
The one-loop order is mirrored by the fact that the Hamiltonian only acts on the sites which are nearest neighbors. The identity operator $\mathrm{I}_{l,l+1}$ leaves the spins invariant, while the permutation operator $P_{l,l+1}$ exchanges the two spins. 

We want to compute the spectrum. This means that we want to solve the Schr\"{o}dinger equation $H|\Psi\rangle = E|\Psi\rangle$. $|\Psi\rangle$ will be some operators of the type \eqref{su2_operator}, and the energy will give us the one-loop anomalous dimension for such operator. The standard approach would require us to list all the $2^L$ states and then, after evaluating the Hamiltonian on such a basis, we should diagonalize it. This is doable for a very short spin chain, not in general for any value $L$. The brute force here does not help, and indeed there are smarter ways as the one found by Bethe in 1931 \cite{Bethe:1931hc}. 

\paragraph{One-magnon sector.}

Let us choose a vacuum of the type
\[
\label{vacuum_spinchain}
|0 \rangle \equiv | \uparrow \uparrow\dots\uparrow \uparrow \dots  \uparrow\rangle \,,
\]
and consider an infinite long spin chain, i.e. $L\rightarrow \infinity$. The vacuum has all spins up and it is annihilated by the Hamiltonian \eqref{H_spinchain}. The choice of the vacuum breaks the initial $\grSU(2)$ symmetry to a $\grU(1)$ symmetry. Consider now the state with one excitation, namely with an impurity in the spin chain:
\[
| x\rangle \equiv | \uparrow \uparrow \dots \uparrow\underbrace{\downarrow}_{x}\uparrow\dots \uparrow\rangle\,.
\]
The excitation, called a \emph{magnon}, is sitting in the site $x$ of the spin chain. The wave function is 
\[
\ket \Psi  = \sum_{x=-\infinity}^\infinity \Psi(x) \ket x\,. 
\]
By computing the action of the Hamiltonian $H$ on $\ket \Psi$, one obtains
\be
H \ket \Psi &=&  \sum_{x=-\infinity}^\infinity \Psi(x) \left (\ket{x} -\ket{x+1}-\ket{x-1} \right)\nln
 &= &
 \sum_{x=-\infinity}^\infinity \left  ( 2\Psi (x) -\Psi (x+1)- \Psi(x+1) \right)\ket{x}\,.
\ee
Let us make an ansatz for the wave-function. Choosing
\[
\label{def_psi_1magnon}
\Psi(x) = e^{i p x} \qquad p\in \Reals\,,
\]
then the Schr\"{o}dinger equation for the one-impurity state reads 
\[
H \ket\Psi= \sum_{x=-\infinity}^\infinity e^{i p x} \left (2- e^{i p} -e^{-i p}\right)\ket x\,.
\]
This means that the energy for the one magnon state is
\[
\label{one_magnon_energy}
E (p)= {\lm\over 8\pi^2} \left( 2- e^{i p} -e^{-i p }\right)=  {\lm\over 2\pi^2} \sin^2 {p\over 2}\,.
\]
This is nothing but a plane wave along the spin chain. 

The spin chain is a discrete system. There is a well defined length scale, which is given by the lattice size, and the momentum is confined in a region of definite length, typically the interval $[-\pi,\pi]$ (the first Brillouin zone). 
An infinite chain might be obtained by considering a chain of length L and assume periodicity. Thus we need to impose a periodic boundary condition on the magnon wave function, which means
\[
\Psi (x+L)= \Psi (x) ~~\Rightarrow ~~ e^{i p L} = 1~~ \Rightarrow p_n={2\pi n\over L} ~~ n\in \Z\,.
\]
These are the {\it Coordinate Bethe Equations} for the one-magnon sector.%
\footnote{In condensed matter physics they are usually called {\it Bethe-Yang} equations.}
They are the periodicity conditions of the spin chain. 

Leaving the spin chain picture, and going back to the gauge theory, the operator $\mathcal O$ in \eqref{su2_operator} is not only periodic but cyclic (due to the trace). For the single magnon this implies that the excited spin must be symmetrized over all the sites of the chain. Thus the total energy vanishes.%
\footnote{This is equivalent to impose $\Psi(x)=\Psi(x+1)$, which gives $e^{i p}=1$.}
%
Indeed, operators of the kind
\[
\mathcal O =\Tr \left (\dots Z Z Z W Z Z \dots\right) 
\]
are chiral primary operators: their dimension is protected and one can see that the cyclicity of the trace means that the total momentum vanishes, which is another way of saying that the energy is zero, cf. \eqref{one_magnon_energy}.

Thus there is no operator in SYM that corresponds to the single magnon state. This is actually true for all sectors, since it follows from the cyclicity of the trace. 

\paragraph{Two-magnon sectors.}

Consider now a state with two excitations, namely two spins down:
\be
&&\ket{ x < y}= \ket{ \uparrow \dots \uparrow \underbrace{\downarrow}_{x} \uparrow\dots \uparrow \uparrow \underbrace{\downarrow}_{y}\uparrow\dots}\,, \nln
&& \ket{\Psi}= \sum_{x<y =-\infinity}^\infinity  \Psi(x,y) \ket{ x < y}\,.
\ee
The Hamiltonian \eqref{H_spinchain} is short-ranged, thus when $x + 1< y$ it proceeds as before for the single magnon state, just that in this case the energy $E$ would be the sum of two magnon dispersion relations. 
The problem starts when $x+1=y$, namely in the contact terms. In this case the Scr\"{o}dinger equation for the wave-function gives
\be
\label{eq_2magnons}
2\Psi(x,x+1)-  \Psi(x-1,x+1)-\Psi(x,x+2) =0 \,.
\ee
It is clear that a wave function given by a simple sum of the two single magnon states as in \eqref{def_psi_1magnon} does not diagonalize the Hamiltonian \eqref{H_spinchain}, but ``almost''. Using the following ansatz%
\footnote{For the case with $x>y$ it is sufficient to exchange the role of x and y.}
\[
\label{2magnon_wf}
\Psi (x,y)= e^{i p x+i q y-i {\dl\over 2}}+ e^{i q x+i p y + i {\dl\over 2}}\qquad x< y\,,
\]
and imposing that it diagonalizes the Hamiltonian,
one finds the value for the {\it phase shift} $\delta$ that solves the equation, namely 
\be
\label{phase_shift}
e^{i\dl (p,q)}= - \frac{ 1-2 e^{i q} +e^{i p+i q}}{1- 2 e^{i p}+e^{i q+\imath p}}= - \frac{\cot{p/2} - \cot{q/2}-2i}{\cot{p/2} - \cot{q/2}+2 i}\,.
\ee
For this phase shift the total energy is just the sum of two single magnon dispersion relations (trivially the ansatz \eqref{2magnon_wf} with the phase shift given by \eqref{phase_shift} solves the case with $x+1< y$). What does this phase shift represent? This is the shift experienced by the magnon once it passes through the other excitation, namely when it {\it scatters} a magnon of momentum $q$. Hence $S(p,q)\equiv e^{i\dl (p,q)}$ is nothing but the corresponding {\it scattering-matrix}. 

We still have to impose the periodic boundary conditions on the wave functions%
\footnote{The wave function is symmetric with respect to x, y.}:
\[
\Psi(0, y)= \Psi (L,y)\,,
\]
which, after substituting the phase shift \eqref{phase_shift} in \eqref{2magnon_wf}, gives
\[
\label{BE_2magnons}
e^{i p L}= e^{-i \dl(p,q) }= e^{i\dl (q,p)}= S(q,p)
\qquad 
e^{i q L}= e^{i\dl (p,q)}= S(p,q) \,.
\]
Again, these are the Coordinate Bethe equations for the $\mathfrak{su}(2)$ sector with two magnons. 

Finally, we need to impose the cyclicity condition, i.e. $p+q=0$, which means that the Bethe equations \eqref{BE_2magnons} are solved for 
\[
p= {2\pi n\over L-1}=-q\,.
\]
The energy becomes 
\[
E= E(p)+E(q)={\lm\over \pi^2} \sin^2 \left({\pi n\over L-1}\right)\,.
\]
Maybe the reader is more familiar to the Bethe Equations expressed in terms of the {\it rapidities}, also called {\it Bethe roots}%
\footnote{In section \ref{chapter:Intro_integrability} the rapidity is denoted with the greek letter $\theta$. Although the notation might seem confusing it is the standard one used in literature.},
namely introducing
\[
\label{rapidity}
u_k= \Half \cot{p_k\over 2}
\]
and using $p=-q$, the phase shift reads
\[
e^{i\dl (u, -u)}= S(u,-u)= -\frac{u-{i\over 2}}{u+{i\over 2}}\,.
\]

\paragraph{K magnon sectors.}

The results of the previous section can be generalized to any number of magnons K (with $K< L$). The Bethe Equations for general $K$ are
\[
\label{BE_Kmagnons}
e^{i p_k L}= \prod_{j\neq k}^K e^{-i \dl(p_k, p_j)} = \prod_{j\neq k}^K S(p_j, p_k)\,.
\]
The energy is a sum of K single particle energies
\[
E = \sum_{k=1}^K E_k= {\lm\over 2\pi^2} \sum_{k=1}^K \sin^2 {p_k\over 2}\,,
\]
and the cyclicity condition is
\[
\prod_{k=1}^K e^{i p_k L}=1\,.
\]
In terms of the rapidities \eqref{rapidity} all these conditions take the maybe more common form of
\be
\label{general_Kmagnons}
&& 1= \left( \frac{u_k+{i\over 2}}{u_k-{i\over 2}}\right)^L \prod_{j=1,k\neq j}^K \frac{u_j-u_k +i}{u_j-u_k-i} \qquad \text{(Bethe Equations)}
\nln
&& E= \sum_{k=1}^K \left( \frac{i}{u_k+{i\over 2}}-\frac{i}{u_k-{i\over 2}}\right) \qquad \text{(energy)}
\nln
&& 1= \prod_{k=1}^K \frac{u_k+{i\over 2}}{u_k-{i\over 2}} \qquad \text{(cyclicity)}\,.
\ee
What have we achieved? The remarkable point is that the Hamiltonian of a (1+1)-dimensional spin chain has been diagonalized by means of the {\it 2-body S-matrix} $S(p,q)$, cf. \eqref{BE_Kmagnons}. Indeed, in order to know the spectrum of $K$ magnons, where $K$ is arbitrary, we only need to solve the Bethe equations and to compute the two-body S-matrix. The K-body problem is then reduced to a 2-body problem, which is an incredible achievement. This does not happen in general. The underlying notion that we are using here is that {\it each magnon goes around the spin chain and scatters only with one magnon each time}. This is possible only for {\it integrable} spin chains, or in general for {\it integrable} models.%
\footnote{There are indeed further assumptions about integrability. We are indeed assuming that the only kind of scattering is elastic, that there is no magnon produced in such scatterings and that the initial and final momenta are the same. We have already used these hypothesizes in equation \eqref{2magnon_wf} for the two-magnon sector 
}
%
We will come back more extensively on this in the next section.

\subsubsection{The full planar $\grp{PSU}(2,2|4)$ ABE}
\label{sec:fullABE}


Here we have shown in details the $\grSU(2)$ sub-sector for the fields in the spin~$\Half$ representation. However, this can be generalized to other representations for the same group, or to other groups (e.g $\grSU(N))$ and also to higher loops. What is really interesting for us, in an AdS/CFT perspective, is that the asymptotic Bethe Equations for the full (planar) $\grp{PSU}(2,2|4)$ group have been written down. This has been done by Beisert and Staudacher \cite{Beisert:2005fw}. They are reported in appendix \ref{app:fullABE}.

At the beginning of the section we explained that the Bethe equations are called ``asymptotic''. ``Asymptotic'' since the Bethe procedure captures the correct behavior of the anomalous dimension only up to $\lm^L$ order for a chain of length L. 
After this order, {\it wrapping} effects have to be taken into account. They reflect the fact that the chain has a {\it finite size}. At the order $n$ in perturbation theory, the spin chain Hamiltonian involves interaction up to $n+1$ sites: $H_{l,l+1,\dots, l+n}$. If the spin chain has total length $L=n+1$, then it is clear that there might be interactions that go over all the spin chain, namely they wrap the chain.%
\footnote{I will come back on the wrapping effects in section~\ref{chapter:NFS}.}
At this point the ABE are no longer valid. In order to compute these finite-size effects, one might proceed with different techniques as the L\"{u}scher corrections~\cite{Luscher:1986pf, Luscher:1985dn},%
\footnote{For generalizations and applications of L\"{u}scher formulas for the computations of finite-size effects we refer the reader to the papers~\cite{Janik:2007wt, Bajnok:2008bm, Bajnok:2009vm}. The four loop anomalous dimension for the Konishi operator computed in \cite{Bajnok:2008bm} has been positively checked against the gauge theory perturbative computation of~\cite{Fiamberti:2007rj}. }
the Thermodynamic Bethe Ansatz (TBA)~\cite{Ambjorn:2005wa}, cf.~\cite{Arutyunov:2009ur, Arutyunov:2009ux, Gromov:2009bc, Bombardelli:2009ns} for very recent results, and the Y-system~\cite{Gromov:2009tv}. These topics currently are one of the main area of research in the context of integrability and AdS/CFT, however here we will not face the problem of finite-size effects%
\footnote{The wording ``finite-size effect`'' should not be confused with what we will illustrate in section \ref{chapter:ABJM}, cf. the discussion therein.}.
The explicit one-loop $\grp{PSU}(2,2|4)$ spin chain Hamiltonian has been derived by Beisert in~\cite{Beisert:2003jj}. This means that the expression of the one-loop dilatation operator for the $\mathcal N=4$ SYM is known. Increasing the loop order usually makes things (and thus also the dilatation operator) sensibly more complicated, cf. e.g.~\cite{Beisert:2003tq}. Moreover, we do not really need the explicit expression of the Hamiltonian, once one has the Bethe equations. Indeed, nowadays we have from the one-loop~\cite{Beisert:2003yb} to the {\it all loop} asymptotic Bethe equations for the planar $\grp{PSU}(2,2|4)$~\cite{Beisert:2005fw}.
%
%



\section{Classical vs. Quantum Integrability}

\label{chapter:Intro_integrability}

The superstring theory on $\ads_5\times \Sphere^5$ can be described by a very special two-dimensional field theory. Indeed, such a theory shows an infinite symmetry algebra. Before discussing such an algebra for the specific case of the superstring we will review other integrable $(1+1)$ field theories, their conserved (local and non-local) charges and finally stress the difference between integrability at classical and quantum level. 

The discovery of an infinite set of conserved charges in two-dimensional classical $\sigma$ models is due to Pohlmeyer \cite{Pohlmeyer:1975nb} and L\"{u}scher and Pohlmeyer~\cite{Luscher:1977rq}. A different derivation of the tower of conserved charges has been given by Brezin {\it et al.} in~\cite{Brezin:1979am}. A very useful review is Eichenherr's paper~\cite{Eichenherr:1981pa}. 


\subsection{Principal Chiral Model}
\label{sec:PCM}

As a prototype to start our discussion with, we consider the so-called Principal Chiral Model (PCM). The following presentation is mostly based on~\cite{Evans:1999mj}.  The PCM is defined by the following Lagrangian:
\[
\label{Lagrangian_PCM}
\mathcal L= {1\over \gamma^2} \rm{Tr} \left(\p_\mu g^{-1}\p^\mu g\right)\,,
\]
where $g$ is a group valued map, $g\, : \Sigma \rightarrow G$ with $\Sigma$ a two-dimensional manifold and $G$ a Lie group. In particular $\Sigma$ is parameterized by $\sigma^\mu=(\tau,\sigma)$. We can think to $\Sigma$ as the string world-sheet. $\gamma$ is a dimensionless coupling constant, the model is conformally invariant. The model (\ref{Lagrangian_PCM}) possesses a $\grp{G}_L\times\grp{G}_R$ global symmetry (simply due to the trace cyclicity) which corresponds to left and right multiplications by a constant matrix, i.e. $\grp{G}_L\times\grp{G}_R:~~g \rightarrow g_{0L} \,g \, g^{-1}_{0R}$. The conserved Noether currents associated to such symmetries are
\[
j^R= - d g g^{-1}
\quad
j^L=+g^{-1} d g\,
,~~~~\text{with}~~~ g j^Lg^{-1}=-j^R\,.
\]
These currents are one-forms and they are also called Maurer-Cartan forms (MC-forms). They are nothing but vielbeins; indeed $j^{(L,R)}$ are $\mathfrak{g}$-valued functions and they span the tangent space for any point $g(\tau,\sigma)$ in $\grp{G}$. We can then write
\be
&& j= j^a t^a =  E^a_M d X^M t^a
\nln
&& j^R_\mu= -\p_\mu g g^{-1} \qquad   j^L_\mu=+g^{-1} \p_\mu g \,
\ee
where $X^M$ denotes the specific parameterization chosen for the M-dimensional group manifold $G$. $t^a$ are the generators of the corresponding Lie algebra $\mathfrak{g}$, which obey the standard Lie algebra relations $[t^a, t^b]=f^{abc}t^c$. 

The Lagrangian \eqref{Lagrangian_PCM} can be written in terms of the right and left currents, namely $\mathcal L=-{1\over \gamma^2} \Tr( j_\mu^L j^{\mu L})=- {1\over \gamma^2} \Tr( j_\mu^R j^{\mu R})$. 
The equations of motion following from \eqref{Lagrangian_PCM} are nothing but the conservation laws for the right and left currents:
\[
\label{eom_PCM}
\p^\mu j_\mu^L=\p^\mu j_\mu^R=0\,.
\]
Moreover, by construction the currents also satisfy the so-called Maurer-Cartan identities
\[
\label{PCM_flatness}
\p_\mu j_\nu^{(R,L)}-\p_\nu j_\mu^{(R,L)} + [j^{(R,L)}_\mu, j^{(R,L)}_\nu]=0\,.
\]
The equation \eqref{PCM_flatness} encodes all the information about the algebraic structure of the model. Also, $j_\mu^{(R,L)}$ can be seen as a two-dimensional gauge field. Then, when one introduces the covariant derivative $D_\mu^{(R,L)}=\p_\mu+[j_\mu^{(R,L)},~~]$, the identity \eqref{PCM_flatness} can be interpreted as a zero-curvature equation. The covariant derivative $D_\mu$ acts on the elements of the Lie algebra $\mathfrak{g}$. 
%
%

\paragraph{Local and non-local conserved charges in PCM.}

The PCM has two different sets of conserved charges: the local and the non-local ones. Both conserved quantities can be obtained from a unique generating functional, the \emph{monodromy matrix}. They correspond to an expansion of the monodromy matrix around different points%
\footnote{There is, indeed, another way of constructing such non-local charges by an iterative procedure, for more details we refer the reader to the original paper \cite{Brezin:1979am}.},
and I will discuss these aspects more extensively below.

First consider the following charges:
\be
\label{non_local_charges_PCM}
&& Q^a_{(0)}= \int_{-\infty}^\infty j_\tau^a (\sigma) d\sigma\,,
\nln
&& Q^a_{(1)}=\int_{-\infty}^\infty j_{\sigma}^a(\sigma) d\sigma -{1\over 2} f^{abc}\int_{-\infty}^\infty d\sigma j_\tau^b(\sigma) \int_{-\infty}^\sigma d\sigma' j_\tau^c(\sigma')\,.
\ee
The first one is local, i.e. it is an integral of local functions, and it is the global right and left symmetry of the model; while the second one is bi-local. The Poisson brackets between $Q^a_{(0)}$ and $Q^a_{(1)}$ generate a series of charges, $Q^a_{(n)}$, which are conserved and which are integrals of non-local functions. Therefore the set of charges generated by $Q^a_{(0)}$ and $Q^a_{(1)}$ are called non-local charges. The basic idea is that such charges show certain ``hidden'' symmetries of the two-dimensional model, not the ones directly seen by dynamical point-particles. 
The conservation laws for $Q^a_{(n)}$ follow directly from the equations of motion (\ref{eom_PCM}).
Note that since the charges $Q^a_{(n)}$ are non-local, they will not commute in general, and they will not be additive when acting on some generic multi-particle state. They are fundamental in order to understand the classical and quantum integrability of the model. In particular when it is possible to extend such charges to the quantum level, they generate a quantum group called Yangian, whose structure yields to the factorizability of the S-matrix.

Beside the charges $Q^a_{(n)}$ there are another type of conserved quantities, which are integrals of local functions of the fields. Such charges are additive on (asymptotic) multi-particle states and since they commute this puts severe constraints on the dynamics, as we will discuss in the section \ref{sec:Smatrix_fact}. The basic idea is that such local charges directly generalize the energy-momentum conservation law to higher spin. Indeed, consider the quantities $\Tr(j_\pm^{(R,L)} j_\pm^{(R,L)} )$, where we have rewritten the currents in the light-cone coordinates $x^{\pm}=\sigma\pm\tau$. From the equations of motion \eqref{eom_PCM} and the Maurer-Cartan identities \eqref{PCM_flatness} it follows that
\[
\label{PCM_local2}
\p_+ \Tr\left (j_-^{(R,L)} j_-^{(R,L)} \right )= \p_- \Tr\left (j_+^{(R,L)} j_+^{(R,L)} \right )=0\,.
\]
This is nothing but the conservation of the PCM energy-momentum tensor. Differentiating the action \eqref{Lagrangian_PCM} with respect to the two-dimensional (world-sheet) metric $g_{\mu\nu}$ one has
\[
T_{\mu\nu}= -{1\over 2\gm^2}\Tr\left( j_\mu j_\nu -\frac{1}{2} g_{\mu\nu} \left( j_\lm j^\lm\right)\right)\,,
\]
and in the light-cone coordinates it becomes $T_{\pm\pm}=  -{1\over 2\gm^2}\Tr\left( j_\pm j_\pm\right)$. In general, we can extend \eqref{PCM_local2} by considering a higher $m$ rank tensor, namely 
\[
\label{PCM_localm}
\p_+ \Tr\left ( (j_-^{(R,L)})^m \right )= \p_- \Tr\left (( j_+^{(R,L)})^m \right )=0\,.
\]
In particular, in order to satisfy the equation \eqref{PCM_localm}, any higher $m$-rank tensor should be associated with the invariant and completely symmetric Casimir tensor $C^{a_1... a_m} t^{a_1}... t^{a_m}$. Note that, for the case $m=2$, the invariant tensor is simply the trace of two generators, i.e. $C^{ab} \propto \dl^{ab}$ (multiplied by a constant numerical factor which depends on the particular normalization of the algebra). Then, the conservation laws \eqref{PCM_local2} and \eqref{PCM_localm} follow, apart from the equations of motion for the currents, also from the algebraic identities which involve the products of symmetric tensors $C^{a_1... a_m} $ and the antisymmetric structure constant $f^{abc}$. The corresponding charges are then
\[
\label{def_localcharge}
q^s_{\pm}= \int_{-\infty}^\infty d\sigma C^{a_1... a_m} j^{a_1}_\pm (\sigma)... j^{a_m}_\pm (\sigma)\,,
\]
where $s$ denotes the Lorentz spin, namely $s=m-1$. The currents in $q^s$ can be the right or left-invariant ones, they will give the same local conservation laws. 

\paragraph{The Lax pair in PCM.}

We have seen that we have currents which are conserved and which are flat, cf. equations \eqref{eom_PCM} and \eqref{PCM_flatness} respectively. At this point, we would like to construct a \emph{flat} linear combination of the currents $j$ themselves. This means that we consider a linear combination with arbitrary coefficients and demand that it should satisfy the equation \eqref{PCM_flatness}:
\[
\label{a_flat}
a_\mu= \alpha j_\mu+\beta \eps_{\mu\nu} j^\nu ~~~\text{such that} ~~ \p_\mu a_\nu-\p_\nu a_\mu +[a_\mu, a_\nu]=0\,.
\]
Since the mixed terms with $\alpha\beta$ are zero, and the terms with the product $\eps\eps$ gives a factor $-1$, the solution for the coefficients are obtained from the equation $\alpha^2-\alpha-\beta^2=0$, explicitly:
\[
\label{sol_flat_a}
\beta= \Half \sinh{\lm}\qquad 
\alpha= \Half \left(1\pm\cosh{\lm}\right)
\]
with $\lm\in\Reals$. This means that there is an entire family of solutions depending on a parameter $\lm$, the \emph{spectral parameter}.%
\footnote{The spectral parameter is usually complex in theories with Euclidean signature.}
The zero-curvature equation for the connection $a$ encodes all the dynamical informations, such as equations of motion and Maurer-Cartan identities. Note that in general $a$ is not conserved, namely it does not satisfy the equations of motion \eqref{eom_PCM}. 

We now explain why we want such connection $a$. The flatness condition for $a$ is associated with a two-dimensional differential system. In particular, for the generic group-valued function $U(\tau,\sigma)$, the compatibility condition for the differential equations 
\be
\label{Lax_representation}
{\p U\over \p\tau} = a_\tau (\lm) U\qquad 
 {\p U\over \p\sigma }= a_\sigma (\lm) U
\ee
gives ${\p^2 U\over \p\tau\p\sigma}={\p^2 U\over \p\sigma\p\tau}$, which corresponds to the zero-curvature equation for the connection $a$, \eqref{a_flat}. 
The system \eqref{Lax_representation} is also called the Lax representation, and for this reason, the two components of the connection $a$ are called the Lax pair. The system \eqref{Lax_representation} is integrable provided that $a$ is flat and the solution for $U$ is given by 
\[
\label{def_U}
U(\mathcal{C}, \,\lm)= {\rm P} \, e^{-\int_{\mathcal{C}} a}\,,
\] 
where $\rm{P}$ denotes the path-order prescription for the generators contained in $a$ and $\mathcal C$ is a path on the world-sheet $\Sigma$. For any initial data, or boundary condition $U(\tau_0,\sigma_0)$, the system \eqref{Lax_representation} has a unique solution given by the operator \eqref{def_U}. This Wilson line operator, which defines the parallel transport along the path $\mathcal C$ with the connection $a$, is called the {\it monodromy matrix}. %

The integrability of the system \eqref{Lax_representation} is guaranteed by the fact that the connection has a zero curvature \eqref{a_flat}, namely that the solution \eqref{def_U} is independent of path deformations. Let $s$ parameterize the path $\mathcal C$.
A small variation of the contour of integration, $\sigma^\mu(s)\rightarrow \sigma^\mu(s)+\dl \sigma^\mu(s)$, produces a variation on the Wilson loop operator according to \cite{Polyakov:1980ca}
\[
\label{def_var_U}
{\dl \over \dl \sigma^\mu(s)}U ={\rm P} \left ( \mathcal{F}_{\mu\nu} \frac{d \sigma^\nu}{ds} e^{\int_{\mathcal{C}} a(s)}\right)\,,
\]
where $\mathcal{F}_{\mu\nu}$ is the field strength for the connection $a$. It is clear that for a flat current, i.e. when $\mathcal{F}_{\mu\nu}=0$, such variation vanishes, namely the Wilson line operator is invariant under continuos path deformations if the connection is flat. 
This is a key point: From the fact that $U$ cannot be deformed, it follows that it might be the proper generating functional for the conserved charges. Considering paths $\mathcal C$ of constant time and looking at small deformations of the contours in the $\tau$ direction, then for a flat connection the Wilson line operator will be invariant under variations of these particular paths, namely under deformations in time. Explicitly
\[ 
\label{def_QdaU}
Q (\lm)= \lim_{\sigma\rightarrow\pm\infty}\, U(\mathcal{C}_0; \lm)\,=\, {\rm P} \, e^{-\int_{-\infty}^\infty a}|_{\tau_0}
\]
where it has been stressed that the contour $\mathcal{C}_0$ is over surfaces of constant time $\tau_0$ and that $\sigma\rightarrow\pm\infty$.%
\footnote{Closed strings require a closed loop and the trace in the definition of U. Moreover one needs to assume a proper behavior  for the currents at the boundary $\sigma\rightarrow\pm\infty$.}
Thus, summarizing, the conservation of the charges $Q (\lm,\tau_0)$ is guaranteed by the flatness of $a$ \eqref{a_flat}. One can easily differentiate $U$, and assuming that the currents fall down to zero at infinity and that $a$ is flat, one will get a vanishing time derivative for $Q (\lm,\tau_0)$. 

The non-local charges which we have discussed above can be obtained as a Taylor expansion around the zero value of the spectral parameter $\lm$. Around $\lm=0$ the expansion of the flat connection $a$ with the minus solution in \eqref{sol_flat_a} is
\[
a_\mu (\lm)\cong {\lm\over 2} \eps_{\mu\nu} j^\nu - {\lm^2\over 4} j_\mu+\Op (\lm^3)\,.
\]
Then defining 
\[
Q(\lm) \equiv 1+\sum_{n=1}^\infinity {(-1)^n\over n!}\lm^n Q^{(n-1)}\,,
\]
one has at the leading order in $\lm$ expanding the exponential in \eqref{def_QdaU}
\be
&& Q^{(0)}= \Half \int_{-\infinity}^\infinity d\sigma j_\tau (\sigma)\,,\nln
&& Q^{(1)}= {1\over 2} \int_{-\infinity}^\infinity d\sigma j_\sigma (\sigma) -{1\over 4} \int_{-\infinity}^\infinity d\sigma \int_{-\infinity}^\sigma d\sigma' [ j_\tau (\sigma) j_\tau (\sigma')]\,. 
\ee
Apart for an irrelevant numerical factor these charges are the same presented above in \eqref{non_local_charges_PCM}.

Some concrete examples of the PCM  are the models with group $G=\grSU(N)$ and the $\grp{O}(4) \sim \grSU(2)\times \grSU(2)$ model. Most relevant for us is the GS type IIB superstring in $\ads_5\times\sphere^5$ in the light-cone gauge with symmetry group $\grp{P}(\grp{SU}(2|2)\times \grSU(2|2))$. This model will be elaborated on in section \ref{chapter:NFS}.


\subsection{Coset model}
\label{sec:coset_model}

We now review some other very special two-dimensional $\sigma$-models, namely those defined on a coset space. The presentation closely follows the paper by Bena {\it et al.}~\cite{Bena:2003wd}. 

For a coset space, the map $g(\tau,\sigma)$ takes values in the quotient space $\grp{G}/\grp{H}$. $\grp{H}$ is a $\grp{G}$-subgroup, called \emph{isotropy group} or \emph{stabilizer} since it is required to leave invariant the G elements. The coset space $\grp{G}/\grp{H}$ corresponds to the identification 
\[
\label{coset_def}
g(\tau,\sigma)\cong g(\tau,\sigma) h(\tau,\sigma)\,, \qquad  h(\tau,\sigma)\in \grp{H}\,.
\]
In some sense we can say that we have ``half'' of the global symmetries compared with the PCM of the previous section \ref{sec:PCM}: what is now left is only the invariance under global left multiplication.
However, now the subgroup $\grp{H}$ plays the important role of gauge group, since each point in every orbit in the target-space is defined up to a \emph{local} transformation, i.e. a gauge transformation, which does not contain any further physical information. For this reason $g(\tau,\sigma)$ is the coset representative. 
Note that we could have used left-multiplication in \eqref{coset_def} to identify different $g$ and then the remaining global symmetry would have been the right one. The forthcoming arguments then run analogously, with some obvious exchange between the left and right sectors.  

It is possible to give a geometric construction for spaces such as $\C P^{n}= \grSU(n+1)/(U(1)\times \grSU(n))$, $\ads_n= \grSO(n-1,2)/ \grSO(n-1,1)$ and $\Sphere^n=\grSO(n+1)/\grSO(n)$. For example, consider the $n$-dimensional sphere $\Sphere^n$ embedded in $\R^{n+1}$. Fixing the north-pole $(0,0,\dots,1)$ we still can have all the rotations in the $n$ transverse directions, namely $\grSO(n)$, which leave the north pole fixed and do not change the points on the sphere $\Sphere^n$. 

As already seen in the previous section, we can introduce the one-forms
\[
\label{MCforms_coset}
J_\mu= J_\mu^a t^a= g^{-1} \p_\mu g \,.
\]
We follow the literature and use capital letters $J_\mu$ for the left-invariant currents and vice versa, small letters $j_\mu$ for the conjugated currents, since now the roles played by the two kinds of Maurer-Cartan forms are very different. 
Indeed, the group $\grp{G}$ acts on the coset representative as a left multiplication $g_0$, thus the currents $J_\mu$ transform according to
\[
J_\mu= g^{-1} \partial_\mu g \rightarrow  (g_0 g)^{-1}\partial_\mu (g_0 g)=  g^{-1} \partial_\mu g\,,
\]
since $g_0$ is constant. Thus the currents are left-invariant, which corresponds to the action of the global symmetry $\grp{G}$. What happens to the MC-forms when we consider the coset identification? This means that an element $g$ will be multiplied by an element of the subgroup $\grp{H}$, which now depends on the world-sheet coordinates $\sigma^\mu$. Replacing $g \rightarrow g h$ in $J$ we obtain the following transformation
\[
\label{gaugetransf_J}
J_\mu \rightarrow  h^{-1} g^{-1} \partial_\mu g h + h^{-1}\partial_\mu h \,. 
\]
The first term transforms covariantly under a local gauge $\grp{H}$ transformation, but not the second term. Considering the conjugate currents
\[
j_\mu= - g J_\mu g^{-1}= - \partial_\mu g g^{-1} \,,
\]
we see that they transform covariantly under global left-multiplication:
\be
&& g\rightarrow g_0\, g \qquad j_\mu \rightarrow g_0\, j_\mu \, g_0^{-1}\,.
\ee
For this reason it is important to distinguish between the left and right sectors, since now the two types of currents are not both conserved anymore as it was in the PCM case \eqref{eom_PCM}, and they transform in different ways under gauge transformations.
Obviously, we could have started defining the coset space by a left-multiplication and inverted the role between ``small'' and ``capital'' currents.

The algebra $\mathfrak{g}$ is split in two sectors with respect to the $\grp{H}$-action: $\mathfrak{g}=\mathfrak{h}\oplus\mathfrak{k}$, where $\mathfrak{k}\equiv \mathfrak{g/h}$ is the orthogonal complement in $\mathfrak{g}$ with respect to $\mathfrak h$. As a consequence, also the left-invariant currents undergo the same split, namely
\[
J= K + H\,,
\]
with obvious notation for the various terms. Thus $H$ is really a connection, a gauge field, while $K$ represents the part of the one-form which transforms covariantly under gauge transformations, i.e. $h^{-1} g^{-1}\p g h$ in \eqref{gaugetransf_J}. Notice that the current
\[
k= - g K g^{-1}
\]
is gauge invariant. 
Finally, the current $j_\mu$ does not have a defined grading, since the rotation with $g$ and $g^{-1}$ mixes the two sectors $\mathfrak{h}$ and $\mathfrak{g/h}$, however one keeps the notation $h$ and $k$ to denote $g H g^{-1}$ and $g K g^{-1}$ respectively. 

The Lagrangian is as for the PCM \eqref{Lagrangian_PCM}
\[
\label{Lagrangian_coset}
\mathcal L= {1\over \gamma^2} \rm{Tr} \left(\p_\mu g^{-1}\p^\mu g\right)
                  = {1\over \gamma^2} \rm{Tr} \left(J_\mu J^\mu \right)
                  = {1\over \gamma^2} \rm{Tr} \left(j_\mu j^\mu \right)\,. 
\]
Since the two tangent spaces $\mathfrak{h}$ and $\mathfrak{k}$ are orthogonal, this leads to the following expression for the $\mathcal L$
\[
\label{L_explicit}
\mathcal L  = {1\over \gamma^2} \rm{Tr} \left(H_\mu H^\mu+ K_\mu K^\mu \right)\,.
\]
The term $\Tr (A_{|\grp{G/H}} B_{|\grp{H}})$ vanishes, as it should, since the trace is a bilinear invariant tensor that respects the structure of the space:
\[
\label{coset_algebra}
[ \mathfrak{k}, \mathfrak{h}] \subseteq \mathfrak{k}
\qquad
[ \mathfrak{h},  \mathfrak{h}] \subseteq  \mathfrak{h}\,.
\]
Indeed, the grading $\mathfrak{g}=\mathfrak{h}\oplus\mathfrak{k}$ means that the generators of one set span the tangent space labelled by $\mathfrak{k}$ and the other complementary set generates $\mathfrak{h}$, and there is no generator left. Thus, the trace between any two elements spanning orthogonal spaces vanishes, since the trace is nothing but a scalar product in this tangent space. 

Since the action \eqref{L_explicit} is gauge invariant, it is clear that one can integrate out the gauge field $H$ so that the only remaining contribution to the currents in $\grp{G/H}$ is
\[
\label{L_coset}
\mathcal L_{\grp{G/H}} = {1\over \gamma^2} \rm{Tr} \left( K_\mu K^\mu \right)\,,
\]
which is again manifestly gauge invariant (recall that $K$ is covariant under local H transformations) and it is naturally defined on the quotient space $\grp{G/H}$. 

Again it follows from the equations of motion that the left-invariant currents are conserved;
they satisfy the usual identity $\p_\mu J_\nu-\p_\nu J_\mu + [J_\mu, J_\nu]=0$. As for the PCM, we can construct the flat linear combination $a$. However, in the coset space we need a further requirement: the space should be symmetric, namely beyond the standard algebraic structure for a coset space \eqref{coset_algebra}, we need also that
\[
\label{def_symmetric_space}
[ \mathfrak{k}, \mathfrak{k} ]\subseteq \mathfrak{h}\,.
\]
This is indeed a necessary and sufficient condition for a bosonic coset space to have a Lax representation~\cite{Eichenherr:1979ci, Eichenherr:1981sk}. Note that other models can still have a Lax representation. The $\ads_5\times \Sphere^5$ superstring case is eloquent in this sense: the bosonic sub-sector, which is strictly the coset $\ads_5\times \Sphere^5$, is a symmetric space. However, its full supersymmetric generalization is not. The corresponding superstring action is not simply $S_{\grp{G/H}}$ but there is a further contribution of the Wess-Zumino-Witten type (WZW)~\cite{Metsaev:1998it} which allows a Lax pair reformulation \cite{Bena:2003wd}. 

In order to construct a flat connection let us consider the projections of the Maurer-Cartan identities over $\mathfrak h$ and $\mathfrak k$. Then $\p_\mu J_\nu-\p_\nu J_\mu +[J_\mu, J_\nu]=0$ gives
\be
&& \p_\mu H_\nu-\p_\nu H_\mu + \left [H_\mu, H_\nu\right] +\left [K_\mu, K_\nu\right] =0
\nln
&& \p_\mu K_\nu-\p_\nu K_\mu + \left [H_\mu, K_\nu\right] +\left [K_\mu, H_\nu\right] =0\,.
\ee
Without the condition \eqref{def_symmetric_space} the commutator $\left [K_\mu, K_\nu\right] $ would have contributed to both the differentials, $dH$ and $dK$
\footnote{As explained in~\cite{Bena:2003wd}, if $\mathfrak{k}$ is a sub-algebra, then the commutator $\left [K_\mu, K_\nu\right] $ sits only in the $\p k$ terms.}.
Using the following identity
\[
\p_\mu l_\nu -\p_\nu l_\mu= - g (\p_\mu L_\nu- \p_\nu L_\mu ) g^{-1} - [l_\mu, j_\nu] - [j_\mu, l_\nu]
\]
valid for any current $L$ and its conjugate $l= -g L g^{-1}$,
one has
\be
 \p_\mu k_\nu -\p_\nu k_\mu + 2[k_\mu, k_\nu]=0\,. 
\ee
In this way, the flat connection corresponding to $a$ in the PCM is just the gauge-invariant one-form $2 k_\mu$, since it is conserved and it is also flat. Then the construction for the monodromy matrix follows exactly the PCM model in section \ref{sec:PCM}.


\subsection{The magic of (1+1)-dimensional theories}
\label{sec:Smatrix_fact}

Something special happens for two-dimensional field theories which have an infinite amount of conserved {\it higher} charges. This is mainly due to the fact that there is only one spatial dimension, and that the charges can be used to reshuffle the amplitudes in scattering processes. 
The role of integrability in constraining the dynamics of the theory was discovered in the late 1970s and early 1980s by Zamolodchikov and Zamolodchikov~\cite{Zamolodchikov:1978xm}, L\"{u}scher~\cite{Luscher:1977uq}, Kulish~\cite{Kulish:1975ba}, Parke~\cite{Parke:1980ki} and by Shankar and Witten~\cite{Shankar:1977cm}.
In order to illustrate this point, we start with a two-dimensional theory with an infinite set of charges, which are integrals of local functions and which are diagonal in one-particle states. The charges are of the kind illustrated in Section \ref{sec:PCM}. 

Let us first introduce some notation and define what we mean by scattering.
We denote the particle state with the wave-function $|A (\theta)\rangle $, where $\theta$ is the \emph{rapidity}, which is defined for a massive field theory%
\footnote{The rapidity can also be introduced for massless theory, but we are indeed interested in massive field theories. }
as
\[
p_{+a}= 2 m_a e^{\theta_a}
\qquad 
 p_{-a}= 2 m_a e^{-\theta_a}\,.
\]
$p_+$ and $p_-$ are the momenta in the light-cone coordinates%
\footnote{The light-cone momenta are defined according to $p_{\pm}=\Half (p_0\pm p_1)$.}. 
Suppose the asymptotic \emph{in}-state is composed of $m$ particles. We can then write
\[
|in\rangle = | A_{a_1}(\theta_1) \dots A_{a_m}(\theta_m) \rangle \,.
\]
The hypothesis is that the particles are described by wave packets with an approximate position for each momentum (for each rapidity) and that all the interactions are short-ranged (since we are discussing massive field theories) such that the $m$-particle state can be approximated by a sum of $m$ single-particle states (the wave packets are far enough apart to be considered single particle states). An asymptotic in-state means that sufficiently backwards in time the $m$ particles do not interact. This imposes a certain ordering in the state, since the particle which is traveling faster must be on the left in order to avoid crossing with all other particles, vice versa the slowest particle should be the first on the right, i.e. 
\[
\label{in_order}
\theta_1 > \theta_2>\dots > \theta_m\qquad \text{for \emph{in} states}\,. 
\]
This also implies the reversed ordering for the \emph{out}-state. Consider as well the asymptotic state containing $n$ particles, namely $n$ independent wave packets 
\[
|out \rangle = | A_{b_1}(\theta_1) \dots A_{b_n}(\theta_n) \rangle \,.
\]
Now the particles should travel without interacting for future times and the slowest particle should be on the left and the particle moving fastest on the right, namely in terms of rapidities 
\[
\label{out_order}
\theta_1< \theta_2 <\dots <\theta_n \qquad \text{for \emph{out} states}\,. 
\]
The letters $a_1,\dots a_m$ and $b_1\dots b_n$ denote any possible set of quantum numbers characterizing the particles. 

The \emph{S-matrix} or \emph{scattering matrix} is by definition the mapping relating the {\it in} and {\it out}-states, namely it is defined by 
\[
| A_{a_1}(\theta_1) \dots A_{a_m}(\theta_m) \rangle = 
S_{a_1\dots a_m}^{b_1\dots b_n} (\theta_1,\dots \theta_m; \theta'_1\dots \theta'_n)
| A_{b_1}(\theta'_1) \dots A_{b_m}(\theta'_n) \rangle \,,
\]
where it is intended to sum over the indices $b_1\dots b_n$, and over the out-going rapidities, which are ordered as explained above. We can also introduce the Faddeev-Zamolodchikov (ZF) notation~\cite{Zamolodchikov:1978xm, Faddeev:1980zy} and write each asymptotic state as a sequence of $A_a (\theta)$'s, remembering that they do not commute and they are ordered in increasing or decreasing rapidity for in or out-state respectively, according to \eqref{in_order} and \eqref{out_order}. Then one can write the state and the S-matrix element in the following way:
\be
&& A_{a_1}(\theta_1) \dots A_{a_m}(\theta_m) 
\nln
&& A_{a_1}(\theta_1) \dots A_{a_m}(\theta_m)  = 
S_{a_1\dots a_m}^{b_1\dots b_n} (\theta_1,\dots \theta_m; \theta'_1\dots \theta'_n)
A_{b_1}(\theta'_1) \dots A_{b_n}(\theta'_n) \,. \nln
\ee
The S-matrix is a unitary operator, namely it should respect the condition (in operator notation)
\[
S\left(\theta_1,\theta_2\right) \, S^\dagger \left( \theta_2,\theta_1\right)= \mathbb{1}\,.
\]
In general one also requires that the S-matrix is invariant under parity transformation (in our case the discrete symmetry which flips the spatial coordinate $\sigma$ to $-\sigma$), time reversal and charge conjugation. In relativistic quantum field theories the S-matrix turns out to be invariant also under the \emph{crossing symmetry}, namely the transformation which exchanges one in-coming particle of momentum $p$ with an out-going anti-particle of momentum $-p$, cf. discussion in section \ref{sec:Smatrix}.

\paragraph{Selection rules.}

Let us now come back to the local charges $q_{\pm }^s$. Since they commute with the momentum operator, for a single particle state we have
\[
\label{q_oneparticle}
q_{\pm}^s |A_{a} (\theta)\rangle= \om_{a}^{(s)} e^{\pm s \theta}  |A_{a} (\theta)\rangle\,,
\]
where $ \om_{a}^{(s)} $ are the corresponding eigenvalues. For $s=0$ and $s=1$ we can think about them as the energy and the momentum. 
However, we are assuming that there exists an infinite number of higher rank local conserved charges, namely we are assuming $s> 1$. 
Suppose now we act with the local conserved charges on the in and out-states. Since the wave packets are well separated and the charges are integrals of local functions, their action on such states is additive, namely
\be
\label{q_instate}
&& q_s  | A_{a_1}(\theta_1) \dots A_{a_m}(\theta_m) \rangle= \nln
&& =\left(  \om_{a_1}^{(s)} e^{s \theta_1}+\dots +\om_{a_m}^{(s)} e^{s \theta_m} \right )  | A_{a_1}(\theta_1) \dots A_{a_m}(\theta_m) \rangle\,.
\ee
Again, just to understand, for $s=0$ the above relation is the energy conservation condition and for $s=1$ the momentum conservation law. 
Obviously we can write the expression above \eqref{q_instate} also for out-going states:
\[
\label{q_outstate}
q_s  | A_{b_1}(\theta'_1) \dots A_{b_m}(\theta'_m) \rangle= 
\left(  \om_{b_1}^{(s)} e^{s \theta'_1}+\dots +\om_{b_m}^{(s)} e^{s \theta'_m} \right )  | A_{b_1}(\theta'_1) \dots A_{b_m}(\theta'_m) \rangle\,.
\]
The charges are conserved during the entire scattering process and they are diagonalized by asymptotic multi-particle states as stated above \eqref{q_instate} and \eqref{q_outstate}. Then for any $m\rightarrow n$ scattering amplitude it must be true that 
\[
\label{s_system}
 \om_{a_1}^{(s)} e^{s \theta_1}+\dots +\om_{a_m}^{(s)} e^{s \theta_m}=  \om_{b_1}^{(s)} e^{s \theta'_1}+\dots +\om_{b_n}^{(s)} e^{s \theta'_n}
 \]
 for all the possible infinite values of $s$. Thus there are $s$ such equations, with $s$ taking infinitely many values. Hence, the only solution for generic values of the in-coming momenta is 
 \[
 \label{sol_inout}
 n=m \qquad  \om_{a_i}^{(s)} =  \om_{b_i}^{(s)}  \qquad \theta_i=\theta'_i \,,
 \]
with $i=1,\dots m$. 
The consequences of the solutions \eqref{sol_inout} are severe for the dynamics of the system. 
\begin{itemize}
\item 
Since $n$ must be equal to $m$ this implies that there cannot be processes where the number of particles changes, namely the number of particles is conserved during the scattering and there cannot be particle production. 
\item The set of in-coming momenta, $\{ p_i\}$ must be equal to the set of out-going momenta $\{ p'_i\}$, or in terms of rapidities $\{\theta_i\}=\{ \theta'_i\}$.
\end{itemize}
However, this does \emph{not} imply that the sets of quantum numbers before and after the scattering $\{a_i\}$ and $\{b_i\}$ should be the same. They can have different values, namely scatterings which lead to changing flavor are still allowed. There is some subtlety, in the sense that one might find solutions to the equations \eqref{s_system} for specific values of the in-coming momenta and for $n\neq m$. However these values turn out to not be physical \cite{Dorey:1996gd}. 
The scatterings which are possible and consistent with the infinite set of charges are the elastic processes.

\paragraph{S-matrix factorization.}

\begin{figure}%
\begin{center}%
\includegraphics[scale=0.3]{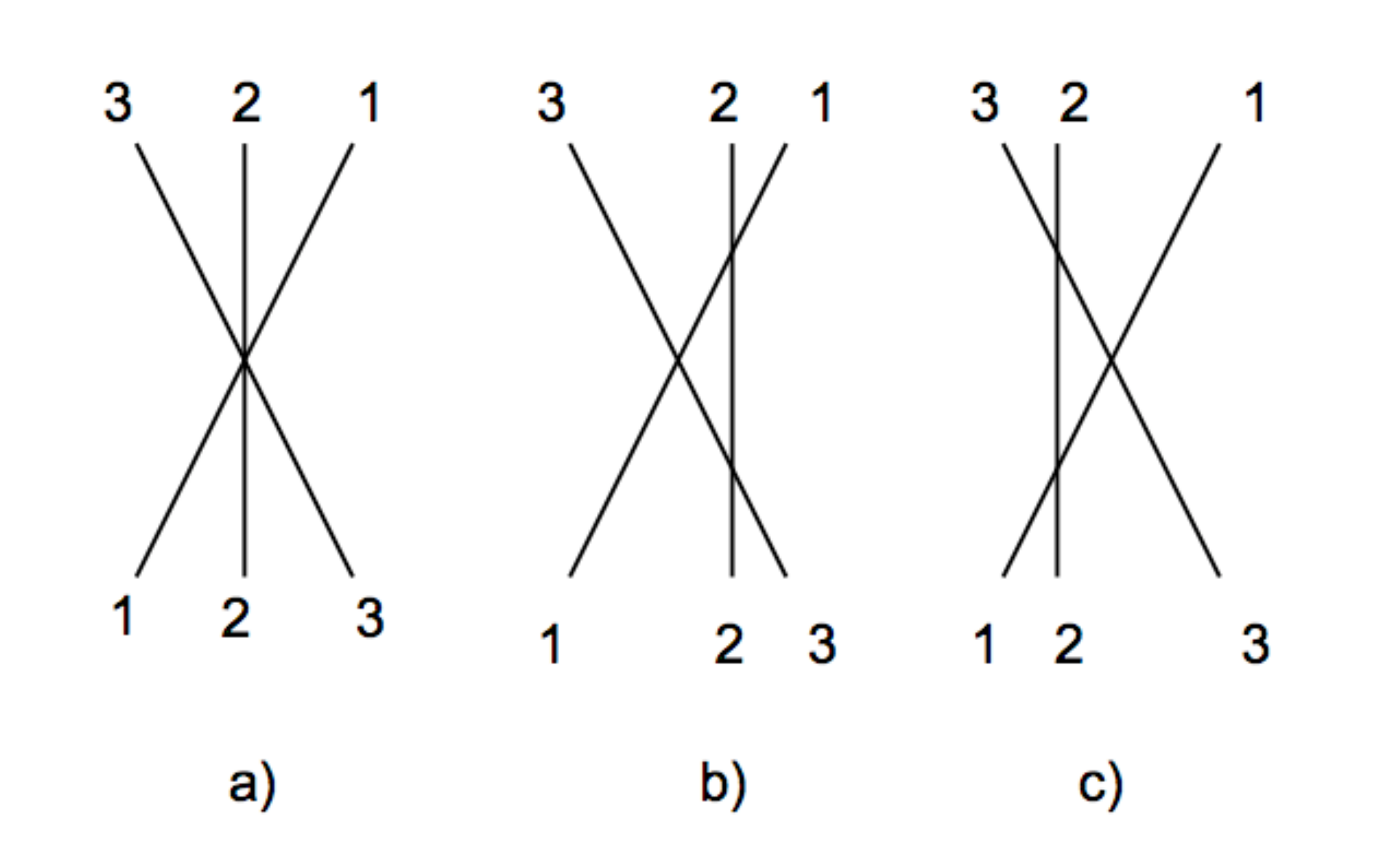}%
\end{center}%
\caption[]{Tree level diagrams for three-body scattering $3\rightarrow 3$.}%
\label{fig:tree_level_3}%
\end{figure}

\begin{figure}%
	\begin{center}%
		\includegraphics[scale=0.6]{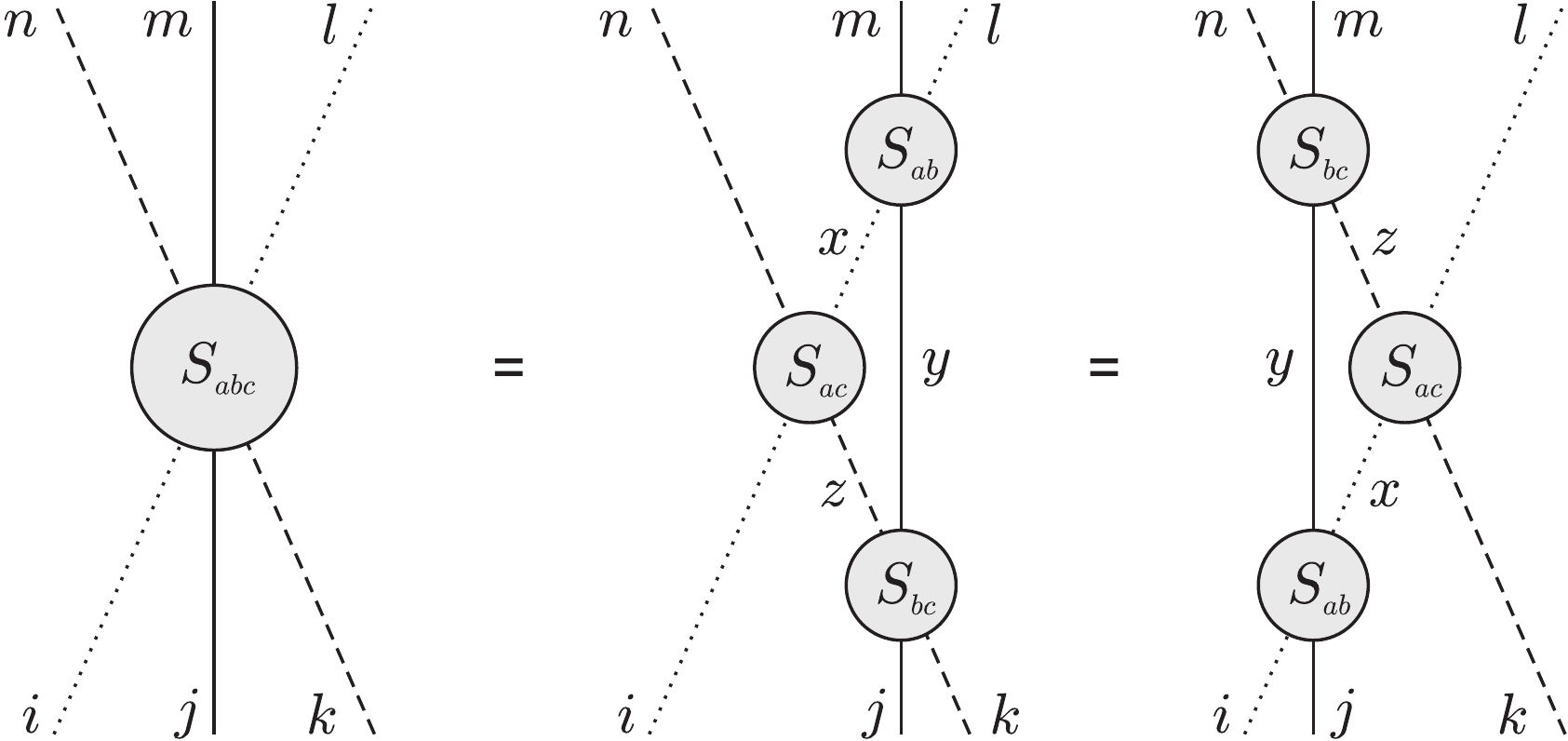}%
	\end{center}%
	\caption[vvv]{Factorization of the three-body S-matrix and Yang-Baxter equation.}%
	\label{fig:factorization}%
\end{figure}

There is still another dynamical constraint which makes the two-dimensional integrability a really powerful tool: the factorizability of the S-matrix. Each wave packet is localized, and we can model it by a gaussian distribution around the position $x_i$ with momentum $p_i$. Acting on such a state with an operator of the type $e^{- \imath c P^s}$ shifts the phase factor by a function depending on the momentum%
\footnote{The argument that we are following is from \cite{Dorey:1996gd}, rigorously we should here use the operator $e^{- i c q_s}$ as it has been done in \cite{Parke:1980ki}. However, since it does not spoil the effectiveness of the argument and it makes a bit ``digestive'' from a technical point of view, we adopt the same technique as in Dorey's paper \cite{Dorey:1996gd}.}:
in particular the position is shifted by $\dl x_i= c s p_i^{s-1}$. 
When the operator acts on an $m$-particle state of the type seen before, namely $m$ times a single particle state, then each localized wave packet is shifted by a different quantity since such shift depends on the wave packet momentum. 
Then, since the asymptotic states are eigenstates for the higher conserved charges and since such charges commute with the S-matrix, we can use them in order to reshuffle the in and out-states. Explicitly one can write 
\[
\label{S&P_commute}
\langle out |S| in\rangle= \langle out | e^{i c P^s} S e^{-i c P^s}| in\rangle\,. 
\]
We can rearrange the wave-packets and make their phase factor change according to their momenta. In order to illustrate the ideas, let us consider the $3\rightarrow 3$ scattering. At tree level we can have three types of diagrams, cf. figure \ref{fig:tree_level_3}.
The first graph $(a)$ visualizes the scattering of three particles at the same point, while the remaining two diagrams, $(b)$ and $(c)$, represent a series of three two-body scatterings. Namely, in the diagram $(b)$, first the particles 2 and 3 meet, collide and \emph{then} the particle 3 collides further with 1 and \emph{then} the particle 2 with 1. Of course we can start with the initial scattering between 1 and 2 and proceed analogously, as in figure \ref{fig:tree_level_3} $(c)$. Now, we use the operator $e^{- i c P^s}$ in order to shift the particle positions as in \eqref{S&P_commute}. However, everything must respect the macro-causality principle, namely it cannot happen that the particle 1 goes out before that also the particle 3 participates in the scattering. Otherwise the corresponding amplitude would just vanish%
\footnote{This argument can also be used to show that processes of the type $2\rightarrow n$ are zero in integrable two-dimensional field theory, since it should always be true that $t_{12}\le t_{23}$, where 1 and 2 are the in-coming particles and $t_{12}$ is the time that occurs for the scattering between 1 and 2, while 3 is the fastest particle among the out-going ones.}.
Namely, nothing can happen between the slowest in-coming particle and the fastest out-going particle before that all the in-coming particles have collided. Now the point is that one can use the higher charges to rearrange the phase shift for the multi particle state, but indeed the diagrams in figure \ref{fig:tree_level_3} only differ by a phase factor. This means that we can use the operators $P^s$ in order to move the lines 1, 2, 3 in Fig. \ref{fig:tree_level_3}  $(a)$, in order to get \emph{any of the two} other graphs in Fig. \ref{fig:tree_level_3}. Hence all the graphs in figure \ref{fig:tree_level_3} are equal. 
This implies that the three-body S-matrix (Fig. \ref{fig:tree_level_3} a) is equal to a sequence of two-body S-matrices (Fig. \ref{fig:tree_level_3}  $(b)$ and $(c)$). This is the meaning of the first equality in Fig. \ref{fig:factorization}, where what we have discussed for the tree-level is extended to generic $n$-loop order. The second equality in figure \ref{fig:factorization} represents the Yang-Baxter equations. They are really non trivial equations, since they fix the flow of indices that we can have in the S-matrix elements.
This is something special which can happen in two dimensions. Indeed, we are using the higher charges to reshuffle the in-coming particle positions. Hence, if their rapidities differ, they will still meet at some point in space. This is not true for the four-dimensional case, where there are still two dimensions where the in-coming particles can completely avoid the scattering. This is the main reason why an integrable theory in 4 dimensions only has a trivial S-matrix, which is stated in the \emph{Coleman-Mandula theorem} \cite{Coleman:1967ad}. In one spatial dimension the particles necessary will meet at some point: They run in the same line there is no way to go out%
\footnote{Parke has proved that the existence of only two higher conserved charges $q_{\pm s}$ with $s>1$ is sufficient for the arguments presented above \cite{Parke:1980ki}.}.

Let us pause here and summarize the previous paragraph. In any (1+1)-dimensional theory with infinitely many local conserved charges, any $n\rightarrow n$ process can actually be known since the corresponding S-matrix element is given by a sequence of ${n (n-1)\over 2}$ two-body S-matrix elements. In many well-understood theories even if the 2-body S-matrix is computed, it is hopeless to compute the three-particle S-matrix. But now we are saying that we do not need it. We can compute any particle number scattering and the corresponding amplitude will be a product of $2\rightarrow 2$ scattering amplitudes. 
Thus, any scattering process involving more than two particles is a sequence of 2 by 2 collisions, which are all elastic and before and after any collision the particles keep on traveling freely. 

Until now we have only discussed the local conserved charges since the arguments in order to run need to use the fact that these objects are additive on multi-particle states. However, in \cite{Iagolnitzer:1977sw} Iagolnitzer gave a more general proof for the S-matrix factorization and for the selection rules. The same is done in L\"{u}scher's paper \cite{Luscher:1977uq} where he proved the relation between non-local charges and S-matrix factorization for the $\grp{O}(n)$ sigma model. For simplicity and for pedagogical reasons we have chosen to use the local charges to simpler visualize the arguments.


\paragraph{Remarks on the $\ads_5\times \Sphere^5$ string world-sheet S-matrix.} 

\begin{figure}%
	\begin{center}%
		\includegraphics[scale=0.5]{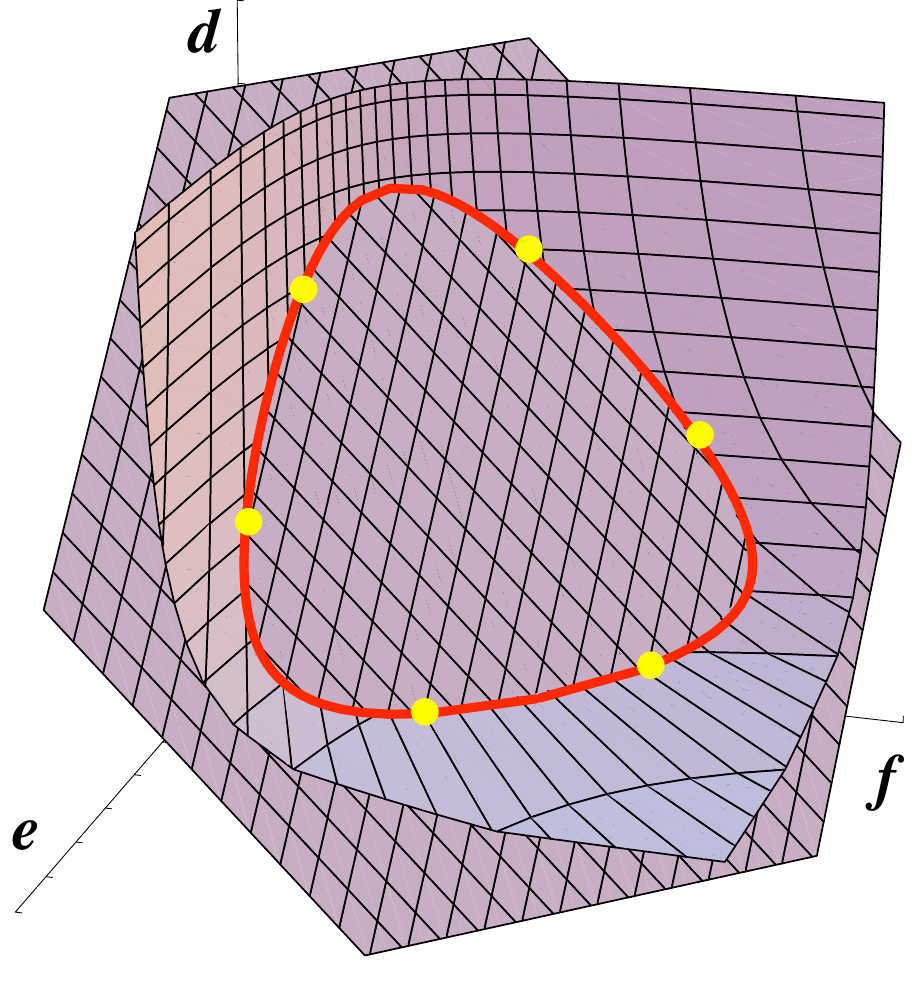}%
	\end{center}%
	\caption[vvv]{Three particle phase space.}%
	\label{fig:phasespace}%
\end{figure}

From the discussion above it is clear that we can use the factorization of the S-matrix and the selection rules (and the Yang-Baxter equations) as a definition for a two-dimensional integrable field theory. It is often really difficult to explicitly construct the (non-local and local) charges and usually it is more useful to know the S-matrix elements.  
This has been studied in~\cite{Puletti:2007hq}, where we have explicitly verified the factorization of the one-loop S-matrix for the near-flat-space limit of the type IIB superstring on $\ads_5\times \Sphere^5$. This is equivalent to state the integrability of the model at leading order in perturbation theory. However, this will be explained in more detail in section \ref{chapter:NFS}. Here we only want to stress once more that these dynamical constraints severely restrict the motion in the phase space. As example consider the $3\rightarrow 3$ process. Any scattering amplitude must respect the energy and the momentum conservation laws. In the light-cone coordinates one has that $p_- p_+= 4 m^2$. Then $p_{\pm}$ can be parameterized as $p_+=2 m a$ and $p_-=2m/a$ and the energy-momentum conservation laws become
\be
\label{enmom_conserv}
&& {1\over a}+{1\over b}+{1\over c}= {1\over d} +{1\over e}+ {1\over f} 
\nln
&& a+b+c= d+e+f
\ee
where the set $(a,b,c)$ is for the in-coming momenta, which are fixed (it is the external input which we give when we start to run our collision), while $(d,e,f)$ is the set of out-going momenta, which are constrained to respect the above equations \eqref{enmom_conserv}. The equations in \eqref{enmom_conserv} describe two surfaces. Without any further conservation law the out-going particles could lie in any point along the curve described by the intersection of the two equations. However, since we have a higher charge and we can impose another equation, there are only {\it six} valid points in all the phase space! These points correspond to the permutations given by the equation $\{ a,b,c\}=\{ d, e, f\}$, see Fig.\ref{fig:phasespace}.
This of course means that we have completely solved the motion. If we have a $4\rightarrow 4$ scattering then we need a fourth higher charge to fix univocally the points in the phase space, and so on. This is the concrete way how the charges manifest themselves. How to get the extra equation, namely how the higher charges actually operate on the phase space, will be discussed in section \ref{chapter:NFS}. There we also explain why we want to show the quantum integrability of the AdS superstring.


\subsection{Quantum charges in PCM and Coset model}
\label{sec:quantum_int}

Until now the discussion has only been at the classical level. Can we generalize the arguments above to the corresponding quantum field theory in a straightforward way? This question is far from trivial: numerous works in the past years ('70s-'80s) have been devoted to understand when integrability survives at the quantum level. However, also the answer is far from being trivial: for the $\grp{O}(n)$ model all the integrability properties survive after quantization~\cite{Luscher:1977uq, Goldschmidt:1980wq, Evans:2004ur}, which is not the case for the $\C \rm{P}^n$ model~\cite{Abdalla:1980jt}. Can we say why? Can we say where and how the troubles are originated? Can we learn something useful for the type IIB string theory? In this section we will try to partly answer these questions.

\paragraph{Quantum non-local charges.}

Going back to the definition of the non-local charges \eqref{non_local_charges_PCM}, one would like to implement such definition at the quantum level. The first trouble which one needs to face is the fact that the currents, and all fields in general, now are promoted to operators. The first term in \eqref{non_local_charges_PCM} now contains a product of two operators. When the two points where the operators are sitting at get closer and closer the currents can interact and give rise to singularities. In quantum field theory any product of operators is in general not-well defined. Also, the second term in \eqref{non_local_charges_PCM} can get renormalized and in general there will be some field renormalization coefficient which can be divergent. 

In order to have a reliable charge definition, it is necessary to slightly modify the expression in \eqref{non_local_charges_PCM}~\cite{Luscher:1977uq}:
\[
\label{bilocal_charge_quantum}
Q^a_{(1)}= Z \int_{-\infty}^\infty j_{\sigma}^a(\sigma) d\sigma -{1\over 2} f^{abc}\int_{-\infty}^\infty d\sigma j_\tau^b(\sigma) \int_{-\infty}^\sigma d\sigma' j_\tau^c(\sigma')\,.
\]

The second step is to compute the short-distance expansion for the current product in \eqref{bilocal_charge_quantum} and see if UV-dangerous terms can come out. This means to compute the operator product expansion (OPE) for the currents:
\[
\label{OPE_general}
j^a_\mu (x-\eps) j^b_\nu (x+\eps) \sim \sum_k C_{\mu\nu}^k(\eps) \mathcal O_k^{ab}(x)\,,
\qquad \eps\rightarrow 0
\]
where the sum $k$ denotes the sum over a basis of operators $\mathcal O_k^{ab}$. The operators $\mathcal O^{ab}_k$ do not depend on the short-distance parameter $\eps$, while the coefficients in the expansion $C_{\mu\nu}^k$ are functions of the coordinates, and thus of $\eps$. 
The problematic terms are linearly (i.e. ${1\over \eps}$) and logarithmically divergent in $\eps$. For example, for the PCM by dimensional analysis and since the currents have conformal dimension 1, we can expect an expansion of the type
\[
j^a_\mu (x-\eps) j^b_\nu (x+\eps) \sim C_{\mu\nu}^{\lm, abc}(\eps) j^c_\lm (x) + D_{\mu\nu}^{\lm\rho,abc}(\eps)  \p_\lm j^c_\rho (x)+\dots\,,
\]
where $C_{\mu\nu}^\lm(\eps)$ behaves as ${1\over \eps}$, just by dimensional analysis. This gives rise to possible logarithmic terms once one integrates. 

For the $\grp{O}(n)$ $\sigma$ model L\"{u}scher showed that the quantum charges are well-defined, they are conserved quantum mechanically and they force the S-matrix to factorize \cite{Luscher:1977uq}. The  same is not true for the $\C P^n$ model, which was investigated by Abdalla {\it et al.} in \cite{Abdalla:1980jt}. The $\C P^n$ model is classically integrable, however at quantum level an anomaly appears in the conservation law for the quantum non-local charges. As before, one needs to study the short-distance expansion for the currents \eqref{OPE_general} and then plug back the OPE in the quantum non-local charge \eqref{bilocal_charge_quantum}. The term responsible for the anomaly in the $\C P^n$ case is the field strength of the currents, namely a dimension two operators, whose corresponding coefficient in \eqref{OPE_general} contains logarithmically and linearly divergent terms. (Notice that the supersymmetric $\C P^n$ is quantum integrable \cite{Abdalla:1981yt}.) 

Can we give some kind of rules, about when or whether we could expect an anomaly in the charge conservation laws? For symmetric coset models of the type discussed in section \ref{sec:coset_model} this issue has been addressed in \cite{Abdalla:1982yd}. If one would like to summarize the results of the paper, one could say that the breaking of integrability at quantum level is related to $\grU(1)$ factor in the denominator of the quotient space, a fact which is confirmed by the $\C \rm{P}^n$ example, where the corresponding field strength gives rise to the anomaly. In some sense in the $\grp{O}(n)$ model there is not a great variety of operators $\mathcal O_k^{ab}$ of dimension 1 and 2 with the proper symmetries required by the model itself in order to be a candidate for the anomaly.%
\footnote{For a more detailed and complete explanation one should say that for Riemannian symmetric coset space the anomaly is forbidden when the sub-algebra $\mathfrak{h}$ is simple, and vice versa it is originated when the sub-algebra contains non-trivial ideals. Roughly speaking we can say that the decomposition of the sub-algebra $\mathfrak{h}$ corresponds to the possible operators $\mathcal O_k^{ab}$ which are the basis in the current OPE \eqref{OPE_general}. }
%


\paragraph{Remarks on the $\ads_5\times \Sphere^5$ superstring case.}

From all this one can understand why it is not so trivial to investigate the quantum integrability for two-dimensional $\sigma$ model, as for example the superstring world-sheet theory. Recall that the super-coset $\ads_5\times \Sphere^5$ is not a symmetric space, thus we cannot extend directly the analysis of \cite{Abdalla:1982yd}. However we can learn much from the $\C \rm{P}^n$ case and with this example in mind we have started to investigate the quantum pure spinor superstring in $\ads_5\times \Sphere^5$ in the papers~\cite{Puletti:2006vb} and~\cite{Puletti:2008ym}. In particular recall the expression for the variation of the monodromy matrix \eqref{def_var_U}, the integrability of the model is strictly related to the tensor $\mathcal F_{\mu\nu}$, cf. section \ref{chapter:PSformalism}.


\section{Green-Schwarz-Metsaev-Tseytlin superstring}

\label{chapter:GS_formalism}


The section is mainly based on the textbooks by \cite{Green:1987sp, Becker:2007zj} and also on the original papers by Green and Schwarz \cite{Green:1983sg, Green:1983wt} for the first part. For the second part I will mainly refer to the work by Metsaev and Tseytlin \cite{Metsaev:1998it} for the super-coset construction of the action, to the paper by Bena, Polchinski and Roiban \cite{Bena:2003wd} for the classical integrability of the GSMT action and finally to the reviews written by Zarembo~\cite{Zarembo:2004hp} and by Arutyunov and Frolov~\cite{Arutyunov:2009ga}.


\subsection{Green-Schwarz action in flat space}
\label{sec:GSflataction}

In the Green-Schwarz (GS) approach the target space supersymmetries are manifest and in some sense the superspace coordinates are treated more symmetrically with respect to the Ramond-Neveu-Schwarz (RNS) formalism.%
\footnote{The RNS formalism is another formulation to describe supersymmetric strings. In this case the supersymmetries are implemented into the theory by means of fields which are spinors on the world-sheet and vectors on the target space. However this approach is not suitable for describing superstrings supported by Ramond-Ramond fluxes, as it is our favorite $\ads_5\times \sphere^5$ superstring.}
In string theory, the embedding coordinates $X^a(\tau,\sigma)$ map the world-sheet $\Sigma$, parameterized by $(\tau, \sigma)$, into the target space. Now the same concept is generalized to the ``fermionic embedding coordinates'' $\theta^I(\tau,\sigma)$. These are spinors on the target space and scalars from a world-sheet point of view.

The GS superstring action in a flat background~\cite{Green:1983wt} is
\be
\label{GS_action_flat}
 && S_{GS, flat} =S_{\rm kin} + S_{\rm WZW} \nln
&&=-{1\over 4\pi\alpha'} \int d^2\sigma \sqrt{-h} h^{\mu\nu}
                                              \big (\partial_\mu X^{a}- i\bar\theta^I \Gamma^{a}\partial_\mu \theta^I\big)
                                              \big(\partial_\nu X_{ a}- i\bar\theta^J \Gamma_{ a}\partial_\nu \theta^J \big)\nln
&&+  {1\over 2\pi\alpha'} \int d^2\sigma  \epsilon^{\mu\nu} \big( -i\partial_\mu X^{a}\sigma_3^{IJ} \bar\theta^I \Gamma_{ a}\partial_\nu \theta^J
                                                   +\bar\theta^1 \Gamma^a \partial_\mu\theta^1\bar\theta^2 \Gamma_a \partial_\nu \theta^2\big) \,. \nln                                         
\ee
$h^{\mu\nu}$ is the world-sheet metric, $X^a$ are the ten embedding coordinates in the flat space $a=0,\dots\, 9$ and $\theta^I$ with $I=1,2$ are the two Majorana-Weyl spinors in ten dimensions%
\footnote{I have dropped the spinorial index $\alpha$.},
with $\sigma_3^{IJ}=diag (1,-1)$. For the specific case of the type IIB superstring, the two fermions have the same chirality, vice versa in type IIA they have opposite chirality, namely
\be
&& \Gamma_{11} \theta^I= \theta^I\quad \text{with} ~~ I=1,2 \quad \text{type IIB}
\nln
&& \Gamma_{11} \theta^{I}=(-1)^{I+1} \theta^{1}\quad \text{with} ~~ I=1,2 \quad \text{type IIA}\,,
\ee
where $\Gamma_{11}=\Gamma^0\Gamma^1\dots\Gamma^9$ and $\Gamma_a$ are the $32\times 32$ $\Gamma$-matrices which satisfy the $\grSO(9,1)$ Clifford algebra:
 \[
 \{ \Gamma_a,\Gamma_b\}= 2 \eta_{ab}~~~\text{ with}~ \eta_{ab}=diag(-1,1,\dots,1)\,.
 \]
 The action \eqref{GS_action_flat} is essentially built of two terms. The first contribution $S_{\rm kin}$ is a $\sigma$-model (the term symmetric in the world-sheet indices). The second line comes from the Wess-Zumino-Witten (WZW) term, i.e. $S_{\rm WZW}$ (the one antisymmetric in the world-sheet indices). I will give more detail on the two terms at the end of the section. 
 
An important feature of the GS action \eqref{GS_action_flat}, which is valid also in curved backgrounds, is the invariance under a local fermionic symmetry, which is called {\it $\kappa$-symmetry}~\cite{Green:1983wt}. Such a symmetry fixes univocally the coefficient in front of the WZW term. The $\kappa$-symmetry allows one to gauge away half of the fermionic degrees of freedom, leaving only the physical ones. Counting the fermionic degrees of freedom, we start with a Dirac fermion in ten dimensions, namely with $2^{D/2}=32$ components. We impose the Majorana-Weyl condition which removes half of the components, leaving only 16 real fermionic degrees of freedom. Finally we can use the $\kappa$-symmetry to reduce the spinor components further, namely to 8. Recalling that we started with two supersymmetries ($I=1,2$), we have in total 16 real independent fermionic degrees of freedom.%
\footnote{The equations of motion, e.g. in the light-cone gauge, remove again half of the spinorial components, namely the real independent components left are 8.}
Furthermore, the action \eqref{GS_action_flat} is invariant under super-Poincar\'{e} transformations and world-sheet reparameterizations.


\subsection{Type IIB superstring on $\ads_5\times \mbox{S}^5$: GSMT action}
\label{sec:GSMT_string}

Before getting to the hearth of the discussion about the AdS superstring action, let me first review certain crucial properties of the $\algPSU(2,2|4)$ algebra. In the next paragraph I will heavily use the results of the two sections \ref{sec:PCM} and \ref{sec:coset_model}.

\paragraph{More on the algebra.}

A notably property of the $\algPSU(2,2|4)$ algebra is its inner automorphism%
\footnote{Such a feature is indeed true for the general algebra $\alg{psu}(n,n | 2n)$~\cite{Berkovits:1999zq}, cf. also \cite{Serganova:1989fj}.}, 
defined by a map $\Omega$ which decomposes the algebra in four subsets. 
Explicitly, we have
\[
\label{Z4grading_alg}
\algPSU(2,2|4)\equiv  \alg g= \alg g_0 +\alg g_1 +\alg g_2 + \alg g_3\,,
\]
and the $\Z_4$-grading is generated by the transformation $\Omega$, where
\be
\Omega(M)= - \Sigma M^{ST} \Sigma^{-1}\,.
\ee
Here
$M$ and $M^{ST}$ are $8\times 8$ super-matrices and $\Sigma$ is the following matrix
\be
\label{Sigma_matrix_def}
\Sigma= \matrto{J}{0}{0}{J}
\quad
\text{where}
\quad
J=\matrto{-i\sigma_2}{0}{0}{-i\sigma_2}\,,
\ee
with $\sigma_2$ the Pauli matrix. 
The subsets $\alg{g}_k$ are the eigenspaces with respect to $\Omega$, namely $\Omega \alg{g}_k= i^k \alg{g}_k$. The $\Z_4$-grading respects the bilinear invariants of the algebra, namely 
\be
[\alg{g}_m,\alg{g}_n] =\alg{g}_{m+n~\text{ mod}~ 4}\,.
\ee
From the above relation we can see the reason why the supersymmetric extension of $\ads_5\times \Sphere^5$ is not a symmetric space, namely $[\alg{g}_1,\alg{g}_1] =[\alg{g}_3,\alg{g}_3] =\alg{g}_{2 }$, cf. \eqref{def_symmetric_space} in section \ref{sec:coset_model}. 
The bilinear invariants can be naturally represented by the super-trace in the algebra space, and we have
\[
\langle T_{m}, T_{n} \rangle = 0 \quad \text{unless} \quad m+n=0 ~~(\text{mod}~ 4)\,.
\]
In particular, the sub-algebra $\alg{g}_0$ is the invariant locus of the $\algPSU(2,2|4)$ algebra and it is the algebra for the gauge group $H$, which in our case is $\grSO(4,1)\times \grSO(5)$. This is a crucial point from the super-coset construction point of view. $\alg{g}_2$ contains all the bosonic generators which are left after modding out the Lorentz generators for $\algSO(4,1)\times \algSO(5)$, namely it contains the translation generators, and it is a ten-dimensional space. Notice that $\mathfrak{g}_2$ is not a sub-algebra.%
\footnote{This $\Z_4$-grading works the same for the $\grp{SU}(2,2|4)$ supergroup, thus one might wonder where the difference is. The point is that the projection $\grp{P}$ removes the identity matrix in the algebra, namely the central charge term. Such a factor is sitting in the bosonic subset $\mathfrak{g}_2$, hence it is equivalent to consider traceless matrices within this subspace.}
Finally $\alg g_1$ and $\alg g_3$ are spanned by the fermionic generators, and the two sectors are related by complex conjugation. 

According to the algebra decomposition (\ref{Z4grading_alg}), also the currents will respect the $\Z_4$-grading. Denoting with $J_m \equiv J_{|\alg{g}_m}$ the projection onto the sub-algebra $\alg{g}_m$, then 
\be
\label{J_grading}
J=J_0+J_2+J_1+J_3\,.
\ee
Notice that $J_1$ and $J_3$ are even since they are contracted with the generators and that the gauge-invariant currents $j$ mix under the $\Z_4$-grading. In the language of the previous section $J_0$ is $H$, cf. section \ref{sec:coset_model}.

\paragraph{Green-Schwarz-Metsaev-Tseytlin action.}

Let me first explain the name for this action. 
In 1998 Metsaev and Tseytlin constructed the world-sheet action for the type IIB superstring on $\ads_5\times\Sphere^5$ from a geometrical point of view based on a super coset approach \cite{Metsaev:1998it}. They use the Green-Schwarz (GS) formalism \cite{Green:1983sg, Green:1983wt}, where the target space supersymmetry is manifest. This is due to the fact that the background, curved and with Ramond-Ramond (RR) fluxes, prevents the use of the Ramond-Neveau-Schwarz (RNS) approach, (cf. section \ref{chapter:PSformalism}).

Recalling how the anti-De Sitter spaces and the spheres are realized:
\[
\ads_5 = {\grSO(4,2)\over \grSO(4,1)} \qquad  \Sphere^5 = {\grSO(6) \over \grSO(5)}\,,
\]
and that the direct product $\grSO(4,2) \times \grSO(6)$ is the bosonic sector for the full $\grPSU(2,2|4)$, thus the supersymmetric generalization of the above relation is
\[
{\grPSU(2,2|4)\over  \grSO(4,1) \times  \grSO(5)}= \text{super} ~(\ads_5 \times \Sphere^5 )\,.
\]
In particular, $g$ maps the string world-sheet $\Sigma$ into the super-coset ${\grPSU(2,2|4)\over  \grSO(4,1) \times  \grSO(5)}$. To be more precise, we should say instead of $\grPSU(2,2|4)$ its corresponding universal covering. The left-invariant Maurer-Cartan forms are defined in the same way as in \eqref{MCforms_coset}:
\[
J_\mu= J_\mu^A T^A= g^{-1} \p_\mu g
\qquad
J_\mu^A= J_{\hat M}^A \p_\mu Z^{\hat M}\,,
\]
where $A$ is the $\algPSU(2,2|4)$ algebraic index, $T^A$ are the corresponding generators, which span the four $\mathfrak g_m$ as in \eqref{J_grading}, $\mu$ is the world-sheet index, $\hat M$ is the ten-dimensional target space index and the embedding coordinates are $Z^{\hat M} = (X^M,\theta^\alpha,\hat\theta^{\hat\alpha})$. Recalling the action for the coset model \eqref{Lagrangian_coset} and considering for simplicity only the bosonic sector, then one easily sees that the one-forms $J_\mu^A$ are indeed nothing but vielbeins, namely%
\footnote{This is indeed an expansion, for example by choosing a specific parameterization on the super-coset the full action can be expanded in the number of fermions, cf. \cite{Metsaev:1998it} and \cite{Kallosh:1998zx}. Here it is meant to illustrate the geometrical meaning of the currents, cf. section \ref{subsec:lightcone_gauge}, equation \eqref{fermionic_exp_lagrangian}.}
\be
\label{kin_term}
 S_{\rm{G/H}} &=&  -{\sqrt{\lambda}\over 4\pi} \int d^2 \sigma \sqrt{-h} \, h^{\mu\nu} \,\rm{STr} \left( J_\mu\, J_\nu\right)_{|\mathfrak{g}_2}=\nln
&=&  -{\sqrt{\lambda}\over 4\pi} \int d^2 \sigma \sqrt{-h}\, h^{\mu\nu} \, J_{\hat M}^A \, J_{\hat N}^B\, \p_\mu Z^{\hat M} \,\p_\nu Z^{\hat N}\, \rm{STr} \left( T^A \, T^B \right)_{|\mathfrak{g}_2}=\nln
&=&
 - {\sqrt{\lambda}\over 4\pi} \int d^2 \sigma \sqrt{-h} h^{\mu\nu} \,\p_\mu X^M \, \p_\nu X^N\, \left( J_M^A\, J_N^B \, g_{AB}\right)_{|\mathfrak{g}_2}+ \text{fermions}= \nln
&=&
 -{\sqrt{\lambda}\over 4\pi} \int d^2 \sigma \sqrt{-h} h^{\mu\nu} \,G_{MN} \p_\mu X^M \, \p_\nu X^N+ \text{fermions}\,.
\ee
As for the bosonic coset model the currents $J_\mu$ are invariant under global $\grPSU(2,2|4)$ left multiplication while under the gauge $ \grSO(4,1) \times  \grSO(5)$ transformations they transform as a connection, cf. section \ref{sec:coset_model}. Moreover, they satisfy the Maurer-Cartan identity $\p_\mu J_\nu-\p_\nu J_\mu+[J_\mu, J_\nu]=0$. 

The kinetic term $S_{\rm{G/H}}$ respects the structure of the bosonic coset model as discussed in section \ref{sec:coset_model}. The fermionic currents enter through a Wess-Zumino-Witten term, namely a closed and exact three form%
\footnote{The closure of the WZW term comes from the Maurer-Cartan identity for the left-invariant currents, while from the fact that the third cohomology group of the superconformal group is trivial follows the exactness for the WZW term~\cite{Berkovits:1999zq, Arutyunov:2009ga}.}:
\[
I_{WZ} \sim \kappa \int_{M_3} d^3\sigma \,\Om_3
\]
with
\[
\label{WZW_3form}
\Om_3= \left( J_2\wedge J_1\wedge J_1- J_2\wedge J_3 \wedge J_3\right)\,,
\]
where the boundary of $M_3$ is the string world-sheet $\Sigma$ which we are integrating over. The form for the WZW term is indeed the only relevant one which is compatible with the invariance under $\grp{SO}(4,1)\times \grp{SO}(5)$ gauge transformations and which has the correct flat space limit. The coefficient $\kappa$ in the expression above is fixed by the the local fermionic symmetry which characterizes the GS formalism. In particular, the values allowed are $\kappa=\pm1$. The exchange of sign is related to a parity transformation in the world-sheet coordinates and to an exchange of the two fermionic sectors $\mathfrak{g}_1$ and $\mathfrak{g}_3$. Once one integrates such a three-form \eqref{WZW_3form}, it gives the antisymmetric term
\[
\label{WZ_action}
S_{WZ}=-{\sqrt{\lm}\over 4\pi}\kappa \int d^2 \sigma \eps^{\mu\nu} J_{\mu,1} J_{\nu,3}\,.
\]

Thus the final action is the sum of the two terms \eqref{kin_term} and \eqref{WZ_action}, namely
\be
\label{action_GSMT}
S_{GSMT}= -{\sqrt{\lambda}\over 4\pi} \int d^2\sigma\, \mathrm{Str}\,\big(\, \gamma^{\mu\nu}\, J_{\mu 2}\, J_{\nu 2}\, +\, \kappa\, \epsilon^{\mu\nu} \,J_{\mu 1}\, J_{\nu 3}\,\big)\,,
\ee
with $\gamma^{\mu\nu}=\sqrt{-h} \, h^{\mu\nu}$.
Summarizing the properties of $S_{GSMT}$ we have that 
\begin{itemize}
\item the bosonic part of $S_{G/H}$ reproduces the standard bosonic coset model on $\ads_5\times \Sphere^5$, cf. equation \eqref{kin_term};
\item the full action \eqref{action_GSMT} is invariant under global $\grp{PSU}(2,2|4)$ invariance;
\item it is also invariant under local $ \grSO(4,1) \times  \grSO(5)$ transformation, 
\item and under the $\kappa$ symmetry,
\end{itemize}
as it has been shown in \cite{Metsaev:1998it}. Finally, in the flat space limit, namely for $R\rightarrow \infinity$, the above action \eqref{action_GSMT} reproduces the GS type IIB superstring in flat space~\eqref{GS_action_flat}. This is indeed how Metsaev and Tseytlin uniquely constrained their ansatz for the action~\cite{Metsaev:1998it}.

%
\paragraph{The classical equations of motion.}

In oder to fix the ideas, let us consider the complex world-sheet coordinates%
\footnote{I will use the same normalization and convention as in \cite{Polchinski:1998rq}. }
$z, \bar z$ given by
\[
z=\sigma^1+i \sigma^2
\qquad
\bar z= \sigma^1-i \sigma^2\,.
\]
For the conventions and more detail we refer the reader to appendix \ref{app:notation}. The GSMT action \eqref{action_GSMT} becomes in the new coordinates
\[
\label{action_GSMT_zzbar}
S_{GSMT}= {\sqrt{\lm}\over 2 \pi} \int d^2 z \, \STr \, \big( \, J_2 \bar J_2 - {\kappa\over 2} \,(J_1 \bar J_3 -\bar J_1 J_3)\big)\,.
\]

In order to derive the equations of motion, one can consider an infinitesimal variation $\xi$ of the $\grPSU(2,2|4)$ coset representative $g$, namely 
\[
g=g \xi \qquad \delta g^{-1}=-\xi g^{-1},
\]
where $\xi=\sum_{i=1}^3 \xi_i$ and $\xi_i \in \mathfrak{g}_i$. This implies that small variations for the currents $J= g^{-1} dg$ satisfy 
\be
\label{first_variation_J}
&& \delta_\xi J_i= \partial \xi_i +[J,\xi]_i
\qquad 
\delta_\xi \bar J_i = \bar \partial \xi_i +[\bar J,\xi]_i\nln
&& \delta J_0=[J,\xi ]_0 
\qquad 
\delta\bar J_0=[\bar J,\xi ]_0\,.
\ee
Plugging such variations \eqref{first_variation_J} in the GSMT action \eqref{action_GSMT_zzbar} and using the Maurer-Cartan identities, one obtains the following equations of motion:
\be
\label{eom_GSMT}
&& D \bar J_2 +\Half (1+\kappa) [J_1,\bar J_1]+\Half (1-\kappa) [J_3, \bar J_3]=0
\nln
&& \bar D J_2 +\Half (1-\kappa) [\bar J_1, J_1]+\Half (1+\kappa) [\bar J_3, J_3]=0
\nln
&& (1-\kappa) [J_2, \bar J_1]- (1+\kappa) [J_1, \bar J_2]=0
\nln
&& (1-\kappa) [J_3, \bar J_2]-(1+\kappa) [J_2, \bar J_3]=0\,,
\ee
where the covariant derivatives are defined as
\[
D\,\, = \p \,\, +[J_0,\,\,]
\qquad
\bar D\,\, = \bar \p \,\, +[\bar J_0,\,\,]\,.
\]

It is clear that the choices $\kappa=\pm 1$ are special values, which definitely simplify the above equations. As an example, for $\kappa=1$ the equations of motion~\eqref{eom_GSMT} become
\be
&& D \bar J_2 + [J_1,\bar J_1]=0
\qquad
\bar D J_2 + [\bar J_3, J_3]=0
\nln
&&  [J_1, \bar J_2]=0
\qquad
 [J_2, \bar J_3]=0\,.
 \ee
These equations should be compared with the ones that will be derived in Berkovits formalism in section \ref{chapter:PSformalism}, cf. eq. \eqref{classicaleom_matter}.


\subsection{Classical integrability for the GSMT superstring action}
\label{sec:classical_integrability_GSMT}

The integrability of the $\ads_5\times\sphere^5$ world-sheet action has been proven at classical level in~\cite{Mandal:2002fs} for the bosonic sector and in~\cite{Bena:2003wd} for the full supersymmetric model by constructing the Lax pair, as I will review in this section. The string integrable structure has been showed also in the work~\cite{Kazakov:2004qf} and in~\cite{Kazakov:2004nh, Beisert:2004ag, Beisert:2005bm, SchaferNameki:2004ik}, which are mostly based on the algebraic curve techniques.%
\footnote{I refer the reader to Zarembo's review~\cite{Zarembo:2004hp} for more detail on this topic.}

In order to have a generating functional%
\footnote{For closed strings, the path in the world-sheet is a closed loop.}
for the (local and non-local) charges and prove that the type IIB superstring in $\ads_5\times \Sphere^5$ is classically integrable, we would like to generalize the construction of the flat connection for the PCM and coset models discussed in section \eqref{sec:PCM} and \eqref{sec:coset_model}. Here we also have the contribution from the fermionic currents, however the arguments run absolutely in the same way~\cite{Curtright:1979am, Bena:2003wd}. Again we can take a linear combination of the gauge invariant currents, namely
\[
a_\mu=\alpha j_{2,\mu} +\beta \eps_{\mu\nu} j_2^\nu + \gm\, k_\mu +\dl \,k'_\mu
\]
where $k_\mu= j_{1,\mu}+j_{3,\mu}$ and $k'_\mu= j_{1,\mu}-j_{3,\mu}$, using the notation of~\cite{Bena:2003wd}. Imposing the zero curvature equation
\[
\p_\mu a_\nu -\p_\nu a_\mu + [a_\mu, a_\nu]=0\,,
\] 
one obtains a system of equations
. The solutions, which give two one-parameter families of flat connections, are~\cite{Bena:2003wd} 
\be
&& \alpha= -2\sinh^2\lm \qquad \beta= \mp 2\sinh \lm \cosh \lm
\nln
&& \gm= 1\pm \cosh \lm \qquad \dl= \sinh \lm\,. 
\ee
Thus, remarkably, the classical GSMT superstring action admits a Lax representation, showing its classical integrability. 
Expanding the coefficients for $\lm=0$ at the leading order, one obtains exactly the Noether currents for the global $\grPSU(2,2|4)$ symmetry~\cite{Kazakov:2004nh}, namely
\[
a_\mu= 2 \,\lm\, \eps_{\mu\nu} j_2^\nu  + \lm\, k'_\mu\,. 
\]
In order to deduce the flat connection one uses the equations of motion and the algebraic identities, but one does not need to fix the $\kappa$ symmetry. However, it has been shown in \cite{Arutyunov:2009ga} that integrability forces the coefficient in front of the WZW term to be fixed to the same values which are allowed by the $\kappa$-symmetry ($\kappa=\pm 1$). This means that the word-sheet action, in order to have the infinite set of conserved charges, should also be $\kappa$ symmetric and vice versa.%
\footnote{Indeed rescaling the WZW term the higher symmetries and the $\kappa$-invariance are broken, (not for the special value $\sigma \rightarrow -\sigma$ which corresponds to the world-sheet parity). }
We will come back to the integrability of classical superstring in the discussion for the Pure Spinor formulation of the type IIB superstring in $\ads_5\times\Sphere^5$ in section \ref{chapter:PSformalism}, and there we will discuss the extension to the quantum theory using the Berkovits formalism.


\section{The Pure Spinor $\ads_5\times\mbox{S}^5$ superstring}
\label{chapter:PSformalism}

\subsection{Motivations}

One of the main advantages of the Green-Schwarz formalism is that the target space supersymmetries are manifest. However, already for the type IIB superstring in flat space, we encounter serious difficulties once we try to quantize the theory. Recalling the kinetic term $S_{\rm int}$ in the GS action $S_{GS, flat}$ in the flat ten-dimensional space in~\eqref{GS_action_flat}, one sees that the kinetic term for the fermions is degenerate:
\[
\partial_\mu X^{a} \bar\theta^J \Gamma^{a}\partial_\mu \theta^J\,.
\]
Indeed, when $\partial_\mu X^a=0$ it simply vanishes. Moreover computing the canonical momenta for the spinors
\[
\rho_K\equiv {\delta \mathcal L\over \delta \dot{\theta}^K}\,,
\]
one obtains a complicated and non-linear function of all the phase-space variables. According to the Dirac classification, the canonical momenta are primary constraints, which can be of first or second class. We can say  that in the latter case the momenta have a non-vanishing bracket with the constraints themselves. When the first and second class of constraints are coupled, one needs to disentangle them and quantize the system introducing the so called Dirac brackets as the new anticommutation relations. In the GS superstring, the two classes of constraints cannot be separated in a covariant way. 
A way of bypassing the problem is to fix the light-cone gauge and quantize the superstring action in this gauge.%
\footnote{This is true for the GS formalism in general, namely for the GS superstring action in a flat and curved space, cf. \cite{Green:1987sp, Becker:2007zj}.}
The light-cone quantization allows one to compute the string spectrum leaving only the physical degrees of freedom, and it is very helpful for example in computing the string energies, cf. section \ref{chapter:NFS}. However, it is not completely satisfactory: one would really like to have a covariant quantization for the string action.%
\footnote{There is actually another alternative approach based on the so called Pohlmeyer reduction. The idea is to reduce the string world-sheet action to an equivalent action containing only the physical degrees of freedom, with equivalent integrable structures and with a manifest two-dimensional Lorentz invariance. We refer the reader to the original paper by Grigoriev and Tseytlin~\cite{Grigoriev:2007bu} and to the work by Mikhailov and Schafer-Nameki~\cite{Mikhailov:2007xr} and references therein for more detail.}

These are the main motivations in order to have a formalism with manifest space-time supersymmetries \emph{and} a full covariant formulation which allows one to quantize the superstring action keeping the ten-dimensional Lorentz symmetry manifest. 

These two aspects are joint in the formalism proposed by Berkovits in \cite{Berkovits:2000fe}, extending and completing a previous idea of Siegel \cite{Siegel:1985xj}: the target-space supersymmetry is manifest and the ten-dimensional Lorentz covariance is also manifest and present in all the stages of the theory.  Obviously there is a price to pay. In order to have a standard fermionic kinetic terms, certain ghost fields have to be introduced (the \emph{pure spinors}), as well as their conjugate momenta. The non-physical degrees of freedom introduced in the theory in this way are later removed through a BRST-like operator Q.

\paragraph{Outline.}

In this section~\ref{chapter:PSformalism} I would like to review some basic notions and concepts about the pure spinor (PS) formalism. I will focus on the type IIB superstring action and on the role of the pure spinors in the context of integrability. Thus, the next section will not be an exhaustive introduction to the pure spinor formalism. For this we refer the reader to the ICTP lectures given by Berkovits in 2000~\cite{Berkovits:2002zk} and Oz in 2008~\cite{Oz:2008zz}. 

In the first part, I will discuss some basic features of the pure spinors and of their space. Then I will formulate the PS action for open strings in flat space. The generalization to closed strings is straightforward, since basically one ``squares'' the ghost fields. 

The second essential step is the formulation of the superstring action in curved backgrounds. I will focus on the $\ads_5\times \sphere^5$ type IIB action, since this is the relevant case for the $\ads_5/\CFT_4$ correspondence. 

At this point, in the context of integrability, we need to discuss the key features of the superstring action. In order, we will see the gauge and BRST invariance of the action at classical \cite{Berkovits:2004jw} and quantum \cite{Berkovits:2004xu} level. Notice that these properties are fundamental to guarantee the consistency of the action also at quantum level. Hence, we will review the classical integrability of the PS type IIB action \cite{Vallilo:2003nx} and the explicit construction of the BRST non-local charges~\cite{Berkovits:2004jw}. Indeed, it turns out that the higher conserved charges have to be BRST invariant. The same steps should be repeated at quantum level. In particular, I will summarize the results of \cite{Berkovits:2004xu} for the BRST invariance of the quantum non-local charges and I will discuss the finiteness of the monodromy matrix at the quantum leading order \cite{Mikhailov:2007mr}. 

Finally, the last section is dedicated to the finiteness of the charges, the absence of anomaly in the variation of the monodromy matrix~\cite{Puletti:2008ym} and the operator algebra at the leading order in perturbation theory~\cite{Puletti:2006vb}. 


\subsection{The pure spinor formalism: basic review}

The pure spinors are world-sheet ghosts $\lambda^\alpha$ which carry a space-time spinor index but they are commuting objects, which are constrained to satisfy the following condition (the \emph{pure spinor constraint}):
\[
\label{def_PS}
\lambda^\alpha \hat{\gamma}_{\alpha\beta}^a \lambda^\beta=0\,,
\]
where $\hat\gamma^a$ are $16\times 16$ $\grSO(9,1)$ gamma matrices in the Majorana-Weyl representation, $a=0,1,\dots, 9$. 
Hence the pure spinors are complex Weyl spinors, however the conjugate $\bar\lambda_\alpha$ never appears in the theory. The canonical momenta to $\lambda^\alpha$ are the ghost fields $\om_\alpha$. The system $(\om_\alpha, \lambda^\alpha)$ is analogue to the $(\beta,\gamma)$ system in string theory, however now the conformal weight is (1,0) and the fields are not free. Their ghost number is $(-1,1)$. 

From the condition (\ref{def_PS}) it follows that the actual independent components in $\lambda$ are 11 and not 16 as one would naively expect. The number of exact degrees of freedom is really important, as we will see, thus we would like to spend some time to explain how to count them. 
For simplicity, we can Wick-rotate $\grSO(9,1)$ to $\grSO(10)$. The space where the pure spinors live is singular in the origin, since the constraint \eqref{def_PS} is degenerate at the point $\lambda=0$ (as well as its variation). It is indeed a cone, and removing the singularities we can describe the space as a ${\grSO(10)\over \grU(5)}$ coset. We can break the $\grSO(10)$ description to $\grU(5)$, according to $\grSO(10) \rightarrow \grSU(5) \times \grU(1)$. The $\grU(5)$ gamma matrices are 
\be
\hat\gamma^{\underline a} = {\hat\gamma^{\underline a} +i \hat\gamma^{{\underline a} +1}\over 2} \qquad \text{with} \,\,\, {\underline a} =1,..., 5
\nln
\hat\gamma_{\underline a}  = {\hat\gamma^{\underline a} -i \hat\gamma^{{\underline a} +1}\over 2} \qquad \text{with} \,\,\, {\underline a} =1,..., 5\,.
\ee
We can interpret $\hat\gamma^{\underline a} $ as a raising operator and $\hat\gamma_{\underline a} $ as a lowering operator. They satisfy the $\algU(5)$-algebra, namely
\[
\{ \hat\gamma^{\underline a} , \hat\gamma^{\underline b} \}=\{ \hat\gamma_{\underline a} , \hat\gamma_{\underline b} \}=0
\qquad
\{\hat\gamma^{\underline a} ,\hat\gamma_{\underline b} \} =\delta^{\underline a} _{\underline b} \,.
\]
Let us define the ground state $u^\alpha_+$ as the state annihilated by all the lowering operators, i.e. $\gamma_{\underline a}  u^\alpha_+=0$ for $a=1,...,5$. Then, acting with the $\grU(5)$ $\gamma$-matrices we can obtain the complete basis of the $\grU(5)$ spinors. In particular, acting with an odd number of $\gamma$-matrices leads to a change of the chirality (since the spinor index will be a lower one, i.e. an antichiral spinor).
Hence, the basis for the spinor $\lambda^\alpha$ is 
\[
u^\alpha_+    \qquad  
(u^{{\underline a} {\underline b} })^\alpha\equiv \left( \gamma^{\underline a} \gamma^{\underline b}  u_+  \right)^\alpha
\qquad
u^\alpha_{\underline a} = \epsilon_{{\underline a} {\underline b} {\underline c} {\underline d} {\underline e} } \left(\gamma^{\underline b} \gamma^{\underline c}  \gamma^{\underline d} \gamma^{\underline e}   u_+\right)^\alpha
\]
and any chiral $\grU(5)$ spinor can be written as 
\[
\lambda^\alpha= \lambda^+ u^\alpha_+ + \lambda_{{\underline a} {\underline b} } (u^{{\underline a} {\underline b} })^\alpha+ \lambda^{\underline a}  u_{\underline a} ^\alpha\,.
\]
Notice that $\lambda^+$ is a $\grU(5)$ singlet, $\lambda_{{\underline a} {\underline b} }$ transforms in the $\bar{\mathbf{10}}$ antisymmetric representation of $\grU(5)$ and $\lambda^{\underline a} $ in the $\mathbf{5}$ one. For an antichiral spinor we have
\[
\om_\alpha= \om_+ \tilde{u}^+_\alpha + \om_{\underline a}  (\gamma^{\underline a}  u_+)_\alpha + \om^{{\underline a} {\underline b} } (u_{{\underline a} {\underline b} })_\alpha
\,,
\]
with $(u_{{\underline a} {\underline b} })_\alpha= \epsilon_{{\underline a} {\underline b} {\underline c} {\underline d} {\underline e}} (\gamma^{\underline c}\gamma^{\underline d}\gamma^{\underline e} u_+)_\alpha$ and $ \tilde{u}^+_\alpha=\epsilon_{{\underline a} {\underline b} {\underline c} {\underline d} {\underline e}}(\gamma^{\underline a} \gamma^{\underline b} \gamma^{\underline c} \gamma^{\underline d} \gamma^{\underline e}  u_+)_\alpha$. 
At this point, one can readily decompose the ten equations (\ref{def_PS}) in the $\grU(5)$ basis and obtain
\footnote{In order to decompose the constraints \eqref{def_PS} some useful identities are 
\[
u_+\gamma^1\gamma^2\gamma^3\gamma^4\gamma^5 u_+=1
\qquad
u_+ \gamma^{\underline a}  \gamma^{\underline b} \gamma^{\underline c}  u_+= u_+ \gamma^{\underline a}  u_+=0\,.
\]
}
\be
\label{PS_constraints_U(5)}
&& \lambda \gamma^{\underline a} \lambda= \lambda^+ \lambda^{\underline a}  +{1\over 8} \epsilon^{{\underline a} {\underline b} {\underline c} {\underline d} {\underline e}}\lambda_{{\underline b} {\underline c}}\lambda_{{\underline d} {\underline e}}=0
\nln
&& \lambda\gamma_{\underline a} \lambda= \lambda^{\underline b}\lambda_{{\underline a} {\underline b}}=0\,,
\ee
with ${\underline a} =1,\dots, 5$. Hence, fixing $\lambda^+\neq 0$, the first equation of \eqref{PS_constraints_U(5)} is solved for $\lambda^{\underline a} = -{1\over 8}(\lambda^+)^{-1}   \epsilon^{{\underline a} {\underline b} {\underline c} {\underline d} {\underline e}}\lambda_{{\underline b} {\underline c}}\lambda_{{\underline d} {\underline e}}$, which automatically solves also the second equation. Thus $\lambda^\alpha$ is a function of eleven complex parameters, namely $\lambda^+ $ and $\lambda_{{\underline a} {\underline b}}$. 
Hence the final parameterization for $\lambda$ is 
\[
\label{PS_U5}
\lambda^+= e^s
\qquad
\lambda_{{\underline a} {\underline b}}= u_{{\underline a} {\underline b}}
\qquad
\lambda^{\underline a}= {1\over 8} e^{-s} \epsilon^{{\underline a} {\underline b} {\underline c} {\underline d} {\underline e}} u_{{\underline b} {\underline c}} u_{{\underline d} {\underline e}}\,.
\]
The fact that the vector $\lambda^{\underline a} $ is redundant in this description, namely the second equation in eq. (\ref{PS_constraints_U(5)}) is identically satisfied (for some constant non-vanishing $\lambda^+$), implies that the corresponding antichiral spinor $\om_{\underline a} $ is defined up to gauge transformation, i.e. $\delta \om_{\underline a}  \sim \eta_{\underline a}  \lambda^+$, with $\eta_{\underline a} $ the gauge parameter. As a consequence, it can be directly set to zero, and choose%
\footnote{We should introduce a normal order constant in $\om$, cf. e.g.~\cite{Berkovits:2002zk} and \cite{Schiappa:2005mk}. However the issues about the normal ordering can be ignored here, because we are only interested in the OPE's involving the ghost Lorentz currents $N$.}
\[
\label{PS_w_U5}
\om_+= e^{-s} \partial t 
\qquad
\om^{{\underline a} {\underline b}}=  v^{{\underline a} {\underline b}}
\qquad
\om_{\underline a} =0\,.
\]  
%
%

Some properties are better shown in the $\grU(5)$ basis, where the ghost fields are free. In particular, it is easier to understand better the origin of the ``correction'' term in the OPE \eqref{true_OPE_PS} between the pure spinor and its conjugate field. 

The ghosts are maps from the two-dimensional world-sheet to the target-space, which is the ten-dimensional flat space in this case. In terms of the free $\grU(5)$ components, the ghost action in a flat background in the conformal gauge is  
\[
\label{PS_action_U5}
S_G= {1\over \pi \alpha'}\int d^2z \big( \partial t\bar\partial s -\Half v^{{\underline a} {\underline b}} \bar \partial u_{{\underline a} {\underline b}}\big)\,.
\]
Hence, the OPE's can be directly read from the above action:
\be
\label{free_field_OPE}
&&
t(z_1) \, s(z_2) \sim \log{(z_1-z_2)}
\nln 
&&
v^{{\underline a} {\underline b}}(z_1) \, u_{{\underline c} {\underline d}}(z_2) \sim {\delta^{[{\underline a} }_{{\underline c} } \delta^{{\underline b} ]}_{\underline d} \over z_1-z_2}\,.
\ee

In the covariant ten-dimensional $\grSO(10)$ notation, the pure spinor action in flat space is
\be
\label{action_PS_flat}
S_G= {1\over \pi\alpha'}\int d^2z \,\om_\alpha\bar \partial \lambda^\alpha \,.
\ee
The two actions \eqref{PS_action_U5} and \eqref{action_PS_flat} describe the pure spinors and the conjugate fields in a flat space (even though in different notations), but the latter contains also the non-physical degrees of freedom. 

Without breaking the $\grSO(10)$ covariance, the OPE is
\[
\label{true_OPE_PS}
\om_\alpha (z_1) \lambda^\beta(z_2) \sim {\delta^\beta_\alpha\over z_1-z_2} -\Half\hat \gamma_a^{\beta+} e^{-s} {(\hat\gamma^a\lambda)_\alpha \over  z_1-z_2}\,.
\]
As before, the + index is the $\mathbf{1}$ spinor component in the $\grU(5)$ notation. The second term in eq. (\ref{true_OPE_PS}) takes care of the fact that, due to the PS condition (\ref{def_PS}), $\om$ is defined only up to gauge transformations 
\[
\label{gauge_transf_w}
\delta \om_\alpha = \Lambda^a (\hat\gamma_a \lambda)_\alpha\,.
\]
This is exactly the same statement above the expressions \eqref{PS_w_U5} in the $\grSO(10)$ notation. Alternatively, we can say that the second term in \eqref{true_OPE_PS} assures that the PS constraint remains valid also when we consider the OPE between $\om$ and the condition (\ref{def_PS}) itself.

Since $\om_\alpha$ is defined only up to gauge transformations \eqref{gauge_transf_w}, it means that it can appear only in gauge covariant combinations, as for example the Lorentz ghost currents
\[
\label{def1_N}
N^{ab}=\Half \om\, \hat{\gamma}^{ab} \lambda\,.
\]
One can see that the second term in eq. \eqref{true_OPE_PS} does not contribute to the OPE between N and $\lambda$ due to the identity $\lambda\hat\gamma^{ab}\hat\gamma^c\lambda=0$:
\[
\label{OPE_Nlambda}
N^{ab}(z_1) \lambda^\alpha \sim \Half (\hat\gamma^{ab} )^\alpha_{\,\,\,\beta} {\lambda^\beta\over z_1-z_2}\,.
\]
%
%
The ghost Lorentz currents satisfy the following OPE
\be
\label{OPE_N}
N^{a b}(z_1) N^{c d}(z_2)\sim
 {\eta^{c[b} N^{a]d}(z_2)-\eta^{d[b} N^{a]c} (z_2)\over z_1-z_2}
-3 {\eta^{ad}\eta^{b c} -\eta^{ac}\eta^{b d}\over (z_1-z_2)^2}\,,
\ee
The OPE's for the Lorentz currents and for $\lambda$ are manifestly covariant. They are most easily computed in the $\grU(5)$ formalism, where all the fields are free. Indeed, decomposing $N^{ab}$ in $(N,N^{\underline a}_{\underline b}, N^{{\underline a} {\underline b}}, N_{{\underline a}{\underline b}})$ and using the free field OPE's (\ref{free_field_OPE}), one can compute the expression (\ref{OPE_N}), cf. \cite{Schiappa:2005mk} for explicit computations.

The fact that the pure spinors have 11 degrees of freedom is essential, because it is what one needs in order to cancel the conformal anomaly. Let us consider the kinetic term for the GS action in flat space. In the conformal gauge, the world-sheet metric is flat. In the $z,\bar z$ coordinates, cf. appendix \ref{app:notation}, the kinetic term of \eqref{GS_action_flat} becomes
\[
\label{matter_action_PS_flat}
S= {1\over  \pi\alpha'}\int d^2 z\, \big(\Half \partial X^a \bar\partial X_a + \rho_\alpha \bar\partial \theta^\alpha\big)\,,
\]
where $\rho_\alpha$ is the canonical momentum%
\footnote{The fermions are Majorana-Weyl spinors in ten dimensions, thus one can directly use the $16 \times 16 $ Dirac matrices $\hat\gamma^a$ instead of the $32\times 32$ $\Gamma^a$ matrices.}
\[
\label{rho}
\rho_\alpha= {i \over 2} \partial X^a (\bar\theta \hat\gamma_a)_\alpha + \bar\theta \hat\gamma^a  \partial \theta (\bar\theta\hat\gamma_a)_\alpha\,.
\]

In the flat ten-dimensional Minkowski space the PS action is given by eq. (\ref{matter_action_PS_flat}) and eq. (\ref{action_PS_flat}). By computing the central charge, the contribution from the matter sector is $c_M=10-32=-22$, from the bosonic and fermionic sector respectively. Thus, the ghosts should contribute to the central charge with $c_G=+22$, in order to cancel the conformal anomaly. 
Indeed, the ghost stress-energy tensor is 
\[
T_{G}= \Half v^{{\underline a} {\underline b}} \partial u_{{\underline a} {\underline b}} +\partial t\partial s +\partial^2 s
\]
and the OPE gives
\[
T_G(z_1) \, T_G(z_2) \sim {\mathrm{dim}_{\C}\mathcal M\over (z_1-z_2)^4}\,,
\]
where $\mathcal M$ is the manifold where the pure spinors live, i.e. their degrees of freedom. 
Eventually, the corresponding central charge is $c_G= 2 \, \mathrm{dim}_{\C}\mathcal M=+22$. 

For completeness, let me write the ghost number operator
\[
J_G= \om_\alpha\lm^\alpha\,.
\]


\paragraph{BRST operator.}

One can define a BRST-like operator%
\footnote{The name BRST means Becchi-Rouet-Stora-Tyutin~\cite{Becchi:1974xu, Becchi:1975nq, Tyutin:1975qk}. }
as
\[
\label{Qps}
Q=\oint \lambda^\alpha d_\alpha\,,
\] 
where $d_\alpha$ is the fermionic constraint 
\[
d_\alpha= \rho_\alpha- {i\over 2}\partial X^a (\bar\theta \hat\gamma_a)_\alpha - \bar\theta \hat\gamma^a \partial \theta(\bar\theta\hat\gamma_a)_\alpha\,.
\]
Q has ghost number 1, thus the physical string states are the elements which are in the cohomology%
\footnote{The BRST cohomology of the nilpotent operator Q~\eqref{Qps} is the space of all equivalent states $\ket v$ which are closed and exact, namely which satisfy $Q\ket v=0$ and which differ by a null state $\ket v= \ket{v'}+Q\ket u$ for some state $\ket u$.}
of Q and have ghost number 1. Q is guaranteed to be nilpotent by the PS constraint \eqref{def_PS}, since
 \[
 Q^2 = \lambda^\alpha \lambda^\beta \{d_\alpha, d_\beta\}\sim \lambda\hat\gamma_a\lambda =0\,.
 \]
In the GS formulation the superstring action was invariant under $\kappa$-symmetry. This symmetry is no longer present and its role is replaced by the BRST symmetry. I will come back on this point when the pure spinor action in curved background will be discussed.

Until now we have discussed the open string action in flat space. We want to deal with closed strings, which means to double the system described above.  Namely, we will have two sets of ghosts $( \om_\alpha,\lambda^\alpha)$ and $(\hat{\om}_{\hat\alpha},\hat \lambda^{\hat\alpha})$ with constraints
\[
\label{PS_constraint_closed}
\lambda \hat\gamma_a\lambda=0 \,,
\qquad
\hat\lambda \hat\gamma_a \hat\lambda =0\,.
\]
They are left and right-moving bosonic spinors, with conformal weight (1,0) and (-1,0). They are described by the following action in a flat background
\[
\label{def_ghost_action_closedstring}
S_{G}= {1\over \pi\alpha'}\int d^2z \,\big( \om_\alpha \bar\partial \lambda^\alpha + \hat{\om}_{\hat{\alpha}} \partial\hat\lambda^{\hat\alpha}\big)\,,
\]
and they give rise to two BRST operators as well
\[
Q=\oint \lambda^\alpha d_\alpha
\qquad
\bar Q= \oint \hat\lambda^{\hat\alpha} \hat d_{\hat\alpha}\,.
\]
Essentially, all the arguments presented above run in the same way.


\subsection{Type IIB superstring on $\ads_5\times \mbox{S}^5$: PS action}

%
%
%

\paragraph{Matter content.}

 
In the matter content we have two contributions~\cite{Berkovits:1999zq}. The first term is the sigma model action on the super-coset, which is $\grPSU(2,2|4)/(\grSO(4,1)\times \grSO(5))$, namely
\[
\label{S_GH_PS}
S_{G/H}= {1\over 2\gamma^2} \int d^2z\, \mathrm{STr} (J_{G/H})^2\,.
\]
$1/\gamma^2$ is the coupling constant, that we will fix at the end.
In section \ref{sec:coset_model} we have explained how to construct the above action. However, the main difference with the bosonic GSMT action \eqref{kin_term} is that now we also include the fermionic currents. %
%
%
Explicitly, \eqref{S_GH_PS} contains 
\[
\sqrt{-h} h^{\mu\nu} \STr \left( J_{2\mu} J_{2\nu}+J_{1\mu}J_{3\nu}+J_{3\mu}J_{1\nu}\right) \,.
\] 

The action \eqref{S_GH_PS} is invariant under gauge $\grp{H}$-transformations and under the global $\grp{G}$-symmetry. Hence, it is naturally defined on the coset space $\grp{G/H}$.
However, this is not sufficient to guarantee a conformal theory.%
\footnote{Here, the world ``conformal'' is referred only to the matter sector or to the $\ads_2 \times \mbox{S}^2$ case.}
For this reason it is necessary to introduce a topological term, such as the WZW term, which is a gauge invariant three-form. As for the GS action, it should be closed and $d$-exact. Writing
\[
\Omega_3=d \mathrm{STr} \big( J_1\wedge J_3   \big) 
\]
one obtains
\[
\label{WZ_PS}
S_{WZ}= {k \over 2\gamma^2} \int d^2 z\,  \mathrm{STr} \big( J_1\wedge J_3   \big) \,. 
\]
The WZW term in (\ref{WZ_PS}) is exactly the same which is in the GSMT action, cf.~\eqref{action_GSMT}. However, here the level $k$ is fixed by requiring the superconformal invariance of the action. The $k$ values which are allowed are $\pm \Half$~\cite{Bershadsky:1999hk}. Recall that the coefficient in front of the WZW term is fixed by the $\kappa$-symmetry in the GS formalism. In the PS approach the term $J_1 J_3$ in (\ref{S_GH_PS}) breaks such a symmetry, but on the other hand it gives the possibility to have a kinetic term for the fermions, (thus to construct a fermionic propagator in the standard way and proceed with a perturbative covariant quantization). Indeed, at the leading order one has:
\[
J_{1\mu} J_{3\nu} \sim \partial_\mu \theta^1_L\partial_\nu \theta^3_R\,. 
\]

Thus the total matter contribution for the PS in the conformal gauge%
\footnote{In the conformal gauge the world-sheet metric is flat, cf. appendix~\ref{app:notation}.}
is
\be
\label{Matter_action_PS}
 S_M= S_{G/H}+S_{WZ}=
{1\over 2\gamma^2} \int d^2z\, \mathrm{Str} \big ( J_2 \bar J_2 + {3\over 2} J_3 \bar J_1 + \Half \bar J_3 J_1 \big)\,. 
\ee
Note that this action corresponds to the choice $k=\half$ and that a change in the sign of the WZW term coefficient leads to exchange $J_1$ and $J_3$. The one-loop beta function for the purely matter sector (i.e. $\ads_2\times \sphere^2$) has been computed in~\cite{Berkovits:1999zq} and showed explicitly that the renormalization of the coupling constant is proportional to $(2 k^2-\Half)$, namely $k$ and $\gamma$ are not renormalized at first quantum order for $k=\pm \Half$. Actually it is believed that it is true to all orders in perturbation theory, \cite{Bershadsky:1999hk}.

\paragraph{Ghost content.}

In order to present the ghost content for the type IIB action in $\ads_5\times \Sphere^5$, let me rewrite the pure spinor conjugate momenta and the constraints in a more suitable and elegant form.%
%
%
We have two types of spinors (they are actually the same since we are discussing type IIB strings, however I will keep distinct the indices for left and right-moving), i.e. $\lambda^\alpha,\,\hat \lambda^{\hat\alpha}$. Then we will have
\[
\label{new_def_lambda}
\lambda_1= \lambda^\alpha T_\alpha\,,
\qquad
\lambda_3= \hat \lambda^{\hat\alpha} T_{\hat\alpha}\,, 
\]
where $T_\alpha$ and $T_{\hat\alpha}$ are the $\mathfrak{g}_1$ and $\mathfrak{g}_3$ generators respectively. 
We are in the $\ads_5\times \Sphere^5$ background, thus the two fermionic sectors can talk to each other. Namely, there exists a matrix $\gamma^{01234}$ in the AdS directions which couples the two indices $\alpha\,, \hat\alpha$. This is nothing but the 5-form Ramond-Ramond flux. We can use such a matrix in order to rewrite the conjugate fields $\om$ as chiral spinors:
\[
\label{new_def_omega}
\omega_{3+}= \omega_\alpha( \gamma^{01234})^{\alpha\hat\alpha} T_{\hat\alpha}
\qquad
\omega_{1-}= \hat\omega_{\hat\alpha} ( \gamma^{01234})^{\hat\alpha \alpha} T_{\alpha}
\,,
\]
where the $\pm$ in $\om$ are meant to stress the conformal weight of the conjugate fields. 
At this point we can rewrite the ghost Lorentz currents as
\[
\label{def_N}
N_{0}= -\{
\omega_{3+}, \lambda_1\}
\qquad 
\bar N_{0} = -\{\omega_{1-},\lambda_3\}
\,,
\]
and one can check using the structure constants for the $\algPSU(2,2|4)$ algebra given in appendix \ref{app:psu_structure_const}, that is indeed the same definition of (\ref{def1_N}). 
The pure spinor constraints~\eqref{PS_constraint_closed} become
\[
\label{new_ps_constraints}
\{\lambda_1,\, \lambda_1\}=0
\qquad 
 \{\lambda_3,\, \lambda_3\}=0\,,
\]
or analogously 
\[
[\lambda_1, N_{0}]=0
\qquad
[\lambda_3, \bar N_{0}]=0\,.
\]
%
%

The pure spinor carries a spinor index, hence, under Lorentz transformations, they vary according to
\be
\label{gauge_transf_PS}
&&\delta_\Lambda \lambda_1= [\lambda_1, \Lambda]
\qquad 
\delta_\Lambda \omega_{3+}= [\omega_{3+}, \Lambda]
\nln
&&\delta_\Lambda \lambda_3= [\lambda_3, \Lambda]
\qquad
\delta_\Lambda \omega_{1-}= [\omega_{1-}, \Lambda]\,,
\ee
where $\Lambda$ is a gauge parameter. 
This implies that the Lorentz ghost currents transform in the following way under local $\grSO(4,1)\times \grSO(5)$ transformations:
\[
\label{gauge_transf_N}
\delta_\Lambda N_{0}= [N_{0}, \Lambda]
\qquad
\delta_\Lambda \bar N_{0}= [\bar N_{0}, \Lambda]\,.
\]
In order to write down the PS action in the AdS background, we need to covariantize the ghost action (\ref{def_ghost_action_closedstring}). Our gauge field is $J_0$, then introducing the covariant derivatives
\[
D= \partial+ [ J_0\,,~~]
\qquad
\bar D=\bar \partial + [\bar J_0\,,~~]\,,
\]
one can rewrite the terms $\om\partial \lm$ as $\om D\lm$. Explicitly:
\be
&&\omega_{3+}\bar D \lambda_1=
 \omega_{3+}\bar\partial \lambda_1 +\omega_{3+}[\bar J_0,\lambda_1]=
 \omega_{3+}\bar\partial \lambda_1-\omega_{3+}[\lambda_1,\bar J_0]=\nln
 &&= \omega_{3+}\bar\partial \lambda_1-\{\omega_{3+},\lambda_1\}\bar J_0 =
 \omega_{3+}\bar\partial \lambda_1+N_{0}\bar J_0 \,.
\ee
The same is true for the other term: $\omega_{1-} D \lambda_3= \omega_{1-}\partial \lambda_3+\bar N_{0} J_0 $. Note that $\lambda_{1,3}$ and $\omega_{1,3}$ are anticommuting objects, since the components $\lambda^{\alpha} , \hat\lambda^{\hat\alpha}$ and $\omega_{\alpha}, \hat\omega^{\hat\alpha}$ commute and they are contracted with the fermionic generators $T_{\alpha}\,, T_{\hat\alpha}$ (vice versa the currents $J_1\,, J_3$ are commuting objects). The pure spinors are local objects (they live on the tangent space), thus they transform non-trivially under local tangent space Lorentz rotations. For this reason, they can couple to the gauge field ($J_0$, $\bar J_0$) and to the constant target space curvature tensor through their currents.

The right and left-moving sectors are mixed, once we write the ghost fields as in (\ref{new_def_lambda}) and in (\ref{new_def_omega}). Indeed, also the Cartan metric mixes the two sectors. It is defined in terms of the bilinear invariant $\STr$, in particular the elements of such metric are:
\be
\label{alg_metric}
&& \STr (T_a T_b)= \eta_{a b}\qquad \STr (T_{[ab]} T_{[cd]})= \eta_{[ab][cd]} \nln
&& \STr (T_\alpha T_{\hat\beta})= \eta_{\alpha\hat\beta} \qquad \STr (T_{\hat\alpha}T_\beta)= \eta_{\hat\alpha\beta}\,,
\ee
where $\{ T_{[ab]} ,\, T_a, \, T_\alpha,\, T_{\hat\alpha}\,\}$ span $\{ \mathfrak{g}_0,\,  \mathfrak{g}_2,\,  \mathfrak{g}_1,\,  \mathfrak{g}_3\, \}$ respectively. 
An explicit representation for the Cartan metric is not necessary here, it strictly depends on the normalization of the structure constants of $\algPSU(2,2|4)$ and of the super-trace, e.g. cf. appendix in~\cite{Puletti:2006vb}. For the moment, it is sufficient to notice that $\eta_{\alpha\hat\beta}$ is proportional to the matrix $\gamma^{01234}$ and $\eta_{[ab][cd]}$ is the combination $\eta_{a[c}\eta_{d]b}$.
Finally the PS action for the $\ads_5\times \sphere^5$ string is 
\be
\label{def_PS_IIB_action}
S_G= {1\over \gamma} \int \rm{STr} \big( 
\omega_{3+}\bar\partial \lambda_1 + N_{0} \bar J_{0} 
+\omega_{1-}\partial \lambda_3 + \bar N_{0} J_{0} 
- N_{0}\bar N_{0}\big) \,.
\ee
The coefficient in front of the coupling between matter and ghost currents, i.e. $N_{0} \bar J_{0} $ and $\bar N_{0} J_{0} $, is fixed by requiring the gauge invariance of the ghost action (\ref{def_PS_IIB_action}). The action (\ref{def_PS_IIB_action}) must be gauge invariant in order to make sense in this coset construction. Further, note that the term $N_{0}\bar N_{0}$ in (\ref{def_PS_IIB_action}) is automatically gauge invariant under the transformations \eqref{gauge_transf_N}. 
The coupling with the space-time connection gives rise to mixed matter-ghost terms ($J_0 \bar N_{0}$ and $\bar J_0 N_{0}$).


\paragraph{Summary.}

Let me summarize the complete action for the type IIB superstring living on $\ads_5\times\Sphere^5$ in the pure spinor formalism~\cite{Berkovits:2000fe, Berkovits:2000yr, Vallilo:2002mh}:
\be
\label{action_PS_total}
 S &=& S_G+ S_M= \nln
 &=&
{1\over \gamma^2} \int d^2z\, \mathrm{Str} \big (\Half J_2 \bar J_2 + {3\over 4} J_3 \bar J_1 + \Quarter \bar J_3 J_1
+ 
\omega_{3+}\bar\partial \lambda_1 + N_{0} \bar J_{0} \nln
&&+
\omega_{1-}\partial \lambda_3 + \bar N_{0} J_{0} 
- N_{0}\bar N_{0}\big) \,.
\ee

The coupling constant is
%
%
\[
{1\over \gamma^2}= {\sqrt{\lm}\over 4\pi}= {R^2\over 4\pi\alpha'}\,.
\]
Note the non-perturbative parity symmetry of the action which exchanges
\[
\label{parity_symmetry}
z\leftrightarrow \bar z 
\qquad\quad \theta\leftrightarrow \hat\theta
\qquad\quad \mathfrak{g}_1\leftrightarrow \mathfrak{g}_3\,.
\]

\paragraph{The classical equations of motion.}

Recall the MC-current definition in terms of the super-coset representative: 
\[
J= g^{-1} dg \quad \text{with} \quad g\in {\grPSU(2,2|4)\over \grSO(4,1)\times \grSO(5)}\,.
\] 
We have already seen how to derive the equations of motion in section \ref{chapter:GS_formalism} for the GSMT string, cf. section \ref{sec:GSMT_string}. We need to consider a small variation $\xi$ of $g$, i.e. $\dl g=g \xi$, $\delta g^{-1}=-\xi g^{-1}$, which gives for the currents the expressions \eqref{first_variation_J}. 
Plugging the variations for the left-invariant currents (\ref{first_variation_J}) in the action (\ref{action_PS_total}) and using the Maurer-Cartan identities
$
\partial \bar J-\bar\partial J + [J,\bar J]=0
$,
provides the following equations of motion for the matter currents
\be
\label{classicaleom_matter}
&& \bar D J_2=[J_3,\bar J_3]+ [N,\bar J_2]- [J_2,\bar N_0]\nln
&&D \bar J_2=-[J_1,\bar J_1]+[N,\bar J_2]- [J_2,\bar N_0]
\nln
&&  \bar D J_3= [N,\bar J_3]- [J_3,\bar N_0] \cr &&
 D \bar J_3=
-[J_1,\bar J_2]-[J_2,\bar J_1]+ [N,\bar J_3]-[J_3,\bar N_0] \cr
&&  \bar D
J_1=[J_3,\bar J_2]+[J_2,\bar J_3]+ [N,\bar J_1]- [J_1,\bar N_0]\cr
&& D \bar J_1= [N,\bar J_1]-[J_1,\bar N_0] \,. 
\ee

By considering a small perturbation for the ghost fields $\delta \lambda$, $\delta \omega$ leads to the equations of motion for the ghost sector:
\be
&&\bar D \lambda_1 - [\bar N_0, \lambda_1]=0
\qquad
\bar D \omega_{3+}- [\bar N_0, \omega_{3+}]=0
\nln
&& D \lambda_3 -[ N_0, \lambda_3]=0
\qquad
 D \omega_{1-}-[ N_0, \omega_{1-}]=0\,.
\ee
From the definition of the Lorentz ghost currents \eqref{def_N} and from the above equations it follows
\be
\label{classicaleom_N}
\bar D N_0+ [N_0, \bar N_0]=0
\qquad 
 D \bar N_0+ [\bar N_0,  N_0]=0\,. 
 \ee

\paragraph{BRST transformations.}

In the context of integrability, a crucial role is played by the BRST operator. 
In the curved $\ads_5 \times\sphere^5$ background it is given by
\be
Q=Q_L+Q_R= \oint \rm{STr} \big(\lambda_1 J_3+\lambda_3\bar J_1\big)\,,
\ee
namely it is made by a right and a left-moving BRST operator, $Q_L=\lambda_1 J_3$ and $Q_R=\lambda_3\bar J_1$.
The BRST operator $Q$ acts by right-multiplication on the coset representative $g(x,\theta,\hat \theta)$~\cite{Berkovits:2004jw}, and the infinitesimal BRST transformations for g are
\be
\label{BRST_transf_g}
\epsilon Q (g)= g\big(\epsilon \lambda_1 +\epsilon\lambda_3\big)
\qquad
\epsilon Q (g^{-1})= -\big(\epsilon \lambda_1 +\epsilon\lambda_3\big) g^{-1}\,,
\ee
where $\epsilon$ is an anticommuting parameter introduced for convenience, since $\lm_1$ and $\lm_3$ are anticommuting bosons. 
For the matter currents it implies 
\be
\label{BRST_transf_J}
&& \eps \,Q (J_m)=\dl_{m+3,0}\partial (\eps\lm_1)+[\, J_{m+3}\,,\, \eps\, \lm_1\, ]+\dl_{m+1,0}\partial (\eps\lm_3)+[\, J_{m+1}\,,\eps\,\lm_3\,]\nln
&& \eps \,Q (\bar J_m)=\dl_{m+3,0}\bar\partial (\eps\lm_1)+[\, \overline J_{m+3}\,,\, \eps\, \lm_1\, ]+\dl_{m+1,0}\bar \partial (\eps\lm_3)+[\, \bar J_{m+1}\,,\eps\,\lm_3\,]\,,\nln
\ee
where we have used the definitions of the MC-currents, the relations \eqref{BRST_transf_g} and then the projection on $\mathfrak{g}_m$, with $m=0,\dots,3$. 

The ghost fields transform under BRST transformations according to~\cite{Berkovits:2004jw}
\be
\label{BRST_transf_PS}
\epsilon Q(\lambda_1)=\epsilon Q(\lambda_3)=0
\qquad 
\epsilon Q(\omega_{3+})=-J_3\epsilon
\qquad
\epsilon Q(\omega_{1-})=-\bar J_1\epsilon\,. 
\ee
From these relations, one obtains the BRST transformations for the ghost currents%
\footnote{The ghost current BRST transformations can be computed recalling the OPE's reported at the beginning of the section, cf. \eqref{OPE_Nlambda}.},
i.e. 
\be
\label{BRST_transf_N}
\eps Q(N_{0})\,= \,[\,J_3,\,\eps\,\lm_1\,]
\qquad 
\eps Q(\bar N_{0})\,= \,[\,\bar J_1,\,\eps\,\lm_3\,]\,.
\ee

As mentioned above, the BRST operator must be nilpotent. The two operators $Q_L$ and $Q_R$ are nilpotent thanks to the pure spinor constraints \eqref{new_ps_constraints}. However Q is nilpotent only up to gauge transformations. Using the PS constraints \eqref{new_ps_constraints}, one can check that 
\[
Q^2 (g)= -g\left\{\lm_1,\lm_3\right\}\,. 
\]
$\left\{\lm_1,\lm_3\right\}$ belongs to the $\mathfrak{g}_0$ sub-algebra, i.e. $\grSO(4,1)\times \grSO(5)$, and thus it parameterizes a gauge transformation. 
As an example, computing the squared BRST transformation for $J_2$%
\footnote{
We have used the pure spinors constraints \eqref{new_ps_constraints} as well as the Jacobi identity
\be
\big\{\big [ J,\lm_{1 (3)}\big ],\lm_{1 (3)}\big\} - \big\{\big [ \lm_{1 (3)}, J \big ],\lm_{1 (3)}\big\} +\big [ \big \{ \lm_{1 (3)}, \lm_{1 (3)}\big\}, J\big ]=0
\ee
which implies that $ \big\{\big [ \lm_{1 (3)}, J \big],\lm_{1 (3)}\big\}$ vanishes.},
one gets
\be
Q^2 (J_2)= - \left[\left\{ \lm_1,\lm_3\right\} ,J_2\right]\,.
\ee
With the same procedure one can compute 
\be
&& Q^2 (N_0)= -\left \{ \lm_1, D\lm_3 -\left [N_0,\lm_3\right]\right\} -\left [N_0, \left \{\lm_1, \lm_3\right\}\right ]
\nln
&& Q^2 (\bar N_{0})= -\left \{ \lm_3, \bar D\lm_1-\left [\bar N_{0},\lm_1\right]\right\} -\left [\bar N_{0}, \left \{\lm_1, \lm_3\right\}\right ]\,.
\ee
Hence, the BRST operator is nilpotent up to classical equations of motion and up to gauge transformations parameterized by $\left\{\lm_1,\lm_3\right\}$~\cite{Berkovits:2004xu}. This is consistent because all the action is invariant under transformations generated by $\grSO(4,1)\times\grSO(5)$. 


\paragraph{The classical BRST and gauge invariance.}

%
\begin{itemize}

\item The action is BRST invariant at classical level. In particular this can be easily shown by applying the BRST transformations (\ref{BRST_transf_J} -- \ref{BRST_transf_N}) to the action (\ref{action_PS_total}). Then the BRST variation coming from the purely matter sector is
\be 
\dl_Q S_m\equiv \eps Q (S_m) = \rm{STr} \left ( \bar J_1 D(\eps\lm_3)+J_3 \bar D (\eps \lm_1)\right)
\ee
which is exactly canceled by the BRST variation of the ghost sector
\be
\dl_Q S_g\equiv \eps Q(S_g) = - \rm{STr} \big(   \bar J_1 D(\eps\lm_3)+J_3 \bar D (\eps \lm_1)\big)\,. 
\ee
\item As already discussed the action is classically gauge invariant, by construction for the matter sector and by covariantization for the ghost sector. 
\end{itemize}


\paragraph{The quantum gauge and BRST invariance.}

We need to consider if these properties survive at quantum level. We want to discuss quantum integrability for type IIB string on $\ads_5\times \Sphere^5$, thus we need to consider whether the quantum PS superstring action is consistent. The statements in~\cite{Berkovits:2004xu} and~\cite{Berkovits:2004jw} are that

\begin{itemize}
\item
The PS action \eqref{action_PS_total} is gauge invariant at quantum level;
\item The PS action \eqref{action_PS_total} is BRST invariant at quantum level.
\end{itemize}
It is worth giving some detail on how this has been shown in \cite{Berkovits:2004xu}.
I will discuss the gauge invariance first and then the BRST invariance.

If there is an anomaly at quantum level, namely if the gauge invariance is broken quantum mechanically it means that there exists a local operator which generates such anomaly. This operator should be local, since the anomaly comes from the short distance behavior of some operator that quantum mechanically becomes ill-defined, cf. section \ref{sec:quantum_int}.
Hence, one can proceed with an engineering construction of such generic operator. Since it is local, it should vanish for global transformations; since it is responsible for the gauge symmetry breaking, it should be in the sub-algebra $\mathfrak{g}_0$.
Then the ansatz is \cite{Berkovits:2004xu}
\be
\dl_\Lm S= \rm{STr}\left( \alpha N_{0}\bar\partial \Lm+\bar \alpha \bar N_{0}\partial \Lm +\beta J_0\bar \partial \Lm+\bar\beta \bar J_0\partial \Lm\right)
\ee
where $\left (\alpha, \bar\alpha,\beta,\bar\beta\right)$ are some arbitrary coefficients and $\Lm$ parameterizes the $\grSO(4,1)\times \grSO(5)$ gauge transformations. Proposing a possible counter-term \cite{Berkovits:2004xu} such as
\be
S_c= - \rm{STr}\big ( \alpha N_{0}\bar J_{0} +\bar\alpha \bar N_{0} J_{0} +\Half (\beta+\bar\beta) J_0\bar J_0\big)
\ee 
is possible to cancel partially the anomaly, and the remaining terms, namely 
\be
\dl_\Lm \left ( S+S_c\right) = \Half \left( \beta-\bar \beta\right) \rm{STr} \left (J_0\bar\partial\Lm-\bar J_0\partial \Lm\right)
\ee
vanish due to the non-perturbative symmetry which exchanges right and left-moving \emph{and} bar and unbar coordinates in the world-sheet, cf. \eqref{parity_symmetry}, and which, in this case, constraints to have $\beta=\bar\beta$. 

The quantum BRST invariance of the action (\ref{action_PS_total}) has been shown in \cite{Berkovits:2004xu}, and the arguments proceed analogously. One constructs an ansatz for the anomalous local operator. In order to relate the terms and thus to reduce the possible linear combination, one can use the classical equations of motion and the Maurer Cartan identities. 
However, one needs to keep in mind that the anomalous terms should be a gauge invariant local ghost number 1 operator. Again, local is due to the short-distance behavior of the operators, gauge invariant since the gauge and BRST transformation commute and finally ghost number 1 since it is a variation generated by the BRST operator. The required properties restrict the possibilities for the coefficients in the linear combination. In this way it is possible to find a local counter-term which exactly cancels the variation. Thus the quantum effective action is BRST invariant. 

There are some points to notice. First, the use of the classical equations of motion and the fact that the BRST operator, as well as the BRST transformations, are always the classical one. Second, since the BRST variation of the effective action can be written as a BRST variation of suitable counter-terms, this  means that the BRST cohomology of gauge invariant local ghost-number 1 operators is trivial, namely they can always be written as a BRST variation of some suitable operator. In this way the BRST transformation of the total action, given by the effective quantum terms plus the counter-terms, is zero~\cite{Berkovits:2004jw, Berkovits:2004xu}. 

This was at the first order in perturbation theory. However the arguments can be extended by induction at any order in perturbation theory \cite{Berkovits:2004xu}. The basic idea is that if one has proved that the effective action is BRST invariant up to order $h^n$, then a possible anomaly would be generated by a local operator of the same type before. Using the fact that the BRST cohomology for such operators is trivial, proves the BRST invariance up to $h^{n+1}$ order, and thus one can go on by induction. Let me stress that we concretely use the classical BRST operator, the classical equations of motion and Maurer-Cartan identities.%
\footnote{
In this reasoning there is indeed some caveat. I will try to explain briefly remanding the reader to \cite{Berkovits:2004xu} for more detailed explanations. The argument works if there are no conserved currents of ghost number 2. Such currents indeed can spoil the nilpotency of Q, since Q has ghost number 1, thus $Q^2$ has ghost number 2 and the existence of some charges of ghost number 2 would in principle generate an anomaly in the nilpotency of the quantum operator Q. However such currents are not present~\cite{Berkovits:2004xu}, implying that Q remains nilpotent at quantum level.}
%

\paragraph{The quantum conformal invariance.}

The action \eqref{action_PS_total} is conformal invariant at quantum level. By means of the background field method (cf. section~\ref{sec:papersPS}) this has been shown to one loop in perturbation theory~\cite{Vallilo:2002mh} and by cohomology arguments to all orders~\cite{Berkovits:2004xu}%
\footnote{The quantum conformal invariance of the pure spinor superstring has been showed also for generic curved backgrounds and for the heterotic string~\cite{Chandia:2003hn, Bedoya:2006ic}.}.
%




\subsection{Classical integrability of the $\ads_5 \times \mbox{S}^5$ PS superstring action}

The classical integrability has been proved by Vallilo in \cite{Vallilo:2003nx} by using the same approach of Bena et al. for the GSMT action \cite{Bena:2003wd}. 
The same Lax pair has been found by Berkovits requiring that the higher charges should be BRST invariant~\cite{Berkovits:2004jw}. The integrability at classical level of the pure spinor action in generic $\ads_n\times \sphere^n$ backgrounds has been studied in~\cite{Adam:2007ws}.

Recall from section \ref{chapter:Intro_integrability} that the existence of a flat connection $a$, namely a connection whose field strength identically vanishes, allows us to construct a not-deformable Wilson-like operator (the monodromy matrix). Its path independence assures the conservation of the corresponding charges. 
Hence, one would like to extend the analysis of Bena et al. to the PS formulation of the $\ads_5\times\sphere^5$ action. 

The zero-curvature equations in the $z,\bar z$ coordinates reads
\[
\label{a_flat_ps}
\p \bar a -\bar \p a - \left[ a, \bar a \right]=0\,.
\]
However it is simpler to work with the left-invariant currents, since they have a well-defined grading. Using $A= - g^{-1} a g$, the flatness condition \eqref{a_flat_ps} becomes
\[
\label{A_flat_ps}
\p \bar A -\bar \p A + \big[ A, \bar A \big] + \big [  J,\bar A\big] + \big[ A, \bar J \big] =0\,,
\]
where $J$ are the MC-currents $J=J_0+\sum_{i=1}^3 J_i$. 

The natural ansatz for $A$ is the linear combination involving {\it all} the possible currents
\[
\label{ansatz_A}
A= \alpha J_2 +\beta J_ 1+\gamma J_3+\delta N_0\,, 
\qquad 
\bar A= \bar \alpha \bar J_2 +\bar \beta \,\bar J_ 1+\bar \gamma \,\bar J_3+\bar \delta \bar N_0\,.
\]
Notice that now also the Lorentz ghost currents participate to the proposed Lax pair. Further, now no antisymmetric combination of the fermionic currents enter, as it was for the GS formulation. The fermionic currents are treated on equal footing with the bosonic ones. 

Plugging the ansatz \eqref{ansatz_A} in the condition \eqref{A_flat_ps} and using the equations of motion \eqref{classicaleom_matter} and \eqref{classicaleom_N}, one obtains for the coefficients the following solutions
\be
\label{sol_vallilo}
&&\alpha= z-1 
\qquad \beta = \pm z^{1/2}-1
\qquad \gamma= \pm z^{3/2}-1
\nln
&&\bar \alpha=z^{-1}-1
\qquad \bar\beta=\pm z^{-3/2} -1 
\qquad \bar\gamma = \pm z^{-1/2}-1
\nln
&&\delta=(1-z^2)
\qquad 
\bar\delta= (z^{-2}-1)\,.
\ee
As it was noted by Vallilo~\cite{Vallilo:2003nx}, the system admits the same solution if we exclude the ghost contributions. Thus, at classical level, the two sectors, matter and ghost, are completely decoupled. This is not true at quantum level, as it can be seen in \cite{Vallilo:2002mh, Mikhailov:2007mr, Puletti:2008ym}. 

\paragraph{The construction of the BRST charges.}

The same result \eqref{sol_vallilo} has been found by Berkovits using a different procedure. Let me sketch this point since it sheds some light, especially in the relations between the non-local charges and the BRST operator. As it is clearly explained in~\cite{Berkovits:2004jw}, such charges are symmetries of the string and can map physical states to physical states, thus they should necessarily respect the symmetries of the  theory, namely they should be BRST invariant (and it follows for the GS formalism that there the conserved non-local charges should be $\kappa$-symmetric). 

The explicit construction of the charges for the type IIB superstring in $\ads_5\times \Sphere^5$ is based in three steps~\cite{Berkovits:2004jw}. 
First, we search for a gauge invariant current $a$, such that 
\[
\label{a_cond}
Q( a)=\partial_\sigma \Lm +[a, \Lm]
\]
for some $\Lm$.
Then, the charges given by
\[
P\left ( e^{-\int_{-\infty}^\infty d\sigma a(\sigma)}\right)
\]
are BRST invariant, since $a$ satisfies (\ref{a_cond}). In order to construct $a$ concretely, one makes an ansatz writing the most general linear combination in terms of all the currents (matter and ghost currents), i.e.
\[
\label{a_ansatz}
a= -g\left (
\delta N_0+\beta J_1+\alpha J_2+\gm J_3
+ \bar\dl \bar N_{0}+\bar\beta \bar J_1+\bar \alpha \bar J_2+\bar \gm\bar J_3\right) g^{-1}\,.
\]
Note that $J_0$ and $\bar J_0$ are not included in the list, since we want a gauge invariant object, for the same reason $a$ is written as a rotation of the left-invariant currents, recall section \ref{chapter:Intro_integrability}. 
First we act with the BRST operator Q on $a$ (\ref{a_ansatz}), and then we impose that $Q(a)$ obtained in this way satisfies (\ref{a_cond}) where $\Lm$ is 
\[
\label{def_Lambda}
\Lm= g\left ( b \lm_1 +\bar b\lm_3\right) g^{-1}\,. 
\]
These constraints fix the coefficients only to certain values. The specific solutions are the same as those found by Vallilo \eqref{sol_vallilo}. Moreover, the remaining coefficients $b$ and $\bar b$ are
\be
b= \pm z^{\Half}-1
\qquad
\bar b =\pm z^{-\Half} -1\,.
\ee
The expansion around the value $z=1$ gives back the first global charge. Namely, for the matter sector is
\[
q\cong (z-1) \int d\sigma j+ \mathcal O(z^2) =  (z-1)\int d\sigma \left( \Half j_1+ j_2+ {3\over 2} j_3 \right)+ \mathcal O(z^2) \,,
\]
with $j=-g J g^{-1}$. 
This is the explicit construction of the charges. However, their existence is related to the fact that the classical BRST cohomology does not contain ghost number 2 states, namely that such states can always be written as BRST variation of certain operators. This is indeed the ultimate condition that guarantees the existence of the higher charges. 


\subsection{Quantum BRST charges and quantum monodromy matrix}

The arguments presented in the previous section are classical. One needs to implement such arguments at quantum level. This has been done in \cite{Berkovits:2004xu} at any order in perturbation theory. 
The argument runs essentially as before. Suppose that we have certain BRST invariant charges at order $h^n$ in perturbation theory, then $\tilde Q(\tilde k^C)= h^{n+1} \Om^C+\mathcal O(h^{n+2})$, where $\tilde Q$ is the BRST operator that generates the classical BRST transformations and their quantum corrections, while $\Om^C$ is some generic integrated local ghost number 1 operator. 
Since the BRST cohomology is trivial for such operators $\Om^C$, namely for local integrated ghost number 1 operators~\cite{Berkovits:2004jw, Berkovits:2004xu}, then it can be always written as a BRST variation of something, namely it can be written as $\Om^C=Q(\int_{-\infty}^\infty d \sigma \Sigma^c(\sigma))$, which means that $\tilde k^c-h^{n+1}\int_{-\infty}^\infty d \sigma \Sigma^c(\sigma)$ is BRST invariant up to order $h^{n+1}$.

\paragraph{Finiteness of the monodromy matrix at the leading order.}

We have discussed until now about the existence of non-local charges and their BRST invariance at quantum level. Nevertheless, this does not tell us whether such quantities remain well defined quantum mechanically! Are these charges finite?

The question is very far from being trivial, since there are examples in which the bilocal charges are not finite and they need to be regularized, cf. section~\ref{sec:quantum_int}. In the pure spinor approach, the question has been initially investigated by A. Mikhailov and S. Schafer-Nameki~\cite{Mikhailov:2007mr}. Indeed what they have explicitly shown is that the monodromy  matrix is well-defined at the leading order in perturbation theory: it does not get renormalized and all the divergences that can pop-up cancel. They have found different types of divergences, namely divergences that go like ${1\over \eps}$ (linear divergences) and logarithmical divergences ($\log\eps$). In a perturbative quantum field theory, the first ones depend on the regularization scheme adopted, while the second ones are independent on the scheme and must be cancelled, also in order to have a consistent quantum conformal invariance. 
Indeed, suppose to have two contours $\mathcal C$ and $\mathcal{C'}$ related by a conformal transformation, namely $\mathcal{C'}=\lm\mathcal C$. Then the monodromy matrices along the two paths have divergences that should be regularized. The independence on the contour and hence the conformal invariance of the monodromy matrices implies that $\Om^{reg}[\mathcal{C}]=\Om^{reg}[\mathcal{C'}]$. On the other side one has that $\Om^{reg}[\mathcal{C}]= \lim_{\eps\rightarrow 0}\left (\Om_\eps [\mathcal{C}]+ C_\eps [\mathcal{C}]\right)$ and by definition $\Om_\eps[\mathcal{C}]=\Om_{\lm\eps}[\mathcal{C'}]$. This forces to have then $\lim_{\eps\rightarrow 0} C_\eps [\mathcal{C}] =\lim_{\eps\rightarrow 0} C_{\lm\eps} [\mathcal{C'}]$ which is not true for the case of logarithmic divergences \cite{Mikhailov:2007mr}.


\subsection{Quantum Integrability}
\label{sec:papersPS}

We go on following the issue about the finiteness of the conserved charges.
We have already explained in section \ref{chapter:Intro_integrability} that the independence on the contour for the monodromy matrix $\Om$ is equivalent to the conservation of the charges. Thus our goal is to move at quantum level and check that the independence on the contour and the zero-curvature equation still yield~\cite{Puletti:2008ym}.

How do we proceed? 
In the first part we show that there cannot exist an anomaly in the deformation of the contour for the monodromy matrix. This is done by using techniques analogous to the ones explained in Berkovits' papers. In the second part we explicitly compute the field strength (\ref{def_F}) and show that all the logarithmic divergent terms disappear to first order in perturbation theory.


\subsubsection{Absence of anomaly}

Before proceeding, I will summarize some of the basic ``ingredients" presented in the previous part of the section. 
Recall that the Lax pair is
\footnote{Note that in Vallilo's notation $\mathcal J$ is given by $J+A$. Here, we use a slightly different parameterization for the one-parameter family of flat connections with respect to the one presented in \cite{Vallilo:2003nx}, cf. \eqref{sol_vallilo}. }
%
\be
\label{my_flatconn}
&& \mathcal J(z)= J_0 +z J_2 + z^{{1\over 2}} J_1 +z^{{3\over 2}} J_3 + (z^2-1) N\nln
&& \bar{\mathcal{J}} (z) = \bar J_0 + {1\over z} \bar J_2 +{1\over z^{{3\over 2}}} \bar J_1 + {1\over z^{{1\over 2}}} \bar J_3 +({1\over z^2}-1)\bar N\,.
\ee
From the BRST transformations for the currents (\ref{BRST_transf_J}) and (\ref{BRST_transf_N}), we can read how the Lax pair varies under the Q action:
\be
&& \eps Q(\mathcal{J})= \partial \left (z^{-\Half} \eps \lm_3+z^{\Half} \eps \lm_1\right) + [\,\mathcal{J}, \, z^{-\Half} \eps \lm_3+z^{\Half} \eps \lm_1]\nln
&& \eps Q(\bar{\mathcal{J}})= \bar \partial \left (z^{-\Half} \eps \lm_3+z^{\Half} \eps \lm_1\right) + [\,\bar{\mathcal{J}}, \, z^{-\Half} \eps \lm_3+z^{\Half} \eps \lm_1]\,,
\ee
where notice that $z^{-\Half} \eps \lm_3+z^{\Half} \eps \lm_1$ is nothing but what we have called $\Lm$ in (\ref{def_Lambda}). The field strength is 
\be
\label{def_F}
\mathcal F^{(1,1)}(z)\equiv \partial \bar{\mathcal{J}}-\partial\mathcal{J} +\left [\,\mathcal{J},\,\bar{\mathcal{J}}\,\right ]
\ee
and it satisfies
\be
\eps Q\left( \mathcal F^{(1,1)}(z)\right) = \left[ \mathcal F^{(1,1)}(z), z^{\half} \lm_1 +z^{-\half} \lm_3\right]\,. 
\ee

Using the equations of motion (\ref{classicaleom_matter} - \ref{classicaleom_N}) as well the Maurer Cartan identities, one can easily show that indeed the Lax pair with components $\mathcal J$ and $\bar{\mathcal{J}}$ given above does satisfy the zero-curvature equation at classical level, i.e. that the field strength 
vanishes 
\be
\mathcal F^{(1,1)}(z)=0\,.
\ee

Let us now investigate the relation between the monodromy matrix and the world-sheet path \eqref{def_var_U}, which I rewrite here for convenience:
\[
\label{def_var_Omega}
{\dl \over \dl x^\mu(s)}\Om ={\rm P} \left ( \mathcal{F}_{\mu\nu} \dot{x}^\nu e^{\oint_{\mathcal{C}} \mathcal J(s)}\right)\,.
\]
Fix a point along the path $\mathcal C$ and consider an infinitesimal deformation on $\mathcal C$, i.e. $x^\mu(s)\rightarrow x^\mu(s)+\dl x^\mu(s)$. Since the deformation is really small, the ``disturbance'' in this $\eps$ path is represented by some operators $\mathcal O$ sitting on it. At higher and higher energies these operators can interact and produce divergences which spoil the contour independence of the monodromy matrix. 

Let us try to engineering construct $\mathcal O$ and then we will see that such an operator cannot indeed exist. 
$\mathcal O$ should be 
\begin{enumerate}
\item local, since as explained we are worried about the short-distance behavior of the currents which are operators and could produce UV divergences;
\item gauge invariant;
\item by dimensional analysis it is expected to have conformal dimension (1,1), this can be seen already in (\ref{def_var_Omega});
\item we have also seen that the charges are BRST invariant, namely that the Wilson loop is BRST invariant classically and quantum mechanically. This implies that $\mathcal O$ should transform according to
\[
\label{BRST_O}
\eps Q (\mathcal{O}^{(1,1)})= \left [\, \mathcal{O}^{(1,1)},\, z^{\Half} \lm_1 +z^{-\Half} \lm_3\,\right]\,,
\]
which corresponds to ask for the BRST closure of $\mathcal O$;
\item finally the operator should have ghost number zero, which follows from the equation (\ref{def_var_Omega}). 
\end{enumerate}
At this point we can write the most general linear combination satisfying the properties from 1)  to 5). Notice that the BRST closure (\ref{BRST_O}) implies that the matter currents $J_1$ and $\bar{J}_3$ are not present in the possible list, because their BRST transformations (\ref{BRST_transf_J}) contain derivatives of ghosts which cannot be reabsorbed by the equations of motion. Moreover the point 2) leads to exclude the gauge currents $J_0$ and $\bar{J}_0$. 
The ansatz for the operator $\mathcal{O}^{(1,1)}$ has been given in~\cite{Puletti:2008ym}, namely
\be
\label{ansatz_O}
\mathcal{O}^{(1,1)}(z) &=& 
A^{2+,2-}(z) [J_2,\bar J_2]
+A^{1+,3-}(z) [J_1,\bar J_3]
+A^{2+,3-}(z) [J_2,\bar J_3]\nln
&+&A^{1+,2-}(z) [J_1,\bar J_2]
+A^{0+,2-}(z) [N_0,\bar J_2]
+A^{2+,0-}(z) [J_2,\bar N_0]\nln
&+&A^{1+,0-}(z) [J_1,\bar N_0]
 +A^{0+,3-}(z) [N_0,\bar J_3]
+A^{0+,0-}(z) [N_0,\bar N_0]\,.\nln
\ee
The coefficients $A$ are arbitrary functions of the spectral parameter $z$ and they are of order $h$, using Berkovits terminology. All the other possible terms are related by classical equations of motion and Maurer-Cartan identities. 
We have to impose the relation (\ref{BRST_O}) to $\mathcal{O}^{(1,1)}(z)$. This is indeed the most strict requirement on $\mathcal{O}^{(1,1)}(z)$ and from this constraint eventually follows the non-existence of such operator $\mathcal{O}^{(1,1)}(z)$: The system of equations for the unknowns $A$ admits only the trivial solution.
Since we have proven that there are no operator obeying to the properties 1) - 5), this excludes the possibility to have an anomaly in the contour deformation of the quantum monodromy matrix.

Actually, by using Berkovits arguments and by recalling that the non-local charges have been proven to be BRST invariant to all orders in perturbation theory, we can extend the validity of our argument to any $n$-loop order ($h^n$).

In some sense order by order in the quantum theory the BRST symmetry fixes the contour in such a way that any small deformation in the path will not produce any anomaly in the monodromy matrix. This is because is really the constraint (\ref{BRST_O}) to rule out the possibility to have an anomaly. This is quite different from the case of quantum $\C P^n$ models~\cite{Abdalla:1980jt}, where there is no such a ``constraining'' symmetry that prevents the model from an anomaly.

\paragraph{Finiteness of the monodromy matrix to all orders.}

Finally, let us to comment about another implication. The authors of \cite{Mikhailov:2007mr} have argued that the independence of the contour for the monodromy matrix leads necessarily to the cancellation of the logarithmically divergent terms in the quantum monodromy matrix. 
Consequently the arguments presented in~\cite{Puletti:2008ym} indicate that since the monodromy remains independent of the contour to all orders in perturbation theory then it is also finite, or better, it is free from logarithmic divergences to all loops.


\subsubsection{The operator algebra}

Our aim in this section is to show and to explain how to proceed with explicit one-loop computations in the pure spinor formalism. In particular we want to explain the computations of the current OPE's and the field strength (\ref{def_F}) and we want to show that $\mathcal F$ is free from logarithmic divergent terms. 
The operator algebra has been derived in~\cite{Puletti:2006vb, Puletti:2008ym} at the leading order.%
\footnote{
The OPE's of the matter current at leading order ${1\over R^2}$ and up to linear term in the currents have been computed in~\cite{Puletti:2006vb}. Such tree-level results were then confirmed in \cite{Mikhailov:2007mr}. A very similar problem was faced in \cite{Bianchi:2006im} by using a Hamiltonian approach. Successively, the OPE's for matter and ghost currents, still at the leading order in perturbation theory, i.e. ${1\over R^2}$, but containing up to contributions quadratic in the currents (up to $V_2$-like insertion or the ``square'' of $V_1$-vertices), have been computed in~\cite{Puletti:2008ym}.}

Since the world-sheet currents are not holomorphic or anti-holomorphic, it is not possible to derive the OPE's by symmetry considerations. They have to be computed perturbatively. The OPE results show indeed the non-holomorphicity of the currents but also that the $\Z_4$-grading of the $\algPSU(2,2|4)$ algebra is preserved. 

Let me sketch the procedure. 
The method used is the background field method~\cite{Berkovits:1999zq, Vallilo:2002mh}, which means that the fields are expanding around a classical solution. The quantum fluctuations around the classical background interact and give rise to new effective interactions. %
\begin{enumerate}
\item
We write each field $\Phi$ as
\[
\Phi = \Phi_{cl}+ \Phi_q\,.
\]
In particular, the group-valued map $g$ is expanded in quantum fluctuations $X$ around a classical solution $\tilde g$, namely 
\[
g= \tilde g e^{\gamma X}\,, \quad \text{with} \,\, X\in \mathfrak{g/g_0}\,,
\]
where $\gamma$ is the parameter of the expansion, namely the (inverse of the) coupling constant in front of the action in (\ref{action_PS_total}). This means that we are considering the limit
\[
R\rightarrow \infinity\,, \quad \text{or equivalently} \quad \gamma\rightarrow 0\,.
\]
The gauge invariance of the (super) coset space can be used to fix the fluctuations in $\mathfrak{g/g_0}$. Hence from the definition of the currents $J=g^{-1}dg$ one can compute their expansion in terms of the fields $X$, i.e.
\be
\label{expansion_J}
&& J_i= \tilde J_i+ 
\gamma \big( \partial X_i+ [\tilde{J}, X]_i \big)
+{\gamma^2\over 2} \big ( \big[ \partial X, X\big]_i +\big[\big[ \tilde{J},X\big],X\big]_i \big)
+\mathcal{O}(\gamma^3)\nln
&& J_0= \tilde J_0+ 
\gamma \big[\tilde{J}, X\big ]_0
+{\gamma^2\over 2} \big (\big[ \partial X, X\big]_0 +\big[ [ \tilde{J},X ],X\big]_0\big)
+\mathcal{O}(\gamma^3)
\ee
where the subscript $i$ denotes the projection into $\mathfrak{g}_i$ and its values are $i=1,2,3$. $\tilde{J}$ is the classical current, i.e. $\tilde J=\tilde{g}^{-1}d \tilde{g}$. The analogous expansion \eqref{expansion_J} holds for the bar components of the currents, with the obvious substitutions $\p\rightarrow \bar\p$ and $\tilde J\rightarrow \bar{\tilde{J}}$.
The same method can be applied to the ghost fields \cite{Vallilo:2002mh, Chandia:2003hn, Bedoya:2006ic}, 
\be
\omega_{3+} \rightarrow \tilde{\om}_{3+}+\gamma \,\om_{3+}
\qquad
\lm_1\rightarrow \tilde{\lm}_1+ \gamma\,\lm_1
\nln
\omega_{1-} \rightarrow \tilde{\om}_{1-}+\gamma\, \om_{1-}
\qquad
\lm_3\rightarrow \tilde{\lm}_3+ \gamma\,\lm_3
\ee
which means that the Lorentz ghost currents transform according to the following expressions
\be
\label{expansion_N}
&& N_0=\tilde N_0+\gamma\, N_0^{(1)} +\gamma^2\, N_0^{(2)}
\nln
&& \bar N_0= \tilde{\bar{N}}_0 +\gamma\, \bar{N}^{(1)}_0 +\gamma^2\, \bar{N}^{(2)}_0\,,
\ee
with 
\be
 && N_0^{(1)}= -\{\om_{3+},\tilde\lm_1\}-\{ \tilde\om_{3+}, \lm_1\}
\qquad
N_0^{(2)}= -\{\om_{3+},\lm_1\}
\nln
&&
 \bar{N}^{(1)}_0= -\{\om_{1-},\tilde\lm_3\}-\{ \tilde\om_{1-}, \lm_3\}
\qquad
\bar{N}^{(2)}_0= -\{\om_{1-},\lm_3\}\,.
\ee
\item
We plug \eqref{expansion_J} and \eqref{expansion_N} in the action \eqref{action_PS_total}, we obtain an effective action%
\footnote{The contribution to the effective action denoted with the letter $\beta$ denotes the ones computed also by Vallilo \cite{Vallilo:2002mh} for the $\beta$-function, while all the other ones have been computed in~\cite{Puletti:2008ym}.},
which gives us the new Feynman diagrams. What is really interesting are the terms quadratic in the quantum fluctuations, $\Phi_q$, since they will give us the diagrams which correct the two-point functions. Explicitly for the matter sector, we have
\[
S_M=S_{M; 0}+ S_{M;\bt}+ S_{M;2}
\]
where $S_{M;0}$ is the classical matter action~\eqref{Matter_action_PS}, $S_{M;\bt}$ is the effective action for the matter contribution used for computing the one-loop $\beta$-function in~\cite{Berkovits:1999zq} and in \cite{Vallilo:2002mh}, while $S_{M;2}$ contains the off-diagonal terms:
\be
\label{action_matter_beta}
    S_{M;\bt} &=& \frac{1}{\pi} \int d^2z \;\textrm{Str}
        \big(
             \bar{\partial} X_1 \p X_3 + \half \bar\partial X_2 \p X_2  \nln
            &-& [\p X_2, X_3] \bar J_3- [\bar\p X_2, X_1] J_1
            - \frac{1}{2}[\p X_3, X_3] \bar J_2-\frac{1}{2} [\bar\p X_1, X_1] J_2 \nln
            &+& \frac{3}{4}[[\bar J_1, X_3], X_1] J_3 + \frac{1}{2}[[\bar J_1, X_2], X_2] J_3 +\frac{1}{4}[[\bar J_1, X_1],X_3] J_3  \nln
            &+&\frac{1}{2}[[\bar J_2, X_2],X_2] J_2 + \frac{1}{4}[[\bar J_2, X_1], X_3] J_2 -\frac{1}{4} [[\bar J_2, X_3], X_1] J_2 \nln
           & -&\frac{1}{4}[[\bar J_3, X_3], X_1] J_1 -\frac{1}{2}[[\bar J_3, X_2], X_2] J_1+  \frac{1}{4}[[\bar J_3, X_1], X_3]J_1
        \big)\nln
\ee
\be
\label{double-ins-mm}
    S_{M;2} &=& \frac{1}{\pi} \int d^2z \;\textrm{Str}
        \big(
         \frac{1}{2} [[\bar J_3,X_1],X_1]J_3+ \half [[\bar J_1,X_3],X_3]J_1 \nln
        &+& \frac{5}{8}[[\bar J_2,X_2],X_1]J_3+ \frac{3}{8}[[\bar J_2,X_1],X_2] J_3
        +\frac{3}{8}[[\bar J_1,X_2],X_3]J_2 \nln
        &+&\frac{5}{8}[[\bar J_1,X_3],X_2]J_2
      - \frac{3}{8}[[\bar J_3,X_2],X_1]J_2 +\frac{3}{8}[[\bar J_3,X_1],X_2]J_2\nln
    &  -&\frac{3}{8}[[\bar J_2,X_3],X_2]J_1+\frac{3}{8}[[\bar J_2,X_2],X_3]J_1
      \big)\,.
\ee
For the ghost sector one has
\[
S_{G}= S_{G;0}+ S_{GM;\bt}+ S_{GM;2}+S_{GM;3}+S_{G;2}\;.
\]
$S_{G;0}$ is the classical ghost action~\eqref{def_PS_IIB_action}, $S_{GM;\bt}$ contributes to the one-loop $\bt$-function~\cite{Vallilo:2002mh}
\be
\label{three-leg-gm}
S_{GM;\bt} &=& \frac{1}{2\pi} \int d^2 z\; \textrm{Str}\big(
                        N_0 [\bar{\partial} X_3, X_1] +  N_0 [\bar{\partial} X_2, X_2] + N_0 [\bar{\partial} X_1, X_3]\nln
                        &+& \bar N_0 [\partial X_3, X_1] + \bar N_0 [\partial X_2, X_2] + \bar N_0 [\partial X_1, X_3] \big)\;,
\ee
and further contributions are contained in
\be
\label{double-ins-mg}
S_{GM;2} &=& \frac{1}{2\pi} \int d^2z\;\textrm{Str}
           \big(
           N_0 [[\bar J_3,X_3],X_2]+  N_0 [[\bar J_3,X_2],X_3]+ \bar N_0 [[J_{2},X_1],X_1]  \nln
           &+& N_0 [[\bar J_2,X_3],X_3] + N_0 [[\bar J_1,X_1],X_2] +N_0 [[\bar J_1,X_2],X_1] \nln
           &+& N_0 [[\bar J_2,X_1],X_1] +\bar N_0 [[J_{3},X_3],X_2]+  \bar N_0 [[J_{3},X_2],X_3] \nln
          & +& \bar N_0 [[J_{2},X_3],X_3] +\bar N_0 [[J_{1},X_1],X_2] + \bar N_0 [[J_{1},X_2],X_1]   \big) \;, 
          \ee
\be
\label{4-leg-vertex-matter-ghost}
S_{GM;3} & =& \frac{1}{\pi} \int d^2z\; \textrm{Str} \big(
          - N_0^{(1)} ([\bar J_3,X_1]+[\bar J_1,X_3]+[\bar J_2,X_2])\nln
                    &-& \bar N_0^{(1)}([J_{3},X_1]+[J_{1},X_3]+[J_{2},X_2])
          \big)\;,
\ee
\[
\label{eff-action-gg}
S_{G;2}= -\frac{1}{\pi} \int d^2z\; \textrm{Str} \big(
           N_0^{(1)}\bar N_0^{(1)}\big)\;.
\]
$S_{G;2}$ is responsible for the interaction between the two types of ghost currents, so we will have also a non-zero OPE between $N$ and $\bar N$.%
\footnote{
In principle the effective one-loop action can have terms such as
\be
S_{GM;4}= \frac{1}{\pi} \int d^2z\; \textrm{Str} \left(
          N_0^{(2)}\tilde{\bar{ J}}_{0} + \bar N_0^{(2)}\tilde J_{0}  \right)\;
\ee
or
\be
S_{G;4}= -\frac{1}{\pi} \int d^2z\; \textrm{Str} \left(
          N_0^{(2)} \tilde{ \bar{ N}}_0 + \bar N_0^{(2)}\tilde N_0  \right)\;,
\ee
which could correct the propagators for the ghost fields. However, since at this order such corrections are not required, we do not enter in the details for the ghost propagators.}
\item
We compute the effective propagators (or two-point functions) according to
\be
\label{inverse_prop_X}
(A+V_1+V_2)^{-1}=
A^{-1}- (A^{-1} V_1 A^{-1})+ (A^{-1} V_1 A^{-1}V_1 A^{-1})- (A^{-1}V_2 A^{-1})+\dots\,,  
\ee
where $A$ represents the kinetic operator $A\sim {1\over 2\pi} \partial\bar\partial$. $V_1$ represents the three-leg vertices with interaction terms of the type $J\cdot \partial$, such as those in \eqref{three-leg-gm} and the second line in \eqref{action_matter_beta}; $V_2$ contains the four-leg diagrams with interactions of the type $J\, J$, such as those contained in \eqref{double-ins-mm}, \eqref{4-leg-vertex-matter-ghost}, \eqref{double-ins-mg} and in the last lines of equation \eqref{action_matter_beta}. Notice that by dimensional analysis $V_1$ has conformal weight 1, while $V_2$ has conformal weight 2, this is why we truncate the expansion to these operators. 
\item Finally, it is possible to compute the current OPE's contracting the quantum fluctuations $\Phi_q$ with the propagators of the previous point \eqref{inverse_prop_X}. 
In particular for the matter currents the OPE's up to order $\gamma^2\sim {1\over R^2}$ are
\be
J^A(x) \, \bar J^B(y)
&\cong &
\langle J^A(x) \, \bar J^B(y)\rangle
+\gamma^2 \big( \langle \partial X^A(x)\bar \partial X^B(y)\rangle
+ \langle \partial X^A(x)[\tilde{\bar{J}}, X]^B(y)\rangle\nln
& +&\langle [\tilde J, X]^A(x)\bar\partial X^B(y)\rangle
+\langle [\tilde J, X]^A(x)[\tilde{\bar{J}}, X]^B(y)\rangle\big )
+...\,,
\ee
where $A$ is a $\algPSU(2,2|4)$ index. 
\end{enumerate}

If we allow ourselves to keep up to dimension-2 operators in the OPE's, as in~\cite{Puletti:2008ym}, then at order ${1\over R^2}$ the ghosts and the matter are coupled and they give rise to the following OPE's 
\be
\label{OPE_NJ}
&& N_0(x)\bar J_i(y)
\cong - {1\over R^2} \left( \langle \{ \om_{3+},\tilde\lm_1\} (x)\bar\p X_i (y)\rangle+\langle \{ \tilde \om_{3+},\lm_1\} (x)\bar\p X_i (y)\rangle\right) +\dots 
\nln
&& \bar N_0(x) J_i(y)
\cong - {1\over R^2} \left( \langle \{ \om_{1-},\tilde\lm_3\} (x) \p X_i (y)\rangle+\langle \{ \tilde \om_{1-},\lm_3\} (x)\p X_i (y)\rangle\right) +\dots \nln
\ee
\be
\label{OPE_NN}
N_0(x) \bar N_0 (y)
&\cong&  {1\over R^2} \big( \langle \{ \om_{3+},\tilde\lm_1\} (x)  \{ \om_{1-},\tilde\lm_3\} (y)\rangle \nln
					&+& \langle \{ \om_{3+},\tilde\lm_1\} (x)  \{ \tilde \om_{1-},\lm_3\} (y)\rangle 
					+ \langle \{ \tilde \om_{3+},\lm_1\} (x)  \{ \om_{1-},\tilde\lm_3\} (y)\rangle \nln
					& +& \langle \{ \tilde \om_{3+},\tilde\lm_1\} (x)  \{ \tilde \om_{1-}, \lm_3\} (y)\rangle \big) +\dots \,.
\ee
All the OPE results are listed in appendix \ref{app:OPE_results}. 

Moreover at this order ${1\over R^2}$ the currents can get renormalized, namely there are loop-diagrams that can contribute. In particular looking at the expansion (\ref{expansion_J}) one sees that the corrections at order  ${1\over R^2}$ contain two quantum fields $X$ which can be contracted. Since they are on the same point, this will give rise to one-loop diagrams, such as tadpoles or self-energy diagrams. Explicitly%
\footnote{Actually, this is true only for the currents in $\mathfrak{g}_2$.
The currents in fermionic subalgebras cannot contribute just because one would have a fermionic and bosonic index contracted together. 
}:
\[
\label{no_contribution}
{1\over R^2} \langle J^{(2)}(x) \rangle= {1\over 2  R^2} \langle \big [ \partial X, X\big] (x) \rangle + {1\over 2  R^2} \langle \big[\big [ \tilde J, X\big], X \big] (x)\rangle\,.
\]
%


\subsubsection{The field strength}

As discussed in~\cite{Puletti:2008ym}, looking at the expression (\ref{def_var_Omega}) the field strength is our prototype for the operator $\mathcal O$. However in (\ref{ansatz_O}) we mod out the redundancy coming from the equations of motion and the Maurer Cartan identities. This means that there might be operators which classically vanish on-shell and which satisfy all the requirements 1) - 5). Obviously, how it can be readily seen, the field strength (\ref{def_F}) has all these features. For this reason we have also explicitly computed the field strength at one-loop showing that all the logarithmic divergences cancel. However, we have not showed the complete vanishing of the field strength, namely that the finite terms also cancel, due to technical difficulties. 

Once we have expanded the left-invariant currents in ${1\over R^2}$, cf. \eqref{expansion_J}--\eqref{expansion_N}, the Lax pair $\mathcal J$ \eqref{my_flatconn} and the field strength $\mathcal F$ \eqref{def_F} will be also expanded consequently:
\[
\mathcal J\rightarrow \tilde{\mathcal{J}} +\gamma \mathcal{J}^{(1)}+\gamma^2 \mathcal{J}^{(2)}+\mathcal{O}(\gamma^3)
\]
\[
\mathcal F^{(1,1)}\rightarrow \tilde{\mathcal{F}} +\gamma \mathcal{F}^{(1)}+\gamma^2 \mathcal{F}^{(2)}+\mathcal{O}(\gamma^3)\,. 
\]
Notice that $ \tilde{\mathcal{J}}$ is the classical flat connection, which means that $ \tilde{\mathcal{F}}=0$. 

One can write the curvature tensor as
\[
\label{normal-order}
\mathcal{F}^{(1,1)}(z)\;=\;
:\mathcal{F}^{(1,1)}(z):
+ \sum_k C_k (\eps)  \mathcal{O}_{k}^{(1,1)}(z) \,.
\]
The symbol $:\;:$ denotes the normal ordering prescription, namely the contribution to $\F$ coming from the internal contractions in the currents~\eqref{no_contribution}, while the sum $\sum_k C_k (\eps) \mathcal{O}_k$ is the operator product expansion (OPE) which, by definition, takes into account the effects of the  operator $\Jflat\Jflat$.
Explicitly, since $\mathcal F^{(1,1)}$ is defined as in equation \eqref{def_F}, in order to compute $\mathcal F^{(2)}$, we need to consider two contributions:
\[
\label{derivatives_Jflat}
\partial \bar{\mathcal{J}}- \bar{\partial} \mathcal{J} \;= \;
:\partial \bar{\mathcal{J}} -\bar{\partial} \mathcal{J} \;:
\]
and
\be
\label{comm-J-flat}
[\mathcal{J}(x),\bar{\mathcal{J}}(y)] \; &=& \;
:[\mathcal{J}(x),\bar{\mathcal{J}}(y)]:+
f_{BC}^A \,\Jflat^B(x) \bar\Jflat^C(y)\,t_A \; =\nln
&=&\,:\,[\mathcal{J}(x),\bar{\mathcal{J}}(y)]\,:
+ \sum_k C_k (\eps)  \mathcal{O}_{k;+-} (\si) \;\;,
\ee
when $x-y\sim \eps$ and $\si\equiv\frac{x+y}{2}$.
Notice that both expressions \eqref{derivatives_Jflat} and \eqref{comm-J-flat} depend on the spectral parameter $z$. In particular, for the commutator \eqref{comm-J-flat} one has
\be
\label{comm-explicit-list}
[\Jflat,\bar \Jflat] &=&
     [J_0,\bar J_0]-[J_0,\bar N]-[N,\bar J_0] + 2[N,\bar N]
    +[J_2,\bar J_2]+[J_1,\bar J_3]+[J_3,\bar J_1]\nln 
    &+& z^{-2}\big([J_0,\bar N]-[N,\bar N]\big)\nln
    &+ & z^{2}\big([N,\bar J_0]-[N,\bar N]\big)\nln
    &+& z^{-1}\big([J_0,\bar J_2]+[J_1,\bar J_1]+[J_2,\bar N]-[N,\bar J_2]\big)\nln
    &+& z^{-3/2}\big([J_0,\bar J_1]+ [J_1,\bar N]-[N,\bar J_1]\big)\nln
    &+& z^{-1/2}\big([J_0,\bar J_3]+[J_2,\bar J_1]+[J_1,\bar J_2]+[J_3,\bar N]-[N,\bar J_3]\big)\nln
    &+& z\big([J_2,\bar J_0]+[J_3,\bar J_3]-[J_2,\bar N]+[N,\bar J_2]\big)\nln
     &+& z^{1/2}\big([J_1,\bar J_0]+[J_2,\bar J_3]+[J_3,\bar J_2]-[J_1,\bar N]+[N,\bar J_1]\big)\nln
    &+& z^{3/2}\big([J_3,\bar J_0]-[J_3,\bar N]+[N,\bar J_3]\big)\;.
\ee
The various sectors labelled by $z^s$ distinguish the different sub-algebras and thus they cannot mix.

The strategy is to calculate the contributions to \eqref{derivatives_Jflat} and \eqref{comm-J-flat} and to show the cancellation of the divergences for each different sector $z^s$. Notice that, in principle, each commutator in \eqref{comm-explicit-list} gives again two types of terms, namely each commutator in \eqref{comm-explicit-list} is written as
\be
\label{single_commutator}
&&[J (x),\bar J (y)]^A\; =\; f^A_{BC}\; J^B (x) \bar J^C (y) + :[J (x),\bar J (y)]^A:\;,\\ \nn
&&[J(x), \bar N(y)]^A =f_{B[ab]}^A J^B(x)\bar N^{[ab]}(y)+ :[J(x), \bar N(y)]^A:\;,\\ \nn
&&[\bar J(x), N(y)]^A=f_{B[ab]}^A \bar J^B(x)N^{[ab]}(y)+ :[\bar J(x), N(y)]^A:\;,\\ \nn
&&[N(x),\bar N(y)]^{[ab]}= f^{[ab]}_{[c_1d_1][c_2 d_2]}N^{[c_1d_1]}(x) \bar N^{[c_2 d_2]}(y)
                           +:[N(x),\bar N(y)]^{[ab]}:\;,
\ee
where again all the first terms are computed from the OPE's while the second is the normal ordered commutator which contributes with terms as in \eqref{no_contribution}.

Finally, summing all the contributions illustrated in this section, and using the OPE results listed in appendix \ref{app:PS}, it has been possible to show that indeed the one-loop field strength $\mathcal{F}^{(2)}$ is free from UV divergences~\cite{Puletti:2008ym}.


\section{$\ads_5/\CFT_4$ as a 2d particle model and the near-flat-space limit}

\label{chapter:NFS}


\subsection{Introduction}

The integrable structures found on both sides of the correspondence allow one to treat the planar AdS/CFT as a two-dimensional particle model. On the gauge theory side, this is due to the correspondence between the $\mathcal N=4$ SYM theory and the one-dimensional spin chain, in particular it follows from the identification between the dilatation operator and the spin chain Hamiltonian, cf. section \ref{chapter:AdSIntro}. We can treat the scatterings of the impurities in the spin chain as collisions among (1+1) dimensional particles and consider the S-matrix for describing all the relevant kinematical observables. In particular, the integrability of the model ensures that each magnon only scatters with another one each time (S-matrix factorization). 

What about the string theory side? There we have a two-dimensional world-sheet description for closed strings in AdS backgrounds. We need to identify which are the elementary excitations \emph{of} the world-sheet which correspond to the spin chain magnons. In this sense the full GSMT formulation might seem hopeless: keeping all the symmetries for the AdS superstring does not help to find the spectral information. However in the (generalized) light-cone gauge the world-sheet theory describes only the physical degrees of freedom of the AdS superstring. And it is in this way that it is possible to interpret the world-sheet excitations as two-dimensional particles. 

Having a theory which describes particles in (1+1) dimensions and which \emph{might} be integrable, means that we can know all the spectrum through the S-matrix, cf. section~\ref{chapter:AdSIntro} and~\ref{chapter:Intro_integrability}. In particular, even without an exact knowledge of the dilatation operator, the (asymptotic) spectrum can be encoded in the Coordinate Bethe Equations, which in turn can be derived from the S-matrix. Naturally this should be true on both sides of the AdS/CFT duality and in fact it turns out that it is the same S-matrix which describes (asymptotically) the collisions of magnons along the (infinitely long) spin chain and of world-sheet excitations (in an infinite volume). 

Historically on the gauge theory side, the S-matrix was initially discussed by Staudacher in \cite{Staudacher:2004tk}. Beisert explained how it is determined by the unbroken symmetries of the model up to an abelian overall phase in \cite{Beisert:2005tm, Beisert:2006qh}. On the string theory side, it was initially discussed by Arutyunov, Frolov and Staudacher in~\cite{Arutyunov:2004vx}, by Klose and Zarembo in \cite{Klose:2006dd} and by Roiban, Tirziu and Tseytlin in \cite{Roiban:2006yc}. Further fundamental works in this direction are the paper by Klose, McLoughlin, Roiban and Zarembo \cite{Klose:2006zd}, where the world-sheet S-matrix is computed to tree level and the papers by Arutyunov, Frolov and Zamaklar \cite{Arutyunov:2006yd} where the S-matrix has been rewritten in a string basis and by Arutyunov, Frolov, Plefka and Zamaklar~\cite{Arutyunov:2006ak} where the symmetries are discussed on the string theory side. Actually, we will use the S-matrix in the near-flat-space limit (NFS) which was computed to one-loop by Klose and Zarembo in~\cite{Klose:2007wq} and to two-loops by Klose, McLoughlin, Minahan and Zarembo in~\cite{Klose:2007rz}. 

There is a key-point in the discussion above. Such a ``S-matrix-program'' \emph{assumes} (quantum) integrability: the kinematical information is obtained by means of the \emph{two-body} S-matrix. As explained in the previous section~\ref{chapter:PSformalism}, proving rigorously the quantum integrability for the type IIB superstring is an incredible hard task probably as much as proving the gauge/string correspondence. But now, after section~\ref{chapter:Intro_integrability}, we know that in two-dimensional field theories the higher conserved charges leave dynamical constraints (particle production, elastic scattering, factorization of the S-matrix) which can be \emph{tested}. For example, this is the strategy used in~\cite{Puletti:2007hq}: Show that all these properties hold up to one-loop for the type IIB superstring in $\ads_5\times\Sphere^5$. 

We should be more precise. 
First point to discuss is that, even fixing the light-cone gauge, the $\sigma$-model described by Metsaev and Tseytlin in~\cite{Metsaev:1998it} is still prohibitive or at least very complicated. For this reason we use for the explicit computation the so-called \emph{near-flat-space} limit, introduced in 2006 by Maldacena and Swanson \cite{Maldacena:2006rv}. We will explain the features of the model in this limit and the corresponding S-matrix. We will also introduce the light-cone gauge and the BMN limit \cite{Berenstein:2002jq}, since we will reuse these notions in section~\ref{chapter:ABJM} discussing the ``new'' gauge/gravity duality. Notice also that we are discussing the S-matrix and the spectrum in the infinite volume limit. 

A second point to stress. We should not be confused about which kind of S-matrix we are discussing. As mentioned at the beginning, we are describing the superstring in AdS spaces from a \emph{world-sheet} point of view. Indeed we have always discussed the integrability of the world-sheet action. The complete kinematical and dynamical information is contained in this very special two-dimensional quantum field theory. In the light-cone gauge the excitations, which are left after gauge-fixing, are only the physical ones. These are \emph{massive} excitations in the string world-sheet. Thus when we talk about and describe the S-matrix on the string theory side, we really mean the \emph{world-sheet} S-matrix, and not the target space S-matrix. It is really the S-matrix which describes the scattering of these particle excitations \emph{on} the string world-sheet. 

On the gauge theory side, it is the same, namely we are dealing with the \emph{internal} S-matrix, adopting the expression used by Staudacher in \cite{Staudacher:2004tk}. This means that we are considering the scattering of magnons, namely the fundamental excitations in the spin chain picture. This should be not confused with the external S-matrix, namely the scattering matrix associated with the collisions of gluons in four dimensional space-time.


\subsection{Light-cone gauge, BMN limit and decompactification limit} 
\label{sec:BMN}

In this section, we explain more concretely what we mean by a two-dimensional particle model from the string theory point of view, introducing the generalized light-cone gauge, the decompactification limit and the fields. 


\subsubsection{Light-cone gauge}
\label{subsec:lightcone_gauge}

In the GS formalism in order to treat the AdS superstring we need to break the (super) Lorentz covariance by imposing the light-cone gauge~\cite{Callan:2004uv, Arutyunov:2004yx, Swanson:2005wz, Arutyunov:2005hd, Arutyunov:2009ga}.
We introduce the $\ads_5\times \Sphere^5$ metric in the global coordinates
\be
\label{ads5s5_global}
ds^2=\underbrace{ - G_{tt}(z) dt^2+G_{zz}(z) dz^2}_{\ads_5}+\underbrace{G_{\varphi\varphi}(y)d\varphi^2 +G_{yy}(y) dy^2}_{\sphere^5}\,,
\ee
with 
\be
&&
G_{tt}(z)= \left( \frac{1+{z^2\over 4}}{1-{z^2\over 4}}\right)^2\,\,,
\qquad
G_{zz}(z)= \frac{1}{\left(1-{z^2\over 4}\right)^2}\,\,,\nln
&&
G_{\varphi\varphi}(y)=\left( \frac{1-{y^2\over 4}}{1+{y^2\over 4}}\right)^2\,\,,
\qquad
G_{yy}(y) =\frac{1}{\left(1+{y^2\over 4}\right)^2}\,\,.
\ee
In $\ads^5$, the coordinates $z^i$ are the four transverse directions and $t$ is the global time; in $\sphere^5$, $y^{i'}$ are the four transverse coordinates and $\varphi$ is the angle along one of the big circle of the 5-sphere. 
The corresponding embedding coordinates, the world-sheet fields, are denoted by
\[
\underbrace{T, ~~ Z_{i}}_{\ads_5}\,,~~~ \underbrace{\phi, ~~Y_{i'}}_{\mbox{S}^5}\qquad \text{with} \quad i\,,i'=1,2,3,4\,.
\]

One can introduce the light-cone coordinates which mix the two $\grp{U}(1)$ directions, in particular to keep the discussion more general we can use the following parameterization
\[
\label{lightcone_coords_general}
X^+= (1-a) T+a\phi
\qquad
X^-= \phi-T\,,
\]
where $a$ is a real number defined between $0\le a\le 1$. The typical values for $a$ are $a=\Half$, which is called the \emph{uniform} gauge, and $a=0$ which is called the {\it temporal} gauge.%
\footnote{Note that names for the different gauge choices are not globally valid.}
There are some simplifications for the different gauge choices, in particular in the next section~\ref{chapter:ABJM} in the context of the $\ads_4/\mbox{CFT}_3$, we will make use of the temporal gauge. Here we will assume the uniform light-cone gauge, which corresponds to the most symmetric choice and has remarkable simplifications in the S-matrix computations. 

The conjugate momenta are defined by $p_M=  {\dl \mathcal L \over \dl \dot x^M}$. Hence inverting the relations \eqref{lightcone_coords_general}, $T=X^+-a X^-$ and $\phi=X^++(1-a) X^-$, the light-cone momenta are
\[
\label{lcmomenta}
p_+\equiv {\dl \mathcal L \over \dl \dot X^+} = p_\phi +p_T\,,
\qquad
p_-\equiv  {\dl \mathcal L \over \dl \dot X^-}  =  - a p_T+(1-a) p_\phi\,.
\]

In the light-cone gauge the target space time (in light-cone coordinates) is identified with the world-sheet time coordinate%
\footnote{This is rigorously true only if the \emph{winding number} is zero, the number of times that the closed string winds along the one-sphere parameterized by the angle $\phi$. In our case we are always discussing closed strings with vanishing winding number.}
and the conjugate momentum to the field $X^-$ is kept constant, namely
\[
\label{lcgauge_intro}
X^+ \overbrace{=}^{!} \tau\,,
\qquad
p_- \overbrace{=}^{!} \text{constant} (\equiv C) \,. 
\]
Notice that this means that the total space-time momentum in the light-cone coordinates is
\be
\label{pminus_length}
P_-  &=& \frac{1}{2\pi\alpha'} \int_{-\pi}^{\pi} d\sigma p_-= {C\over \alpha'}  \,,
\nln
P_-& =&\frac{1}{2\pi\alpha'}  \int_{-\pi}^{\pi} d\sigma p_-= \frac{1}{2\pi\alpha'}  \int_{-\pi}^{\pi} d\sigma  \left(- a p_T+(1-a) p_\phi\right)
      = a E +(1-a) J\,.
\ee
The first line in \eqref{pminus_length} says that the total space-time light-cone momentum $P_-$ measures the world-sheet circumference, which we have chosen to parameterize with $-\pi\le\sigma\le \pi$. However, we could have integrated between the interval $[-s, s]$ after rescaling the world-sheet coordinate $\sigma$, and nothing would have changed in the first line, a part from the appearance of the constant $2s$. Thus $P_-$ is related to the string length. Notice that we have set $R=1$, but it can be easily restored by multiplying the results in \eqref{pminus_length} by $R^2$.

Let us now comment on the second line in \eqref{pminus_length}. By definition, $P_-$ is related to the $\grp{U}(1)$ charges which are the energy, conjugated to the global time in $\ads$, and the angular momentum $J$, conjugated to the angle for the $\sphere^5$-equator. Since this is important, let me stress that we have
\[
\label{U1_charges}
E= -\frac{1}{2\pi\alpha'} \int_{-\pi}^{\pi} d\sigma p_T\,,
\qquad 
J= \frac{1}{2\pi\alpha'}\int_{-\pi}^{\pi} d\sigma p_\phi\,.
\]
Notice that for the temporal gauge ($a=0$) the total space-time light-cone momentum $P_-$ is the angular momentum $J$. Finally for $P_+$ we have
\[
P_+=  \frac{1}{2\pi\alpha'}\int_{-\pi}^{\pi} d\sigma p_+ = \frac{1}{2\pi\alpha'} \int_{-\pi}^{\pi} d\sigma\left( p_\phi + p_T\right)= J-E\,.
\]

Even though we have fixed the light-cone gauge, there is still some choice left: there is still the reparameterization invariance for the world-sheet coordinates. Closed strings are parameterized by $\tau$ which can take any real values and by $\sigma$ which takes values in the $S^1$ circle, since by definition the string is closed. Then topologically the closed string world-sheet is a cylinder.
In particular, this implies that when we shift the coordinate $\sigma$ along the circle by a constant, the physics we are describing should not change. In other words the total momentum along the word-sheet spatial direction (namely the operator which generates the translation in $\sigma$) should vanish. This is the so-called \emph{level matching condition}: the total world-sheet momentum should vanish. Physical closed strings must be level-matched. 
The reparameterization invariance with respect to the world-sheet coordinates is encoded in the Virasoro constraints. Namely we have to impose that the energy-momentum tensor for the superstring world-sheet vanishes: 
\[
\label{Virasoro_constraints_general}
T_{\mu\nu}\equiv S_{\mu\nu}  -\half \gamma_{\mu\nu} \gamma^{\lm\rho}S_{\lm\rho} = 0\,,
\]
where the definition for $ S_{\mu\nu}$ comes from recalling that the GSMT AdS superstring world-sheet Lagrangian is $\mathcal L=\mathcal L_{kin}+\mathcal L_{WZW}$, cf. section \ref{sec:GSMT_string}, i.e.
\be
\label{fermionic_exp_lagrangian}
\mathcal L_{\rm{kin}} &=& -\half \gamma^{\mu\nu} S_{\mu\nu }= -\half \gamma^{\mu\nu} \STr \left (J_\mu J_\nu \right)_{|\mathfrak{g}_2} \nln
&=& -\half \gamma^{\mu\nu}  S_{\mu\nu}^{(0f)} -\half \gamma^{\mu\nu} S_{\mu\nu}^{(2f)}+\dots\nln
\mathcal L_{\rm{WZ}} &=& \mathcal L_{\rm{WZ}}^{(2f)}+\dots\,.
\ee
We have expanded in the inverse powers of the string tension ($\frac{1}{\sqrt{\lm}}$) and for each \emph{loop} in the number of fermions. The string world-sheet metric is $\gamma^{\mu\nu}$ with determinant $-1$ and defined by $\gamma^{\mu\nu}=\sqrt{- h}h^{\mu\nu}$. 

We can concentrate on the bosonic sector for simplicity. In this case $S_{\mu\nu}$ is simply given by $S_{\mu\nu}^{(0f)}= \p_\mu X^M \p_\nu X^N G_{MN}$ and the Virasoro constraints read 
\[
\label{Virasoro_constraints_bos}
T_{\mu\nu}^{bos}= \p_\mu X^M \p_\nu X^N G_{MN}- \half \gamma_{\mu\nu} \gamma^{\lm\rho}\p_\lm X^M \p_\rho X^N G_{MN}=0\,.
\]
One can define the conjugate momenta as
\[
p_M= -\gamma^{\tau \mu} G_{MN}\p_\mu X^N
\]
which is only another way of rewriting the functional derivative $\frac{\dl \mathcal L}{\dl \dot X^N}$ for the bosonic sector. Then one has
\[
\dot X^M= -{1\over \gamma^{\tau\tau}} G^{MN}p_N -{\gamma^{\tau\sigma}\over  \gamma^{\tau\tau}} {X^M}'\,,
\]
where the world-sheet metric basically plays the role of a Lagrange multiplier as it can be seen also rewriting the Hamiltonian and the Lagrangian, i.e.   
\be
\mathcal L &=& -{1\over 2 \gamma^{\tau\tau}} G^{MN} p_M p_N +{1\over 2 \gamma^{\tau\tau}} G_{MN} {X^M}' {X^N}'
\nln
\mathcal H &=& p_M \dot{X}^M -\mathcal L= -{1\over 2 \gamma^{\tau\tau}}\left( G^{MN} p_M p_N +G_{MN} {X^M}' {X^N}' \right)-{\gamma^{\tau\sigma}\over \gamma^{\tau\tau}}p_M {X^M}' \,.
\ee
Thus the Virasoro constraints just become
\[
\label{Virasoro}
 G^{MN} p_M p_N +G_{MN} {X^M}' {X^N}' =0
 \qquad
p_M {X^M}' =0\,.
\] 
The standard procedure is to solve the second Virasoro constraints in \eqref{Virasoro} in order to find ${X^-}'$ and substitute it back in the first constraint $ G^{MN} p_M p_N +G_{MN} {X^M}' {X^N}' =0$. 
In particular one finds 
\be
p_M {X^M}'= p_- {X^-}'+ p_I {X^I}'\, =0 \rightarrow {X^-}'= - \frac{1}{C}  p_I {X^I}'\,
\ee
with the index $I=1,\dots\,, 8$ labeling the transverse directions, i.e. $I=(i, i')$. Thus ${X^-}'$ is a function of the physical transverse fields, which are periodic in $\sigma$. Indeed, in the light-cone gauge, ${X^-}'$ measures the density of the variation of the fields along the $\sigma$ direction, namely it measures the world-sheet momentum density. Then, once one integrates the second constraint in \eqref{Virasoro}, we recognize in it the level-matching condition. 

Plugging back the solution for ${X^-}'$ in the first constraint in \eqref{Virasoro}, one obtains a quadratic equation for $p_+$, that can be solved by
\[
\label{H_lc}
\mathcal H_{lc}= -p_+\,,
\] 
where $\mathcal H_{lc}$ is the light-cone world-sheet Hamiltonian density. Again $p_+$ is now only a function of the transverse coordinates and momenta, once that all the gauges are imposed and the constraints are solved. The equation \eqref{H_lc} tells us that the time evolution in the world-sheet coincides with the time evolution in the target space as it should be, since we have chosen to identify the two time coordinates $X^+$ and $\tau$. 
The world-sheet Hamiltonian is then
\[
\label{Hamiltonian_def}
H=  \frac{1}{2\pi\alpha'} \int_{-\pi}^\pi d\sigma \mathcal H_{lc}\,.
\]
In particular since the Hamiltonian density does not depend on constants related to gauge choices, it does not depend on $P_-$. The length of the circumference $P_-$, (or the angular momentum $J$ in the temporal gauge), enters only trough the interval of integration in \eqref{Hamiltonian_def}. This implies that in fact one can rescale the boundary of integration by $\pi\rightarrow \pi P_-/\sqrt{\lm}$, (or by $\pi\rightarrow \pi J/\sqrt{\lm}$ in the temporal gauge).  
The equation \eqref{H_lc} has also another important consequence. Rewriting $p_+$ from the equation \eqref{lcmomenta}, as a consistency condition one has
\[
H=  \frac{1}{2\pi\alpha'}\int_{-\pi}^\pi \mathcal H_{lc}=-  \frac{1}{2\pi\alpha'}  \int_{-\pi}^\pi p_+= -  \frac{1}{2\pi\alpha'}  \int_{-\pi}^\pi \left(p_T+ p_\phi\right)= E-J\,,
\]
where we used the definitions for the $\grp{U}(1)$ charges in \eqref{U1_charges}. 


\paragraph{The fields.}

After gauge-fixing the type IIB Lagrangian, we are left with 8 bosonic and 8 fermionic degrees of freedom. The bosons correspond to the transverse directions in $\ads_5\times\sphere^5$. The initial symmetry $\grp{PSU}(2,2|4)$ is broken by the gauge-choice. In particular for the bosonic sector we have killed the directions $T$ and $\phi$ in favor of $Y$ and $Z$. Thus the manifest bosonic symmetries left are
\[
\grp{SO}(4,2)\times \grp{SO}(6)\rightarrow \grp{SO}(4)\times \grp{SO}(4)\,.
\]
The light-cone gauge preserves the $\grSO(4)\times \grSO(4)$ symmetry. However in the BMN limit, the unbroken symmetry group is enhanced to $ \grp{SO} (8)$, but not in the NFS limit, where the quartic interactions break $\grp{SO} (8)$ into two copies of $\grp{SO} (4)$, cf. sections \ref{sec:BMN_details} and \ref{sec:NFS}, respectively. 
The indices $i,i'$, with $i,i'=1,2,3,4$ carried by the fields $Z$ and $Y$ respectively can be rewritten in terms of spinorial indices thanks to the Pauli matrices~\cite{Klose:2006zd}, namely each group $\grp{SO}(4)$ can be decomposed as two copies of $\grSU(2)$:
\[
\grp{SO}(4)\sim \left( \grSU(2)\times \grSU(2)\right)/ \Z_2\,. 
\]
Notice that one $\grp{SO}(4)$ comes from the AdS isometry. It represents what is left from the conformal group after gauge-fixing. The second $\grp{SO}(4)$ comes from the sphere isometry, corresponding to what is left from the R-symmetry. Thus the two copies of $\grSU(2)$ contained in $\grp{SO}(4,2)$ are the Lorentz symmetry group while the other two $\grSU(2)$'s contained in $\grp{SO}(6)$ describe the flavor symmetry of the model. 

In terms of the fields this means that the embedding coordinates can be rewritten as bi-spinors 
\[
\label{bispinor_notation}
Z_{\alpha \dot\alpha}= (\sigma_i)_{\alpha\dot\alpha} Z^i
\qquad
Y_{a\dot a}=  (\sigma_{i'})_{a\dot a} Y^{i'}\,,
\]
where the $\sigma$ matrices are $\sigma_i= \sigma_{i'}=\left( \mathbb{1},\imath \overrightarrow \sigma\right)$ and the indices are $a=1,2$, $\dot a=\dot 1,\dot 2$, $\alpha=3,4$, $\dot\alpha=\dot 3,\dot 4$. 
The fermions mix between the two different sectors:
\[
\Psi_{a\dot\alpha}\qquad \Upsilon_{\alpha\dot a}\,,
\]
and one can rewrite all the fields as a $4\times 4$ matrix
\be
\label{matrix_fields}
\matrto{Y_{a\dot a}}{\Psi_{a\dot\alpha}}{\Upsilon_{\alpha\dot a}}{Z_{\alpha \dot\alpha}}\,. 
\ee
The fields transform in the bifundamental $(\mathbf{2}|\mathbf{2})^2$ representation of $\grp{PSU}(2|2)_L\times \grp{PSU}(2|2)_R$. The left and right group acts along the columns and the rows of the matrix \eqref{matrix_fields} respectively. Notice that in the matrix notation above, the first block diagonal corresponds to $\sphere^5$. 

Finally, the two $(\mathbf{2}|\mathbf{2})$ indices can be rearranged in the super-indices $A=(a,\alpha)$ and $\dot A=(\dot a, \dot \alpha)$, where $a, \dot a$ are even and $\alpha,\dot \alpha$ are odd.


\subsubsection{Decompactification limit}
\label{subsec:decompactification_limit}

We have seen that we can rescale the interval of integration in $\sigma$ by a factor depending on the total light-cone momentum $P_-$. Consider now the limit 
\[
P_-\rightarrow \infinity\,.
\]
This means that the world-sheet action is an integral between $-\infinity$ and $+\infinity$, namely for the spatial world-sheet coordinate it means $\sigma\in \Reals$. Equivalently, we can say that instead of considering closed strings we are discussing open strings, whose world-sheet has the topology of a plane. 

Why would one like to consider such a limit? The point is that in this  \emph{decompactification limit}  the world-sheet becomes an infinite plane and it makes sense to introduce asymptotic states (as the ones we discussed in section \ref{sec:Smatrix_fact})
 and the S-matrix for the world-sheet excitations.
It is worth noticing that on the gauge theory side the decompactification limit corresponds to gauge-invariant operators with very large \emph{R}-charge ($J$). 


\subsubsection{The BMN limit}
\label{sec:BMN_details}

The name ``BMN'' stays for the authors of \cite{Berenstein:2002jq}: Berenstein, Maldacena and Nastase. Another fundamental work in this direction is the paper by Gubser, Klebanov and Polyakov~\cite{Gubser:2002tv}. The terms BMN limit and {\it plane-wave} limit will be used as synonyms. The plane wave limit of the $\ads_5\times\sphere^5$ type IIB superstring action was found in~\cite{Metsaev:2001bj} by Metsaev and in~\cite{Metsaev:2002re} by Metsaev and Tseytlin.

The $\ads_5\times \sphere^5$ metric in global coordinates can be rewritten as
\be
\label{ads5s5_BMN}
ds^2 = R^2\, \big( -  dt^2 \cosh^2{\rho}+d\rho^2 +\sinh^2\rho d\Omega_3^2 
                              + d\phi^2 \cos^2\theta +d\theta^2+\sin^2\theta d{\Omega'}^2_3\,\big)\,,
\ee
where the explicit dependence in the radius $R$ is restored
\footnote{Recall the relation $T= {R^2\over 2\pi\alpha'}= {\sqrt{\lm}\over 2\pi}$, namely ${R^2\over \alpha'}=\sqrt\lm$, cf. section \ref{chapter:Intro}. In the previous section we set $R=1$ while now we set $\alpha'=1$.}.
The metric is the same as in \eqref{ads5s5_global} after transforming the coordinates according to
\[
\cosh\rho= \frac{1+{z^2\over 4}}{1-{z^2\over 4}}
\qquad
\cos\theta= \frac{1+{y^2\over 4}}{1-{y^2\over 4}}\,.
\]

We will deal with an infinitely boosted string along the $\sphere^5$ equator parameterized by $\phi$. Such a string carries a very large angular momentum $J$. One can treat it semi-classically and consider small fluctuations around the classical null geodesic of the point-like string which is described by $\rho=\theta=0$. By dimensional analysis one has that $J\sim R^2$, thus it is equivalent to consider the large radius limit ($R\rightarrow\infinity$) of the $\ads_5\times \sphere^5$ background (Penrose limit).

It is useful to rescale the coordinates for the choice $a=0$ according to 
\be
\label{ppwave_coordinates}
t\rightarrow x^+
\qquad
\varphi\rightarrow x^++{x^-\over R^2}
\qquad
z^i \rightarrow {z^i\over R}
\qquad 
y^{i'} \rightarrow {y^{i'}\over R}\,.
\ee
Notice that $X^+$ is dimensionless, $X^-$ has length dimension 2 while the transverse coordinates have dimension 1. Plugging back the coordinate transformations \eqref{ppwave_coordinates} in the metric \eqref{ads5s5_BMN} and taking the large $R$ limit one obtains
\[
\label{ppwave_metric}
ds^2\cong 2 d x^+ d x^- + d z^2 +d y^2 - \left( z^2+y^2\right) (dx^+)^2+ \mathcal O(1/R^2)\,.
\]
This is the Penrose limit of $\ads_5\times \sphere^5$ space, which is equivalent to the plane-wave geometry seen by a very fast particle. 

The Ramond-Ramond (RR) flux survives the Penrose limit, thus we need to impose the light-cone gauge in order to study the fate of our string:
\[
\label{lcgauge}
X^+=\tau
\qquad 
p_-=\text{constant}\,.
\]
Notice that after the rescaling \eqref{ppwave_coordinates} the $\grU(1)$ charge corresponding to the angular momentum $J$  gets also rescaled by a factor $R^2$, namely now we have
\[
\label{momenta_rescaled}
P_-= {J\over R^2}
\qquad
P_+ = J-E\,.
\]
The limit we are considering is
\[
R\rightarrow\infinity
\qquad
J\rightarrow \infinity
\qquad 
P_-= \text{fixed}
\qquad 
E-J=\text{fixed}\,,
\]
and we will neglect all the terms of order $\mathcal O(1/R^2)$.
Notice that ${\lm\over J^2}\sim {R^4\over J^2}$ and $P_-$ plays the role of an effective parameter. For example, recalling that at the leading order the bosonic Lagrangian is  $\mathcal L=-\half S_{\mu\nu}^{(0f)}= -\half \p_\mu X^M \p_\nu X^N G_{MN}$ and plugging in $G_{MN}$ the plane-wave metric \eqref{ppwave_metric}, one obtains at the leading order
\be
 \mathcal L_{B,BMN} &=& \half \sum_{i=1}^4 \left\{ ({Z'}^i)^2 +(Z^i)^2- (\dot{Z}^i)^2 \right\}
      + \half \sum_{i'=1}^4 \left\{ ({Y'}^{i'})^2 +(Y^{i'})^2 - (\dot{Y}^{i'})^2\right\} \,,\nln
 \mathcal H_{B,BMN}&=& \half \sum_{i=1}^4 \left\{ ({Z'}^i)^2 +(Z^i)^2\right\} + \half \sum_{i'=1}^4 \left\{ ({Y'}^{i'})^2 +(Y^{i'})^2\right\} \,.
\ee
We have distinguished between $Y$ and $Z$ coordinates just to make contact with the notation used in the previous section, but indeed they should be treated on equal footing. The above Hamiltonian describes a {\it free} system of 8 bosonic {\it massive} fields. It is straightforward to introduce the fermions, in particular at the leading order we will have only bilinear fermionic terms ($\mathcal L_{kin}^{(2f)}$). After gauge-fixing the local fermionic $\kappa$ symmetry, only the $\grSO(8)$ spinors survive and they also acquire mass from the RR flux (the term is contained in the covariant derivative).

After expanding in Fourier modes the bosonic (and the fermionic) fields, the quantized Hamiltonian
\be
H_{B,pp}= \sum_{n=-\infinity}^\infinity \om_n  \sum_{I=1}^8 (a^I_n)^\dagger (a^I_n)
\ee
describes 8 different kinds of free oscillators, completely decoupled and with unit mass.%
\footnote{Since the 8 modes have the same dispersion relation and they are not really distinguished, we have recollected all together. If one includes the fermions then it is a free $(8|8)$ harmonic oscillator systems.}%

The BMN dispersion relation is relativistic, namely
\[
\label{BMN_energy}
\om_n^2= 1+ k^2 = 1+\left ({n\over \alpha' P_-}\right)^2 = 1+\left({n R^2\over \alpha' J}\right)^2\,,
\]
which is valid for fermions and bosons. Notice that since the theory is free the S-matrix is trivially the identity. 

Let us consider the first non trivial case%
\footnote{One level-matched oscillator, e.g. $(a^I_{-n})^\dagger|0\rangle$, implies $n=0$ and thus zero energy.},
namely a string state where only two level-matched oscillators are excited, i.e. $ (a^I_n)^\dagger \,  (a^I_{-n})^\dagger|0\rangle$. The corresponding energy is 
\[
2 \om_n = 2\sqrt{ 1+\left({n R^2\over \alpha' J}\right)^2}\simeq 2 + \left({n R^2\over \alpha' J}\right)^2 + \mathcal O \big({\lm\over J^2}\big)\,.
\]

It is possible to consider the same limit also on the gauge theory side. The corresponding spin chain carries operators with an infinite R-charge ($J$) and the dispersion relation computed gives the same result \eqref{BMN_energy}. 
In section \ref{sec:BE}, we have analyzed the dispersion relation for an operator such as
\be
\label{example_BMN}
\Tr \left ( Z^{L-K} W^K \right)\,. 
\ee
In the particular case where $K=2$, we have computed $E_{K=2}= {\lm\over \pi^2} \sin^2 \left( {\pi n\over L-1}\right)$ where the quantized momentum for the magnons is $\pm p= \pm {2\pi n\over L-1}$. $L$ is the spin chain length and the {\it R}-charge is $J=L-K$. Let us consider the small momentum limit $p\rightarrow 0$, or equivalently the large $L$ limit, then 
\[
E_{K=2}\cong {\lm n^2\over L^2} \cong \left({R^2 n \over \alpha' J}\right)^2\,,
\]
where we have made all the factors explicit to facilitate the comparison with the formula \eqref{BMN_energy}, namely ${R^4\over {\alpha'}^2 J^2}={\lm\over J^2}$, and we are using the fact that $J\sim L\rightarrow \infinity$ while $K\sim \mathcal O(1)$. Indeed the two dispersion relations match exactly, recalling that now the scaling dimension is $\Delta= J+2 +\gamma$ and the string energy is $E= \Delta-J$, where $J$ is just the bare scaling dimension. 
Thus, $E_{K=2}$ gives the first ${\lm\over J^2}$ correction to the string energy $E$ and to the anomalous dimension $\Delta-J$. Hence, the plane-wave string is dual to a single trace operator with infinite {\it R}-charge.%
\footnote{We should really match the $I$ directions of the oscillators $(a^I_n)^\dagger$ with the operators of  \eqref{example_BMN}, namely we should match the other quantum numbers to identify operators and oscillators.}
%

\paragraph{The BMN scaling.}

Notice that on the string side the BMN limit means $\lm\rightarrow \infinity$ and $J\rightarrow \infinity$, but the ratio $\lm'\equiv{\lm\over J^2}$ is kept fixed. One might wonder what happens if we consider $\lm'$ as a small effective parameter. This is the so-called BMN scaling, where an expansion in $\lm'$ gives the sub-leading terms to the dispersion relation:
\be
\label{BMN_scaling_E}
E = J+ J\left[ \left( \sum_{l=0}^\infinity {a_1^{(l)}\over J^l}\right )\lm'
+ \left(  \sum_{l=0}^\infinity {a_2^{(l)}\over J^l}\right) {\lm'}^2+\dots\right]\,.
\ee
Notice that it is a joint expansion%
\footnote{ In section \ref{sec:paperIII}, in the context of $\ads_4/\CFT_3$, I will come back on the BMN scaling and on the near-BMN strings, namely on those string configurations close to the plane-wave (BMN) limit, where ${1\over J}$ corrections are taken into account. }
 in $\lm'$ and ${1\over J}$.

The coefficient $a_n^{(l)}$ gives the $n$-th term in the $\lm'$ expansion at $l$ loop order in the string $\sigma$ model, i.e. $(\frac{1}{\sqrt \lm})^{l-1}$, with $n=1,2,\dots$ and $l=0,1,2,\dots$. 
The relation \eqref{BMN_scaling_E} was initially understood by Frolov and Tseytlin in~\cite{Frolov:2003qc} and there are many examples in literature, mostly due to Frolov and Tseytlin%
\footnote{A partial list of the fundamental works on spinning strings at classical and one-loop level is~\cite{Frolov:2002av, Frolov:2003qc, Frolov:2003tu, Frolov:2003xy,Frolov:2004bh, Gubser:2002tv, Arutyunov:2003uj, Arutyunov:2003za, Park:2005ji, SchaferNameki:2006gk}. These are different configurations with respect to those considered in this work, we will only consider expansions around the BMN geodesic. For more detailed references we refer the reader to Tseytlin's review~\cite{Tseytlin:2003ii}.},
where for strings with very large (multi)-spins their energy scales according to \eqref{BMN_scaling_E}. I refer the reader to Tseytlin's review~\cite{Tseytlin:2003ii} and references therein. 

On the gauge theory side, it is also possible to organize the scaling dimension in the same kind of expansion, where here $\lm\ll 1$, $J\rightarrow \infinity$ and the ratio $\lm'$ is small, namely 
\be
\label{BMN_scaling_D}
\Delta =  J
+ J\left[ \left( \sum_{l=0}^\infinity {c_1^{(l)}\over J^l}\right )\lm'
+ \left(  \sum_{l=0}^\infinity {c_2^{(l)}\over J^l}\right) {\lm'}^2+\dots\right]\,.
\ee
Here, the $l$ loop term in the coefficients $c_n^{(l)}$ corresponds to terms of order $\lm^l$. 

The BMN scaling opens the possibility of a direct comparison between gauge and string theory, since it offers a window where the two perturbative regimes overlap. Hence the proposal is that the two series of coefficients in \eqref{BMN_scaling_E} and \eqref{BMN_scaling_D} should match:
\[
\label{proposed_equality}
a_n^{(l)} \overbrace{=}^{?} c_n^{(l)} \qquad \text{with} ~ n=1,2,\dots ~ \text{and}~ l=0,1,\dots\,.
\]
The computations of the near-BMN and Frolov-Tseytlin strings~\cite{Kazakov:2004qf, Kruczenski:2003gt} showed an agreement with the gauge theory predictions~\cite{Gross:2002su, Santambrogio:2002sb, Beisert:2003xu, Beisert:2003ea, Engquist:2003rn, Beisert:2005mq, Freyhult:2005fn, Hernandez:2005nf, Beisert:2005bv, SchaferNameki:2005tn, Gromov:2005gp, SchaferNameki:2005is} up to one and two-loop order, cf. also the works~\cite{Arutyunov:2003rg, Arutyunov:2004xy} where the matching was verified also for the infinite commuting conserved charges.
However, at three loops the proposed equality \eqref{proposed_equality} breaks down: The explicit three-loop computation of the near-BMN strings~\cite{Callan:2004ev, Callan:2004uv, Callan:2003xr}, i.e. $a_3^{(1)}$, and of the spinning strings~\cite{Arutyunov:2004xy} showed a mismatch with the gauge theory predictions coming from the Bethe ansatz~\cite{Serban:2004jf, SchaferNameki:2005tn, Beisert:2005cw}, (``three loop discrepancy''). 

The physical reason for such a disagreement, as initially pointed out by Serban and Staudacher~\cite{Serban:2004jf} and then by Beisert, Dippel and Staudacher~\cite{Beisert:2004hm}, is that we are really comparing two different perturbative regions, where the order of the limits, which have been used to construct the expressions \eqref{BMN_scaling_E} and \eqref{BMN_scaling_D}, matters.
On the string theory side, one firstly sends $J\rightarrow \infinity$ and then expands in small $\lm'$, vice versa, on the gauge theory side the first step is the perturbative expansion in small $\lm$ and secondly in the large {\it R}-charge $J$. The two limits do not commute and thus the results for the string energy and for the anomalous dimension coefficients, i.e. $a_n^{(l)}$ and $c_n^{(l)}$, will not necessarily match. In particular, the gauge theory perturbative computation neglects wrapping effects, as discussed at the end of section~\ref{sec:BE}. Thus, one should re-sum the corresponding Feynman diagrams (namely the series in $\lm, J$) in order to correctly compare the two BMN scalings~\eqref{BMN_scaling_E} and \eqref{BMN_scaling_D}. 
I will come back on the three-loop disagreement in section \ref{sec:dressingphase}.


\subsection{The near-flat-space limit}
\label{sec:NFS}

The curved background ($\ads^5\times \sphere^5$) as well as the RR fluxes give rise to interactions in the world-sheet. The spectrum that we want to compute is the spectrum in the presence of such intricate effects. In order to perform concrete computations we need some simplifications.

In 2006 Maldacena and Swanson proposed an interesting truncation of the AdS superstring action~\cite{Maldacena:2006rv}. The remarkable feature of such a model, (Near-flat-space model, NFS) is that even though more treatable than the original MT action, it is still capable of containing interesting physics. In particular we will see that it interpolates between two regimes as the BMN limit and the giant magnon regime.

The region we are discussing is the strong coupling region, namely the region where the 't Hooft coupling is very large, i.e. $\lm \rightarrow \infinity$. The momentum $p$ of the single excitation (magnon)%
\footnote{Notice that now $p$ is the conjugate momentum to the world-sheet coordinates, since it is the momentum carried by the magnons. This $p$ should not be confused with the space-time light-cone momenta of the previous section \ref{sec:BMN}.},
 can be chosen to scale in different ways and this will give different regimes. In particular, scaling $p$ as $\sqrt{\lm}$, when $\lm\gg 1$ one obtains the BMN limit, where the theory is a free massive $(8|8)$ theory and the S-matrix is trivial, cf. section \ref{sec:BMN_details}. Keeping p fixed to some constant value (it can take periodic values), the regime covered is dominated by the \emph{giant magnon}~\cite{Hofman:2006xt}, which is a solitonic solution of the two-dimensional world-sheet theory. In this region the theory is highly interacting. The scaling considered by Maldacena and Swanson is something in between these two regions, namely $p$ scales as $\lm^{-1/4}$.
\begin{figure}
	\begin{center}
		\includegraphics[scale=0.3]{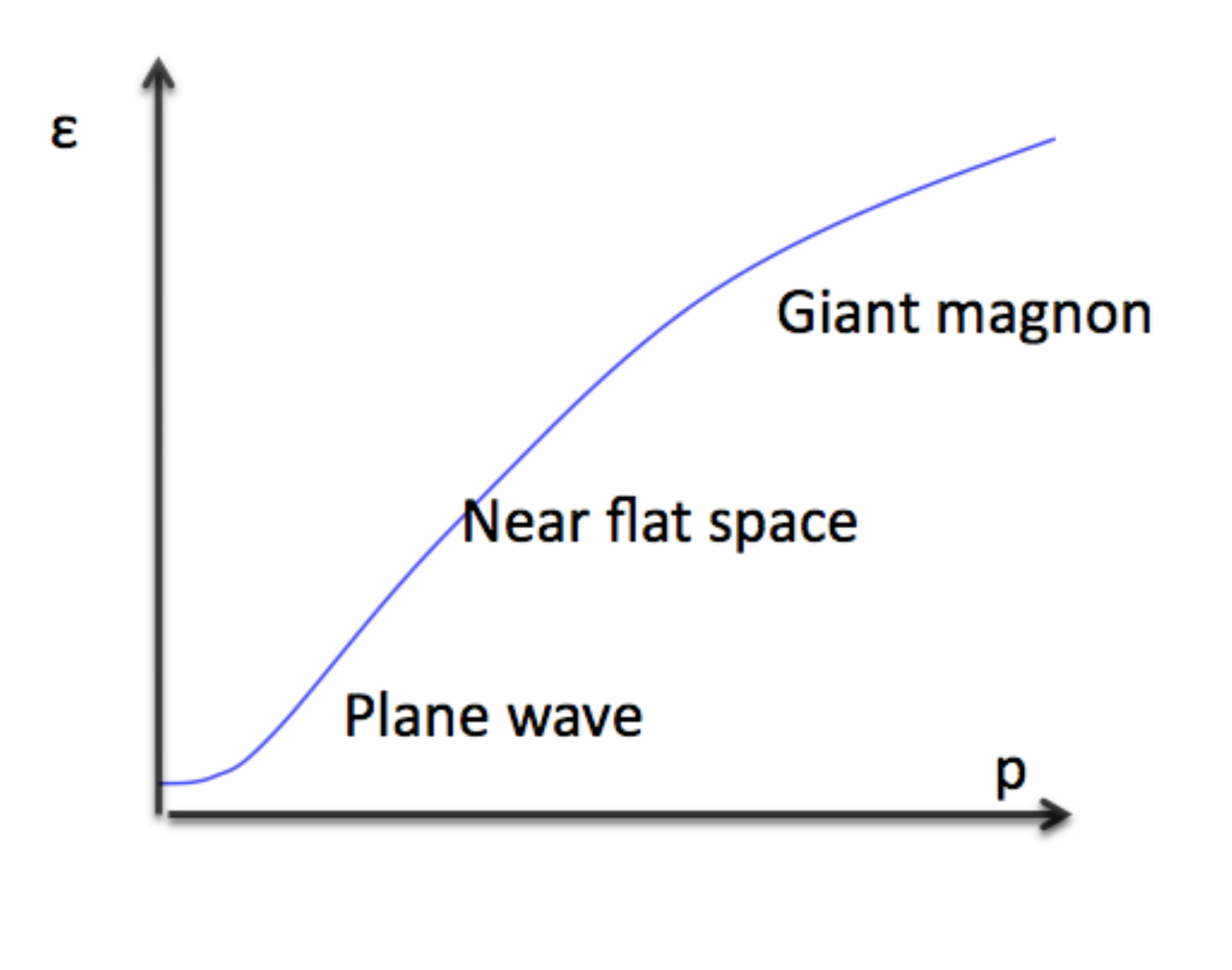}
	\end{center}
	\caption[vvv]{Near-flat-space limit. It interpolates between the plane wave regime and the giant magnon regime. The diagram shows the energy as a function of the momentum. In the plane wave limit the momentum p scales as $\sim \lm^{-\Half}$, in the NFS region $p\sim \lm^{-\Quarter}$ and finally in the giant magnon regime p is a constant (an angle).}
	\label{fig:NFS_diagram}
\end{figure}

The magnon dispersion relation is%
\footnote{See section \ref{sec:Smatrix}.}%
\be
\label{magnon_dispersion_rel}
 E(\lm, p)= \sqrt{1+\frac{\lm}{\pi^2}\sin^2 {p\over 2}}\,.
\ee
Introducing $g$
%
%
 and rescaling the momenta as
\[
g^2 \equiv {\lm\over 8 \pi^2}
\qquad
p {\sqrt{g}\over \sqrt{2}} \equiv k\,,
\]
in the strong coupling limit ($\lm\gg 1$) one obtains the following expansion for the energy
\be
\label{disp_rel_exp}
 E(\lm, p) \sim \sqrt{2 g} k-{1\over \sqrt{2 g}} \left( {k^3\over 6}- {1\over 2 k}\right) +\dots \,.
\ee
The first term is the free energy in the plane wave limit, where the particles have an ultra-relativistic dispersion relation. The other two terms are the ones which characterize the near-flat-space limit and they correspond to keeping up to the second order term in the expansion of the sine function in \eqref{magnon_dispersion_rel}. Namely now we are keeping the sub-leading  corrections in the momentum dependence of $ E$. This is really the region corresponding to the near-flat-space limit, cf. figure \ref{fig:NFS_diagram}. The dispersion relation is not relativistic, not in the exact sense, and it represents some deviation from the Fermi surface. The velocity $v={d E\over d k}$ turns out to depend on the momentum $k$ and the scattering between two excitations carrying different momenta will be non-trivial.%
\footnote{At the leading order the velocity is just the speed of light, namely $v\sim \sqrt{2 g}$, with $g \rightarrow\infinity$. }
Notice that for the giant magnon the dispersion relation%
\footnote{The momentum in the giant magnon regime takes values between 0 and $2\pi$, since it is interpreted as the angle where the open string endpoints sit in the $\Sphere^5$ equator.}
reads $E\sim  { \sqrt{\lm}\over\pi}\sin{p\over 2}$.


\paragraph{The NFS action.}

The form for the near-flat-space Lagrangian used here is the one presented in~\cite{Klose:2007wq}, where the world-sheet coordinates and the fermions are rescaled by 
\be
\label{Kostya_rescaling}
\sigma^{\pm}\rightarrow \gm^{\pm\half} m \sigma^{\pm}
\qquad
\psi_{\pm} \rightarrow \gm^{\mp\Quarter} m^{-\half} \psi_{\pm}\,,
\ee
where $\gamma$ (half inverse string tension) is a power-counting parameter
\[
\gamma={\pi\over \sqrt\lm}\,.
\]
Indeed Maldacena-Swanson action in~\cite{Maldacena:2006rv} does not depend anymore on any dimensionless or dimensional parameter. The embedding coordinates are also rescaled by $\Half$ in order to bring the action in the canonical form for the kinetic and mass terms. Finally, after the rescaling \eqref{Kostya_rescaling}, the $\psi_+$ fermions are integrated out since they only enter quadratically in the action, for more details we refer the reader to~\cite{Klose:2007wq} or to the appendix contained in~\cite{Puletti:2007hq}. Hence the final version for the near-flat-space model is 
\be
\label{NFS_action}
\mathcal L_{NFS} &=&
\Half (\p Y)^2-{m^2 \over 2} Y^2 +
\Half (\p Z)^2-{m^2 \over 2} Z^2
+ {i \over 2}\psi {\p^2+m^2\over \p_-} \psi\nln
&+&\gamma \left(Y^2-Z^2\right) [ (\p_- Y)^2+ (\p_- Z)^2]
+i \gamma \left (Y^2-Z^2 \right)\psi\p_-\psi\nln
&+&i \gamma\psi \left( \p_- Y^{i'}\Gm^{i'}+\p_- Z^{i}\Gm^{i}\right) \left( Y^{j'}\Gm^{j'}-Z^j\Gm^j\right)\psi\nln
&-&{\gamma\over 24} \left(\psi \Gm^{i'j'}\psi  \psi \Gm^{i'j'}\psi -\psi \Gm^{ij}\psi \psi \Gm^{ij}\psi \right)\,,
\ee
where $\psi$ are only the $\psi_-$ components of the original spinors.

Let us summarize and stress once more what the NFS truncation concretely implies. We are considering the following rescaling for the world-sheet excitation momenta
\[
p_{\pm}\lm^{\pm \Quarter}= \text{fixed}
\]
which implies that
\[
p_+\rightarrow 0
\qquad
p_- \rightarrow \infinity\,, ~~~ \text{when} ~~~  \lm\rightarrow \infinity\,.
\]
Hence the NFS limit is a decoupling limit, which factorizes the left and right moving sectors of the AdS string by suppressing the right-moving modes. Further, notice that the truncation breaks the two-dimensional Lorentz invariance of the action.

The NFS model inherits the symmetry of the original GSMT superstring in the light-cone gauge, i.e. $\grp{P} \left(\grSU(2|2)\times \grSU(2|2)\right)$. However, as mentioned at the beginning, the quartic interactions break $\grp{SO}(8)$ to $\grp{SO}(4)\times \grp{SO}(4)$, as it can be seen in the Lagrangian \eqref{NFS_action}, where there is a relative sign for the interactions with four bosonic fields. 

The NFS model has been useful most in the simplification of the S-matrix, such as for example, to test the dressing phase at two loops~\cite{Klose:2007rz} or to verify the factorization of the S-matrix~\cite{Puletti:2007hq}. The key point is that the interactions which appear in \eqref{NFS_action} are at most quartic interactions, and in this sense they make our life easier.


\subsection{The S-matrix}
\label{sec:Smatrix}

In section~\ref{chapter:Intro_integrability} we have presented the S-matrix as a unitary operator mapping asymptotic in and out states. In section~\ref{chapter:AdSIntro} we have introduced the Coordinate Bethe Equations for the Heisenberg spin chain, written in terms of the phase shift. Naturally the phase shifts are nothing but the S-matrix elements for the Heisenberg model. Now it is time to recollect the two pictures. We have already explained that there is \emph{one} S-matrix for the planar asymptotic AdS/CFT. In a certain sense the derivation of the S-matrix gives a theoretical background for the Bethe Ansatz equations. 

We want to discuss the S-matrix for the full (asymptotic) $\grPSU(2,2|4)$ model. We are going to skip many details and this presentation is far from being a rigorous derivation, for which we refer the reader to the original papers~\cite{Beisert:2005tm, Klose:2006zd, Arutyunov:2006yd,Arutyunov:2006ak}. Nevertheless we want to make some comments and illustrate the results.


\subsubsection{Introduction}
\label{sec:Intro_Smatrix}

\paragraph{The symmetries and Beisert's derivation.}

First we need to discuss which are the symmetries of the S-matrix. On the gauge theory side, the initial global symmetry is broken by the choice of the spin chain vacuum. The unbroken symmetry left is $\grp{P}\left( \grSU(2|2) \times \grSU(2|2)\right) $, whose corresponding algebra is $\algPSU(2|2) \oplus \algPSU(2|2) \ltimes \Reals$. The two copies of  $\algPSU(2|2)$ share the same central extension $\mathfrak{C}$ (this is the meaning of the symbol $\ltimes$) which is nothing but the energy. Considering only one sector of the full $\algPSU(2|2) \oplus \algPSU(2|2)\ltimes \Reals$, the fields transform in the $(\mathbf{2}|\mathbf{2})$ bifundamental representation.%
\footnote{We have shown that on the string theory side the fields form a $(\mathbf{2}|\mathbf{2})^2$ super-multiplet of the $\algPSU(2|2) \oplus \algPSU(2|2) \ltimes \Reals$ super-algebra. Obviously, the same happens on the gauge theory side, even though we did not show it explicitly.}
However, in this representation the algebra requires a central charge with semi-integer values $\pm \half$~\cite{Beisert:2005tm}. This cannot be, since we know that the dispersion relation depends continuously on the coupling constant ($\lm$), as for example, it can be seen in the BMN limit, cf. section~\ref{sec:BMN_details}. The apparent contradiction is solved by introducing two other central charges such that the enlarged algebra%
\footnote{Such centrally extended algebra is indeed unique~\cite{Beisert:2005tm}.}
becomes $\algPSU(2|2) \ltimes \Reals^3$, or extensively $\algPSU(2|2) \oplus \algPSU(2|2) \ltimes \Reals^3$. The new central charges $\mathfrak P$ and $\mathfrak K$ are unphysical and they play the role of a momentum and its complex conjugated. ``Unphysical'' means that they should vanish on physical gauge invariant states. It might seem that they have been introduced \emph{ad hoc} but indeed they are responsible for changing the length of the spin chain by removing or adding a background field in the chain~\cite{Beisert:2005tm}. For this reason the spin chain is said to be dynamical: its length is not fixed.%
\footnote{Cf. the interesting paper~\cite{Beisert:2003ys} by the same author on dynamical spin chain for the subsector $\algSU(2|3)$.}

Focusing on one sector, the $\algPSU(2|2) \ltimes \Reals^3$ algebra is spanned by the $\grSU(2)\times \grSU(2)$ generators $\mathfrak{L}^\alpha_{~\beta}$ and $\mathfrak{R}^a_{~b}$ and by the supercharges $\mathfrak{Q}^\alpha_{~~a}$, $\mathfrak{S}^b_{~\beta}$ through the following relations
\be
\label{algebra}
&& \left[ \mathfrak{R}^a_{~b}\,, \mathfrak{J}^c \, \right] = \dl^c_b \, \mathfrak{J}^a -\Half \dl^a_b\, \mathfrak{J}^c
\qquad 
 \left[ \mathfrak{L}^\alpha_{~\beta}\,, \mathfrak{J}^\gamma\, \right] = \dl^\gamma_\beta \, \mathfrak{J}^\alpha -\Half \dl^\alpha_\beta \, \mathfrak{J}^\gamma
 \nln
 && 
 \left \{ \mathfrak{Q}^\alpha_{~~a}\,, \mathfrak{S}^b_{~\beta}\, \right\}= \dl^b_a\, \mathfrak{L}^\alpha_{~\beta} + \dl^\alpha_\beta \,\mathfrak{R}^b_{~a} + \half \mathfrak{C}\,  \dl^b_a\, \dl^\alpha_\beta
 \nln
 &&
 \left \{ \mathfrak{Q}^\alpha_{~~a}\,, \mathfrak{Q}^\beta_{~~b} \,\right\}= \eps^{\alpha\beta} \,\eps_{ab}\, \mathfrak{P}
 \qquad
 \left \{ \mathfrak{S}_{~\alpha}^a\,, \mathfrak{S}_{~\beta}^b\, \right\}= \eps_{\alpha\beta}\, \eps^{ab}\,  \mathfrak{K}\,,
 \ee 
where $\mathfrak{J}^\gamma$ and $\mathfrak{J}^c$ are generic generators and $\mathfrak C$, $\mathfrak P$, $\mathfrak K$ are the central extensions corresponding to the energy and the momenta respectively. The same relations hold for the other $\algPSU(2|2) $ sector just replacing undotted with dotted indices. One of the main result is the derivation of the central charges, in particular of the dispersion relation
\[
\label{disp_rel}
\mathfrak{C}= \sqrt{1+8 g^2\sin^2 {p\over 2}}\,,
\]
where the coupling constant%
\footnote{The definition of $g^2$ is not uniform: in literature it is possible to find also $g^2={\lm \over 16 \pi^2}$.}
is 
\[
g^2 ={\lm\over 8\pi^2}\,.
\]
The dispersion relation~\eqref{disp_rel} has been conjectured by Beisert, Dippel and Staudacher in~\cite{Beisert:2004hm}, but Beisert showed that its specific functional dependence is constrained by the symmetry algebra, even though in order to determine the dependence on the coupling constant $g^2$ one needs to use the BMN limit, for example~\cite{Beisert:2005tm}.


Under the full symmetry algebra $\algPSU(2|2) \oplus \algPSU(2|2) \ltimes \Reals^3$ the two-body S-matrix undergoes a group factorization, namely we can rewrite the total scattering operator as 
\be
\label{group_fact}
\Smatrix= \smatrix_{\grPSU(2|2)} \otimes \smatrix_{\grPSU(2|2)}\,. 
\ee
$\Smatrix$ is an operator which acts on the vector space given by the tensor product of single particle vector spaces, explicating the indices we can write
\be
\Smatrix ~ :~ && \rm{V}_a \otimes \rm{V}_{b} \rightarrow \rm{V}_a \otimes \rm{V}_{b}\nln
&& | \Phi_{A\dot A}(a)\, \Phi_{B \dot B} (b) \rangle \rightarrow | \Phi_{C\dot C}(a)\, \Phi_{D \dot D} (b) \rangle \Smatrix_{A\dot A B \dot B}^{C \dot C D\dot D} (a,b)\,,
\ee
where the $a,b$ are the particle momenta.
Thus the group factorization leads to the expression
\[ \label{eqn:Smat-2p}
 \Smatrix_{A\dot{A}B\dot{B}}^{C\dot{C}D\dot{D}}(a,b) =
 (-)^{\abs{\dot{A}}\abs{B} + \abs{\dot{C}}\abs{D}} \,
 S_0(a,b) \,
 \smatrix_{AB}^{CD}(a,b) \,
 \smatrix_{\dot{A}\dot{B}}^{\dot{C}\dot{D}}(a,b) \; .
\]
Actually this is a graded tensor product according to the statistic of the indices, namely $\abs{A}$ is $0$ and $1$ for even and odd indices respectively. The group factorization in \eqref{group_fact} turns out to be true whenever the symmetry group is a direct product of two groups \emph{and} the Yang-Baxter equations are satisfied~\cite{Ogievetsky:1987vv}. 

In order to compute the S-matrix elements we must write down the action of the $\smatrix_{\grPSU(2|2)}$ on two-particle states where the fields are in the fundamental representation and ask for the invariance of the S-matrix under the algebra generators. 
Let us call the superfield in the $\mathbf{(2|2)}$ fundamental representation as $\chi_A$, where $A$ is the super-index $A=(a,\alpha)$ discussed previously, namely $\chi_A=(\phi^a, \psi^\alpha)$, with $a=1,2$ and $\alpha=3,4$. The $\algPSU(2|2) \ltimes\Reals^3$ generators in \eqref{algebra} act on $\chi_A$ according to
\be
\label{rep_theory}
&& \mathfrak{R}^a_{~b}\ket{\phi^c}= \delta^c_b\ket{\phi^a}-\half \delta^a_b \ket{\phi^c}\,,
\qquad
\mathfrak{R}^a_{~b}\ket{\psi^\alpha}=0\,,
\nln
&& \mathfrak{L}^\alpha_{~\beta}\ket{\phi^c}= 0\,,
\qquad
\mathfrak{L}^\alpha_{~\beta}\ket{\psi^\gamma}= \delta^\gamma_\beta\ket{\psi^\alpha}-\half \delta^\alpha_\beta\ket{\psi^\gamma}\,,
\nln
&& \mathfrak{Q}^\alpha_{~~a} \ket{\phi^b}=\mathrm{a}\, \delta^b_a\ket{\psi^\alpha}\,,
\qquad
\mathfrak{Q}^\alpha_{~~a} \ket{\psi^\beta}=\mathrm{b}\,\epsilon^{\alpha\beta}\epsilon_{ab}\ket{\phi^b Z^+}\,,
\nln
&& \mathfrak{S}^a_{~\alpha}\ket{\phi^b}= \mathrm{c}\,\epsilon^{ab}\epsilon_{\alpha\beta}\ket{\psi^\beta Z^-}\,,
\qquad
 \mathfrak{S}^a_{~\alpha}\ket{\psi^\beta}= \mathrm{d}\,\delta^\beta_\alpha\ket{\phi^a}\,.
\ee
From the fulfillment of the algebra \eqref{algebra} the coefficients $\mathrm{a, b,c,d}$ turn out to be
\[
\mathrm{a}= {\sqrt{g}\over 2^{{1\over 4}}} \gamma\,,
\quad
\mathrm{b}= -\frac{\sqrt{g}}{2^{{1\over 4}}\gamma}\,,
\quad 
\mathrm{c}=i {\sqrt{g}\over 2^{{1\over 4}}} \gamma {1\over x^-} \,,
\quad
\mathrm{d}= -i {\sqrt{g}\over 2^{{1\over 4}}\gamma} (x^+-x^-)\,,
\]
with $\gamma= \sqrt{i (x^--x^+)}$ and $e^{ip}=\frac{x^+}{x^-}$. 
In \eqref{rep_theory} $Z^{\pm}$ represent the insertion ($Z^+$) and the removal ($Z^-$) of a background field in the spin chain. 

The two-body S-matrix \eqref{group_fact} acts on the two-particle states $\ket{\chi_A\, \chi_B}$ as
\be
\label{smatrix_elem}
&& \smatrix  \ket{\phi_1^a \phi_2^b}  =A_{12} \ket{\phi_2^{\{ a} \phi_1^{b\}}} +B_{12}  \ket{\phi_2^{[ a} \phi_1^{b]}} +\Half C_{12} \eps^{ab} \eps_{\alpha\beta} \ket{\psi_2^\alpha \psi_1^\beta Z^-}\qquad\quad
\\ \nn
&& \smatrix \ket{\psi_1^\alpha \psi_2^\beta}= D_{12} \ket{\psi_2^{\{\alpha}\psi_1^{\beta\}}}+E_{12} \ket{\psi_2^{[\alpha}\psi_1^{\beta]}}+\Half F_{12}  \eps_{ab} \eps^{\alpha\beta} \ket{\phi_2^a \phi_1^bZ^+}
\\ \nn
&& \smatrix \ket{\phi_1^a \psi_2^\beta} = G_{12} \ket{\psi_2^\beta\phi_1^a}+H_{12} \ket{\phi_2^a\psi_1^\beta}
\\ \nn
&& \smatrix \ket{\psi_1^\alpha \phi_2^b} =K_{12} \ket{\psi^\alpha_2\phi^b_1}+L_{12} \ket{\phi^b_2\psi_1^\alpha}\,,
\ee
with $p_1\equiv 1$ and $p_2\equiv 2$. The ten coefficients are functions of the particle momenta. 
In order to compute the arbitrary coefficients $A_{12},\dots, L_{12}$ we impose the invariance of the S-matrix under the algebra, i.e.
\[
[\mathfrak{J}_1 + \mathfrak{J}_2\,, \smatrix_{12}]=0\,,
\]
as well as unitarity condition and Yang-Baxter equations (which are automatically satisfied).%
\footnote{The central charges are computed by acting with the algebra generators on single particle states in the fundamental representation, cf.~\cite{Beisert:2005tm, Arutyunov:2006yd}.}
In this way, the matrix elements are univocally determined~\cite{Beisert:2005tm} up to an overall abelian phase which we have indicated with $ S_0(a,b) $ and which will be discussed later in section \eqref{sec:dressingphase}: 
\be
\label{smatrix_coeff}
&& A_{12} =S_0(1,2) {x_2^+ -x_1^-\over x_2^- -x_1^+}
\qquad 
D_{12}=-S_0 (1,2)
\\ \nn
&& B_{12} =S_0(1,2) {x_2^+ -x_1^-\over x_2^- -x_1^+} \big( 1-2{ 1-\frac{1}{x_1^+ x_2^-}\over 1-\frac{1}{x_1^+ x_2^+}} \frac{x_2^- -x_1^-}{x_2^+-x_1^-}\big)
\\ \nn
&& C_{12}=S_0 (1,2)\frac{2\gamma_1 \gamma_2}{x_1^+ x_2^+} \frac{1}{1-\frac{1}{x_1^+ x_2^+}}\frac{x_2^--x_1^-}{x_2^--x_1^+}
\\ \nn &&
 E_{12}= -S_0(1,2) \big (1-2 \frac{1-\frac{1}{x_1^-x_2^+}}{1-\frac{1}{x_1^- x_2^-}} \frac{x_2^+-x_1^+}{x_2^--x_1^+}\big)
\\ \nn &&
F_{12}=-S_0(1,2)\frac{2}{\gamma_1\gamma_2 x_1^- x_2^-}\frac{(x_1^+-x_1^-)(x_2^+-x_1^+)}{1-{1\over x_1^-x_2^-}} \frac{x_2^+-x_1^+}{x_2^--x_1^+}
\\ \nn &&
G_{12}= S_0(1,2) \frac{x_2^+-x_1^+}{x_2^--x_1^+}
\qquad
H_{12}= S_0(1,2) {\gamma_1\over \gamma_2} \frac{x_2^+-x_1^-}{x_2^--x_1^+}
\\ \nn
&& K_{12} = S_0(1,2) {\gamma_2\over \gamma_1} \frac{x_1^+-x_1^-}{x_2^--x_1^+}
\qquad
L_{12}= S_0 (1,2) \frac{x_2^- -x_1^-}{x_2^- -x_1^+}\,,
\ee
where $\gamma_p= |x_p^- -x_p^+|^\half$ and 
\[
x_{p}^{\pm} ={\pi\over \sqrt{\lm}} \frac{e^{\pm ip/2}}{\sin{p\over 2}} \left(1+\sqrt{1+{\lm\over \pi^2}\sin^2 {p\over 2}}\right)\,. 
\]


\paragraph{On the string theory side.}

What about the string theory side? Does everything translate automatically in a string language? From the previous section~\ref{sec:BMN}, we have learned that in order to construct the world-sheet S-matrix we need to decompactify the world-sheet. 

However, in order to study the scattering between string excitations that we can interpret as particles for a two-dimensional theory, we actually need to relax the level-matching condition. The ``particles'' can travel along the world-sheet and collide with an arbitrary momentum. In this way it makes sense to compute the scattering amplitude, and thus the S-matrix elements for such particles. 

In the paper \cite{Arutyunov:2006ak} Arutyunov, Frolov, Plefka and Zamaklar showed that the actual world-sheet symmetry algebra for the $\ads_5\times \sphere^5$ light-cone string not level-matched (and decompactified) is $\mathfrak{psu}(2|2) \oplus \mathfrak{psu}(2|2) \ltimes  \Reals^3$ (\emph{off-shell} algebra). Relaxing the level-matching condition is equivalent on the gauge theory side to opening the spin chain, because the string level-matching condition is equivalent to the cyclicity of the trace. This is another way of saying that the operators are no longer gauge invariant, namely that two extra unphysical central charges can appear ($\mathfrak{K}$, $\mathfrak{P}$).
In the same paper the unphysical central charges $\mathfrak{P}$ and $\mathfrak{K}$ have been computed in terms of string fields, and they turn out to be proportional to the world-sheet momentum which should vanish for physical (i.e. level-matched) states. 
%

In \cite{Arutyunov:2006yd} the world-sheet S-matrix has been rewritten in a string basis. This essentially means that the scattering matrix elements have been deduced by requiring the fulfillment of the Zamolodchikov-Faddeev (ZF) algebra. This is the algebra that we have briefly presented in section~\ref{chapter:Intro_integrability}. Such an algebra takes into account the effects of the interactions in the commutation relation for the free oscillators (i.e. creation and annihilation operators). The symbols $A_a (\theta)$ introduced in section \ref{sec:Smatrix_fact} are not the creation and annihilation operators, since now we have an interacting field theory and we cannot use the free field picture for the oscillators. The interactions affect the free oscillators algebra, but on the other hand for integrable field theories the structure of the Hilbert space is preserved (this is really the job of integrability!). Hence, there must be a non-trivial operator which modifies and takes care of the algebra such that the Hilbert space is preserved. This operator is nothing but the S-matrix and the corresponding algebra is the ZF one, as we discussed in section \ref{sec:Smatrix_fact}. 

Concretely, one needs to impose for the scattering matrix elements the invariance under the off-shell symmetry and physical constraints such as
\begin{itemize}
\item unitarity condition
\item CPT invariance
\item crossing symmetry
\footnote{The crossing symmetry is usually present in relativistic quantum field theories and it relates the exchange between particles and anti-particles. Here we are dealing with a non-relativistic theory, however since the two-dimensional Lorentz invariance is spontaneously broken, it might hold also in this case. This has been proposed by Janik \cite{Janik:2006dc}. Such a symmetry constraints the phase factor $S_0$,  cf. section~\ref{sec:dressingphase}.}
\item Yang-Baxter equations\,. 
\end{itemize}

The basis for the two-particle states in which the S-matrix elements satisfy all the properties listed above as well as the ZF algebra (by construction) is what is called the canonical string basis%
\footnote{It is not exactly the same basis in which the spin-chain S-matrix~(\ref{smatrix_elem}, \ref{smatrix_coeff}) has been written. Local transformations which change the two-body basis can change the matrix elements without leading to any actual change in the physical information. However in the new basis the S-matrix might not respect the standard ZF algebra, but rather a ``twisted'' ZF algebra. Namely the standard ZF relation is multiplied by a local operator which does not modify the vacuum. This is what happens to the spin chain S-matrix derived by Beisert. For a more precise relation between the two basis (spin chain and string) we refer the reader to the paper \cite{Arutyunov:2006yd}.}.

In \cite{Klose:2006zd} Klose, McLoughlin, Roiban and Zarembo derived the perturbative tree-level S-matrix by considering a slightly different perspective. The key-point is requiring the invariance of the two-body S-matrix with respect to the {\it Hopf algebra}. 
%
%
The action of the $\algPSU(2|2)$ symmetry generator is non-local. The charges generated indeed are non-local expressions and they are not additive, cf. section \ref{chapter:Intro_integrability}. Thus, when they act on multi-particles states they do not follow the standard Leibniz rule, but rather the so called coproduct, which characterizes the Hopf algebra. This simply means that when one rearranges the order of the fields on the world-sheet the non-locality of the symmetry generators creates a ``disturbance'' which is reflected in a non-trivial coproduct from an algebraic point of view.


\paragraph{The three-body S-matrix.}

The three-body S-matrix acts on the triple tensor product of single-particle states and it is defined by the relation
\be
\Smatrix ~ :~ && \rm{V}_a \otimes \rm{V}_{b} \otimes \rm{V}_{c} \rightarrow \rm{V}_a \otimes \rm{V}_{b} \otimes \rm{V}_{c}\nln
  && \Smatrix | \Phi_{A\dot A}(a) \Phi_{B\dot B}(b) \Phi_{C\dot C}(c)\rangle  = \ket{\Phi_{D\dot D}(a) \Phi_{E\dot E}(b) \Phi_{F\dot F}(c)} \Smatrix_{A\dot A B\dot B C \dot C}^{D\dot D E \dot E F \dot F} (a,b,c) \, .
\ee
The Yang-Baxter equations now read
\be
 \label{eqn:Smat-factorization}
 \tilde \Smatrix_{A\dot A B\dot B C \dot C}^{D\dot D E \dot E F \dot F}(a,b,c)
 &=& \sum_{X\dot X, Y \dot Y, Z \dot Z}
 \tilde \Smatrix_{X\dot X Y \dot Y}^{D\dot D E\dot E}(a,b)\,
 \tilde \Smatrix_{A\dot A Z\dot Z}^{X\dot X F\dot F}(a,c)\,
 \tilde \Smatrix_{B\dot B C\dot C}^{Y\dot Y Z\dot Z}(b,c)\nln
& =& \sum_{X\dot X, Y \dot Y, Z \dot Z}
 \tilde \Smatrix_{ Y \dot Y Z \dot Z}^{E\dot E F\dot F}(b,c)
 \tilde \Smatrix_{X\dot X C\dot C}^{D\dot D Z\dot Z}(a,c)
 \tilde \Smatrix_{A\dot A B\dot B}^{X\dot X Y\dot Y}(a,b)
 \, ,
\ee
where the graded matrix elements are 
\[
\tilde \Smatrix_{A\dot A B\dot B}^{C\dot C D\dot D} = (-)^{\abs{A}\abs{\dot{A}}\abs{B}\abs{\dot{B}}} \Smatrix_{A\dot A B\dot B}^{C\dot C D\dot D}
\]
and 
\[
\tilde \Smatrix_{A\dot A B\dot B F\dot F}^{C\dot C D\dot D E\dot E} = (-)^{\abs{A}\abs{\dot{A}}\abs{B}\abs{\dot{B}} + \abs{B}\abs{\dot{B}} \abs{F}\abs{\dot{F}}  + \abs{F}\abs{\dot{F}}\abs{A}\abs{\dot{A}} } \Smatrix_{A\dot A B\dot B F\dot F}^{C\dot C D\dot D E\dot E} \,.
\]
Notice that each element $\Smatrix_{A\dot A B\dot B}^{C\dot C D\dot D}$ decomposes according to the group factorization \eqref{eqn:Smat-2p}. 

What we are really interested in is the number of degrees of freedom of the three-body S-matrix. Each field is in the fundamental representation $\mathbf{4}$ of $\algPSU(2|2)\ltimes \Reals^3$, i.e. $\Box$. The three body S-matrix is an invariant unitary operator on their triple tensor product which decomposes in two irreducible representations, each with dimension $\mathbf{32}$~\cite{Puletti:2007hq}. In terms of the super-Young tableau%
\footnote{For a more technical and comprehensive discussion the reader can consult~\cite{Beisert:2006qh} and references therein.}
this means
\[
  \raisebox{-0.9mm}{\includegraphics[scale=0.7]{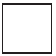}} \;\otimes\;
  \raisebox{-0.9mm}{\includegraphics[scale=0.7]{yang-00.pdf}} \;\otimes\;
  \raisebox{-0.9mm}{\includegraphics[scale=0.7]{yang-00.pdf}} \;=\;
  \raisebox{-2.8mm}{\includegraphics[scale=0.7]{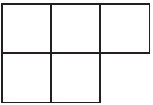}} \;\oplus\;
  \raisebox{-4.6mm}{\includegraphics[scale=0.7]{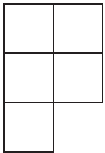}} \; .
\]
Taking also the other $\algPSU(2|2)$ factor into account, then the three-particle S-matrix is a sum of four projectors~\cite{Puletti:2007hq}
\[ \label{eqn:S-matrix-proj}
  \Smatrix
  = C_1 \,
  P_{\lrbrk{\raisebox{-0.1mm}{\includegraphics[scale=0.2]{yang-10.pdf}},
            \raisebox{-0.1mm}{\includegraphics[scale=0.2]{yang-10.pdf}}}}
  + C_2 \,
  P_{\lrbrk{\raisebox{-0.1mm}{\includegraphics[scale=0.2]{yang-10.pdf}},
            \raisebox{-0.7mm}{\includegraphics[scale=0.2]{yang-01.pdf}}}}
  + C_3 \,
  P_{\lrbrk{\raisebox{-0.7mm}{\includegraphics[scale=0.2]{yang-01.pdf}},
            \raisebox{-0.1mm}{\includegraphics[scale=0.2]{yang-10.pdf}}}}
  + C_4 \,
  P_{\lrbrk{\raisebox{-0.7mm}{\includegraphics[scale=0.2]{yang-01.pdf}},
            \raisebox{-0.7mm}{\includegraphics[scale=0.2]{yang-01.pdf}}}}
  \; .
\]
This means that the three particle S-matrix is constrained by the symmetries up to four scalar functions $C_i$, which depend on the incoming momenta and which are the eigenvalues of the corresponding projectors $P_i$. In order to determine them, one needs to compute the scattering amplitudes for the four eigenstates, namely for the highest weight states. These are~\cite{Puletti:2007hq}:
\begin{align}
\label{eqn:hw-states}
  & \Smatrix  \,\ket{    Y_{1\dot{1}}(a) \,    Y_{1\dot{1}}(b) \,    Y_{1\dot{1}}(c) }
  = C_1(a,b,c)\,\ket{    Y_{1\dot{1}}(a) \,    Y_{1\dot{1}}(b) \,    Y_{1\dot{1}}(c) } \; , \\
  & \Smatrix  \,\ket{ \Psi_{1\dot{3}}(a) \, \Psi_{1\dot{3}}(b) \, \Psi_{1\dot{3}}(c) }
  = C_2(a,b,c)\,\ket{ \Psi_{1\dot{3}}(a) \, \Psi_{1\dot{3}}(b) \, \Psi_{1\dot{3}}(c) } \; , \nn \\
  & \Smatrix  \,\ket{ \Bsi_{3\dot{1}}(a) \, \Bsi_{3\dot{1}}(b) \, \Bsi_{3\dot{1}}(c) }
  = C_3(a,b,c)\,\ket{ \Bsi_{3\dot{1}}(a) \, \Bsi_{3\dot{1}}(b) \, \Bsi_{3\dot{1}}(c) } \; , \nn \\
  & \Smatrix  \,\ket{    Z_{3\dot{3}}(a) \,    Z_{3\dot{3}}(b) \,    Z_{3\dot{3}}(c) }
  = C_4(a,b,c)\,\ket{    Z_{3\dot{3}}(a) \,    Z_{3\dot{3}}(b) \,    Z_{3\dot{3}}(c) } \; . \nn
\end{align}
I will come back on the highest weight states \eqref{eqn:hw-states} in the section \ref{sec:paperII}. 


\subsubsection{The dressing phase}
\label{sec:dressingphase}

%
%
%
%

The three-loop disagreement, discussed at the end of section \ref{sec:BMN_details}, pushed the research in the direction of the so called {\it dressing phase}. 

Searching for Bethe equations that fulfill the BMN scaling \eqref{BMN_scaling_D} to all orders leads Beisert, Dippel and Staudacher~\cite{Beisert:2004hm} to modify the rapidity and the dispersion relation, as mentioned in section \ref{sec:Smatrix}%
\footnote{Let us focus on the $\algSU(2)$ sector and on the gauge theory side. Beyond the one-loop order, the model describing the $\algSU(2)$ sector is not anymore the Heisenberg spin chain discussed in section \ref{sec:BE}. Serban and Staudacher proposed to incorporate such a sub-sector into the Inozemtsev spin chain~\cite{Serban:2004jf}. However it breaks the BMN scaling beyond the three loops. The Inozemtsev model is formulated in terms of rapidity and charges which are not the same of the Heisenberg model, obviously. }.
 Indeed, the specific functional form for the energy, and in general for the higher conserved charges, as well as for the rapidity depends on the model we are considering. The BDS proposal for the rapidity, which turned out to be correct, is
\[
\label{rapidity_corrected}
u(p)= \Half \cot \left({p\over 2}\right )\sqrt{1+8 g^2 \sin^2 \left( {p\over 2}\right)}\,,
\]
where the coupling constant $g$ is related to the 't Hooft coupling by
$
g^2={\lm\over 8 \pi^2}
$.
The dispersion relation is only one of the infinite tower of higher charges that an integrable model possesses, and they are modified according to 
\[
\label{charges_spinchain}
\mathbf{q}_{r+1}(p)= g^{-r}\,\, \frac{2\sin{(\Half r p)}}{r}\,\,\left( \frac{\sqrt{1+8 g^2\sin^2 ({p\over 2})}-1}{2 g \sin({p\over 2})}\right)^r\,. 
\]
Notice that the first charge $\mathbf{q}_{1}(p)$ is the momentum $p$, while the second one is the single magnon energy, i.e. $\mathbf{q}_{2}(p)= {1\over g^2} \left( \sqrt{1+8 g^2 \sin^2 ({p\over 2})}-1\right)$. 
The total charge is defined by
\[
\mathbf{Q}_r= \sum_{k=1}^K \mathbf{q}_{r}(p_k)\,,
\]
where $K$ is the total number of magnons.%
\footnote{For the $\algSU(2)$ Heisenberg model the higher charges are given by \mbox{$\mathbf{q}_r(p)=  {2^r\over r-1}\sin(\half (r-1)p)\sin^{r-1}({p\over 2})$} and the rapidity is given by the formula \eqref{rapidity}. For the $r=2$ case, one finds the single magnon energy~\eqref{one_magnon_energy} discussed in section \ref{sec:BE}.}
The BMN limit result can be found by considering the string energy $\dl E=g^2 \mathbf{Q}_2$. 

We have discussed until now the Bethe equations in the spin chain context, let us move back to the string theory side. Kazakov, Marshakov, Minahan and Zarembo (KMMZ) proposed the string Bethe equations (a set of non linear integral equations) in order to describe the classical string $\sigma$-model~\cite{Kazakov:2004qf}. 
%
One would like to generalize (and discretize) such equations in order to capture also quantum string effects. Since the elementary excitations are the same on both sides of the duality, it seemed reasonable to introduce a phase in the S-matrix and thus in the Bethe equations without modifying the BDS dispersion relation~\cite{Arutyunov:2004vx}. This phase shift is part of the scalar factor (the {\it dressing phase}) that cannot be determined by the symmetry algebra, but rather it can be obtained by using the crossing relation%
\footnote{Recall the footnote about the crossing symmetry in section \ref{sec:Smatrix}.}.
 The initial step in the direction of determining the phase factor and the quantum string Bethe equations has been done in \cite{Arutyunov:2004vx} by Arutyunov, Frolov and Staudacher for the $\algSU(2)$ sub-sector. 
The AFS phase has been deduced in such a way that it reproduces the thermodynamic or continuum limit of the string KMMZ Bethe equations.
Explicitly, for $K$ impurities and for the $\algSU(2)$ sub-sector, the Bethe equations formally are still 
\[
e^{i L p_k} =\prod_{j=1, j\neq k}^K S(u_j, u_k)\,,
\]
but now the S-matrix acquires an extra phase:
\be
\label{AFS_phase}
 S(u_j, u_k) &=& \frac{u_k-u_j+i}{u_k-u_j-i}\times\nln
& \times & \exp{ \left(2 i \sum_{r=2}^\infinity ({g^2\over 2})^{r} (\mathbf{q}_{r}(p_k) \mathbf{q}_{r+1}(p_j)-\mathbf{q}_{r+1}(p_k) \mathbf{q}_{r}(p_j))\right)}\,, \nln
\ee
where the charges are the ones in \eqref{charges_spinchain}.

This is not the end of the story for the dressing phase, but rather the beginning: The AFS represents the leading quantum correction to the Bethe equations
%
%
%
and to the S-matrix.
The phase in \eqref{AFS_phase} can be generalized by shifting the S-matrix according to
\[
\exp{2i \theta (p_k, p_j)}= \exp{2 i \sum_{r=2}^\infinity \sum_{{s=1+r \atop s+r=odd}}^\infinity c_{r,s}(g) \left( \mathbf{q}_{r}(p_k) \mathbf{q}_{s}(p_j)-\mathbf{q}_{s}(p_k) \mathbf{q}_{r}(p_j)\right)}\,.
\]
The coefficients $ c_{r,s}(g)$ are expanded in the strong coupling limit according to 
\[
 c_{r,s}(g)=\left({g^2\over 2}\right)^r\sum_{n=0}^\infinity  c_{r,s}^{(n)} g^{-n}\,.
 \]
 We see that the AFS phase is obtained by substituting $c_{r,s}^{(0)}=\dl_{s,r+1}$. The first quantum coefficient $c_{r,s}^{(1)}$ has been deduced by Hernandez and Lopez (HL)~\cite{Hernandez:2006tk}, cf. also~\cite{Beisert:2005cw}, the all-loop strong coupling limit was discovered by Beisert, Hernandez and Lopez \cite{Beisert:2006ib}, i.e. $c_{r,s}^{(n)} $ for all $n \ge 0$, and finally, the full series at strong and weak coupling has been found by Beisert, Eden and Staudacher (BES) in~\cite{Beisert:2006ez}.  
Nowadays, there have been numerous tests for the BES proposal: From the world-sheet point of view up to two-loops~\cite{Roiban:2007jf} and in the near-flat-space limit~\cite{Klose:2007rz}; at weak coupling by direct gauge theory computations~\cite{Bern:2006ew} and up to four loop in the $\grSU(2)$ sector~\cite{Beisert:2007hz}. Other important tests which confirm the BES result have been given in the works~\cite{Gromov:2007ky, Gromov:2007cd, Freyhult:2006vr}. Finally in~ \cite{Arutyunov:2006iu} it has been shown that the HL dressing phase satisfies Janik's equation~\cite{Janik:2006dc}.


\subsubsection{The S-matrix in the NFS limit}
\label{sec:NFS_Smatrix}

We want now to consider the world-sheet S-matrix in the NFS limit. One might wonder whether the NFS truncation is consistent or not, namely if the S-matrix computed directly from the action \eqref{NFS_action} is the same matrix obtained taking the NFS limit from the original world-sheet S-matrix. This was investigated by Klose and Zarembo for the one-loop order \cite{Klose:2007wq} and then to two loops by Klose, McLoughlin, Minahan and Zarembo in \cite{Klose:2007rz}. Indeed, even if we truncate and decouple the right and the left moving sectors, saying that the right modes are faster, it might be that the left moving particles can reappear in the interactions, if we have enough time to wait. Then they might give contributions in loop diagrams at quantum level. 

In the near-flat-space limit the S-matrix elements are given by 
\[ 
 \Smatrix_{A\dot{A}B\dot{B}}^{C\dot{C}D\dot{D}}(a,b) =
 (-)^{\abs{\dot{A}}\abs{B} + \abs{\dot{C}}\abs{D}} \,
 S_0(a,b) \,
 \smatrix_{AB}^{CD}(a,b) \,
 \smatrix_{\dot{A}\dot{B}}^{\dot{C}\dot{D}}(a,b) \; .
\]
The arguments $a \equiv p_{a-}$ and $b \equiv p_{b-}$ are the minus components of the particle light-cone momenta. Up to order $\mathcal O(\gamma^4)$ corrections, the prefactor $S_0$ can be written as
\[ \label{eqn:Smat-pre}
 S_0(a,b) = \frac{\,\,e^{
 \frac{8i}{\pi} \gamma^2 \, \frac{a^3 b^3}{b^2 - a^2}
  \lrbrk{1-\frac{b^2 + a^2}{b^2 - a^2} \, \ln\frac{b}{a}}
  }}
  {1+\gamma^2 \, a^2 b^2 \lrbrk{\frac{b + a}{b - a}}^2} \; .
\]
The matrix part is usually parametrized as follows
\begin{align} \label{eqn:Smat-mat}
  \smatrix_{\lAA\lBB}^{\lCC\lDD} & = A \,\delta_\lAA^\lCC \delta_\lBB^\lDD
                                   + B \,\delta_\lAA^\lDD \delta_\lBB^\lCC \, , &
  \smatrix_{\lAA\lBB}^{\lcc\ldd} & = C \,\levi_{\lAA\lBB} \levi^{\lcc\ldd} \, , &
  \smatrix_{\lAA\lbb}^{\lCC\ldd} & = G \,\delta_\lAA^\lCC \delta_\lbb^\ldd \, , &
  \smatrix_{\lAA\lbb}^{\lcc\lDD} & = H \,\delta_\lAA^\lDD \delta_\lbb^\lcc \, , \\
  \smatrix_{\laa\lbb}^{\lcc\ldd} & = D \,\delta_\laa^\lcc \delta_\lbb^\ldd
                                  + E \,\delta_\laa^\ldd \delta_\lbb^\lcc \, , &
  \smatrix_{\laa\lbb}^{\lCC\lDD} & = F \,\levi_{\laa\lbb} \levi^{\lCC\lDD} \, , &
  \smatrix_{\laa\lBB}^{\lcc\lDD} & = L \,\delta_\laa^\lcc \delta_\lBB^\lDD \, , &
  \smatrix_{\laa\lBB}^{\lCC\ldd} & = K \,\delta_\laa^\ldd \delta_\lBB^\lCC \, , \nn
\end{align}
where the exact coefficient functions are given by
\begin{align} \label{eqn:Smat-coeffs}
  A(a,b) & =  1 + i\gamma \, a\,b\,\frac{b-a}{b+a} \, , &
  B(a,b) & = -E(a,b) = 4i\gamma\, \frac{a^2\,b^2}{b^2-a^2} \, , \nn \\
  D(a,b) & =  1 - i\gamma \, a\,b\,\frac{b-a}{b+a} \; , &
  C(a,b) & =  F(a,b) = 2i\gamma\, \frac{a^{3/2}\,b^{3/2}}{b+a} \, , \\
  G(a,b) & =  1 + i\gamma \, a\,b \, , &
  H(a,b) & =  K(a,b) = 2i\gamma\, \frac{a^{3/2}\,b^{3/2}}{b-a} \, , \nn \\[3mm]
  L(a,b) & =  1 - i\gamma \, a\,b \, . \nn
\end{align}
Notice that the S-matrix elements \eqref{eqn:Smat-coeffs} are exact in the NFS limit, apart from the dressing phase $S_0$ \eqref{eqn:Smat-pre} which is expanded up to order $\gamma^3$. Moreover it turns out that the two-dimensional Lorentz invariance is restored in the NFS model, since they depend on the difference of momenta.  



\subsection{The world-sheet S-matrix factorization}
\label{sec:paperII}

We have already stressed that, from the beginning of the section up to now, we are assuming to deal with a quantum integrable system. Surely this is a suitable hypothesis, which have lead to immense progresses and there have been a vast quantity of indirect checks about the validity of this hypothesis. But notice that on the string theory side perturbative computations beyond the leading order are still extremely difficult to perform. Remarkable in this sense the two-loop computations of the world-sheet scattering amplitudes in the NFS limit~\cite{Klose:2007rz}.

Can we give a proof that the $\ads_5\times \sphere^5$ superstring is quantum integrable at least in the planar limit? The word ``proof'' might discourage. However the NFS model offers us a good region where we can test many of the assumed working hypotheses, among them quantum integrability. The NFS Lagrangian \eqref{NFS_action} is not so terrible and the S-matrix is not trivial in this region. This is an incredible good window in the strong coupling limit where we can \emph{directly} face the important and non-trivial issue of quantum integrability. 
Hence the goal of~\cite{Puletti:2007hq} is to check for the first time in a very explicit and direct way that the NFS model is quantum integrable at one-loop. This strongly supports the hypothesis of a quantum integrable field theory describing the AdS superstring. 

The strategy adopted in~\cite{Puletti:2007hq} is to verify the presence (or the absence) of the dynamical constraints which define an integrable two-dimensional field theory: absence of particle production, elastic scatterings, S-matrix factorization. We have focused on a $3 \rightarrow 3$ scattering. Concretely we have compared two sets of data. On the first set (the ``experimental data''), we compute the $3\rightarrow 3$ scattering amplitudes which follow from the Feynman diagrams of the corresponding NSF action \eqref{NFS_action}. On the second set (the ``theoretical data''), we have computed the three-particle S-matrix which would follow assuming the quantum integrability of the model, namely the three-particle S-matrix which is given by the Yang-Baxter equations as a product of two-particle S-matrix elements, i.e. \eqref{eqn:Smat-factorization}. The computations are done perturbatively up to one-loop. 
The scattering amplitude is defined by
\[
 \label{eqn:defamp}
  \Amp(a,b,c,d,e,f) = \braket{A_{b_3}(f) A_{b_2}(e) A_{b_1}(d)}{A_{a_1}(a) A_{a_2}(b) A_{a_3}(c)}_{\mathrm{connected}} \; 
\]
and the process considered is the generic $3\rightarrow 3$ scattering
\footnote{Recall the ordering and the ZF algebra introduced in section \ref{chapter:Intro_integrability}.}
\[
\label{generic_3to3_scatt}
A_{a_1}(a) A_{a_2}(b) A_{a_3}(c)\rightarrow A_{b_1}(f) A_{b_2}(e) A_{b_3}(d)\,.
\]
Notice that we are dealing with \emph{connected} diagrams, since the disconnected diagrams trivially factorize. The S-matrix elements and the scattering amplitudes are related by
\[ 
\label{eqn:exp-amp}
   \Amp(a,b,c,d,e,f) = \sum_{\sigma(d,e,f)} \Samp_\sigma(a,b,c)\; \delta_{ad}\, \delta_{be} \,\delta_{cf} \, ,
\]
where $\sigma(d,e,f)$ are all the permutations of the outgoing momenta. An explicit example among the highest-weight state \eqref{eqn:hw-states} is illustrated in appendix \ref{app:NFS_example}.

The results of~\cite{Puletti:2007hq} show that the two sets of data agree completely: The tree-level and one-loop scattering amplitudes indeed factorize as in equations \eqref{eqn:exp-amp} and the S-matrix elements $ \Samp_\sigma(a,b,c)$ precisely match the three-body S-matrix computed by the Yang-Baxter equations \eqref{eqn:Smat-factorization}. The formula \eqref{eqn:exp-amp} means that the amplitudes give rise to the phase space showed in figure \ref{fig:phasespace} in section \ref{chapter:Intro_integrability}. 

Since the three-body S-matrix is constrained by the symmetries up to four scalar functions $C_i$, cf. equation \eqref{eqn:hw-states}, it is sufficient to compute the scattering amplitudes for the four processes which correspond to the highest weight states \eqref{eqn:hw-states}, namely which correspond to the eigenstates of the three-body S-matrix. Showing the factorization for these four scattering amplitudes means \emph{proving} the factorization of the \emph{entire} three-particle S-matrix to one-loop order. A proof in a ``mathematical sense'' would require to re-sum all the perturbative series and to show the factorization of any $n\rightarrow n$ scattering amplitudes. Not trivial at all. 

Notice that here in the $3\rightarrow 3$ scattering
\begin{itemize}
\item tree-level order means $\gamma^2 \sim {1\over \lm^\Half}$ 
\item one-loop order means $\gamma^3\sim {1\over \lm^{{3\over 2}}}$. 
\end{itemize}

Actually we have computed further scattering amplitudes involving mixed states between fermions and bosons, in order to confirm the supersymmetries of the NFS model. 

According to section \ref{chapter:Intro_integrability} this means that there must exist a higher conserved charge. How does such charge manifest itself? How do the selection rules and the factorization come from Feynman diagram computations? First recall that each Feynman graph contains already the energy and momentum conservation. In computing the scattering amplitudes one can realize that in the phase space points, where the set of incoming momenta is equal to the set of outgoing momenta, the internal propagators go on-shell and diverge. Namely for a $3\rightarrow 3$ scattering the internal propagators may go on-shell (since in the internal diagrams they might run two incoming momenta and one out-going momentum which have different signs, thus in the point where the in-coming momenta are equal to the out-going one this clearly diverges). They must be regularized and this is done by using the $i \eps$ prescription, namely each mass is shifted by $\pm i\eps$ in order to move the singularities on the complex plane. The residues are then computed with \cite{Dorey:1996gd, Arefeva:1974bk}
\[ 
\label{eqn:principle-value-formula}
  \frac{1}{\vec{p}^2 - m^2 \pm i \eps} = \mathcal{P} \frac{1}{\vec{p}^2 - m^2} \mp i \pi \delta(\vec{p}^2 - m^2) \; ,
\]
where $\mathcal{P}$ stands for the principal value prescription. The term with the principal value takes care of the singularities, namely skipping such delicate points in the integration we can brutally apply the energy-momentum conservation which makes the corresponding amplitudes vanish, after summing over all the equivalent diagrams. What is left is only the term in \eqref{eqn:principle-value-formula} with the extra $\delta$ function, ``extra'' since the Feynman diagrams already come with two-delta functions from the energy-momentum conservation. These three $\delta$-functions combine together and force the out-going momenta to be equal to one of the in-coming momenta, cf. \eqref{eqn:exp-amp}. The resulting phase space is as in figure \ref{fig:phasespace} in section \ref{chapter:Intro_integrability}. 

What about the $2\rightarrow 4$ amplitudes? The crucial point is that now the internal propagators will never be on-shell, since all the momenta flowing there have the same sign. Then we can forget the $i \eps$ regularization and proceed with standard brute force computations. Summing all the amplitudes the result turns out to vanish. This indeed corresponds to the fact that we are not in the ``famous six points'' of the phase space. More details can be found in appedix \ref{app:NFS_example}.



\section{\texorpdfstring{The $\ads_4/\mbox{CFT}_3$}{\ads_4/\mbox{CFT}_3} duality }
\label{chapter:ABJM}



We now leave the $\ads_5/\CFT_4$ duality. But we do not leave the gauge/string duality. 
In 2008 A.~Aharony, O.~Bergman, D.~Jafferis and J.~Maldacena (ABJM) proposed a new conjecture where the world-volume theory of a stack of M2-branes probing a $\C^4/\Z_k$ singularity is a three-dimensional conformal field theory~\cite{Aharony:2008ug}%
\footnote{The ABJM paper comes after plenty of works on multiple M2-branes. I will not go into detail and leave the curious reader to consult the work~\cite{Aharony:2008ug} and references therein.}. 
I will refer to this as the ABJM or the $\ads_4/\CFT_3$ conjecture, in the next section it will be clear why. The work has opened a huge amount of possibilities. Indeed, considering the impressing results due to the integrability properties of the planar $\ads_5/\CFT_4$ duality, it is natural to try to export the same techniques (and hopefully the same progresses) in the new correspondence. 
There are numerous features that are shared by the two gauge/string dualities, but there are also important aspects which are different and which make things quite intriguing and far from being obvious.

\subsection{Introduction}

The $\ads_4/\mbox{CFT}_3$ states a duality between a three-dimensional conformal field theory and an M-theory on eleven dimensions. Let us start from the gauge theory side. It is constructed by two Chern-Simons (CS) theories, each one with a $\grU(N)$ gauge group, coupled with bifundamental matter. However the level of the gauge group is different in the two cases: we have indeed $\grU(N)_{k}\times\grU(N)_{-k}$. The theory is conformally invariant at classical and quantum level and it possesses $\mathcal N=6$ supersymmetries. It contains two parameters%
\footnote{There is also a generalization, known as ABJ theory~\cite{Aharony:2008gk}, where the gauge group is $\grU(N)_k\times \grU(M)_{-k}$. It seems that, also in this case, the theory manifests integrable structures in the planar limit~\cite{Bak:2008vd}.
}:
the gauge group rank, $N$, and the level of the algebra $k$. Both parameters assume integer values. However, it is possible to form a continuous parameter $\lm={N\over k}$, that will play the role of the 't Hooft coupling, and that will interpolate between the string and the gauge theory side. In the large $N$ and $k$ limits, $\lm$ is continuous. In particular, the large $N$ limit corresponds to the planar limit of the CS-matter theory.  
Essentially, for the CS-matter theory ${1\over k}$ plays the same role as it was for $\gym^2$ in SYM theory, cf. section \ref{sec:ABJM_theory}. 

The gravity dual describes a stack of $(Nk)$ M2-branes on a flat space. In particular the M-branes probes the orbifold%
\footnote{An orbifold is a coset $\grp{G}/\grp{H}$ where $\grp{H}$ is a group of discrete symmetries~\cite{Polchinski:1998rq}.}
$\Reals^8/\Z_{k}$. The near-horizon geometry is given by M-theory on $\ads_4\times \sphere^7/\Z_k$. Notice that it is an eleven-dimensional space. 
Due to the $\Z_k$ action, it is natural to write the sphere $\sphere^7$ as an $\sphere^1$ fibration over $\C P^3$: roughly speaking we can say that $\sphere^7/\Z_k\cong \C P^3 \times \sphere^1/\Z_k$. 
The radius of the circle $\sphere^1$ depends on $k$ and the effect of the orbifold is to reduce the volume by a factor $k$. In particular when $k$ is very large, effectively the space is ten-dimensional, i.e. $\ads_4\times \C P^3$. Explicitly, the circle radius is given by $R_{\s^1} \sim {(Nk)^{1/6}\over k}$. Thus, when such radius is very large, namely when $N\gg k^5$, then the theory is strongly coupled and the proper description is in terms of the M-theory. Vice versa, when the radius is very small, i.e. $N\ll k^5$, then it can be effectively used a description in terms of IIA superstrings living on $\ads_4\times \C P^3$ with RR fluxes. More details are given in appendix \ref{sec:preliminaries}. 

The two parameters $N$ and $k$, which describe the number of M2-branes and the order of the orbifold group, are contained in the effective string tension and in the string coupling.
They are given by
\[
\label{string_parameters}
T={R^2\over 2\pi \alpha'}= 2^{5/2}\pi \left({N\over k}\right)^\half \qquad g_s= \left(32\pi^2 {N\over k^5}\right)^{1/4}\,.
\]
The specific relations and the ugly numerical factors in \eqref{string_parameters} are obtained analyzing the supergravity regime, cf. appendix \ref{sec:preliminaries}. Again, from the behavior of the string coupling, we can see that for $N\gg k^5$, i.e. $g_s\gg 1$ the string description fails, we need to use the full M-theory formulation, while for $N\ll k^5$ ($g_s\ll 1$) the ``weak coupling'' string limit is a good approximation. 
Notice that again the effective tension goes like the square root of the 't Hooft coupling, namely $T\sim \sqrt\lm$. The string coupling in terms of $\lm$ reads as $g_s= (32 \pi^2 {\lm\over k^4})^{\Quarter}=  (32 \pi^2 {\lm^5\over N^4})^{\Quarter}$, cf. table \ref{tab:AdS/CFT_comparison}. 
\begin{table}
\begin{center}
\begin{tabular}{|c c c|}
\hline
                 $\ads_5/\CFT_4$  & &$\ads_4/\CFT_3$ \\ \hline
                 IIB on $\ads_5\times\sphere^5$& AdS side  &  IIA on $\ads_4\times \C P^3$  \\
                $ \mathcal N=4$ SYM in 4d &  CFT side       &  $\mathcal N=6$ CS-matter in 3d \\ \hline
                $\lm=g_{\rm YM}^2 N$ & 't Hooft coupling & $\lm={N\over k}$\\
                $T={R^2\over 2\pi\alpha'}={\sqrt{\lm}\over 2\pi} $ & String tension & $T={R^2\over 2\pi \alpha'}= 2^{5/2}\pi \sqrt{\lm} $ \\
                $ g_s={g^2_{\rm YM}\over 4\pi}$ & String coupling & $g_s= \left(32\pi^2 {N\over k^5}\right)^{1/4}$\\ \hline
               $\grSU(N)$ & gauge group & $\grU(N)_k \times \grU(N)_{-k}$ \\
$\grPSU(2,2|4)$& global symmetry &$\grp{OSp}(6|4)$\\
$\ads_5\times\sphere^5= {\grSO(4,2)\over \grSO(4,1)} \times {\grSO(6)\over \grSO(5)}$& bosonic subgroup& $\ads_4\times \C P^3= {\grSO(3,2)\over \grSO(3,1)} \times {\grSU(4)\over \grU(3)}$   \\ \hline
   \end{tabular}
\end{center}
\caption{Summarized comparison between the two gauge/string dualities.}
\label{tab:AdS/CFT_comparison}
\end{table}

From now on, we are going to consider only a specific region for the gravity side of the correspondence: the string regime. This means that for us $N$ and $k$ are very large and in particular are such that $N\ll k^5$ or $1\ll \lm\ll k^4$. 

Also the $\ads_4/\CFT_3$ is a weak-coupling duality.


\subsection{The ABJM $\mathcal N=6$ Chern-Simons theory}
\label{sec:ABJM_theory}

The $\mathcal N=6$ Chern-Simons theory in three dimensions is described by the following Lagrangian
\be
\label{def_Lagrangian_CS}
\mathcal L&=& {k\over 4\pi} \Tr\big\{ \eps^{\mu\nu\lm} \big( A_\mu \p_\nu A_\lm +{2\over 3} A_\mu A_\nu A_\lm 
                                             - \hat A_\mu \p_\nu \hat A_\lm -{2\over 3} \hat A_\mu \hat A_\nu \hat A_\lm\big)\nln
                                             &+&D_\mu Y_A^\dagger D^\mu Y^A
                                             +{1\over 12} Y^A Y^\dagger_A Y^B Y^\dagger_B Y^C Y^\dagger_C
                                             +{1\over 12} Y^A Y^\dagger_B Y^B Y^\dagger_C Y^C Y^\dagger_A\nln
                                             &- &{1\over 2} Y^A Y^\dagger_A Y^B Y^\dagger_C Y^C Y^\dagger_B
                                             +{1\over 3} Y^A Y^\dagger_B Y^C Y^\dagger_A Y^B Y^\dagger_C
                                            -\Half Y^\dagger_A Y^A \bar\psi^B \psi_B\nln
                                            &+& Y^\dagger_A Y^B \bar\psi^A \psi_B
                                            +\Half \bar\psi^A Y^B Y^\dagger_B \psi_A
                                            - \bar\psi^A Y^B Y^\dagger_A \psi_B
                                            +i \bar \psi^A \gamma^\mu D_\mu \psi_A\nln
                                            &+& \Half \eps^{ABCD} Y_A^\dagger \bar \psi_{cB} Y^\dagger_C \psi_D
                                            -\Half \eps_{ABCD} Y^A \bar \psi^B Y^C C \psi_c^D
                                            \big\}\,. 
\ee
The gauge group is $\grU(N)_{k}\times\grU(N)_{-k}$, where the subscripts denote the level of the algebra. The relative sign is reflected in the two Chern-Simons contributions in \eqref{def_Lagrangian_CS}, which describe the two gauge fields $A_\mu$ and $\hat A_\mu$. The Lorentz index $\mu$ runs between 0 and 2, i.e. $\mu=0,1,2$, since the theory is three-dimensional.  
The gauge field $A$ transforms in the adjoint representation of $\grU(N)_k$ and it is a singlet with respect to the second $\grU(N)_{-k}$. Vice versa the field $\hat A_\mu$ is a singlet for $\grU(N)_k$ and transforms in the adjoint of $\grU(N)_{-k}$. 

The fields $Y^A$ and $Y_A^\dagger$ are eight scalars, the index $A$ is an $\grSU(4)$ index, namely $A=1,2,3,4$. This is not the original form of~\cite{Aharony:2008ug}, but rather we use the formulation given in~\cite{Minahan:2008hf, Benna:2008zy}, such that the scalars grouped into $\grSU(4)$ multiplet make {\it R}-symmetry manifest. They transform in the fundamental representation of $\grSU(4)$, i.e $\mathbf 4$ and $\mathbf{\bar 4}$ respectively. Moreover, they transform in the bifundamental representation of the gauge group: $(N,\bar N)$ and $(\bar N, N)$ respectively. The explicit components of the scalars are%
\footnote{The fields $A_a$, $B_{\dot b}$ and their Hermitian conjugates $A^\dagger_a$, $B^\dagger_{\dot b}$ are components of the super-potential
\[
\label{super_W}
W= {2\pi\over k} \Tr \eps^{ab} \eps^{\dot a \dot b} \left( A_{a} B_{\dot a} A_b B_{\dot b}\right)\qquad
\text{with}~~ a, b=1,2 ~, ~\dot a, \dot b=\dot 1, \dot 2\,. 
\]
Writing in terms of the super-potential $W$ \eqref{super_W} makes the flavor $\grSU(2)\times\grSU(2)$ symmetry manifest (but not the {\it R}-symmetry). }
 \[
Y^A= ( A_1, A_2, B^\dagger_{\dot 1}, B^\dagger_{\dot 2})
 \qquad
 Y_A^\dagger= ( A^\dagger_1, A^\dagger_2, B_{\dot 1}, B_{\dot 2})\,.
 \]

Furthermore, the fields $A_a$ transform as an $\grSU(2)$ doublet and the same is true for the $B_{\dot a}$'s, as the notation indicates. Hence, there is an $\grSU(2)\times\grSU(2)\in \grSU(4)$ subsector, which is indeed closed and which is given by $Y^1,Y^2$ and $Y_3^\dagger, Y_4^\dagger$. 

The covariant derivatives are
\[
D_\mu \Phi=\p_\mu \Phi +A_\mu  \Phi - \Phi  \hat A_\mu
\qquad
D_\mu  \Phi^\dagger=\p_\mu  \Phi^\dagger +\hat A_\mu  \Phi^\dagger- \Phi^\dagger A_\mu\,.
 \]
 
The scaling dimension of the scalars $Y$ is $\Delta_0= \Half$, while for the derivatives is $\Delta_0=1$. Furthermore the scalars transform in the trivial representation of $\grSO(3)$, while the covariant derivatives transform in spin 1 representation of $\grSO(3)$ and in the trivial one of $\grSU(4)$. 
 
Finally, the fermions $\Psi^\dagger_A$ and $\Psi^A$ are the $\mathbf{4}$ and $\mathbf{\bar 4}$ multiplets in the spinorial representation of $\grSO(6)$, and they also transform in the $\grU(N)_{k}\times\grU(N)_{-k}$ bifundamental representation. The fermions $ \psi_c^A$ are the charge conjugated fields and they are given by $ \psi_c^A= C \gamma_0 \psi_A^\star$ in terms of the charge conjugation matrix $C$ and $\gamma_\mu$ are the Dirac matrices in three dimensions. They transform in the spin $\Half$ representation of $\grSO(3)$.                                     

The action corresponding to \eqref{def_Lagrangian_CS} is invariant under a CP transformation: the parity changes the sign of the Chern-Simons action which is compensated by the exchange of the gauge fields $A_\mu$ and $\hat A_\mu$. 
%
\paragraph{Symmetries and algebra.}

The theory is conformal and supersymmetric. In particular it possesses $\mathcal N =6$ supersymmetries, which is {\it not} the maximal number of supersymmetries that one can have in three dimensions. 
We already see the first difference with the $\ads_5/\CFT_4$ duality. The supercharges transform in the vector representation of $\grSO(6)\cong \grSU(4)$. I will write the 24 odd generators as $Q_{\alpha I}$ and $S^I_\alpha$ where the spinorial index is $\alpha=1,2$ and the $\grSO(6)$ label is $I=1,\dots, 6$.
Actually, for $k=1,2$ the supersymmetries are enhanced to $\mathcal N=8$, and thus the {\it R}-symmetry is lifted to $\grSO(8)$~\cite{Benna:2008zy}. We will not consider these two cases, since as already mentioned, for us $k$ takes very large values.      

The conformal group in three dimensions is $\grSO(3,2)$. The generators are the Lorentz generators $L_{\mu\nu}$, which are in total three, i.e. $\mu=0,1,2$, the three translation generators $P_\mu$, the dilatation generator $D$ and the three special conformal transformations $K_\mu$. 

The {\it R}-symmetry group is $\grSO(6)\cong \grSU(4)$ with 15 generators, $R_{IJ}$, $I,J=1,\dots, 6$, as we discussed in the $\mathcal N=4$ SYM case in section \ref{chapter:AdSIntro}.       

The direct product $\grSO(3,2)\times \grSU(4)$ corresponds to the bosonic subgroup of $\grp{OSp}(6|4)$. Thus the full global symmetry group of the CS-matter theory is $\grp{OSp}(6|4)$. 


The string states and the gauge theory primary operators will organize themselves as $\mathfrak{osp}(6|4)$ multiplets and they will be characterized by the quantum numbers labeling the bosonic sub-sectors. In particular, these are
\[
\label{charges_ads4}
\left ( \Delta= E, S, J_1,J_2, J_3\right)\,.
\]
The first two charges, i.e. $\Delta (E)$ and $S$, are the Cartan generators of the $\grSO(2)\times \grSO(3)$ maximally compact sub-sector%
\footnote{Actually we are splitting the group $\grSO(3,2)$ according to an Euclidean signature.}
of the full conformal group. 
Notice that in the first entry of \eqref{charges_ads4} we have summarized the content of the gauge/gravity correspondence. The scaling dimension $\Delta$ and the string energy $E$ are the only charges which depend on the coupling constant $\lm$: $\Delta(\lm, N)= E(\lm, N)$.
The last three charges $J_1,J_2, J_3$ are the eigenvalues corresponding to the $\grSU(4)$ Cartan generators. I have indicated with $J_1$ and $J_2$ the two generators of the $\grSU(2)\times\grSU(2)$ sub-sector mentioned before. 

%
%
\paragraph{The symmetries on the string theory side.}

Let us see how the global symmetries are realized on the string scenario. 
The IIA superstring lives on $\ads_4\times \C P^3$. The isometry group of $\ads_4$ is indeed $\grSO(3,2)$. As for the previous case, $E$ is the charge corresponding to global time translation and $S$ is the spin in the AdS space. In other words, according to the splitting of $\grSO(3,2)\rightarrow \grSO(2)\times \grSO(3)$ and to the isomorphisms $\grSO(2)\cong \grU(1)$, $\grSO(3)\cong \grSU(2)$, $E$ is the eigenvalue for the $ \grU(1)$ charge, while $S$ is the spin generator of $\grSU(2)$. Thus once more, the conformal group enters on the string theory side as a symmetry of the background. The same is true also for the projective space $\C P^3$: the corresponding isometry group is $\grp{SU}(4)$. Notice that in $\C P^3$ there are two 2-spheres $\sphere^2$ embedded. They corresponds to the $\grSU(2)\times\grSU(2)$ sub-sector on the gauge theory side. Thus, $J_1$ and $J_2$ represent the total angular momenta in each sphere $\sphere^2$.



\subsection{Spin chains and anomalous dimension}

We want to study the correlation functions of primary operators in the ABJM theory. This means that we want to compute the anomalous dimension for such operators, cf. section \ref{sec:spinchain}. Can we use the spin chain picture also in this case? 

We can repeat the arguments for the $\ads_5/\CFT_4$ duality and represent a local gauge invariant single trace operator via spin chain and study the corresponding quantum mechanical model. In particular the spin chain Hamiltonian will be the mixing matrix, and its eigenvalues will be the anomalous dimensions. Once more this was done for the first time by Minahan and Zarembo in~\cite{Minahan:2008hf}. 

Let us consider the $\grSU(4)$ scalar sector. A prototype of the operator that we want to study is 
\[
\label{op_ABJM}
\mathcal O = C^{B_1 B_2\dots B_L}_{A_1 A_2 \dots A_L} \Tr \left( Y^{A_1} Y^\dagger_{B_1} Y^{A_2}  Y^\dagger_{B_2}  \dots Y^{A_L}   Y^\dagger_{B_L}\right) \,,
\]
where $C^{B_1 B_2\dots B_L}_{A_1 A_2 \dots A_L}$ is a generic tensor. We have to insert a field transforming in the $\mathbf 4$ representation in one site of the spin chain, and the next neighbor has to be a field in the $\mathbf{\bar 4}$ representation, since we want a gauge invariant operator and the matter is in the bifundamental representation, as we discussed in the previous section. In this way, the gauge group indices are correctly multiplied. 
Hence, the operator $\mathcal O$ \eqref{op_ABJM} can be represented as an {\it alternating} spin chain. 
This also implies that now the leading order spin chain Hamiltonian involves the next-nearest neighbors, in other words it starts with two-loop interactions ($\sim \lm^2$). Notice that the length of the chain corresponding to the local operator $\mathcal O$ \eqref{op_ABJM} is $2L$. 

When the tensor $C^{B_1 B_2\dots B_L}_{A_1 A_2 \dots A_L}$ gives a symmetric and traceless combination of the scalars in \eqref{op_ABJM}, then the operator $\mathcal O$ is a chiral primary, and its scaling dimension is protected.

The $\grSU(4)$ 2-loop spin chain Hamiltonian is \cite{Minahan:2008hf}
\[
\label{ABJM_spinchain_H}
\Gamma^{(2)}={\lm^2\over 2} \sum_{l=1}^{2L} H_{l,l+1,l+2}= {\lm^2\over 2} \sum_{l=1}^{2L} \left ( 2- 2P_{l,l+2}+ P_{l,l+2} K_{l,l+1}+ K_{l,l+1}P_{l,l+2} \right)\,,
\]
where $P_{l,l+2} $ is the permutation operator and $K_{l,l+1}$ is the trace operator. 

In~\cite{Minahan:2008hf} the scalar $\grSU(4)$ sector was shown to be integrable at leading order (two-loops). The result was also found in~\cite{Bak:2008cp} and in~\cite{Gaiotto:2008cg}. In~\cite{Minahan:2009te} and in~\cite{Zwiebel:2009vb} the two-loop spin chain Hamiltonian for the  entire $\grp{OSp}(6|4)$ group has been constructed and showed that it is integrable. 
The result was also found in~\cite{Bak:2008cp} and in~\cite{Gaiotto:2008cg}.

As before, we can exploit integrability by applying the techniques learned in section \ref{chapter:AdSIntro} in order to compute the anomalous dimensions for single trace local gauge invariant operators. The leading order Bethe Ansatz (ABE) where constructed for the scalar sector in~\cite{Minahan:2008hf} and for the full $\grp{OSp}(6|4)$ group in~\cite{Minahan:2009te}.  Afterwards, N. Gromov and P. Vieira proposed the Bethe Ansatz equations for the entire $\grp{OSp}(6|4)$ group and at {\it all loop} order~\cite{Gromov:2008qe}.

There are already important data available from the string world-sheet computations, in particular for the spinning and rotating strings at one-loop~\cite{Alday:2008ut, McLoughlin:2008ms, Krishnan:2008zs, McLoughlin:2008he}. From these computations it emerges an apparent disagreement with the Bethe ansatz predictions at the next-leading order of the strong coupling limit for the function $h(\lm)$, cf. equation~\eqref{disp_rel_ABJM}. In particular, the string world-sheet computations suggest a one-loop correction entering in $h(\lm)$. However, this is (partially) understood as due to the employment of different regulation schemes among the string theory computations and the Bethe ansatz computations~\cite{McLoughlin:2008he, Bandres:2009kw}. 
%
%


\subsubsection{The $\grSU(2)\times \grSU(2)$ spin chain}

Let us focus on the $\grSU(2)\times \grSU(2)$ bosonic sector. This is a nice testing ground since it is a closed subsector and probably the simplest one. 
Recall that it is generated by the scalars $A_{1,2}$ and $B_{\dot 1, \dot 2}$, i.e. $Y^{1,2}$ and $Y_{3,4}^\dagger$. 

We want to calculate the anomalous dimension $\gamma$ for operators such as 
\[
\label{op_SU2_ABJM}
\mathcal O= C^{a_1 a_2 \dots a_L}_{b_1 b_2 \dots b_L} \Tr \left ( A_{a_1} B_{b_1} A_{a_2} B_{b_2}\dots  A_{a_L} B_{b_L}\right)\,.
\]
The choice of the vacuum
\[
\label{vacuum_ABJM}
\Tr\big( Y^1 Y_3^\dagger \big)^J \equiv \Tr \left( A_1 B_{\dot 1}\right)^J
\]
breaks the initial global symmetry. In particular what is left is an $\grSU(2|2) \times \grU(1)$ symmetry. 
Looking at the Hamiltonian \eqref{ABJM_spinchain_H}, one can see that in this sub-sector the trace operator $K_{l,l+1}$ does not contribute, thus the Hamiltonian reduces to 
\[
\label{Hspinchain_ABJM}
\Gamma^{(2)}_{|\grSU(2)}={\lm^2\over 2} \sum_{l=1}^{2L} H_{l,l+1,l+2}= {\lm^2\over 2} \sum_{l=1}^{2L} \left ( 2- 2P_{l,l+2} \right)\,.
\]
If one remembers section \ref{sec:spinchain}, one will recognize that the Hamiltonian \eqref{Hspinchain_ABJM} is nothing but (two times) the Heisenberg Hamiltonian of section \ref{sec:spinchain}. Thus, we have two separate $XXX_{\half}$ spin chains, one corresponding to the odd sites and the other to the even ones~\cite{Minahan:2008hf}. However, they are not completely decoupled since we have a unique cyclicity condition, which will couple the momenta for the two spin chains. Notice that each spin chain has $L$ sites. 

Recalling the Bethe Ansatz equations for the Heisenberg spin chain in section \ref{sec:BE}, it is straightforward to write down the $\algSU(2)\times \algSU(2)$ Bethe Ansatz equations, essentially they are the same:
\be
\label{su2_ABE_ABJM}
&&
E= 4\lm^2 \left [ \sum_{i=1}^{K_1} \sin^2 {p^{(1)}_i\over 2}+ \sum_{i=1}^{K_2} \sin^2 {p_i^{(2)}\over 2}\right]
\nln
&& e^{i p_k^{(a)} L}= \prod_{j=1,j\neq k}^{K_a} S(p^{(a)}_j,p^{(a)}_k)
\nln
&& \sum_{i=1}^{K_1}p^{(1)}_i+ \sum_{i=1}^{K_2} p^{(2)}_i=0\,.
\ee
$K_1$ and $K_2$ are the magnon numbers in the odd and even sites of the chain, respectively; the superscript $a=1,2$ selects the odd or the even sites. The S-matrix is the same as in section \ref{sec:BE}, namely $S(p_j,p_k)= - {1+e^{i(p_k+p_j)}-2 e^{ip_k}\over 1+e^{i(p_k+p_j)}-2 e^{ip_j} }$.

\paragraph{Symmetries and S-matrix.}

The choice of the vacuum \eqref{vacuum_ABJM} breaks the initial global $\grp{OSp}(6|4)$ symmetry to the $\grSU(2|2)$ symmetry. Once more, the algebra that realizes the integrable structure of the model is the centrally extended $\algSU(2|2)$ algebra. Although now, we have only one copy. 
Analyzing the bosonic sector, we see that the initial symmetries are broken into 
\[
\grSO(3,2)\times \grSU(4)\rightarrow \grSO(2) \times \grSO(3) \times \grSU(2)_{\mathit{R}}  \times \grSU(2)_{\mathit{R}}\,. 
\]
$\grSO(3)\cong \grSU(2)$ is the group of the space-time rotations; one of the two $\grSU(2)_{{\it R}}$ groups is broken by the vacuum choice.
Thus the direct group $\grSU(2)\times \grSU(2)_{{\it R}}$ gives the bosonic subgroup of $\grSU(2|2)$ (with  $\grU(1)$ central extension). 

The full S-matrix has been constructed in~\cite{Ahn:2008aa}. It has been deduced through the ZF algebra, cf. section~\ref{sec:Smatrix}. It has already passed some consistency checks, at two loops at weak coupling~\cite{Ahn:2009zg} and at tree-level at strong coupling~\cite{Zarembo:2009au}. It reproduces the all-loop Bethe Ansatz equations conjectured in~\cite{Gromov:2008qe}. 

The one particle state forms a $(\mathbf 2| \mathbf 2)$ fundamental representation of the centrally extended $\algSU(2|2)$ algebra. The dispersion relation obtained by the BPS condition (or shortening condition), cf. \ref{sec:Smatrix}, is 
\[
\label{disp_rel_ABJM}
\mathfrak{C}= \sqrt{\Quarter + h(\lm) \sin^2 {p\over 2}}\,. 
\]
In $\ads_5/\CFT_4$ the dispersion relation \eqref{disp_rel} is the same (with $h(\lm)\sim\lm $) at strong and weak coupling limit, as we saw, for example, by studying the BMN limit in section \ref{sec:BMN_details}.
However now things are different. Recall that the shortening condition and, more in general, symmetry arguments fix the form of the dispersion relation only up to a scalar function $h(\lm)$. The specific behavior of such a function in the UV or IR regime enters as an input and, for example, it can be fixed by a comparison with the BMN limit. There is no reason why this should be the same at strong and weak coupling limit. For the $\ads_5/\CFT_4$ duality it happens. But this is not true now in the ABJM conjecture. At the weak coupling ($\lm\ll 1$) the authors of~\cite{Minahan:2008hf, Gaiotto:2008cg, Nishioka:2008gz} have found that $h(\lm)\sim 4 \lm^2$:
\[
\mathfrak{C}= \sqrt{\Quarter+ 4\lm^2 \sin^2 {p\over 2}}\qquad \text{when} \quad \lm \ll 1\,.
\]
However, at the strong coupling ($\lm\gg 1$) the results of~\cite{Gaiotto:2008cg, Grignani:2008is, Nishioka:2008gz} give a different behavior: $h(\lm)\sim 2\lm$, cf. section \ref{sec:Penrose_limit_ABJM}:
\[
\mathfrak{C}= \sqrt{\Quarter+2\lm \sin^2{p\over 2}} \qquad \text{when} \quad \lm \gg 1\,.
\]
The violation of the BMN scaling already at the leading order might be due to a lack of supersymmetries.


\subsection{Integrability on the string theory side}
\label{sec:stringside_integrability}


Let us move to the string theory side: the type IIA superstring leaving on $\ads_4\times \C P^3$. The background can be written as a bosonic quotient space, namely
\[
\ads_4 = {\grSO(3,2)\over \grSO(3,1)} \qquad \C P^3 =\frac{\grSU(4)}{\grU(3)}\,,
\]
which is the bosonic subgroup of $\grp{OSp}(6|4)$. Hence the super-coset approach \'a la GSMT, cf. section \ref{sec:GSMT_string}, can be employed in this case for the formulation of the type IIA string action~\cite{Arutyunov:2008if, Stefanski:2008ik}. There are certain subtleties. In the initial GS superstring action there are in total 32 fermionic degrees of freedom, while now they are 24. Thus part of the $\kappa$-symmetries must be fixed in order to adjust the number of fermions, in particular half (8) of such local fermionic symmetries are gauged away~\cite{Arutyunov:2008if}.

Arutyunov and Frolov have proved the classical integrability of the type IIA string $\sigma$ model on $\grp{OSp}(6|4)/\grp{SO}(3,1)\times \grU(3)$ in \cite{Arutyunov:2008if} by constructing the Lax pair as it was done for the $\ads_5\times \sphere^5$ case~\cite{Bena:2003wd}, cf. section \ref{sec:classical_integrability_GSMT}. However, the fact that the superspace $\ads_4\times \C P^3$ is not a super-coset, implies that the classical integrability has been rigorously showed only for a sub-sector of the full complete $\ads_4\times \C P^3$ background~\cite{Gomis:2008jt}.


\subsubsection{The BMN limit}
\label{sec:Penrose_limit_ABJM}

Recalling what we have learned about the BMN limit (especially on the string theory side) in section~\ref{sec:BMN_details}, we will analyze the IIA string on the projective space in an analogous manner. Let us consider a string with a very large angular momentum%
\footnote{On the gauge theory side this corresponds to primary local operators with a very large {\it R}-charge (or alternatively very long spin chain with a finite number of impurities), cf. section~\ref{sec:BMN_details}.}
 $J$ in $\C P^3$.
As we discussed in~\ref{sec:BMN_details}, this limit is equivalent to consider the string moving in the background obtained by taking the Penrose limit ($R\rightarrow \infinity$) of the original geometry, which is now $\ads_4\times \C P^3$. Remember that, by dimensional analysis, the very large $R^2$ limit is the same as the very large $J$ limit. 
The string is excited along the global time direction $t$ in $\ads_4$ and it is rotating very fast in $\C P^3$. Thus we proceed by computing a perturbative expansion around the classical trajectory (the point-like string configuration). 

The Penrose limit has been computed in \cite{Gaiotto:2008cg, Nishioka:2008gz, Arutyunov:2008if, Grignani:2008is}, expanding the motion in very similar null geodesics. However, I will mostly refer to the decoupling limit used by Grignani, Harmark and Orselli~\cite{Grignani:2008is}, which is based on the work~\cite{Bertolini:2002nr} for the $SU(2)$ sector of $\ads_5\times S^5$.

The $\ads_4\times \C P^3$ space is described by
\begin{equation}
\label{adscp} ds^2 = \frac{R^2}{4} ds^2_{\ads_4}+ R^2 ds_{\C P^3}^2\,,
\end{equation}
with the unit metric written as
\be
\label{cp3_metric} 
&& ds^2_{\ads_4}= \left( - \cosh^2 \rho dt^2 +
d\rho^2 + \sinh^2 \rho d\hat{\Omega}^2_2 \right) \nln
&& 
ds_{\C P^3}^2 = \frac{1}{4} d\psi^2 + \frac{1-\sin
\psi}{8} d\Omega_2^2 + \frac{1+\sin \psi}{8} d{\Omega_2'}^2 + \cos^2
\psi ( d \delta + \omega )^2\,. \nln
\ee
%
%
%
%
%
The one-form $\om$ in \eqref{cp3_metric} is given by
\begin{equation}
\omega = \frac{1}{4} \sin \theta_1 d\varphi_1 + \frac{1}{4} \sin
\theta_2 d\varphi_2\,,
\end{equation}
and $d\Omega_2^2$ and $d{\Omega_2'}^2$ parameterize the two spheres $\sphere^2$ embedded in $\C P^3$, in particular we have that:
\[
\label{spheres}
d\Omega_2^2= d\theta_1^2 + \cos^2\theta_1 d\phi_1^2
\qquad
d{\Omega_2'}^2=d\theta_2^2 + \cos^2\theta_2 d\phi_2^2\,.
\]
Thus, the ten embedding coordinates on $\ads_4 \times \C P^3$ are:
\[
\underbrace{t,\, \rho ,\, \hat\Om_2}_{\ads_4}
\qquad
\underbrace{ \psi\,, \dl\,, \theta_1\,, \phi_1\,, \theta_2\,, \phi_2}_{\C P^3}
\]
We want to make two operations at this point:
\begin{itemize}
\item We want to select the $\grSU(2)\times \grSU(2)$ sector;
\item we want to take the Penrose limit, cf. equation \eqref{ppwave_coordinates}.
\end{itemize}
This implies that we have to choose a null geodesic such that the only excited coordinates lie in the projective space (a part the time direction), i.e. $R_t \times \sphere^2\times \sphere^2$. Secondly, the coordinates should be rescaled in order to take the infinite radius limit. 

The coordinates which are suitable in order to select the $\grSU(2)\times \grSU(2)$ sector~\cite{Grignani:2008is}, are
\begin{equation}
\label{su2_change_coord}
t' = t \qquad \chi = \delta - \frac{1}{2} t\,. 
\end{equation}
This gives the following metric for $\ads_4\times \C P^3$
\be
\label{adscp2} 
 ds^2  &=&  - \frac{R^2}{4} {dt'}^2 ( \sin^2 \psi  +
\sinh^2 \rho
) + \frac{R^2}{4}( d\rho^2 + \sinh^2 \rho d\hat{\Omega}_2^2 ) \nln
& +&  R^2 \big[ \frac{d\psi^2}{4} + \frac{1-\sin \psi}{8} d\Omega_2^2
+ \frac{1+\sin \psi}{8} d{\Omega_2'}^2 \nln
&+&  \cos^2 \psi ( dt' + d \chi
+ \omega )( d \chi + \omega )\big]\,. 
\ee
In section \ref{chapter:NFS} we have introduced the $\grU(1)$ charges in equation \eqref{U1_charges}, analogously here we have%
\footnote{The symbol $=$ should be properly read as a prescription here. }
\[
\label{charges_ABJM}
\Delta = i \p_{t}
\qquad 
J = -{i\over 2} \p_\dl\,. 
\]
After the change of coordinates \eqref{su2_change_coord}, by the chain rule%
\footnote{The inverse transformations of \eqref{su2_change_coord} are $t=t'$ and $\dl=\Half t' +\chi$.}
%
 the charges become
\begin{equation}
\label{newch} 
E \equiv \Delta - J = i \partial_{t'} 
\qquad 2 J = - i
\partial_\chi\,.
\end{equation}

Let us rescale the coordinates according to
\begin{equation}
\label{rescaling_coord}
v = R^2 \chi \spa x_1 = R \varphi_1 \spa y_1 = R \theta_1 \spa x_2 =
R \varphi_2 \spa y_2 = R \theta_2 \spa u_4 = \frac{R}{2} \psi\,,
\end{equation}
and transform the transverse coordinates in $\ads_4$ with $u_1$, $u_2$ and $u_3$ defined by the relations
\begin{equation}
\label{u_coords}
\frac{R}{2} \sinh \rho = \frac{u}{1 - \frac{u^2}{R^2}} \spa
\frac{R^2}{4} ( d\rho^2 + \sinh^2 \rho d\hat{\Omega}_2^2 ) =
\frac{\sum_{i=1}^3 du_i^2}{(1-\frac{u^2}{R^2}  )^2} \spa u^2 =
\sum_{i=1}^3 u_i^2\,. 
\end{equation}
Explicitly, the metric \eqref{adscp2} in the new coordinates~\eqref{rescaling_coord} and \eqref{u_coords},
becomes
\be
\label{adscp3} 
ds^2  &=& -  {dt'}^2 \left( \frac{R^2}{4} \sin^2
\frac{2u_4}{R} + \frac{u^2}{(1 - \frac{u^2}{R^2})^2} \right) +
\frac{\sum_{i=1}^3 du_i^2}{(1-\frac{u^2}{R^2}  )^2} + du_4^2\nln
&+&
\frac{1}{8}\left(\cos\frac{u_4}{R}-\sin\frac{u_4}{R}\right)^2\left(
dy_1^2+\cos^2\frac{y_1}{R}d
x_1^2\right)\nln
&+&\frac{1}{8}\left(\cos\frac{u_4}{R}+\sin\frac{u_4}{R}\right)^2\left(
dy_2^2+\cos^2\frac{y_2}{R}d x_2^2\right)
\nln
&+& R^2\cos^2\frac{2u_4}{R} \left[ dt' + \frac{dv}{R^2}
+\frac{1}{4}\left(\sin \frac{y_1}{R} \frac{d x_1}{R} + \sin
\frac{y_2}{R} \frac{d x_2}{R}\right)\right] \times 
\nln
&& \times \left[\frac{dv}{R^2}
+\frac{1}{4}\left(\sin \frac{y_1}{R} \frac{d x_1}{R} + \sin
\frac{y_2}{R} \frac{d x_2}{R}\right)\right]\,. 
\ee
%

At the leading order, the $R \rightarrow \infty$ limit of the metric~\eqref{adscp3} leads to the plane-wave metric given by:
\begin{equation}
\label{ppwavemetric_ABJM} 
ds^2 =  dv dt'  + \sum_{i=1}^4 ( du_i^2 - u_i^2
{dt'}^2 ) + \frac{1}{8} \sum_{i=1}^2 ( dx_i^2 + dy_i^2 + 2 dt' y_i
dx_i )\,.
\end{equation}
The light-cone coordinates in this metric are $t'$ and $v$:
one should read
\[
 t'\rightarrow X^+\qquad v\rightarrow X^-
 \]
 in order to use the results of section \ref{chapter:NFS}. 
In essence, this is equivalent to consider the following classical configuration for the string
\[
\rho=0 \qquad \theta={\pi\over 4} \,.
\]

After the rescaling \eqref{rescaling_coord}, also the $\grU(1)$ charge $J$ in \eqref{charges_ABJM} gets rescaled according to 
\begin{equation}
\frac{2 J}{R^2} = - i \partial_v\,. 
\end{equation}
This is equivalent to $P_-$ in equation \eqref{momenta_rescaled} in the case $a=0$.

\paragraph{The light-cone gauge.}

We need to fix the light-cone gauge if we want to quantize the string Hamiltonian, since there are Ramond-Ramond fluxes and they survive the Penrose limit, cf. equations \eqref{lcgauge_intro} and \eqref{lcgauge}. Explicitly:
\[
\label{lcgauge_ABJM}
t'=c\tau \qquad p_v = \text{constant}\,,
\]
where the constant is fixed by the computation%
\footnote{The constant is fixed through the relation $2J= {1\over 2\pi\alpha'}\int_0^{2\pi} d\sigma p_\chi= {1\over 2\pi\alpha'}\int_0^{2\pi} d\sigma {\dl \mathcal L\over \dl \dot\chi}$.}
of the canonical momentum $p_v={\dl \mathcal L\over \dl \dot v} $ and gives
\[
\label{cconstant}
c={4 J\over R^2}\,.
\]
This will be used as our expansion parameter and it corresponds to $P_-$ of section \ref{sec:BMN_details}, cf. equation \eqref{momenta_rescaled}. 

After solving the Virasoro constraints \eqref{Virasoro}, the bosonic light-cone Hamiltonian computed according to \eqref{H_lc} in the background \eqref{ppwavemetric_ABJM} gives 
\be
\label{densityH_bosons}
&&c \mathcal H_{B,pp}=
\sum_{a=1}^2 (p_{x_a}\dot x_a+ p_{y_a} \dot y_a)+
\sum_{i=1}^4 p_{u_i} \dot u_i -\mathcal L_{B,pp}=\\ \nn
&&=\sum_{a=1}^2\, [ 4 p_{x_a}^2 +4 p_{y_a}^2 +{1\over 16} {x'}^2_a +{1\over 16} {y'}_a^2 -c p_{x_a} y_a +{c^2 \over 16} y_a^2
]
 +\frac{1}{2} \sum_{i=1}^4 [p_{u_i}^2 +{u'}_i^2 + c^2 u_i^2]\,.
\ee
The quantization of the coordinates%
\footnote{The details about the normalization and the explicit expression for the bosonic modes are in appendix \ref{subsec:quantizationABJM}.}
leads to the following free%
\footnote{The same is obtained for the plane-wave fermionic spectrum~\cite{Astolfi:2009qh}:
\be
\label{Hppwave_ferm_final}
H_{F,pp}= {1\over c} \sum_{n\in\Z} \sum_{b=1}^4 \om_n F_n^{(b)}
+{1\over c} \left\{ \sum_{n\in\Z}  \sum_{b=1}^2 \left(\Om_n +{c\over 2}\right) \tilde F_n^{(b)}
                           + \sum_{n\in\Z}  \sum_{b=1}^2 \left(\Om_n -{c\over 2}\right) \tilde F_n^{(b)}\right\}\,,
\ee
with dispersion relations $\om_n=\sqrt{n^2+{c^2\over 4}}$, $\Om_n=\sqrt{n^2+c^2}$ and the number operators $F_n= d^\dagger_n d_n$ and $\tilde F_n= b^\dagger_n b_n$.  We avoid to write the spinorial indices. }
bosonic Hamiltonian
\begin{equation}
 \label{penspectrum}
c H_{\rm free} = \sum_{i=1}^4 \sum_{n\in \Z}\Om_n\,
\hat{N}^i_n
+\sum_{a=1}^2\sum_{n\in \Z}
\left(\om_n - \frac{c}{2}\right) M_n^a
+\sum_{a=1}^2\sum_{n\in \Z} \left(\om_n+
\frac{c}{2}\right) N_n^a\,,
\end{equation}
with the number operators $\hat{N}^i_n = (\hat{a}^i_n)^\dagger
\hat{a}^i_n$, $M_n^a = (a^a)^\dagger_n a^a_n$ and $N_n^a =
(\tilde{a}^a)^\dagger_n \tilde{a}_n^a$, and with the level-matching
condition
\begin{equation}
\label{levelm}
 \sum_{n\in \Z}n \left[\sum_{i=1}^4
\hat{N}^i_n+\sum_{a=1}^2 \left(M_n^a + N_n^a\right)\right]
 = 0\,. 
\end{equation}
The dispersion relations are
\[
\label{def_omeghe}
\Om_n= \sqrt{n^2+c^2}\qquad \om_n=\sqrt{\frac{c^2}{4}+n^2}\, .
\]
This plane-wave Hamiltonian \eqref{penspectrum} describes 8 bosonic (and 8 fermionic) degrees of freedom. But there are some surprises. 

The dispersion relations \eqref{def_omeghe} of the plane-wave Hamiltonian show that, firstly, we have two different sets of excitations, and secondly, that in both cases the dispersion relations do not match the gauge theory result. 
As it is clear from \eqref{penspectrum} and \eqref{Hppwave_ferm_final}, the masses, which appear there, are different. 
We have obtained four bosons with mass $m=\Half$, the {\it light-modes} and four with mass $m=1$, the {\it heavy-modes}. The same is true for the fermions. 
The $(4|4)$ {\it light} multiplet corresponds to the transverse coordinates of $\C P^3$, $(x_1,y_1,x_2, y_2)$, namely to the two spheres $\sphere^2$, \eqref{spheres}, after the rescaling \eqref{rescaling_coord}. These elementary excitations correspond to those seen on the gauge theory side.
In particular for the light-modes, after using \eqref{cconstant}, the energies are
\[
{1\over c} \om_n=\sqrt{{n^2\over c^2}+{1\over 4}}= \sqrt{\Quarter + {2\pi^2\lm\over J^2} n^2}\,.
\]
This is consistent with the dispersion relation discussed in the previous section:
\[
\sqrt{\Quarter + 2 \lm \sin^2{p\over 2}}\,. 
\]
The bosonic heavy modes correspond to the transverse directions $(u_1,u_2,u_3,u_4)$ and they are not observed on the gauge theory side. Actually their role is distinct, this fact is not visible in the BMN limit. Indeed, the coordinate $u_4$ plays a special role. The other coordinates $(u_1\,, u_2\,, u_3)$ are rotated by the group $\grSO(3)$ and they correspond to the derivatives on the gauge theory side. 

Hence, there is an apparent mismatch on the number of the elementary impurities which appear on gauge and string theory side. This was resolved by Zarembo in~\cite{Zarembo:2009au} where he showed the fate of the heavy world-sheet modes. They are not elementary world-sheet excitations. They disappear from the spectrum: Once the leading quantum corrections in the propagator are taken into account, it is possible to see that the pole corresponding to heavy modes is indeed above the threshold for the light-mode pair productions. They are absorbed in the continuum and thus ``invisible'' from the gauge theory point of view.


\subsection{The near-BMN corrections}
\label{sec:paperIII}

%
%


Let us take a step forward in the study of the BMN regime on the string theory side: We want to compute the leading (${1\over J}$) quantum corrections to the string energies, for a certain class of string configurations, following~\cite{Astolfi:2008ji}. This method was proposed by Callan {\it et al.} in \cite{Callan:2004uv, Callan:2003xr} for the $\ads_5\times \sphere^5$ superstring. For other methods used to compute the ${1\over J}$ corrections in the $\ads_5/\CFT_4$ context we refer the reader to the papers~\cite{Arutyunov:2004yx, Arutyunov:2005hd, Frolov:2006cc}.

Summarizing what we have seen in the previous section, our starting point is a light-cone gauged string moving on $t\in \ads_4$ and $\sphere^2\times \sphere^2\in \C P^3$ with a very large angular momentum $J$ in the projective space. We are at strong coupling limit $\lm\gg 1$ and also $J$ (or $R$) is very large, however the ratio $\lm'={\lm\over J^2}$ is kept fixed. This $\lm'$ becomes an effective parameter to explore the spectrum beyond the Penrose limit. 

In particular we want to make a joint expansion in large $J$ {\it and} in small $\lm'$, cf. what we have discussed in section \ref{sec:BMN_details} about the BMN-scaling. In a certain sense, we are saying that the angular momentum is very large but yet finite.%
\footnote{From this, it follows the name finite-size corrections. They should not be confused with the finite size corrections which enter by considering the strings in a finite volume and which are exponentially small. This kind of corrections are not captured by the ABE, thus we will not deal with them, cf. the discussion in section \ref{sec:BE}. The finite-size corrections which are discussing here are near-BMN corrections, and indeed, I will use the two expressions as synonyms.}

Since by dimensional analysis the ${1\over J}$ corrections are equivalent to the ${1\over R^2}$ corrections, the finite-size corrections can be computed by including higher order terms in the inverse of the curvature radius, i.e. up to ${1\over R^2}$. 

Let us focus on the bosonic sector of the type IIA $\ads_4 \times \C P^3$ superstring. Thus, the discussion of the section \ref{subsec:lightcone_gauge} applies directly here. 
All the relevant formulas are written in section \ref{subsec:lightcone_gauge}, let me just recall the main expressions. 
The starting point is the bosonic action%
%
%
\[
S= {1\over 2\pi \alpha'} \int d\sigma^2 \mathcal L\quad \text{with}\quad \mathcal L= -\Half \gamma^{\mu\nu} G_{MN} \p_\mu X^M \p_\nu X^N\,,
\]
and the two Virasoro constraints \eqref{Virasoro}. Solving the second one of \eqref{Virasoro} in favor of ${X^-}'$ ($v'$ in~\cite{Astolfi:2008ji}) gives the light-cone Hamiltonian density
\[
\label{H_lc_paperIII}
\mathcal H_{lc}= -p_{t'}\,.
\] 
Notice that $p_+$ of section \ref{chapter:NFS} is $p_{t'}$ in the notation of~\cite{Astolfi:2008ji}. 

The crucial step is that everything is consistently expanded up to order ${1\over R^2}\sim {1\over J}$. In the curvature radius expansion, the leading term in \eqref{H_lc_paperIII}, i.e. the term of order $\mathcal O(1)$, is the BMN limit ($\mathcal H_{\rm free}$); the next-leading terms are the new contributions that, once they are quantized, will give us the quantum corrections to the string BMN spectrum, i.e. $\mathcal H_{\rm int}$:
\[
\mathcal H_{lc}= \mathcal H_{\rm free}+\mathcal H_{\rm int}\,.
\]
Notice that $\mathcal H_{\rm free}$ reduces to \eqref{densityH_bosons} in the bosonic sector, which is the sector we are interested in. 

With respect to the $\ads_5$ case, one of the surprising properties of the interacting Hamiltonian $\mathcal H_{\rm int}$ is that it contains also three-leg vertices. It is indeed built of two contributions:
\[
 \mathcal H_{{\rm int}}= \CH^{(1)}_{\rm int}+\CH^{(2)}_{\rm int}\,,
 \]
\begin{itemize}
\item
at order ${1\over R}$ it is cubic and it contains three fields (the heavy mode corresponding to $u_4$ and two light-modes corresponding to two of the four $\sphere^2\in \C P^3$ coordinates), i.e. $\mathcal H_{\rm int}^{(1)}$;
\item at order ${1\over R^2}$ it is quartic (the relevant terms for us are the ones with all the transverse $\grSU(2)\times \grSU(2)$ coordinates), $\mathcal H_{\rm int}^{(2)}$. 
\end{itemize}
Explicitly, we have: 
\begin{equation}\label{ch1}
\CH^{(1)}_{\rm int}=\frac{u_4}{8 R c}\left[(\dot x_1)^2-(\dot
x_2)^2+(\dot y_1)^2-(\dot
y_2)^2-(x_1')^2+(x_2')^2-(y_1')^2+(y_2')^2\right]
\end{equation}
and
\begin{align}\label{ch2}
&\CH^{(2)}_{\rm int}=\frac{1}{128 R^2 c^3}\left[4\left(\dot x_a
x_a'+\dot y_a
y_a'\right)^2-\left(\left(x'_a\right)^2+\left(y'_a\right)^2+\left(\dot{x}_a\right)^2+
\left(\dot{y}_a\right)^2\right)^2\right]\cr&+\frac{1}{48 R^2
c}\left[3  \left(\left((\dot x_1
)^2-(x'_1)^2\right)y_1^2+\left((\dot x_2
)^2-(x'_2)^2\right)y_2^2\right)+c\left(\dot x_1 y_1^3+\dot x_2
y_2^3\right)\right]+\dots\,
\end{align}
where the dots are for terms that are irrelevant in the computation of the spectrum of string states belonging to the $SU(2)\times SU(2)$ sector, $c$ is the constant defined in \eqref{cconstant}, the index $a=1,2$ labels the two copies of $\grSU(2)$ and with $\dot x$ and $x'$ we mean the derivative with respect to the world-sheet coordinate $\tau$ and $\sigma$ respectively. 

The classical interacting Hamiltonian $\mathcal H_{{\rm int}}$ must be quantized%
\footnote{However, since $\mathcal H_{{\rm int}}$ is derived classically, there is a normal ordering ambiguity. We choose to fix the constant of normal ordering to zero, by consistency with the zero vacuum energy. 
},
cf. \ref{subsec:quantizationABJM}, and used to compute the energy corrections via standard perturbation theory, namely
\begin{equation}
\label{energycorrection}
E_{s,t}^{(2)}=\langle s, t|H^{(2)}_{\rm
int}|s,t\rangle+\sum_{|i\rangle}\frac{\left|\langle i|H^{(1)}_{\rm
int}|s,t\rangle\right|^2}{E^{(0)}_{|s\rangle,|t\rangle}-E^{(0)}_{|i\rangle}}\,,
\end{equation}
where $|i\rangle$ is a suitable intermediate state. Notice that $E^{(0)}$ is the pp-wave energy, $E^{(1)}$ vanishes and $H^{(1)}_{\rm int}$ is the integral of $\mathcal H_{{\rm int}}$ over $\sigma$.
In concrete terms, in \eqref{energycorrection} we need to insert some specific state: We investigate two specific string configurations with two impurities in both cases.
One state contains two world-sheet excitations sitting on the same sphere $S^2\in \C P^3$ (the state $\ket s$):
\begin{equation}\label{s}
|s\rangle = (a_n^1)^\dagger (a_{-n}^1)^\dagger|0\rangle\,.
\end{equation}
The second case we consider, is when the two world-sheet excitations are on the two different 2-spheres $\grSU(2)$ (the state $\ket t$):
\begin{equation}\label{t}
|t\rangle = (a^1_n)^\dagger (a^2_{-n})^\dagger|0\rangle\,.
\end{equation}
Both terms $\mathcal H_{\rm int}^{(1)}$ and $\mathcal H_{\rm int}^{(2)}$ contribute at order ${1\over J}$, in particular, for example for the state $\ket t$ one has~\cite{Astolfi:2008ji}
\begin{eqnarray}
 \label{mixingspectrum1}
\langle t| H^{(2)}_{\rm int}|t\rangle
 &=&-\frac{\left[n^2+\left(\omega_n-\frac{c}{2}\right)^2\right]^2+
4n^2\left(\omega_n-\frac{c}{2}\right)^2}{R^2 c^3 \omega_n^2}\nln
&\simeq&
-\frac{4 n^4 \pi^4 \lambda'^2}{J} +\frac{16 n^6 \pi^6
\lambda'^3}{J}+\CO\left(\lambda'^4\right)\,,
\end{eqnarray}
\be\label{mixingspectrum2}
\sum_{|i\rangle}\frac{\left|\langle i|H^{(1)}_{\rm
int}|t\rangle\right|^2}{E^{(0)}_{|t\rangle}-E^{(0)}_{|i\rangle}}
=\frac{1}{
R^2c}\sum_p\frac{\left[\left(\omega_{p+n}-\frac{c}{2}\right)\left(\omega_{n}-\frac{c}{2}\right)-(p+n)n\right]^2}
{\omega_{p+n}\omega_n\Omega_p\left(\omega_{p+n}-\omega_n-\Omega_p\right)}
+\frac{\left[\left(\omega_{n}-\frac{c}{2}\right)^2-n^2\right]^2}{R^2
c^3\omega^2_{n}}\,.
\ee
Notice that the cubic Hamiltonian contribution contains divergent terms which we regularize with the $\zeta$-function. Thus, summing the two contributions \eqref{mixingspectrum1} and \eqref{mixingspectrum2}, one obtains
\be\label{Et}
E_t^{(2)} &=&-\frac{[n^2+(\omega_n-\frac{c}{2})^2]^2+
4n^2(\omega_n-\frac{c}{2})^2}{R^2 c^3
\omega_n^2}+\frac{[(\omega_{n}-\frac{c}{2})^2-n^2]^2}{
R^2c^3\omega^2_{n}} \quad\quad \nln
&\simeq & -\frac{64 n^6
\pi^6\lambda'^3}{J}+\CO(\lambda'^4)
\ee
It is interesting that for the state $|t\rangle$ the first finite-size correction appears at the order $\lambda'^3$. Notice also that there is no $\ads_5$ analogous for the state $\ket t$. 
Analogously it can be done for the state $\ket s$~\cite{Astolfi:2008ji}:
\be
\label{Es}
E_s^{(2)}&=&-2\frac{[(\omega_n-c )(4n^2-c^2)-c^2
\omega_n]}{R^2 c^3 \omega_n}-
\frac{[(\omega_{n}-\frac{c}{2})^2+n^2]^2}{R^2
c\,\omega^2_{n}\Omega^2_{2n}}
-\frac{[(\omega_{n}-\frac{c}{2})^2-n^2]^2}{R^2
c^3\omega^2_{n}}\nln
&\simeq &\frac{8 n^2 \pi^2 \lambda'}{J}-\frac{64
n^4 \pi^4 \lambda'^2}{J}+\frac{448 n^6
\pi^6\lambda'^3}{J}+\CO(\lambda'^4)
\ee

\paragraph{Comparing with the Bethe Ansatz equations and with the Landau-Lifshitz model.}

From a spin chain picture, the $\grSU(2)\times \grSU(2)$ light excitations correspond to the insertions of two fundamental magnons such as $A_1 B_2$, $A_1, B^\dagger_{\dot 2}$, $A_2, B_1$ and $B^\dagger_2, B_1$ in the spin chain. We can pictorially think to the case $\ket s$ as two down spins in the same $XXX_{\half}$ chain and to the case $\ket t$ as each spin down for each chain%
\footnote{This picture should not be taken too much seriously: the chains are the same just involving odd and even sites, indeed there is one trace condition.}.
In this way it has been possible to see that, in the case $\ket t$, the dressing phase contribution is responsible for the interactions between the two spin chains since the S-matrix contribution is trivial in this case. The results of~\cite{Astolfi:2008ji} have been confirmed in~\cite{Sundin:2008vt}.

The energies up to order ${1\over R^2}$ obtained with the above finite-size procedure are compared with the strong coupling limit of the Bethe equations proposed in~\cite{Gromov:2008qe}. The $\grSU(2)\times \grSU(2)$ Bethe Ansatz equations are written in~\cite{Astolfi:2008ji} by following the $\ads_5/\CFT_4$ example, and here are reported in equation \eqref{su2_ABE_ABJM}. In particular, at this order, the dressing phase is a direct generalization of the AFS phase~\eqref{AFS_phase} with the substitution $g^2\rightarrow h(\lm)$, cf. section \ref{sec:dressingphase}.
Furthermore, in the concrete computation it has been used the strong coupling leading order value for the function $h(\lm)$, namely $h(\lm)=2\lm$. 

We have also used another approach in order to compute the energy corrections to the string configurations considered: the so called Landau-Lifshitz (LL) model. This is a low-energy effective model that was initially developed in the $\ads_5/\CFT_4$ case by Kruczenski~\cite{Kruczenski:2003gt}. It has the advantage to be free from divergences and to be well-defined at leading quantum level. For a nice review we refer the reader to the paper~\cite{Tseytlin:2004xa}  and for examples in the $\ads_5\times \sphere^5$ context we refer to the works~\cite{Minahan:2005qj, Minahan:2005mx, Kruczenski:2004kw, Kruczenski:2004cn}. 

The final result contained in~\cite{Astolfi:2008ji} is the complete matching between the energy corrections computed with these three different techniques.


\section{Summary and Conclusions}

\label{chapter:epilogue}

The work is devoted to review the study of the string integrability in the context of the AdS/CFT dualities. The integrable structures which emerge on both sides of the $\ads_5/\CFT_4$ correspondence, manifest themselves with an infinite set of conserved charges. These infinite ``hidden'' symmetries solve, at least in principle, the model and provide us with a formidable tool for exploring the string/gauge correspondence. 

The exposition starts with the $\ads_5/\CFT_4$ correspondence. Its gravity side, namely the type IIB superstring action in $\ads_5\times \sphere^5$, can be formulated in two approaches: the Green-Schwarz-Metsaev-Tseytlin (GSMT) formalism and the Berkovits (pure spinor) formalism. The latter allows one to proceed perturbatively to a manifestly covariant quantization of the string action. Using the pure spinor approach we could analyze the operator algebra of the left-invariant currents which are the main ingredient in the construction of the string action. This has been done by computing the operator product expansion (OPE) of the left-invariant currents at the leading order in perturbation theory (i.e. ${1\over R^2}\sim {1\over \sqrt{\lm}}$) and up to terms of conformal dimension 2. This confirms the $\Z_4$-grading of the full $\algPSU(2,2|4)$ algebra, which is the AdS/CFT global symmetry, as well as the non-holomorphicity of the currents. We have then investigated the quantum integrability of the type IIB $\ads_5\times \sphere^5$ superstring. Its proven classical integrability does not automatically imply that such a property survives at quantum level, as the example of the $\C P^n$ model teaches us. In the first order formalism, the integrability is related to the existence of a Lax pair, namely a flat connection, which guarantees the independence of the contour for the monodromy matrix (the functional generating the infinite tower of conserved charges) and thus the conservation of the charges. We have studied the variation of the monodromy matrix under a small path deformation at the leading order in perturbation theory and in the pure spinor approach. We could give a direct and explicit check that indeed its path-independence holds at quantum level and that it remains free from UV logarithmic divergences. A crucial ingredient in this computation are the OPE's mentioned above. 

Employing the GSMT light-cone gauged type IIB superstring action, one can interpret the world-sheet elementary excitations as two-dimensional particles and construct the corresponding S-matrix by assuming that the model is quantum integrable. We have explicitly verified that such a scattering matrix factorizes as it should be for a two-dimensional integrable quantum field theory. For this computation we have exploited the near-flat space truncation of the full string $\sigma$-model up to one-loop, which means $\sim {1\over \lm^{3/2}}$ for the three-particle scatterings considered. 

Finally we have turned our attention to the $\ads_4/\CFT_3$ correspondence. We have considered the gravity dual given by the type IIA superstring in $\ads_4\times \C P^3$. In the GS formalism we have examined near-BMN string configurations with a large angular momentum $J$ in $\C P^3$. For the bosonic $\grSU(2)\times \grSU(2)$ closed sector we have then calculated the first quantum correction, namely ${1\over J}\sim {1\over R^2}$, to the corresponding string energies. The obtained values have been positively checked against the conjectured all-loop Bethe ansatz predictions.

\section*{Acknowledgments}

The first person whom I would like to thank is my supervisor Kostya Zarembo. Thanks for being a guidance, for all the stimulating discussions and advices, and especially for your capacity to ``see and love the Physics'' and to transmit such amazing feelings. 
I am very grateful to Davide Astolfi, Gianluca Grignani, Troels Harmark, Thomas Klose, Olof Ohlsson Sax and Marta Orselli for the valuable collaborations on the works on which this review is partially based on. I am also indebted to Andrei Mikhailov for all the numerous and precious discussions. I would also like to thank Lisa Freyhult, Ulf Lindstr\"{o}m, Joseph Minahan, Antti Niemi, Maxim Zabzine and all the other members of the Department of Theoretical Physics at Uppsala University for helpful discussions and suggestions.
%
Finally, I thank Joel Ekstrand, Malin G\"{o}teman, Thomas Klose, Johan K\"{a}ll\'{e}n, Olof Ohlsson Sax, Kostya Zarembo and the committee Gleb Arutyunov, Marcus Berg, Rikard Enberg, Lisa Freyhult, Arkady Tseytlin and Niclas Wyllard for reading the various parts of this manuscript and for all the important suggestions and comments.


\begin{appendix}
\section{Appendix}

\label{chapter:appendix}


\subsection{Notation}
\label{app:notation}

\paragraph{Complex coordinates.}

The conventions are the same as used by Polchinski in chapter 2 of~\cite{Polchinski:1998rq}. The $z, \bar z$ coordinates are defined according to: 
\[
z=\sigma^1+i\sigma^2
\qquad
\bar z=\sigma^1 -i\sigma^2\,.
\]
The derivatives are
\[
\p_{z}= \Half(\p_1-i \p_2)
\qquad 
\p_{\bar z}= \Half (\p_1+i \p_2)\,.
\]
Notice that for the Maurer-Cartan forms I use $J\equiv J_{z}$ and $\bar J \equiv J_{\bar z}$. In paper \cite{Puletti:2008ym} they are also indicated with $J_+$ and $J_-$ respectively. 
The two-dimensional metric is 
\[
\eta_{z\bar z}=\eta_{\bar z z}=\Half
\qquad
\eta^{z\bar z}= \eta^{\bar z z}=2\,,
\]
where all the other components are zero. 
The Levi-Civita tensor is defined by $\eps^{12}=-\eps^{21}=+1$. In the Minkowski world-sheet the $\eps $ tensor is defined as $\eps^{01}= -\eps^{10}=+1$. 
In particular, we use the prescription $\sigma^2=i\sigma^0$ for Wick-rotating the coordinates.
Finally, the measure in the $z,\bar z$ coordinate is $d^2z= 2 d\sigma^1 d\sigma^2$.


\section{The $\ads_5/\CFT_4$ duality: the full planar ABE}
\label{app:fullABE}

For completeness, here we report the Asymptotic Bethe equations for the planar $\ads_5/\CFT_4$ \cite{Beisert:2005fw}:
\be
\label{fullABE}
&& 1= e^{i (p_1+\dots +p_{K_4})}= \prod_{j=1}^{K_4} \frac{x_{4j}^+}{x_{4j}^-}
\\ \nn
&& 1=\prod^{K_2} \frac{u_{1k}-u_{2j}+i/2}{u_{1k}-u_{2j}-i/2} \prod_{j=1}^{K_4}\frac{2-g^2/x_{1k}x^{+}_{4j}}{2-g^2/x_{1k}x^{-}_{4j}}
\\ \nn
&& 1= \prod_{j=1,j\neq k}^{K_2} \frac{u_{2k} -u_{2j}-i}{u_{2k} -u_{2j}+i}\prod_{j=1}^{K_3} \frac{u_{2k} -u_{3j}+i/2}{u_{2k} -u_{3j}- i/2}
             \prod_{j=1}^{K_1} \frac{u_{2k} -u_{1j}+i/2}{u_{2k} -u_{1j}- i/2}
\\ \nn
&& 1= \prod^{K_2} \frac{u_{3k} -u_{2j}+i/2}{u_{3k} -u_{2j}- i/2}\prod_{j=1}^{K_4}\frac{x_{3k}-x_{4j}^+}{x_{3k}-x_{4j}^-}
\\ \nn
&& 1= \left(\frac{x_{4k}^-}{x_{4k}^+}\right)^L\prod_{j=1,j\neq k}^{K_4}
            \left( \frac{x^+_{4k}-x^-_{4j}}{x^-_{4k}-x^+_{4j}} \frac{2-g^2/x_{4k}^+ x_{4j}^-}{2-g^2/x_{4k}^- x_{4j}^+}e^{2i \theta(x_{4k}, x_{4j})}\right) 
            \\ \nn
            && \qquad \times \prod^{K_1}\frac{2-g^2/x_{4k}^- g_{1j}}{2-g^2/x^+_{4k}x_{1j}}\prod_{j=1}^{K_3} \frac{x_{4k}^--x_{3j}}{x_{4k}^+-x_{3j}}
              \prod_{j=1}^{K_5}\frac{x_{4k}^- -x_{5j}}{x_{4k}^+-x_{5j}} \prod_{j=1}^{K_7} \frac{2-g^2/x_{4k}^- x_{7j}^-}{2-g^2/x_{4k}^+ x_{7j}}
\\ \nn
&& 1= \prod_{j=1}^{K_6} \frac{u_{5k}-u_{6j}+i/2}{u_{5k}-u_{6j}-i/2}\prod_{j=1}^{K_4} \frac{x_{5k}-x_{4j}^+}{x_{5k}-x_{4j}^-}
\\ \nn
&& 1= \prod_{j=1, j\neq k}^{K_6} \frac{u_{6k}-u_{6j}-i}{u_{6k}-u_{6j}+i} \prod_{j=1}^{K_5}  \frac{u_{6k}-u_{5j}+i/2}{u_{6k}-u_{5j}-i/2}
            \prod_{j=1}^{K_7} \frac{u_{6k}-u_{7j}+i/2}{u_{6k}-u_{7j}-i/2}
\\ \nn
&&  1=\prod_{j=1}^{K_6}  \frac{u_{7k}-u_{6j}+i/2}{u_{7k}-u_{6j}-i/2}\prod_{j=1}^{K_4} \frac{2-g^2/x_{7k}x^+_{4j}}{2-g^2/x_{7k}x^-_{4j}}           
 \,.
\ee
The Bethe roots are $(u_{1k},u_{2k},u_{3k},u_{4k},u_{5k},u_{6k},u_{7k})$ corresponding to the excitation numbers $(K_1,K_2, K_3, K_4, K_5, K_6,K_7)$ and the {\it rapidity map} is defined by
\[
x(u)=\Half u \left(1+ \sqrt{1- {2g^2 \over u^2}}\right), \qquad g^2 ={\lm \over 8 \pi^2}
\]
with
$x^{\pm }(u)\equiv x(u\pm {i\over 2})$. The dressing phase is
\be
 \theta (u_k, u_j) = \sum_{r=2}^\infinity \sum_{{s=1+r,\atop s+r=\text{odd}}}^\infinity c_{r,s} (g)\left( q_r (u_k) q_{s} (u_j) -q_r (u_j) q_{s} (u_k)\right)
\ee
where the coefficients are~\cite{Beisert:2006ez}
\be
 c_{r,s}^{(n)}= \frac{(-1)^n\zeta(n)}{2\pi^n \Gamma(n-1)} (r-1) (s-1) \frac{\Gamma\big({1\over 2}(s+r+n-3)\big)\Gamma\big( {1\over 2} (s-r+n-1)\big)}{\Gamma\big({1\over 2}(s+r-n+1)\big)\Gamma\big({1\over 2}(s-r-n+3)\big)}\,.
\ee
In particular at strong coupling they have been discussed in section \ref{sec:dressingphase}.


\section{Pure spinor formalism}
\label{app:PS}

\subsection{The $\mathfrak{psu}(2,2|4)$ structure constants}
\label{app:psu_structure_const}

The non-vanishing structure constants for the $\mathfrak{psu}(2,2|4)$ super-algebra are the following:
\be
&& f^{[\underline{ab}]}_{\al\bth}=\fraz(\gm^{\underline{ab}})_{\al}^{\;\;\gm}\dl_{\gm\bth}\qquad
f^{[\underline{a'b'}]}_{\al\bth}=-\fraz(\gm^{\underline{a'b'}})_{\al}^{\;\;\gm}\dl_{\gm\bth}\nln
&& f^{\al}_{[cd]\bt}=-f^{\al}_{\bt[cd]}=\fraz(\gm_{cd})_{\bt}^{\;\;\al}\qquad
f^{\alh}_{[cd]\bth}=-f^{\alh}_{\bth[cd]}=\fraz(\gm_{cd})_{\bth}^{\;\;\alh}\nln
&& f^{a}_{\al\bt}=f^{a}_{\bt\al}=\gm^{a}_{\al\bt}\qquad
f^{\bth}_{a\bt}=-f^{\bth}_{\bt a}=-(\gm_a)_{\bt\gm}\dl^{\gm\bth}\nln
&& f^{a}_{\alh\bth}=f^{a}_{\alh\bth}=\gm^{a}_{\alh\bth}\qquad
f^{\al}_{a\alh}=-f^{\al}_{\alh a}=(\gm_a)_{\alh\bth}\dl^{\al\bth}\nln
&& f^{[\underline{ef}]}_{\ula \ulb}=-f^{[\underline{ef}]}_{\underline{b} \ula}=\dl^{[\underline{e}}_{\underline{a}}\dl^{\underline{f}]}_{\underline{b}}\qquad
f^{[\underline{e'f'}]}_{\underline{a' b'}}=-f^{[\underline{e'f'}]}_{\underline{b'}
\underline{a'}}=-\dl^{[\underline{e'}}_{\underline{a}'}\dl^{\underline{f}']}_{\underline{b}'}\nln
&& f^{e}_{[cd] b}=-f^{e}_{ b[cd]}=\eta_{ b[ c}\dl^{e}_{ d]}\nln
&& f^{[gh]}_{[cd][ef]}=\eta_{ce}\dl^{[
g}_{d}\dl^{ h]}_{ f}-\eta_{ c f}\dl^{[ g}_{d}\dl^{ h]}_{ e}+\eta_{ d f}\dl^{[ g}_{c}\dl^{h]}_{e}-\eta_{d e}\dl^{[ g}_{c}\dl^{h]}_{f}\qquad\qquad \,.
\ee
The bosonic indices are $a=0,\dots, 9$ labeling $\mathfrak{g}_2$, with $a=(\ul a,\underline{a}')$, where $\ul a=0,\dots,4$ labels the $\ads_5$ directions and $\underline{a}'=5,\dots,9$ labels the $\sphere^5$ directions, and $[ab]$ labeling $\mathfrak{g}_0$. The fermionic indices are $\alpha,\hat \alpha$ for $\mathfrak{g}_1$ and $\mathfrak{g}_3$ respectively.


\subsection{OPE results}
\label{app:OPE_results}

The results listed here are from~\cite{Puletti:2008ym}. Notice the different notation: here $\mathfrak{g}_{1(3)}$ corresponds to $\mathfrak{g}_{3(1)}$ of~\cite{Puletti:2008ym}.   
The symbol $\;\widetilde{}\;$ is omitted, however all the currents in the R.H.S. are classical and there is an overall factor $1/R^2$ also omitted.
It is convenient to perform the OPE's in the symmetric point $\si\equiv (x+y)/2$, i.e. $J(x)J(y)=\sum C(x-y)\mathcal O(\si)$. $v$ and $\bar v$ are defined as $v\equiv x-y$ and $\vbar \equiv \bar x-\bar y$, respectively.


\subsubsection{$J J$}

\paragraph{$J_0J_2$}
\be
\label{ope-j0j2}
           && J^{[ab]}(x) \bar J^a(y)=\\ \nn
&&          = f^{a[ab]}_b\big(\frac{J^b}{\bar v}+\frac{v}{2\bar v}\p J^b+\half \bar\p J^b\big)
            + f^{[ab]}_{\al\alh}f^{\alh a}_\bt
                  \big( J^\al J^\bt\frac{v}{\bar v}
                      - J^\al \bar J^\bt\log\gm|v|^2\big) \\ \nn
&&          + f^{[ab]}_{b c}f^{a c}_{[cd]}J^b\big(N^{[cd]}\frac{v}{\bar v}+\bar N^{[cd]}\log\gm|v|^2\big)
\ee

\be
\label{ope-jbar0j2}
&& \bar J^{[ab]}(x) J^a(y)=\\ \nn
&&        =  f^{a [ab]}_b \big(\frac{\bar J^b}{v}+
                             \half \p \bar J^b +
                             \frac{\vbar}{2v}\pbar \bar J^b \big)
            +f^{[ab]}_{\al\alh}f^{\al a}_{\bth}\big(\bar J^{\alh}\bar J^{\bth}\frac{\vbar}{v}
                                                   -\bar J^{\alh}J^{\bth}\log\gm|v|^2\big)\\ \nn
&&          -f^{[ab]}_{b c}f^{c a}_{[cd]}\bar J^b \big(N^{[cd]}\log\gm|v|^2+ \bar N^{[cd]}\frac{\vbar}{v}\big)
\ee


\paragraph{$J_0J_1$}
\be
\label{ope-j0jbar3}
&& J^{[ab]}(x)\bar J^\al(y)=\\ \nn
&& =  f^{\al[ab]}_{\bt}\big(\frac{J^\bt}{\vbar}+\frac{v}{2\vbar}\p J^\bt+\half\pbar J^\bt\big)
 + f^{\alh\al}_{[cd]}f^{[ab]}_{\bt\alh}J^\bt\big(\bar N^{[cd]}\log \gm|v|^2 +N^{[cd]}\frac{v}{\bar v}\big)
 \\ \nn &&
\ee

\be
\label{ope-j3jbar0}
&& J^\al(x)\bar J^{[ab]}(y)=\\ \nn
&&    f^{[ab]\al}_{\bt}\big(\frac{\bar J^\bt}{v}
                                -\half\p \bar J^\bt
                                -\frac{\vbar}{2v}\pbar \bar J^\bt\big)\\ \nn
&&   +f^{[ab]}_{ab}f^{b\al}_{\bth}\big(J_-^a \bar J^{\bth}\frac{\vbar}{v}
                                   -J^{\bth} \bar J^a\log\gm|v|^2\big)
     +f^{[ab]}_{\bth\gm}f^{\gm\al}_a\big(J^a \bar J^{\bth}\log\gm|v|^2
                                       -\bar J^{\bth} \bar J^a\frac{\vbar}{v}\big)\\ \nn
&&   +f^{[ab]}_{\bt\hat{\gm}}f^{\hat{\gm}\al}_{[\lm\rho]}\bar J^\bt\big(N^{[\lm\rho]}\log\gm|v|^2+\bar N^{[\lm\rho]}\frac{\vbar}{v}\big)
\ee

\paragraph{$J_0 J_3$}

\be
\label{ope-jbar0j1}
&&  \bar J^{[ab]}(x)J^{\alh}(y)=\\ \nn
&&  = f^{\alh [ab]}_{\bth}\big(\bar J^{\bth}\frac{1}{v}+
                           \half \p \bar J^{\bth}+
                           \frac{\vbar}{2v}\pbar \bar J^{\bth}\big)+
  f^{[ab]}_{\al\bth}f^{\al\alh}_{[\lm\rho]}\big( \bar J^{\bth}N^{[\lm\rho]} \log\gm|v|^2+
                                            \bar J^{\bth} \bar N^{[\lm\rho]} \frac{\vbar}{v}  \big)\\ \nn &&
\ee
\be
\label{ope-j0jbar1}
&&  J^{[ab]}(x)\bar J^{\alh}(y)=\\ \nn
&& = f^{\alh [ab]}_{\bth}\big(\frac{J^{\bth}}{\vbar}+
                           \frac{v}{2\vbar}\p J^{\bth}+
                           \half \pbar J^{\bth}\big)
  + f^{[ab]}_{ab}f^{b\alh}_\bt \big(J^a J^{\bt}\frac{v}{\vbar}
                                    -J^a \bar J^{\bt}\log\gm|v|^2\big)\\ \nn
&&  + f^{[ab]}_{\bt\hat{\gm}}f^{\alh\hat{\gm}}_a \big(-J^a J^{\bt}\frac{v}{\vbar}
                                          +J^\bt \bar J^a\log\gm|v|^2\big)
  + f^{[ab]}_{\al\bth}f^{\al\alh}_{[\lm\rho]}J^{\bth}\big(N^{[\lm\rho]}\frac{v}{\vbar}+ \bar N^{[\lm\rho]}\log\gm|v|^2\big)\\ \nn &&
\ee


\paragraph{$J_0 J_0$}

\be
\label{ope-j0jbar0}
&&  J^{[a_1b_1]}(x)\bar J^{[a_2b_2]}(y) =\\ \nn
&& = \big( f^{[a_1b_1]\lm}_a f^{[a_2b_2]}_{b\lm}J^a \bar J^b
  +f^{\bt[a_1b_1]}_\al f^{[a_2b_2]}_{\bt\bth}J^\al \bar J^{\bth}
  +f^{\bth[a_1b_1]}_{\alh} f^{[a_2b_2]}_{\bt\bth}J^{\alh}\bar J^{\bt}\big)\log\gm|v|^2\\ \nn &&
\ee


\paragraph{$J_1 J_1$}

\be
\label{opej-j3j3}
&& J^\al(x)\bar J^\bt(y)=\cr
&&  f^{\al\bt}_a \frac{J^a}{\bar v}
    + f^{\al\bt}_a \big(\frac{v}{2\bar v} \p J^a +\half \pbar J^a-\half \p \bar J^a \log\gm|v|^2 \big)\cr
&&  -\half \log\gm|v|^2 \big(f^\bt_{[ab]\dl}f^{\al[ab]}_\gm -f^\al_{[ab]\dl}f^{\bt[ab]}_\gm\big)\bar J^\gm J^\dl \cr
&&  +\half \log\gm|v|^2 \big(f^{\al\gm}_a f^\bt_{\gm[ab]}-f^{\bt\gm}_a f^\al_{\gm[ab]}\big)
      \big(N^{[ab]}\bar J^a+ \bar N^{[ab]}J^a\big)\cr
&&  -f^{\bt\gm}_a f^\al_{\gm[ab]}\bar J^a\big(N^{[ab]}\log\gm|v|^2+\bar N^{[ab]}\frac{\bar v}{v}\big)
    +f^\al_{a\alh}f^{\alh\bt}_{[ab]} J^a\big(N^{[ab]}\frac{v}{\bar v}+\bar N^{[ab]}\log\gm|v|^2\big)  \cr
&&    \ee

\paragraph{$J_3 J_3$}

\be
\label{ope-jbar1j1}
&&  \bar J^{\alh}(x)J^{\bth}(y)=\\ \nn
&&    = f^{\alh\bth}_a\big( \frac{1}{v}\bar J^a+
                        \half \p \bar J^a
                        -\half\pbar J^a \log\gm|v|^2
                        +\frac{\vbar}{2v}\pbar \bar J^a\big)\\ \nn
&&    +\half \log\gm|v|^2 \big(f^{\alh}_{[ab]\hat{\dl}}f^{\bth [ab]}_{\hat{\gm}}-f^{\bth}_{[ab]\hat{\dl}}f^{\alh[ab]}_{\hat{\gm}}\big)
      \bar J^{\hat{\gm}}J^{\hat{\dl}}\\ \nn
&&    +\half \log\gm|v|^2 \big( f^{\alh\hat{\gm}}_a f^{\bth}_{\hat{\gm}[ab]}-f^{\bth\hat{\gm}}_a f^{\alh}_{\hat{\gm}[ab]}\big)
                       \big(N^{[ab]}_+ \bar J^a + \bar N^{[ab]}J^a\big)\\ \nn
&&    +f^{\alh}_{a\al}f^{\al\bth}_{[ab]}\bar J^a\big(\bar N^{[ab]}\frac{\vbar}{v}+N^{[ab]}\log\gm|v|^2\big)
      -f^{\alh}_{a\al}f^{\al\bth}_{[ab]}J^a\big(\bar N^{[ab]}\log\gm|v|^2+N^{[ab]}\frac{v}{\vbar}\big)\\ \nn &&
\ee

\be
\label{ope-j1jbar1}
&&   J^{\alh}(x)\bar J^{\bth}(y)=\\ \nn
&&   = f^{\alh\bth}_a\big( \frac{\bar J^a}{v}
                        -\half \p \bar J^a
                        +\half \pbar J^a\log\gm|v|^2
                        -\frac{\vbar}{2v}\pbar \bar J^a\big)\\ \nn
&&   = +\half \log\gm|v|^2 \big(f^{\alh}_{[ab]\hat{\dl}}f^{\bth [ab]}_{\hat{\gm}}-f^{\bth}_{[ab]\hat{\dl}}f^{\alh[ab]}_{\hat{\gm}}\big)
      \bar J^{\hat{\gm}}J^{\hat{\dl}}\\ \nn
&&    +\half \log\gm|v|^2 \big( f^{\alh\hat{\gm}}_a f^{\bth}_{\hat{\gm}[ab]}-f^{\bth\hat{\gm}}_a f^{\alh}_{\hat{\gm}[ab]}\big)
                       \big(N^{[ab]}\bar J^a + \bar N^{[ab]}J^a\big)\\ \nn
&&    -f^{\bth}_{a\al}f^{\al\alh}_{[ab]}\bar J^a\big(\bar N^{[ab]}\frac{\vbar}{v}+ N^{[ab]}\log\gm|v|^2\big)
      +f^{\bth}_{a\al}f^{\al\alh}_{[ab]}J^a\big(\bar N^{[ab]}\log\gm|v|^2 +  N^{[ab]}\frac{v}{\vbar}\big)\\ \nn &&
\ee

\paragraph{$J_2J_1$}

\be
\label{ope-j2jbar3}
&& J^a(x) \bar J^\al(y)=\\ \nn
&&=   f^{\al a}_{\alh}\big(\frac{J^{\alh}}{\vbar}
                            +\frac{v}{2\vbar}\p J^{\alh}
                            +\half\pbar J^{\alh}
                            -\half \p \bar J^{\alh}\log\gm|v|^2\big)\\ \nn
&&    +f^{a\gm}_{\alh}f^\al_{\gm[ab]}\bar J^{\alh}\big(N^{[ab]}\log\gm|v|^2+ \bar N^{[ab]}\frac{\vbar}{v}\big)
      -f^{a\gm}_{\alh}f^\al_{\gm[ab]} J^{\alh}\big(N^{[ab]} \frac{v}{\vbar}+ \bar N^{[ab]}\log\gm|v|^2\big) \\ \nn
&&     + \textbf{R}^{a\al}_{+-}\log\gm|v|^2
\ee

\be
\label{ope-j3jbar2}
&& J^{\al}(x)\bar J^a(y)=\\ \nn
&&   f^{a\al}_{\alh}\big(\frac{J^{\alh}}{\vbar}
                      +\frac{v}{2\vbar}\p J^{\alh}
                      +\half\pbar J^{\alh}
                      -\half \p \bar J^{\alh}\log\gm|v|^2\big)\\ \nn
&&    +f^{\al\gm}_b f^a_{\gm\bt}\big(\bar J^b J^\bt \log\gm|v|^2
                                -J^b J^\bt\frac{v}{\vbar}
                                -\bar J^b \bar J^\bt\frac{\vbar}{v}
                                +J^b \bar J^\bt \log\gm|v|^2\big)\\ \nn
&&    +f^{\al b}_{\alh}f^a_{b[ab]} \bar J^{\alh}\big(N^{[ab]}\log\gm|v|^2+  \bar N^{[ab]}\frac{\vbar}{v}\big)
      -f^{\al b}_{\alh}f^a_{b[ab]} J^{\alh}\big(N^{[ab]}\frac{v}{\vbar}+  \bar N^{[ab]}\log\gm|v|^2\big)\\ \nn
&&     +\textbf{R}^{\al a}_{+-}\log\gm|v|^2
\ee
The tensor $\textbf{R}^{a\al}_{+-}$ is a symmetric tensor and it contains all the terms coming from the diagram computed from the vertices \eqref{double-ins-mg} and \eqref{double-ins-mm}. They diverge logarithmically however these type of insertions being symmetric are just cancelled when we take the sum of the commutator between $J_+^a(x) J_-^\al(y)$ and $J_+^{\al}(x)J_-^a(y)$.

\paragraph{$J_3 J_2$}
The same structure as before for the case $J_2 J_1$ appears here.

\be
\label{ope-j1jbar2}
&&  J^{\alh}(x)\bar J^a(y)=\\ \nn
&&     = f^{\alh a}_\bt \big(-\frac{\bar J^\bt}{v}
                      -\half \pbar J^\bt\log\gm|v|^2
                      +\half \frac{\vbar}{v}\pbar \bar J^\bt
                      +\half \p \bar J^\bt\big)\\ \nn
&&     +f^a_{\bt\gm}f^{\gm\alh}_{[ab]} \bar J^\bt\big(N^{[ab]}\log\gm|v|^2 +  \bar N^{[ab]}\frac{\vbar}{v}\big)
       +f^{\alh}_{[ab]\hat{\gm}}f^{\hat{\gm} a}_\bt J^\bt\big(N^{[ab]}\frac{v}{\vbar}+  \bar N^{[ab]}\log\gm|v|^2\big)\\ \nn
&&     +\textbf{R}^{\alh a}_{+-}\log\gm|v|^2
\ee

\be
\label{ope-j2jbar1}
&& J^a(x)\bar J^{\alh}(y)=\\ \nn
&&       =f^{ a\alh}_\bt \big(-\frac{\bar J^\bt}{v}
                               +\half \frac{\vbar}{v}\pbar \bar J^\bt
                               +\half \p \bar J^\bt
                              -\half \pbar J^\bt\log\gm|v|^2\big)\\ \nn
&&      +f^a_{\hat{\gm}\bth}f^{\alh\hat{\gm}}_b \big(\bar J^{\bth}J^b \log\gm|v|^2
                                    -\bar J^b \bar J^{\bth}\frac{\vbar}{v}
                                    -J^{\bth} \bar J^b\frac{v}{\vbar}
                                    +\bar J^b J^{\bth}\log\gm|v|^2\big)\\ \nn
&&       +f^{ab}_{[ab]}f^{\alh}_{b\bt}J^\bt\big(N^{[ab]}\frac{v}{\vbar}+ \bar N^{[ab]}\log\gm|v|^2\big)
        +f^{ab}_{[ab]}f^{\alh}_{b\bt}\bar J^{\bt}\big(N^{[ab]}\log\gm|v|^2 + \bar N^{[ab]}\frac{\vbar}{v}\big)\\ \nn
&&         +\textbf{R}^{\alh a}_{+-}\log\gm|v|^2
\ee
Again $\textbf{R}^{\alh a}_{+-}$ is the same kind of tensor as before, it comes from the same vertices \eqref{double-ins-mg} and \eqref{double-ins-mm}, with the replacement $\al\rightarrow\alh$.

\paragraph{$J_2J_2$}

\be
\label{ope-j2j2bar}
&& J^a(x)\bar J^b(y)=\\ \nn
&&       = -f^{ab}_{[ab]}\big(\frac{N^{[ab]}}{\vbar}+\frac{\bar N^{[ab]}}{v}\big)\\ \nn
&&         +\half f^{ab}_{[ab]}\big( -\frac{v}{\vbar}\p N^{[ab]}+\pbar N^{[ab]}(-1+\log\gm|v|^2)
                                             +\frac{\vbar}{v}\pbar \bar N^{[ab]} +\p \bar N^{[ab]}(1-\log\gm|v|^2)\big)\\ \nn
&&        -f^{a\lm}_{[a_1b_1]}f^b_{\lm[a_2b_2]}\big(N^{[a_1b_1]} N^{[a_2b_2]} \frac{v}{\vbar}
                                                                + \bar N^{[a_1b_1]} \bar N^{[a_2b_2]} \frac{\vbar}{v}\big)\\ \nn
&&        -f^{a\lm}_{[a_1b_1]}f^b_{\lm[a_2b_2]}\big( N^{[a_1b_1]}\bar N^{[a_2b_2]}\log\gm|v|^2
                                                                + \bar N^{[a_1b_1]} N^{[a_2b_2]} \log\gm|v|^2\big)\\ \nn
&&        +f^{\gma}_{\bth}f^b_{\gm\al}\big(\bar J^{\bth}J^{\al}+J^{\bth} \bar J^\al \big) \log\gm|v|^2
          +f^a_{\alh\bth}f^{\bth b}_\bt J^{\alh}J^\bt \frac{v}{\vbar}
          +f^b_{\al\bt}f^{\bt a}_{\alh}\bar J^\al \bar J^{\alh} \frac{\vbar}{v}\\ \nn
&&        -\half \big(f^{a[ab]}_\lm f^b_{[ab]\rho}+ f^{b[ab]}_\lm f^a_{[ab]\rho}\big) \bar J^\lm J^\rho \log\gm|v|^2\\ \nn
&&        -\half\big(f^{a\hat{\gm}}_\al f^b_{\hat{\gm}\bth}+ f^{b\hat{\gm}}_\al f^a_{\hat{\gm}\bth}\big)\bar J^\al J^{\bth} \log\gm|v|^2
          +\half\big(f^{a\gm}_{\bth} f^b_{\gm\al}+ f^{b\gm}_{\bth} f^a_{\gm\al}\big)\bar J^{\bth} J^\al \log\gm|v|^2
\ee


\paragraph{$J_3J_1$}

\be
\label{ope-j1j3bar}
&& J^{\alh}(x) \bar J^\bt(y)=\\ \nn
&&     = f^{\alh\bt}_{[ab]}\big(\frac{N^{[ab]}}{\vbar}+\frac{\bar N^{[ab]}}{v}\big)\\ \nn
&&      +\half f^{\alh\bt}_{[ab]}\big( \frac{v}{\vbar}\p N^{[ab]} +\pbar N^{[ab]}(1-\log\gm|v|^2)
                                           -\frac{\vbar}{v}\pbar\bar N^{[ab]} -\p\bar N^{[ab]}(1-\log\gm|v|^2)\big)\\ \nn
&&      +f^{\alh\gm}_{[a_1b_1]}f^{\bt}_{\gm[a_2b_2]}\big(N^{[a_1b_1]} N^{[a_2b_2]}\frac{v}{\vbar}
                                                                + \bar N^{[a_1b_1]} \bar N^{[a_2b_2]}\frac{\vbar}{v}\big)\\ \nn
&&      +f^{\alh\gm}_{[a_1b_1]}f^{\bt}_{\gm[a_2b_2]}\big( N^{[a_1b_1]} \bar N^{[a_2b_2]} \log\gm|v|^2
                                                                + \bar N^{[a_1b_1]} N^{[a_2b_2]}\log\gm|v|^2\big)\\ \nn
&&      +\frac{1}{4}\big(3 f^{\alh}_{[ab]\bth}f^{\bt[ab]}_\al -f^\bt_{a\bth}f^{\alh a}_\al\big)
                     \bar J^\al J^{\bth} \log\gm|v|^2
       +\frac{1}{4}f^{[ab]}_{\al\bth}f^{\alh\bt}_{[ab]} \bar J^{\bth} J^\al \log\gm|v|^2\\ \nn
&&        +\frac{1}{4}f^{\alh\bt}_{[ab]}f^{[ab]}_{ab}\bar J^a J^b \log\gm|v|^2
\ee

\paragraph{$J_1 J_3$}

\be
\label{ope-j3j1bar}
&& J^\bt(x) \bar J^{\alh}(y)=\\ \nn
&&       = f^{\alh\bt}_{[ab]}\big(\frac{N^{[ab]}}{\vbar}+\frac{\bar N^{[ab]}}{v}\big)\\ \nn
&&       +\half f^{\alh\bt}_{[ab]}\big( \frac{v}{\vbar}\p N^{[ab]} +\pbar N^{[ab]}(1-\log\gm|v|^2)
                                           -\frac{\vbar}{v}\pbar \bar N^{[ab]} -\p \bar N^{[ab]}(1-\log\gm|v|^2)\big)\\ \nn
&&      +f^{\bt\hat{\gm}}_{[a_1b_1]}f^{\alh}_{\hat{\gm}[a_2b_2]}\big(N^{[a_1b_1]} N^{[a_2b_2]} \frac{v}{\vbar}
                                                                + \bar N^{[a_1b_1]} \bar N^{[a_2b_2]} \frac{\vbar}{v}\big)\\ \nn
&&      +\big(f^{\bt\gm}_a f^{\alh}_{\gm b}-f^{\alh}_{a\gm}f^{\gm\bt}_b\big)\bar J^a J^b \log\gm|v|^2
        -f^\bt_{a\hat{\gm}}f^{\hat{\gm}\alh}_b J^a J^b \frac{v}{\vbar}
        -f^{\alh}_{a\gm}f^{\bt\gm}_b \bar J^a \bar J^b \frac{\vbar}{v}\\ \nn
&&      -f^{\bt a}_{\bth}f^{\alh}_{a\al}\bar J^{\bth}J^\al \log\gm|v|^2
        +f^\bt_{\bth a}f^{a\alh}_\al J^{\bth} J^\al \frac{v}{\vbar}
        +f^{\alh}_{\al a}f^{\bt a}_{\bth}\bar J^\al \bar J^{\bth}\frac{\vbar}{v}
        +f^{\alh}_{\al a}f^{\bt a}_{\bth}J^{\bth} \bar J^\al \log\gm|v|^2\\ \nn
&&      +\frac{1}{4}\big( 3f^{\alh}_{[ab]\bth}f^{[ab]\bt}_\al -f^\bt_{a\bth}f^{a\alh}_{[ab]}\big)
                        \bar J^\al J^{\bth} \log\gm|v|^2\\ \nn
&&     -\frac{1}{4}f^{\alh\bt}_{[ab]}f^{[ab]}_{ab}\bar J^a J^b \log\gm|v|^2
       -\frac{1}{4}f^{[ab]}_{\al\bth}f^{\alh\bt}_{[ab]}\bar J^{\bth} J^{\al} \log\gm|v|^2
\ee


\subsubsection{$J N$}

\be
\label{ope-j2nbar}
 J^a(x)\bar N^{[ab]}(y)=
      f^a_{b[a_1b_1]}f^{[ab][a_1b_1]}_{[a_2b_2]}\bar N^{[a_2b_2]}J^b\log\gm|v|^2
\ee
\be
\label{ope-j2barn}
\bar J^a(x)N^{[ab]}(y)=
     f^a_{b[a_1b_1]}f^{[ab][a_1b_1]}_{[a_2b_2]}N^{[a_2b_2]}\bar J^b\log\gm|v|^2
\ee
\be
\label{ope-j3nbar}
J^\al(x)\bar N^{[ab]}(x)=
 f^\al_{[a_1b_1]\bt}f^{[a_1b_1][ab]}_{[a_2b_2]}J^\bt \bar N^{[a_2b_2]}\log\gm|v|^2
\ee
\be
\label{ope-jbar3n}
\bar J^\al(x)N^{[ab]}(y)=
f^\al_{[a_1b_1]\bt}f^{[a_1b_1][ab]}_{[a_2b_2]}\bar J^\bt N^{[a_2b_2]}\log\gm|v|^2
\ee
\be
\label{ope-jbar1n}
\bar J^{\alh}(x)N^{[ab]}(y)=
 f^{[ab][a_1b_1]}_{[a_2b_2]}f^{\alh}_{\bth [a_1b_1]}\bar J^{\bth} N^{[a_2b_2]}\log\gm|v|^2
\ee
\be
\label{ope-j1nbar}
J^{\alh}(x)\bar N^{[ab]}(y)=
 f^{[ab][a_1b_1]}_{[a_2b_2]}f^{\alh}_{\bth [a_1b_1]}J^{\bth} \bar N^{[a_2b_2]}\log\gm|v|^2
\ee

\subsubsection{$ N N$}
\be
\bar N^{[ab]}(x)N^{[cd]}(y)
= f^{[ab]}_{[a_1b_1][a_2b_2]}f^{[cd][a_2b_2]}_{[a_3b_3]}N^{[a_3b_3]} \bar N^{[a_1b_2]}\log\gm|v|^2
\ee


\section{The S-matrix factorization in the NFS limit: an example}
\label{app:NFS_example}

The results in this section are from~\cite{Puletti:2007hq}. In order to show how the factorization of the three-body S-matrix works in the near-flat-space limit at the leading order, among the highest weight states \eqref{eqn:hw-states} we consider the following scattering process
\[
\label{ex_hwstate}
Y_{1\dot{1}}(a) Y_{1\dot{1}}(b) Y_{1\dot{1}}(c)
     \rightarrow Y_{1\dot{1}}(d) Y_{1\dot{1}}(e) Y_{1\dot{1}}(f)\,.
 \] 
In particular, the S-matrix element $C_1$ in \eqref{eqn:hw-states} can be extracted from
\[ \label{eqn:C1}
C_1(a,b,c) \sum_{\sigma(d,e,f)} \delta_{ad} \, \delta_{be} \, \delta_{cf}
=
\bra{Y_{1\dot{1}}(f) \, Y_{1\dot{1}}(e) \, Y_{1\dot{1}}(d)}
\, \Smatrix \,
\ket{Y_{1\dot{1}}(a) \, Y_{1\dot{1}}(b) \, Y_{1\dot{1}}(c)} \; .
\]
Recalling the NFS action~\eqref{NFS_action} and the relation \eqref{bispinor_notation} which allows one to write the fields $Y_i$ with $i=1,\dots,4$ as bispinors, in the $\algSO(4)^2$ notation, the amplitude \eqref{eqn:C1} reads
\be
\label{eqn:fourampl}
\bra{Y_{1\dot{1}}\,Y_{1\dot{1}}\,Y_{1\dot{1}}}
\,\Smatrix \,
\ket{Y_{1\dot{1}}\,Y_{1\dot{1}}\,Y_{1\dot{1}}}  &=&
 \bra{Y_1\,Y_1\,Y_1}\,\Smatrix\,\ket{Y_1\,Y_1\,Y_1}
-\bra{Y_1\,Y_1\,Y_1}\,\Smatrix\,\ket{Y_1\,Y_4\,Y_4} \nln
&-&\bra{Y_1\,Y_1\,Y_1}\,\Smatrix\,\ket{Y_4\,Y_1\,Y_4}
-\bra{Y_1\,Y_1\,Y_1}\,\Smatrix\,\ket{Y_4\,Y_4\,Y_1} \; ,
\ee
where the momentum arguments are as in \eqref{eqn:C1}.

\subsection{Feynman diagram computation}

For practical purposes the amplitudes from Feynman diagrams are more easily computed in the $\algSO(4)^2$ notation. In order to show how the factorization emerges from Feynman graphs, we will illustrate the computation only for the process
\[
\label{scattering_ex}
Y_1 (a) Y_1 (b) Y_1 (c)\rightarrow Y_1 (d)Y_1 (e)Y_1 (f)\,,
\]
which is contained in \eqref{eqn:fourampl}. The remaining scattering amplitudes are completely analogous.

Recalling \eqref{eqn:defamp}, the amplitude is defined as
\[ 
\label{eqn:defamp_ex}
  \Amp(\pin) \equiv \Amp(a,b,c,d,e,f) = \braket{Y_1(f) Y_1(e) Y_1(d)}{Y_1(a) Y_1(b) Y_1(c)}_{\mathrm{connected}} \; .
\]

\paragraph{Tree-level.} 
\begin{figure}
\begin{center}
\includegraphics[scale=0.8]{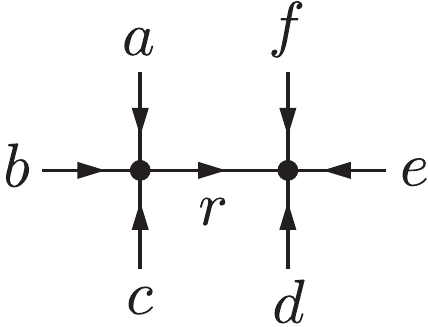}
\end{center}
\caption[vvv]{Tree-level diagram. The label $r$ counts the derivatives acting onto the internal propagator.}
\label{fig:tree}
\end{figure}
At tree-level the amplitude \eqref{eqn:defamp_ex} is computed from diagrams of the kind drawn in Fig. \ref{fig:tree}. For the process \eqref{eqn:defamp_ex} we find
\[ 
\label{eqn:tree-amplitude}
  \Amp^{\mathrm{tree}}(\pin) = -i\gamma^2 \, \frac{1}{\sqrt{64abcdef}} \, \frac{1}{2!\,3!^2} \sum_{\sigma(\pou)} F(\pou) \, \prop_0(\pou) \; ,
\]
where
\[
  F(\pin) = 16 \, \bigbrk{ a^2 + b^2 + c^2 + a b + b c + c a } \, \bigbrk{ d^2 + e^2 + f^2 + de + ef + fd } \; .
\]
and $\prop_0$ is the tree-diagram propagator
\[ 
\label{eqn:prop0}
 \prop_0(\pin) = \frac{\delta^2(\vecpin)}{(\vec{a}+\vec{b}+\vec{c})^2 - m^2 + i \eps} \; .
\]
The sum in \eqref{eqn:tree-amplitude} is taken over all permutations of $\pou \equiv (a,b,c,-d,-e,-f)$. Since the summand is symmetric in $(a,b,c)$ and in $(d,e,f)$ and under the exchange $(a,b,c)\leftrightarrow(d,e,f)$, one can restrict the sum to permutations under which the summand is not symmetric (there are 10 such permutations) and drop the factor $\frac{1}{2!\,3!^2}$. The first fraction in \eqref{eqn:tree-amplitude} originates from the wave-function normalization of the external particles.

After performing the sum in \eqref{eqn:tree-amplitude}, one can use energy-momentum conservation to show that the amplitude indeed vanishes if the sets of in- and out-momenta are different:
\[
 \Amp^{\mathrm{tree}}(\pin) = 0 \quad\mbox{for}\quad \{a,b,c\} \not= \{d,e,f\} \; .
\]
At the points where $\{a,b,c\} = \{d,e,f\}$ the amplitude becomes divergent when $\eps\to0$ in \eqref{eqn:prop0}. The divergences originate from terms where the momentum of the internal propagator $\prop_0$ is equal to one of the external momenta and therefore goes on-shell. The divergence is of $\delta$-function-type and its residue can be extracted by means of the principal value formula \eqref{eqn:principle-value-formula}. In the sum \eqref{eqn:tree-amplitude} the principal value terms cancel because of energy-momentum conservation and we are left with an additional $\delta(\vec{p}^2-m^2)$-function which sets the internal momentum $\vec{p}$ of the corresponding diagram on-shell. The factorized form \eqref{eqn:exp-amp} arises from combining this $\delta$-function with the overall energy-momentum conservation $\delta^{(2)}(\vecpin)$ contained in \eqref{eqn:prop0}. For the case at hand we obtain
\[ \label{eqn:tree-amplitude-result}
  \Amp^{\mathrm{tree}}(\pin) = - \frac{\gamma^2}{4 m^4} \, \frac{a b c \, G(a,b,c)}{(a+c)(c-b)(b-a)}
                                 \sum_{\sigma(d,e,f)} \delta_{ad} \, \delta_{be} \, \delta_{cf}
\]
with
\[
  G(a,b,c) = 16 \bigsbrk{ 2 a b c (a-b+c) + a^3 (b-c) + b^3 (a+c) + c^3 (b-a) } \; .
\]
This result is a special case of \eqref{eqn:exp-amp}, where the coefficients 
are actually independent of the permutation $\sigma$ which is due to the fact that all involved fields are of the same flavor.

\paragraph{One-loop.}

\begin{figure}
\begin{center}
\subfigure[bubble]{\includegraphics[scale=0.6]{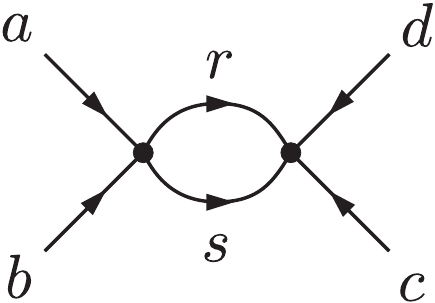}\label{fig:bubble}}\qquad
\subfigure[dog]{   \includegraphics[scale=0.6]{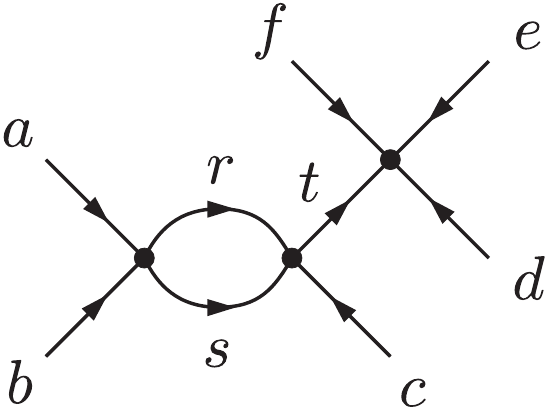}   \label{fig:dog}}   \qquad
\subfigure[sun]{   \includegraphics[scale=0.6]{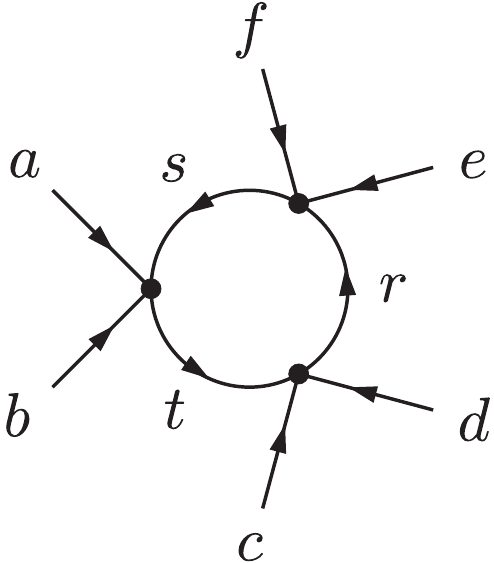}   \label{fig:sun}}
\end{center}
\caption[vvv]{One-loop diagrams. The labels $r$, $s$ and $t$ count the derivatives acting onto the corresponding internal propagator.}
\label{fig:one-loop}
\end{figure}

The one-loop amplitude is given by two sets of diagrams, the ``dogs'' (Fig. \ref{fig:dog}) and the ``suns'' (Fig. \ref{fig:sun}),
\[ \label{eqn:1loop-dog-sun}
  \Amp^{\mathrm{1-loop}}(\pin) = \Amp^{\mathrm{dog}}(\pin) + \Amp^{\mathrm{sun}}(\pin) \; .
\]
As before, the explicit results is for the sample process \eqref{scattering_ex}.

In the case of sun diagrams (Fig. \ref{fig:sun}), it has been possible to reduce the diagrams as a linear combination of tree-level diagrams $\prop_r$ multiplied by bubbles $\bubble_{rs}$ (see Fig. \ref{fig:bubble}) by means of certain cutting rules~\cite{Kallen:1965}.%
\footnote{The tree-level diagram $\prop_r$ is given by
\[ \label{eqn:tree}
 \prop_r(\pin) = \frac{(a+b+c)^r \, (2\pi)^2 \delta^2(\vecpin)}{(\vec{a}+\vec{b}+\vec{c})^2 - m^2 + i \eps} = (a+b+c)^r \prop_0(\pin)\; .
\]
The bubble diagram in Fig. \ref{fig:bubble} is defined by
\[ \label{eqn:bubble}
  \bubble_{rs}(a,b) = \int \frac{d^2\vec{k}}{(2\pi)^2} \frac{k^r \, (a+b-k)^s }{[\vec{k}^2 - m^2 + i\eps][(\vec{a}+\vec{b}-\vec{k})^2 - m^2 + i\eps]} \; .
\]
For more details cf. \cite{Puletti:2007hq}.
}
All bubbles are finite, but -- exactly as at the tree-level -- the propagators in $\prop_r$ become divergent when its momentum goes on-shell. It is clear that there is a potentially divergent propagator also in the dog diagrams (Fig. \ref{fig:dog}). The poles can again be extracted using the principal value formula \eqref{eqn:principle-value-formula}. The partial one-loop amplitudes for \eqref{scattering_ex} are
\begin{align}
\label{eqn:oneloop-sample-dog}
\Amp^{\mathrm{dog}}(\pin) & = + \Ramp(\pin)
-\!\!\sum_{\sigma(a,b,c)} \frac{4 i \gamma^3 a^3 b^2 c}{\lrabs{a^2-c^2}}\; \frac{a^2+b^2}{a^2-b^2}\;
\biggsbrk{ \frac{a^4 + 2a^3b + 10a^2b^2 +2ab^3 + b^4}{a^2-b^2} \\
& \hspace{30mm} - \frac{4i}{\pi} \frac{ab}{\lrabs{a^2-b^2}}\lrbrk{a^2-b^2 + (a^2+b^2)\ln\frac{b}{a}} }
  \times\!\!\sum_{\sigma(d,e,f)}\delta_{ad}\delta_{be}\delta_{cf} \; , \nn 
\end{align}
\begin{align}
\label{eqn:oneloop-sample-sun}
\Amp^{\mathrm{sun}}(\pin) & = - \Ramp(\pin)
-\frac{8i\gamma^3a^2b^2c^2}{(a^2-b^2)(b^2-c^2)(c^2-a^2)} \\[2mm]
& \hspace{20mm} \times \biggsbrk{\, 24a^2b^2c^2 + a^4 b^2 + a^2 b^4 + a^4 c^2 + a^2 c^4 + b^4 c^2 + c^2 b^4 \nn \\
& \hspace{25mm} - a^3 b^2 c + a^2 b^3 c + a^3 b c^2 + a b^3 c^2 + a^2 b c^3 - a b^2 c^3 }
  \times\!\!\sum_{\sigma(d,e,f)}\delta_{ad}\delta_{be}\delta_{cf} \; . \nn
\end{align}
In these expressions, $\Ramp(\pin) = R^{\mathrm{dog}}(\pin) \, \delta^{(2)}(\vecpin) = R^{\mathrm{sun}}(\pin) \, \delta^{(2)}(\vecpin)$ is a function with support on the phase space which drops out in the final one-loop amplitude \eqref{eqn:1loop-dog-sun}. $R^{\mathrm{dog}}(\pin)$ and $R^{\mathrm{sun}}(\pin)$ are rational functions multiplied by logarithms of various ratios of the momenta. They are non-singular for $a \not= b \not= c \not= a$ and the cancelation happens upon energy-momentum conservation between terms with the same momentum flowing through the bubble.



\paragraph{Summary of the results.}

Thus for the scattering process \eqref{ex_hwstate} the total connected amplitudes \eqref{eqn:fourampl} are 
\begin{align}
  \Amp^{\mathrm{tree}}(\pin) & =
  -4\,\gamma^2\,a\,b\,c
\biggsbrk{
a\frac{(a+b)(a+c)}{(a-b)(a-c)} +
b\frac{(a+b)(b+c)}{(a-b)(b-c)} +
c\frac{(a+c)(b+c)}{(a-c)(b-c)}
} \\
& \hspace{15mm} \times \sum_{\sigma(d,e,f)}\delta_{ad}\delta_{be}\delta_{cf} \; , \nn \\
\Amp^{\mathrm{dog}}(\pin) & = + \Ramp(\pin)
-\!\!\sum_{\sigma(a,b,c)} \frac{4i\gamma^3 a^3 b^2 c}{a^2-b^2}
\lrabs{\frac{a+c}{a-c}}
\biggsbrk{
\frac{(a+b)^3}{a-b} \\
&\hspace{30mm} -\frac{4i}{\pi} \frac{ab}{\lrabs{a^2-b^2}}
\lrbrk{a^2-b^2+\lrbrk{a^2+b^2}\ln\frac{b}{a}}
} \times \sum_{\sigma(d,e,f)}\delta_{ad}\delta_{be}\delta_{cf} \; , \nn \\
\Amp^{\mathrm{sun}}(\pin) & = - \Ramp(\pin)
+8i\,\gamma^3\,a^2 b^2 c^2 \, \frac{(a+b)(a+c)(b+c)}{(a-b)(a-c)(b-c)}
\times \sum_{\sigma(d,e,f)}\delta_{ad}\delta_{be}\delta_{cf} \; .
\end{align}

\subsection{S-matrix computation}
\label{sec:essential-processes}

We turn now to the S-matrix elements. Specifically, we verify the factorization by calculating the triple product of two-particle S-matrices according to \eqref{eqn:Smat-factorization} and showing that this product agrees with the computed three-particle amplitudes. To this end we split the tree-level and one-loop S-matrix elements as follows
\[
\begin{split} \label{eqn:Smat-split}
\Smatrix^{(0)} & = \Smatrix^{11\gamma^2} + \Smatrix^{1\gamma\gamma} \; , \\
\Smatrix^{(1)} & = \Smatrix^{11\gamma^3} + \Smatrix^{1\gamma\gamma^2} + \Smatrix^{\gamma\gamma\gamma} \; ,
\end{split}
\]
where the superscripts indicate the perturbative order of the three factors in \eqref{eqn:Smat-factorization}. For instance, $\Smatrix^{1\gamma\gamma^2}$ refers to all terms in the triple product that originate from taking the zeroth order in $\gamma$ from one of the three two-particle S-matrices, the first order from one of the remaining S-matrices and the second order from the final S-matrix.

The first terms in the equations \eqref{eqn:Smat-split} describe processes where one of the particles does not take part in the interaction. These are precisely the terms that correspond to \emph{disconnected} Feynman diagrams. Since we omitted them in the computation in \eqref{eqn:defamp}, we have to discard these terms here, too. We were allowed to disregard these contributions because their factorization is trivial.



Using the near-flat-space S-matrix from section \ref{sec:NFS_Smatrix} in the factorization equation \eqref{eqn:Smat-factorization}, we find for the three-particle S-matrix element governing this process:
\[ \label{eqn:11full}
\Smatrix_{1\dot{1}}^{\mathrm{full}}(a,b,c) =
\bigeval{S_0(A+B)^2}_{(a,b)} \,
\bigeval{S_0(A+B)^2}_{(a,c)} \,
\bigeval{S_0(A+B)^2}_{(b,c)} \; ,
\]
where the relevant coefficients from \eqref{eqn:Smat-coeffs} are
\[ \label{eqn:A+B}
\bigeval{(A+B)}_{(a,b)} = 1 + i\gamma a b \frac{b+a}{b-a} \; .
\]
This sum corresponds to those terms in the two-particle S-matrix which symmetrize two bosonic indices. Expanding the matrix element \eqref{eqn:11full} in $\gamma$, one finds the prediction for the connected tree-level amplitude
\be
\Smatrix_{1\dot{1}}^{1\gamma\gamma}= -4\,\gamma^2\,a\,b\,c
\lrsbrk{
a\,\frac{(a+b)(a+c)}{(a-b)(a-c)} +
b\,\frac{(a+b)(b+c)}{(a-b)(b-c)} +
c\,\frac{(a+c)(b+c)}{(a-c)(b-c)}
}
\ee
and the two pieces of the one-loop amplitude
\be
&& \Smatrix_{1\dot{1}}^{1\gamma\gamma^2}=
-\!\!\sum_{\sigma(a,b,c)} \frac{4i\gamma^3 a^3 b^2 c}{a^2-b^2}
\lrabs{\frac{a+c}{a-c}}
\biggsbrk{
\frac{(a+b)^3}{a-b}
-\frac{4i}{\pi} \frac{ab}{\lrabs{a^2-b^2}}
\lrbrk{a^2-b^2+\lrbrk{a^2+b^2}\ln\frac{b}{a}}
} \,,\nln
&& \Smatrix_{1\dot{1}}^{\gamma\gamma\gamma} =
8i\,\gamma^3\,a^2b^2c^2 \frac{(a+b)(a+c)(b+c)}{(a-b)(a-c)(b-c)} \; .
\ee
These results match those from the Feynman diagrams.


\section{The $\ads_4/\mbox{CFT}_3$ duality: Preliminaries}
\label{sec:preliminaries}


\subsection{Reducing the M-theory background to $\ads_4\times{\s^7\over k}$.}

The near-horizon limit of the M2-brane solution is
$\ads_4\times\s^7$, namely
\[
ds^2={L^2\over 4} ds^2_{AdS_4}+L^2 ds_{S^7}^2\,,
\]
where $L$ is curvature radius for the eleven-dimensional target-space. 

We choose four complex coordinates to parameterize $\s^7$ such that $\sum_{i=1}^4 |X_i|^2=1$ \cite{Nishioka:2008gz}, i.e.
\be
&& X_1=\cos\theta \cos{\theta_1\over 2} e^{\imath (\chi_1+\varphi_1)/2}\qquad
   X_2=\cos\theta \sin{\theta_1\over 2} e^{\imath (\chi_1-\varphi_1)/2}\nln
&& X_3=\sin\theta \cos{\theta_2\over 2} e^{\imath (\chi_2+\varphi_2)/2}\qquad
   X_4=\sin\theta \sin{\theta_2\over 2} e^{\imath (\chi_2-\varphi_2)/2}\,,\nln
\ee
with $0\le \theta \le \pi/2$, $0\le\chi_i\le 4\pi$, $0\le\varphi_i\le 2\pi$ and $0\le \theta_i\le \pi$ for $i=1,2$. Then, the metric on the sphere $\sphere^7$ is
\be
\label{metric_app_s7}
 ds_{S^7}^2 = \sum_{i=1}^4 dX_i d\bar X_i 
 &=& d\theta^2
+{1\over 4} \cos^2 \theta \big\{ \big(d\chi_1+\cos\theta_1 d\varphi_1\big)^2 +d\theta_1^2 +\sin^2\theta_1 d\varphi_1^2\big\}\nln
& + &{1\over 4}\sin^2\theta \big\{ \big(d\chi_2+\cos\theta_2 d\varphi_2\big)^2 +d\theta_2^2 +\sin^2\theta_2 d\varphi_2^2\big\}\,. 
\ee
With the change of coordinates $\chi_1 =2y+2\dl$, $\chi_2=2y-2\dl$ and implementing the orbifold condition according to $y\sim y+{2\pi\over k}$, the metric \eqref{metric_app_s7} becomes
\be
 ds_{S^7}^2
&=& ds^2_{\C\mathbb{P}^3}+(A+dy)^2
  =d\theta^2
+{1\over 4}\cos^2 \theta d\Om_1^2
+{1\over 4}\sin^2\theta d\Om_2^2\nln
&+&(A+dy)^2
 +4 \cos^2\theta\sin^2\theta \big( d\dl +\Quarter \cos\theta_1 d\varphi_1 -\Quarter \cos\theta_2 d\varphi_2\big)^2
\,,
\ee
with
\[
 d\Om_1^2= d\theta_1^2 +\sin^2\theta_1 d\varphi_1^2
 \qquad
d\Om_2^2 =d\theta_2 +\sin^2\theta_2 d\varphi_2^2
\]
and
\[
\label{oneformA}
A= \big(\cos^2\theta-\sin^2\theta\big)d\dl
  +\Half \cos^2\theta\cos\theta_1 d\varphi_1
  +\Half \sin^2\theta \cos\theta_2 d\varphi_2\,.
\]
Thus the total eleven-dimensional metric is
\[
\label{metric11}
ds^2_{11}= {L^2\over 4}ds^2_{AdS_4} + L^2ds^2_{S^7}=
          L^2(\quarter ds^2_{AdS_4}+ds^2_{\C\mathbb{P}^3})+(A+dy)^2\,.
\]
In order to find the dilaton in terms of the other parameters $k, L$, we can compare \eqref{metric11} with the standard eleven-dimensional supergravity metric~\cite{Aharony:2008ug}
\[
\label{metric11dilaton}
ds^2_{11}= e^{-2\phi/3} ds^2_{IIA} + e^{4\phi/3} (d\tilde y+\tilde A)^2
\]
with $\tilde y\sim \tilde y+2\pi$.
Thus, comparing \eqref{metric11} and \eqref{metric11dilaton} (in unit where $\al'=1$), one finds
\be
&& e^{2\phi}= {L^3\over k^3}\nln
&& ds^2_{IIA}= {L^3\over k}(\quarter ds^2_{AdS_4}+ds^2_{\C\mathbb{P}^3})
   \equiv R^2 (\quarter ds^2_{AdS_4}+ds^2_{\C\mathbb{P}^3})\,.
\ee
Hence, summarizing the results, we have that
\[
R^2\equiv{L^3\over k}= k^2 e^{2\phi}\qquad
e^{\phi}={R\over k}\,.
\]
%
%
%
%

In order to make contact with what we have found in this appendix and with the results in \cite{Astolfi:2008ji}, we shift the variables as
\[
\label{shift}
\theta_1\rightarrow \theta_1-\frac{\pi}{2}
\qquad
\theta_2\rightarrow \theta_2+\frac{\pi}{2}\,.
\]
With this change of coordinates we obtain the same metrics used in the main text of this review and in \cite{Astolfi:2008ji}.

\paragraph{The fluxes.}

The type IIA superstring on $\ads_4 \times \C P^3$ is supported by two Ramond-Ramond fluxes $F_{(2)}$ and $F_{(4)}$. They are given by
\be
e^\phi F_{(2)}= R dA
\qquad\quad
e^\phi F_{(4)}={3R^3\over 8} \eps_{AdS_4}\,.
\ee


\subsection{Mode expansion for the bosonic fields}
\label{subsec:quantizationABJM}

The mode expansion for the bosonic fields can be written as
\begin{equation}
u_i (\tau,\sigma ) = i \frac{1}{\sqrt{2}} \sum_{n\in \Z}
\frac{1}{\sqrt{\Omega_n}} \Big[ \hat{a}^i_n e^{-i ( \Omega_n \tau -
n \sigma ) } - (\hat{a}^i_n)^\dagger e^{i ( \Omega_n \tau - n \sigma
) } \Big]\,,
\end{equation}
\begin{equation}
\label{zmode} z_a(\tau,\sigma) = 2 \sqrt{2} \, e^{i\frac{ c
\tau}{2}} \sum_{n \in \Z} \frac{1}{\sqrt{\omega_n}} \Big[ a_n^a
e^{-i ( \omega_n \tau - n \sigma ) } -  (\tilde{a}^a)^\dagger_n e^{i
( \omega_n \tau - n \sigma ) } \Big]\,,
\end{equation}
where $\Omega_n=\sqrt{c^2+n^2}$, $\omega_n=\sqrt{\frac{c^2}{4}+n^2}$
and we defined
$z_a(\tau,\sigma)=x_a(\tau,\sigma)+iy_a(\tau,\sigma)$. The canonical
commutation relations $[x_a(\tau,\sigma),p_{x_b}(\tau,\sigma')] =
i\delta_{ab} \delta (\sigma-\sigma')$,
$[y_a(\tau,\sigma),p_{y_b}(\tau,\sigma')] = i\delta_{ab}\delta
(\sigma-\sigma')$ and $[u_i(\tau,\sigma),p_j(\tau,\sigma')] =
i\delta_{ij} \delta (\sigma-\sigma')$ follow from
\begin{equation}
\label{comrel} [a_m^a,(a_n^b)^\dagger] = \delta_{mn} \delta_{ab}\spa
[\tilde{a}_m^a,(\tilde{a}_n^b)^\dagger] = \delta_{mn}
\delta_{ab}\spa [\hat{a}^i_m,(\hat{a}^j_n)^\dagger] = \delta_{mn}
\delta_{ij}\,.
\end{equation}

\end{appendix}


\small

\bibliographystyle{nb}
\bibliography{References}

\end{document}